\documentclass[11pt,a4paper]{article}
\usepackage{jheppub}
\usepackage[T2A,T1]{fontenc}
\usepackage[cp1250]{inputenc}
\usepackage{amsmath} %na rovnice
\usepackage{amssymb} %na fraktur
\usepackage{braket}
\usepackage{slashed}
\usepackage{bbm}
\usepackage{comment}
\usepackage{yfonts}
\usepackage{simplewick}
\usepackage[boxsize=0.7em,centertableaux]{ytableau}
\usepackage{empheq}
\usepackage{tikz}
\usepackage{subcaption}
\usetikzlibrary{matrix,arrows}
\usetikzlibrary{decorations.markings, decorations.pathreplacing}

\allowdisplaybreaks

\numberwithin{equation}{section}

\DeclareMathOperator{\Tr}{Tr}

\DeclareMathOperator{\res}{res}

\DeclareMathOperator{\hook}{hook}

\title{On Bethe equations of 2d conformal field theory}

\abstract{We study the higher spin algebras of two-dimensional conformal field theory from the perspective of quantum integrability. Starting from Maulik-Okounkov instanton R-matrix and applying the procedure of algebraic Bethe ansatz, we obtain infinite commuting families of Hamiltonians of quantum ILW hierarchy parametrized by the shape of the auxiliary torus. We calculate explicitly the first five of these Hamiltonians. Then, we numerically verify that their joint spectrum can be parametrized by solutions of Litvinov's Bethe ansatz equations and we conjecture a general formula for the joint spectrum of all ILW Hamiltonians, based on results of Feigin, Jimbo, Miwa and Mukhin.

There are two interesting degeneration limits, the infinitely thick and the infinitely thin auxiliary torus. In one of these limits, the ILW hierarchy degenerates to Yangian or Benjamin-Ono hierarchy and the Bethe equations can be easily solved. The other limit is singular but we can nevertheless extract local Hamiltonians corresponding to quantum KdV or KP hierarchy. Litvinov's Bethe equations in this local limit provide an alternative to Bethe ansatz equations of Bazhanov, Lukyanov and Zamolodchikov, but are more transparent, work at any rank and are manifestly symmetric under triality symmetry of $\mathcal{W}_{1+\infty}$. Finally, we illustrate analytic properties of the solutions of Bethe equations in minimal models, in particular for Lee-Yang CFT. The analytic structure of Bethe roots is very rich as it captures the representation theory of $\mathcal{W}_N$ minimal models.}

\author[a]{Tom\'{a}\v{s} Proch\'{a}zka,}
\author[b]{Akimi Watanabe}
\affiliation[a]{Institute of Physics AS CR \\ Na Slovance 2, Prague 8, Czech Republic}
\affiliation[b]{Department of Physics, The University of Tokyo \\ 7-3-1 Hongo, Bunkyo-ku, Tokyo 113-0033, Japan}
\emailAdd{prochazkat@fzu.cz}
\emailAdd{awatanabe@hep-th.phys.s.u-tokyo.ac.jp}

\begin{document}

\maketitle

%\tableofcontents

%\setcounter{footnote}{0}

\section{Introduction}

Two-dimensional conformal field theory owes its solvability to infinite-dimensional symmetry algebra, the Virasoro algebra,
\begin{equation}
\label{virasoroalgebra}
[L_m,L_n]=(m-n)L_{m+n} + \frac{(m-1)m(m+1)}{12} c \delta_{m+n}.
\end{equation}
In favourable situations, such as for so-called minimal models, the whole Hilbert space of the theory decomposes into a finite number of irreducible representations of two commuting copies of the Virasoro algebra. The surprising feature of the representations of the Virasoro algebra in contrast to representations of complex semisimple algebras is the fact that the analogue of Cartan subalgebra is only one-dimensional, spanned by a single generator, $L_0$. All the other generators $L_m, m \neq 0$ act as ladder operators. As consequence of this, the spectrum of $L_0$ is somewhat uninteresting, with all eigenvalues being of the form $\Delta+n, n \geq 0$ where $\Delta$ is the $L_0$ eigenvalue of the lowest weight vector. The eigenspaces of $L_0$ are highly degenerate, the degeneracy going to infinity as $n \to \infty$.

By contrast, there is a larger class of solvable two-dimensional quantum field theory models, whose solvability is due to existence of infinitely many commuting higher conserved quantities \cite{Zamolodchikov:1987jf,Zamolodchikov:1989hfa,Bazhanov:1996aq,Mussardo:1992uc}. Examples of such models are the integrable deformations of 2d CFTs, where one starts with a 2d conformal field theory in the UV and deforms it by a relevant operator. Perhaps the most famous example of such a model is the 2d Ising model in transverse magnetic field \cite{Zamolodchikov:1989fp} related to the exceptional Lie algebra $E_8$. The question is if two-dimensional QFTs with conformal symmetry can be understood from the point of view of this more robust framework of integrability. More concretely, we can ask if are there other quantities besides $L_0$ that would form an infinite dimensional commutative algebra and what their spectrum is. Such quantities have been constructed long time ago by \cite{Sasaki:1987mm,Eguchi:1989hs}. Just as $L_m$ are Fourier modes of the local field $T(z)$, the stress-energy tensor, we can consider zero Fourier mode of the composite local fields such as $(TT)(z)$ and other fields of higher dimension. The requirement of commutativity determines this family of operators uniquely up to overall normalization \cite{Bazhanov:1994ft}. The first few members are \footnote{Here the double dots denote the oscillator ordering where Fourier modes $L_m$ with larger $m$ are on the right. The discussion of conformal normal ordering and its relation to Fourier modes is summarized in appendix \ref{apmodes}.}
\begin{align}
\label{blzintegrals}
I_1 & = T_0 = L_0 - \frac{c}{24} \\
I_3 & = (TT)_0 = L_0^2 + 2\sum_{m>0} L_{-m} L_m - \frac{c+2}{12} L_0 + \frac{c(5c+22)}{2880} \\
\nonumber
I_5 & = (T(TT))_0 + \frac{c+2}{12} (\partial T \partial T)_0 \\
& = \sum_{m_1+m_2+m_3=0} : L_{m_1} L_{m_2} L_{m_3} : + \sum_{m>0} \left( \frac{c+11}{6} m^2 - 1 - \frac{c}{4} \right) L_{-m} L_m \\
\nonumber
& + \frac{3}{2} \sum_{m>0} L_{1-2m} L_{2m-1} - \frac{c+4}{8} L_0^2 + \frac{(c+2)(3c+20)}{576} L_0 - \frac{c(3c+14)(7c+68)}{290304}.
\end{align}

\paragraph{Classical limit}
Before proceeding, it is useful to mention the classical limit of this integrable structure which is well known in the context of the KdV integrable hierarchy \cite{Zabrodin:2018uwz,harnad2021tau}. Consider a second order ordinary differential operators of the form
\begin{equation}
L^2 \equiv \partial_x^2 + u(x)
\end{equation}
which is just the one-dimensional Schr\"odinger operator with potential $-u(x)$\footnote{$u(x)$ should be chosen from a suitable class of functions, for example periodic functions or functions rapidly decaying at infinity, but for our purposes this choice is not relevant as the discussion is mostly algebraic.}. One may ask if there are any deformations of $u(x)$ which do not change the spectrum of the Sch\"odinger operator. Clearly one such example is the rigid translation, but in fact there are infinitely many other deformations preserving the spectrum. These deformations are organized in the form of the KdV hierarchy of partial differential equations. The first non-trivial of these is the Korteweg-de Vries equation
\begin{equation}
\label{kdvequation}
4\partial_t u = 6u \partial_x u + \partial_x^3 u.
\end{equation}
As easily shown in the theory of integrable hierarchies, given a potential $-u(x)$, its arbitrary finite time evolution according to \eqref{kdvequation} has the same spectrum. There exist infinitely many such differential equations and all their associated flows commute with each other and are isospectral, i.e. preserve the spectrum of our Schr\"odinger operator. Furthermore, one can endow the space of potentials with a structure of a Hamiltonian system. For this one introduces a Poisson bracket on the space of potentials
\begin{equation}
\label{kdvpoisson}
\{ u(x), u(y) \} = -\partial_x^3 \delta(x-y) - 4u(x) \partial_x \delta(x-y) - 2\partial_x u(x) \delta(x-y)
\end{equation}
and finds that the equations of KdV hierarchy are generated by Hamiltonians which are integrals of local densities such as
\begin{align}
H_1 & = \frac{1}{2} \int u(x) dx \\
H_3 & = \frac{1}{8} \int u(x)^2 dx.
\end{align}
These two Hamiltonians generate the rigid translations and the flow of \eqref{kdvequation} in the sense that
\begin{align}
\left\{ H_1, u(x) \right\} & = \partial_x u(x) \\
\left\{ H_3, u(x) \right\} & = \partial_t u(x)
\end{align}
and similarly for all higher equations of KdV hierarchy. Since all KdV flows are Hamiltonian flows, the Hamiltonians themselves are conserved quantities, i.e. they are spectral invariants. Now we see how this classical story parallels the discussion of conserved quantities in Virasoro algebra \eqref{blzintegrals}. The Poisson bracket \eqref{kdvpoisson} written in terms of the Fourier modes of $u(x)$ is the classical version ($c \to \infty$ limit) of Virasoro algebra \eqref{virasoroalgebra} and the quantum conserved quantities correspond classically to KdV Hamiltonians. The study of spectra of $I_j$ is therefore the quantum analogue of the spectral problem of one-dimensional Schr\"odinger potentials and its associated classical KdV hierarchy.

\paragraph{Spectrum and BLZ Bethe ansatz equations}
Despite their natural definition and universal appearance in all 2d CFTs, the simulaneous diagonalization of this collection of conserved quantities is notoriously difficult. Since in irreducible representations of the Virasoro algebra the $L_0$ eigenspaces are finite-dimensional, the diagonalization of $I_j$ is in principle reduced to a diagonalization of a finite dimensional matrices. What is missing are simple closed-form formulas for their joint spectra. A significant progress in this direction was achieved in a series of papers \cite{Bazhanov:1994ft,Bazhanov:1996dr,Bazhanov:1998dq} where the authors used the connection of Virasoro algebra to classical KdV hierarchy of differential equations and constructed quantum analogues of the monodromy matrices. Traces of these are generating functions of both local and non-local conserved quantities associated to Virasoro algebra. These generating functions satisfy non-trivial functional equations (such as the analogue of Baxter's $TQ$ relations) as well as non-linear integral equations.

The next important development happened in the context of ODE/IM correspondence \cite{Dorey:2001uw,Dorey:2007zx}. The authors noticed that the same functional equations found earlier by BLZ in the context of Virasoro algebra were appearing in the context of ordinary differential equations in the complex domain in connection with their spectral and monodromy data. First, the discussion was focused on the lowest weight state of the Virasoro representation, but in \cite{Bazhanov:2003ni} BLZ managed to generalize this correspondence to arbitrary excited states in Virasoro representations. The resulting prescription for diagonalizing higher conserved charges in the representations of Virasoro algebra is as follows: we start with the second order differential operator
\begin{equation}
-\partial_z^2 + \frac{\epsilon^2(\Delta+\frac{\epsilon^2}{4}-\frac{1}{2})}{z^2} + \frac{1-\sum_{j=1}^M \gamma_j}{z} + \sum_{j=1}^M \left( \frac{2}{(z-z_j)^2} + \frac{\gamma_j}{z-z_j} \right) + \lambda z^{\epsilon^2-2},
\end{equation}
where $\Delta$ is the conformal dimension of the primary, $\epsilon$ is Nekrasov-like parameter parametrizing the central charge,
\begin{equation}
c = 13 - 6(\epsilon^2 + \epsilon^{-2})
\end{equation}
and $\lambda$, $z_j$ and $\gamma_j$ are so far free parameters. The coefficient of $z^{-1}$ is chosen to fix the scaling symmetry acting on the $z$ coordinate. For generic values of $\Delta$ and $\epsilon$ we have a differential operator with irregular singularities at $z=0$ and at $z = \infty$. To describe the Virasoro excited states at level $M$ (in other words $L_0$ eigensubspace with eigenvalue $\Delta+M$), one allows for $M$ additional regular singular points $z=z_j$ such that the monodromy of the solutions around these singular points is trivial (i.e. $z_j$ are apparent singularities and we can write two linearly independent Frobenius solutions around $z=z_j$ of exactly the same form as around any regular point). The requirement of absence of monodromy at these $M$ points for all $\lambda$ determines all $\gamma_j$ and further imposes $M$ algebraic equations that need to be satisfied by $M$ parameters $z_j$,
\begin{equation}
\label{blzbethe}
\sum_{k \neq j} \frac{z_j(\epsilon^4 z_j^2 - (\epsilon^2-2)(2\epsilon^2+1)z_j z_k + (\epsilon^2-1)(\epsilon^2-2) z_k^2)}{(z_j-z_k)^3} = (1-\epsilon^2) z_j - \epsilon^4 \Delta.
\end{equation}
Generically there are as many solutions of these equations as there are partitions of $M$ \cite{Masoero:2018rel,Conti:2020zft,Conti:2021xzr} which agrees with the number of states in generic Virasoro representation at weight $M$. This gives a nice particle-like interpretation of the states in Virasoro representations: we can think of states at level $M$ as describing states of $M$ indistinguishable particles. Each particle has associated $z_j \in \mathbbm{C}$ and the equations \eqref{blzbethe} can be thought of as being Bethe equations describing allowed collections of Bethe roots $z_j$ corresponding to quantum states, analogously to the situation in XXX spin chain or other quantum mechanical integrable models with finite number of degrees of freedom. All the local higher spin charges are in turn determined as symmetric polynomials of Bethe roots, e.g.
\begin{align}
\label{blzintegralsev}
I_1 & \leadsto \Delta+M -\frac{c}{24} \\
\label{blzintegralsev2}
I_3 & \leadsto (\Delta+M)^2 + \frac{(\epsilon^2-2)(2\epsilon^2-1)}{4\epsilon^2} (\Delta+M) - \frac{4(\epsilon^2-1)}{\epsilon^4} \sum_{j=1}^M z_j + \frac{c(5c+22)}{2880} \\
\nonumber
I_5 & \leadsto (\Delta+M)^3 + \frac{6\epsilon^4-17\epsilon^2+6}{8\epsilon^2} (\Delta+M)^2 - \frac{12(\epsilon^2-1)(\epsilon^2+2)}{\epsilon^4(3\epsilon^2+2)}(\Delta+M) \sum_j z_j \\
& + \frac{(\epsilon^2-2)(2\epsilon^2-1)(18\epsilon^4-59\epsilon^2+18)}{192\epsilon^4} (\Delta+M) + \frac{24(\epsilon^2-1)^2}{\epsilon^6(3\epsilon^2+2)} \sum_j z_j^2 \\
\nonumber
& - \frac{(\epsilon^2-1)(\epsilon^2+2)(6\epsilon^4-17\epsilon^2+2)}{2\epsilon^6(3\epsilon^2+2)} \sum_j z_j -\frac{c(3c+14)(7c+68)}{290304}.
\end{align}

Despite the fact that Bethe equations \eqref{blzbethe} solve the original problem of diagonalizing the local higher spin charges $I_j$, they have their drawbacks. First of all, the Virasoro algebra is a rank $2$ member of a larger family of vertex operator algebras, so called $\mathcal{W}_N$ algebras, all of which can be embedded in two-parametric family called $\mathcal{W}_\infty$ \cite{Gaberdiel:2012ku,Prochazka:2014gqa,Linshaw:2017tvv}. The representation theory of $\mathcal{W}_\infty$ simplifies considerably if we add additional Heisenberg algebra $\widehat{\mathfrak{u}(1)}$, resulting in algebra called $\mathcal{W}_{1+\infty}$. The bases of representation spaces of $\mathcal{W}_{1+\infty}$ are labeled by combinatorial objects such as $N$-tuples of Young diagrams or 3d Young diagrams (plane partitions). It is not obvious from the structure of \eqref{blzbethe} how to generalize these to other $\mathcal{W}_N$ algebras. In \cite{Bazhanov:2001xm} the construction of monodromy matrices and $Q$-operators was generalized to $\mathcal{W}_3$. In \cite{Masoero:2019wqf} the authors considered the generalization of system \eqref{blzbethe} based on work of Feigin-Frenkel \cite{Feigin:2007mr} whose general construction allows in principle to write the analogue of \eqref{blzbethe} for all $N > 2$. But all these systems of Bethe equations become increasingly complicated as the rank increases and in particular at fixed level the number of unknowns and equations increases with the rank $N$. Furthermore, they do not manifest the discrete symmetries (in particular the triality symmetry) of the algebra.

\paragraph{Litvinov's Bethe ansatz equations}

In this work we consider another system of Bethe ansatz equations introduced in \cite{Litvinov:2013zda},
\begin{equation}
\label{betheequationsintro}
q \prod_{l=1}^N \frac{x_j+a_l-\epsilon_3}{x_j+a_l} \prod_{k \neq j} \frac{(x_j-x_k+\epsilon_1)(x_j-x_k+\epsilon_2)(x_j-x_k+\epsilon_3)}{(x_j-x_k-\epsilon_1)(x_j-x_k-\epsilon_2)(x_j-x_k-\epsilon_3)} = 1, \qquad j=1,\ldots,M.
\end{equation}
Here $\epsilon_j$ parameters are Nekrasov-like parameters parametrizing the rank and central charge of $\mathcal{W}_\infty$ algebra. Parameters $a_j$ on the other hand parametrize the weights of the lowest weight vector (generalizing $\Delta$ in the case of Virasoro). Finally, $q$ is a deformation parameter to be discussed later. There are several important differences between \eqref{blzbethe} and \eqref{betheequationsintro}. First of all, the interaction between individual Bethe roots in \eqref{blzbethe} is additive while in \eqref{betheequationsintro} it is multiplicative of Heisenberg XXX type and controlled by
\begin{equation}
\varphi(x) = \frac{(x+\epsilon_1)(x+\epsilon_2)(x+\epsilon_3)}{(x-\epsilon_1)(x-\epsilon_2)(x-\epsilon_3)}
\end{equation}
which is the structure constant of $\mathcal{W}_{1+\infty}$ \cite{Tsymbaliuk:2014fvq,Prochazka:2015deb}. It has also a natural interpretation of a scattering phase (see for instance the TBA equations for elliptic Calogero-Moser system of \cite{Nekrasov:2009rc}). The first product in \eqref{betheequationsintro} controls interaction of Bethe roots with background fields which in our situation are just the lowest weights of $\mathcal{W}_{1+\infty}$ representation.

A new ingredient is the twist parameter $q$. It can be interpreted in several ways and has an important role in the following. From point of view of the representation theory, it controls which infinite dimensional abelian subalgebra of $\mathcal{W}_{1+\infty}$ we are diagonalizing, i.e. which integrable structure we are considering. For generic values of $q$ the associated Hamiltonians are those of \emph{intermediate long wave} equation and its hierarchy \cite{satsuma1979internal,lebedev1983generalized,Litvinov:2013zda,Bonelli:2014iza,buryak2018simple,saut2019benjamin}. The limits $q \to 0$ and $q \to \infty$ correspond to Yangian limits (Benjamin-Ono limits), where the diagonalized abelian subalgebra is the abelian subalgebra manifest in the Yangian presentation of the algebra \cite{Tsymbaliuk:2014fvq,Prochazka:2015deb}. The advantage of this subalgebra is that it is easily diagonalized in terms of combinatorics of Young diagrams (or plane partitions). This is reflected in \eqref{betheequationsintro} by the fact that $q \to 0$ or $q \to \infty$ limit Bethe roots become zeros of numerators or denominators, i.e. Litvinov's Bethe equations factorize into linear factors and can be easily solved. Disadvantage of the Yangian limit is the fact that the higher conserved quantities are non-local from the CFT point of view, i.e. they cannot be written as zero Fourier modes of local currents. To diagonalize the local charges (quantum KdV or KP charges studied by BLZ), one needs to consider the limit $q \to 1$ which is however very singular. One of the aims of this article is to study concretely the behaviour of various quantities in the local limit and extract information about the local charges. There are other roles of parameter $q$. From spin chain point of view it is simply the twist parameter, controlling shape of the auxiliary space. Concretely, $q$ parametrizes the complex structure of the auxiliary Riemann surface which is introduced as a regulator of the trace over auxiliary space. This trace would be otherwise not defined since in our situation the auxiliary space is infinite dimensional. Finally, from the point of view of numerical solution of equations \eqref{betheequationsintro}, $q$ appears as a very natural homotopy parameter, allowing us to smoothly connect the solvable Yangian limit with the physically interesting local BLZ limit. As we will see, there is very interesting analytic structure of solutions of \eqref{betheequationsintro} over $q$-plane.

\subsection{Summary of the paper}

\begin{figure}
\begin{center}
\includegraphics[scale=0.6]{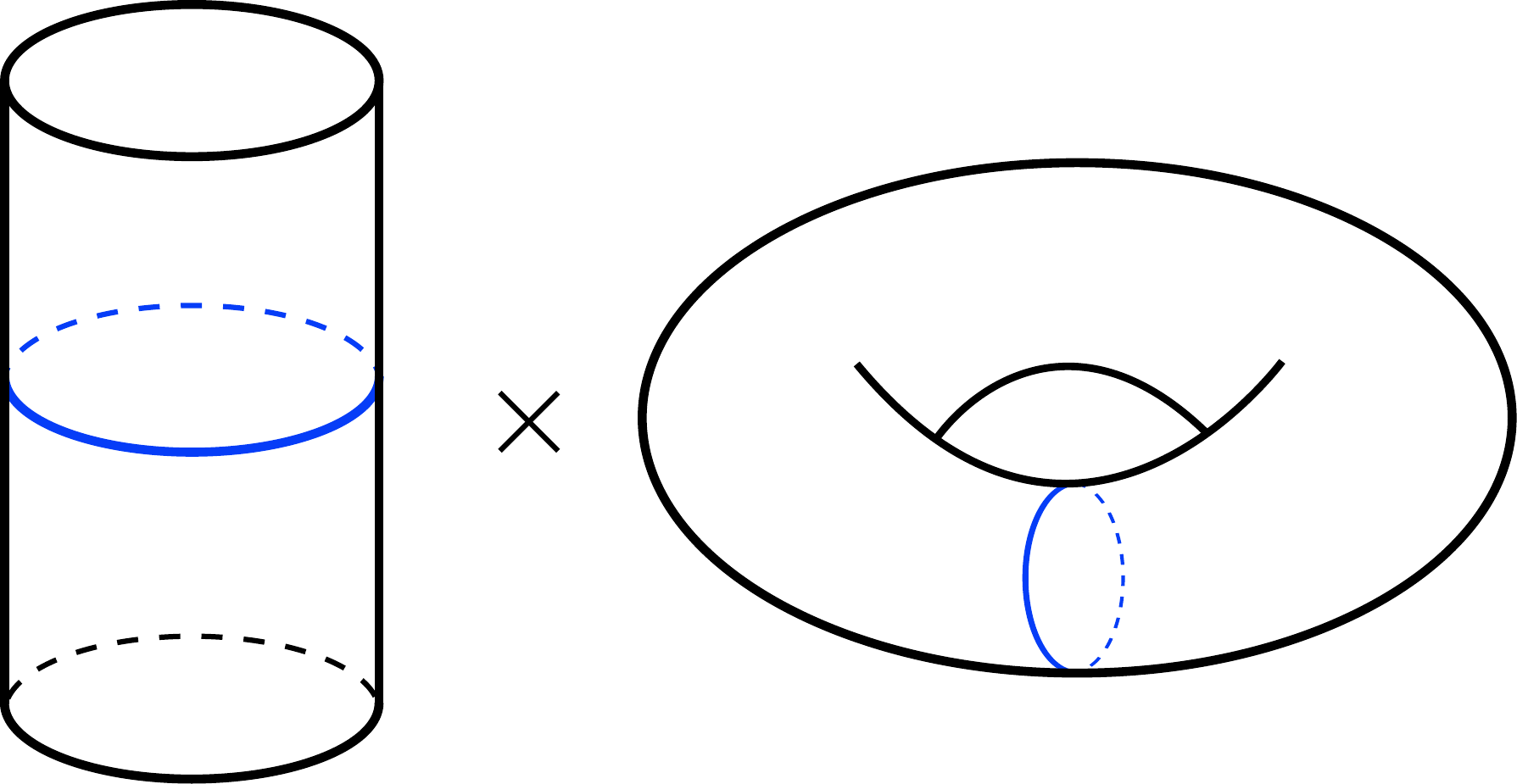}
\end{center}
\caption{$\mathcal{R}$-matrix as a defect (blue) coupling two chiral CFTs, the quantum space (left) is considered to be on a cylinder while the auxiliary space (right) is torus with shape controlled by parameter $q$. The family of Hamiltonians acting on the cylinder on the left is controlled by the shape of the auxiliary torus on the right. In our construction, the torus comes with a choice of a cycle, so it makes sense to talk about thick or thin torus. By a thin torus we mean a torus glued from a long cylinder equipped with a cycle. The limits of infinitely thick or thin torus correspond to Yangian-Benjamin-Ono family (glued from a long cylidner) or Korteweg–De Vries-Kadomtsev–Petviashvili-Bazhanov-Lukyanov-Zamolodchikov family of Hamiltonians (glued from a short cylinder). For generic shapes $q$ we get Hamiltonians from the family of intermediate long wave equation.}
\label{figcyltor}
\end{figure}

The aim of this work is to apply the procedure of algebraic Bethe ansatz \cite{Faddeev:1996iy,Nepomechie:1998jf} to explicitly construct the first few commuting quantum ILW Hamiltonians and to study their spectra in terms of Litvinov's Bethe ansatz equations as well as their local $q \to 1$ limit.

More concretely, we start from Maulik-Okounkov instanton R-matrix \cite{Maulik:2012wi,Smirnov:2013hh,Zhu:2015nha,Prochazka:2019dvu} for a pair of free bosons. Working in the large spectral parameter expansion, we fuse these and calculate first few coefficients of the universal $\mathcal{R}$-matrix, which has the form of the exponential of integral of local currents in two copies of $\mathcal{W}_{1+\infty}$. In the next step we specialize the auxiliary space to be the one of single free boson and calculate the trace to get the transfer matrix acting in the quantum space. Since the free boson Fock space is infinite dimensional, we have to regularize the trace. Following \cite{Litvinov:2013zda}, we choose the regulator to be of the form $q^{L_0}$. This choice effectively puts the auxiliary boson on a torus with modular parameter $q$, without spoiling the commutativity of the resulting ILW Hamiltonians. This setup is illustrated in Figure \ref{figcyltor}. The left part of the diagram represents the quantum space which we identify with a conformal field theory with $\mathcal{W}_{1+\infty}$ symmetry living on a cylinder. The right part of the diagram shows the auxiliary space which is a torus. We will mostly restrict to the situation where the theory associated with the auxiliary space is a single free boson. The $\mathcal{R}$-matrix is an operator acting on product of both spaces so we can think of it as being a codimension $2$ defect inserted along the product of two blue circles. This is somewhat similar to the construction of quantum transfer matrices in \cite{Bazhanov:1994ft,Bazhanov:1996dr,Bazhanov:1998dq}, but in their setup, the auxiliary space is a finite dimensional representation space of underlying quantum group $\mathcal{U}_q(\mathfrak{sl}(2))$, \footnote{The authors of \cite{Bazhanov:1994ft,Bazhanov:1996dr,Bazhanov:1998dq} also consider infinite dimensional auxiliary vector spaces but these do not have an interpretation of another copy of QFT - these representations can be thought of as highest weight representations of continuous spin.} while in our setup the underlying symmetry algebra is $\mathcal{W}_{1+\infty}$ or equivalently the Yangian of $\widehat{\mathfrak{gl}}(1)$.

The most technical part of the present article is the evaluation of the trace. We calculated first five non-trivial ILW Hamiltonians. Assuming the validity of Litvinov's conjectured Bethe ansatz equations, i.e. the fact that the spectra of ILW Hamiltonians can be written in terms of symmetric functions of solutions of \eqref{betheequationsintro}, we are able to determine exact expressions for eigenvalues of all five ILW Hamiltonians that we found. These are in turn matched to a conjectural formula analogous to a formula by Feigin, Jimbo, Miwa and Mukhin \cite{Feigin:2016pld} that was proven in the related context of quantum toroidal algebra. Our expressions for the spectra of the first five ILW Hamiltonians perfectly agree with the conjectural formula.

In the next part, we study carefully the local ($q \to 1$) BLZ/KdV/KP limit of Bethe ansatz equations. This limit is quite singular, because it is exactly the limit of removal of the regulator. Both ILW Hamiltonians as well as most solutions of Bethe ansatz equations become singular in this limit. But surprisingly the singularities in Bethe roots can be associated to excitations of Heisenberg subalgebra of $\mathcal{W}_{1+\infty}$. The remaining Bethe roots, finite in $q \to 1$ limit, are associated to more interesting $\mathcal{W}_\infty$ subalgebra. Using our explicit expressions for the first five ILW Hamiltonians, we show how to disentangle the Heisenberg part from $\mathcal{W}_\infty$ conserved quantities and give formulas for conserved quantities $I_2$ and $I_3$ generalizing \eqref{blzintegrals} from Virasoro algebra to $\mathcal{W}_\infty$.

We conclude by some examples of solutions of Bethe ansatz equations. We study both analytically and numerically the solutions at lower levels in Lee-Yang model (chiral algebra associated to $(A_1,A_2)$ Argyres-Douglas model) and in the first unitary minimal $\mathcal{W}_4$ model. The analytic structure of Bethe roots as functions of parameter $q$ is extremely rich, as it captures the representation theory of $\mathcal{W}_N$ minimal models.

\section{Instanton R-matrix}

In this section, we will review the construction of Maulik and Okounkov's instanton R-matrix \cite{Maulik:2012wi,Zhu:2015nha,Prochazka:2019dvu,Litvinov:2020zeq}. We will follow the notation and conventions of \cite{Prochazka:2019dvu}. Consider a collection of $N$ independent free bosons $J_j(z)$ ($\hat{\mathfrak{gl}}(1)$ Heisenberg vertex operator algebras) with OPE
\begin{equation}
J_j(z) J_k(w) \sim \frac{\delta_{jk}}{(z-w)^2} + reg.
\end{equation}
where $j$ or $k$ label the free bosons, i.e. take values from $1$ to $N$. To each of these currents we associate the Miura operator
\begin{equation}
\label{simplemiura}
\alpha_0 \partial_z + J_j(z)
\end{equation}
where $\alpha_0$ is a fixed complex number. Composing these first order differential operators results in $N$th order differential operator with coefficients that are local quantum fields. There are no ordering ambiguities when we multiply the fields since the $J_j(z)$ coming from the different factors of the product commute. The subalgebra generated by the coefficients of $N$th order differential operator turns out to be closed under operator product expansions. It is a product of diagonal $\hat{\mathfrak{gl}}(1)$ Heisenberg algebra with $\mathcal{W}_N$ algebra and the construction we just reviewed is the simplest free field representation of $\mathcal{W}_N$ \cite{Fateev:1987zh, luk1988quantization, Bouwknegt:1992wg}. In the case of $N=2$ (i.e. starting with just two free bosons) the algebra $\mathcal{W}_2$ goes under the name of Virasoro algebra. Decoupling the diagonal $\hat{\mathfrak{gl}}(1)$ current in this case gives a representation of Virasoro algebra in terms of one free boson $J_1-J_2$ and in the classical limit this is the well-known Miura transformation (studied originally in the context of KdV hierarchy).

Note that the composition of elementary Miura operators \eqref{simplemiura} to construct $\mathcal{W}_N$ algebra is analogous to the construction of $\mathfrak{gl}(M)$ Yangian from its level $1$ representation. Fusing $N$ of these using the Yangian coproduct, we get level $N$ representation of $\mathfrak{gl}(M)$ Yangian. In the context of integrable spin chains $N$ is related to the length of the spin chain. In CFT $N$ is related to the maximal spin of the generating fields of the higher spin symmetry algebra. Yangian of $\mathfrak{gl}(M)$ is obtained by fusion of general number of level $1$ representation (corresponding to spin chains of arbitrary length) and the analogous notion in CFT is the algebra $\mathcal{W}_{1+\infty}$ which has generators of all spins \cite{Gaberdiel:2012ku, Prochazka:2014gqa}. All $\mathcal{W}_N$ algebras can be obtained by truncation from $\mathcal{W}_{1+\infty}$. The construction goes through essentially unmodified if we replace $\hat{\mathfrak{gl}}(1)$ by $\hat{\mathfrak{gl}}(M)$ in which case we obtain a matrix-valued $\mathcal{W}_{1+\infty}$ algebra (Yangian of affine $\mathfrak{gl}(M)$) \cite{Eberhardt:2019xmf,Rapcak:2019wzw}.

In order to construct $\mathcal{W}_N$ generators as differential polynomials in $\hat{\mathfrak{gl}}(1)$ currents $J_j$, we need to multiply the elementary operators \eqref{simplemiura} in certain order. The resulting algebra is independent of the order, but the way it sits in the Fock space of $N$ free bosons depends on the order of the elementary Miura factors. Maulik-Okounkov $\mathcal{R}$-matrix is the similarity transformation that connects these embeddings \cite{Maulik:2012wi,Zhu:2015nha}. Since any permutation is a composition of transpositions, it is enough to study the case of $N=2$. In this case, $\mathcal{R}$ is an operator acting on product of two $\hat{\mathfrak{gl}}(1)$ Fock spaces. Up to an overall normalization, it is uniquely defined by the property
\begin{equation}
\label{instrdef}
\mathcal{R} (\alpha_0 \partial + J_1) (\alpha_0 \partial + J_2) = (\alpha_0 \partial + J_2) (\alpha_0 \partial + J_1) \mathcal{R}.
\end{equation}
Commuting the derivatives to the right, we see that this is equivalent to
\begin{equation}
\mathcal{R} (\alpha_0^2 \partial^2 + \alpha_0 (J_1+J_2) \partial + (J_1 J_2) + \alpha_0 J_2^\prime) \mathcal{R}^{-1} = (\alpha_0^2 \partial^2 + \alpha_0 (J_1+J_2) \partial + (J_1 J_2) + \alpha_0 J_1^\prime).
\end{equation}
Since $\mathcal{R}$ does not depend on the coordinate of the local fields (it is expressible purely in terms of their Fourier modes), in order for this equation to be satisfied the coefficients of each power of $\partial$ must be equal. The coefficients of $\partial^2$ agree trivially. From the coefficients of $\partial^1$ we see that $\mathcal{R}$ must commute with the total current $J_1+J_2$. This means that it has to be constructed out of modes of the difference current
\begin{equation}
\label{jminus}
J_- = \alpha_0 (J_1 - J_2).
\end{equation}
The remaining restriction therefore comes from the equality of coefficients of $\partial^0$. Written in terms of $J_-$ (and simplifying using the commutativity with $J_1+J_2$), this amounts to
\begin{equation}
\label{rmatrixeqn}
\mathcal{R} \left[ \frac{1}{2} (J_- J_-) + \alpha_0^2 \partial J_- \right] \mathcal{R}^{-1} = \frac{1}{2} (J_- J_-) - \alpha_0^2 \partial J_-
\end{equation}
This is the main equation that defines Maulik-Okounkov $\mathcal{R}$-matrix in terms of the local fields. Given an action of $\mathcal{R}$ on the vacuum of $J_-$ Fock space (which determines the overall normalization of $\mathcal{R}$), this equation determines level by level the action of $\mathcal{R}$ on all other states in the Fock space.

One can also solve the equation \eqref{rmatrixeqn} perturbatively in the large spectral parameter \cite{Zhu:2015nha}. In CFT, the spectral parameter is identified with the zero modes of the $\hat{\mathfrak{gl}}(1)$ currents \cite{Prochazka:2019dvu} which are central. For this reason, in \cite{Zhu:2015nha,Prochazka:2019dvu} the zero mode of $J_-$ was treated differently than the remaining modes. Various terms in perturbative expansion of $\mathcal{R}$ were written in terms of oscillators not involving the zero mode of $J_-$. In the analysis that we will do here we will keep this zero mode. Therefore, we will introduce the spectral parameter $u$ by performing a constant shift (which is an automorphism of the Heisenberg algebra)
\begin{equation}
\label{instspecparam}
J_-(z) \to J_-(z)+ \alpha_0 u.
\end{equation}
This introduces certain redundancy which we can use as consistency check of our calculations. The equation \eqref{rmatrixeqn} becomes
\begin{equation}
\label{rmatrixcomm}
\mathcal{R} \left[ \alpha_0 u J_- + \frac{1}{2} (J_- J_-) + \alpha_0^2 \partial J_- \right] = \left[ \alpha_0 u J_- + \frac{1}{2} (J_- J_-) - \alpha_0^2 \partial J_- \right] \mathcal{R}
\end{equation}
This is the equation that we can solve perturbatively at large $u$. It turns out that the large $u$ expansion of $\mathcal{R}$ simplifies if we take the logarithm of $\mathcal{R}$. We therefore write
\begin{equation}
\label{largeuexpelr}
\mathcal{R}(u) = \exp \left[ \frac{r^{(1)}}{\alpha_0 u} + \frac{r^{(2)}}{\alpha_0^2 u^2} + \ldots \right] \equiv \exp r(u)
\end{equation}
Here $r^{(j)}$ turn out to be the zero modes of certain \emph{local} currents constructed out of $J_-$ (this is the reason for taking the logarithm of $\mathcal{R}$) \cite{Litvinov:2013zda}.

At the leading order $u^0$ we need to solve
\begin{equation}
\left[ r^{(1)}, J_- \right] +2\alpha_0^2 \partial J_- = 0
\end{equation}
whose solution is
\begin{equation}
r^{(1)} = -\frac{1}{2} (J_- J_-)_0.
\end{equation}
This is a unique solution that is a zero mode of a local operator of dimension $2$. If we relax the requirement of dimension $2$, we can add to $r^{(1)}$ a linear combination of the zero mode of $J_-$ and of the identity operator (which are both central). This freedom in turn can be absorbed into $u$-dependent normalization of $\mathcal{R}$ which is not fixed by \eqref{rmatrixeqn}.

At the next order $u^{-1}$ we have
\begin{equation}
\label{rmatrixeqno2}
\left[ r^{(2)}, J_- \right] + \frac{1}{2} \left[ r^{(1) 2}, J_- \right] + \left[ r^{(1)}, \frac{1}{2} (J_- J_-) \right] + \alpha_0^2 \left\{ r^{(1)}, \partial J_- \right\} = 0
\end{equation}
which simplifies to
\begin{equation}
\left[ r^{(2)}, J_- \right] + \left[ r^{(1)}, \frac{1}{2} (J_- J_-) \right] = 0
\end{equation}
with solution
\begin{equation}
r^{(2)} = \frac{1}{6} (J_- (J_- J_-))_0
\end{equation}
(which is again unique zero mode of a local field up to addition of a constant or of the zero mode of $J_-$). Note that all terms in \eqref{rmatrixeqno2} that were not commutators canceled. This had to be the case as follows from the ad-exp identity
\begin{equation}
\label{bosonicradexp}
e^{\left[ r, \cdot \right]} \left( \alpha_0 u J_- + \frac{1}{2} (J_- J_-) + \alpha_0^2 \partial J_- \right) = \alpha_0 u J_- + \frac{1}{2} (J_- J_-) - \alpha_0^2 \partial J_-.
\end{equation}
Order $u^{-2}$ terms of this equation are
\begin{multline}
\left[ r^{(3)}, J_- \right] + \frac{1}{2} \left[ r^{(1)}, \left[ r^{(2)}, J_- \right] \right] + \frac{1}{2} \left[ r^{(2)}, \left[ r^{(1)}, J_- \right] \right] + \frac{1}{6} \left[ r^{(1)}, \left[ r^{(1)}, \left[ r^{(1)}, J_- \right] \right] \right] + \\
+ \left[ r^{(2)}, \frac{1}{2} (J_- J_-) + \alpha_0^2 \partial J_- \right] + \frac{1}{2} \left[ r^{(1)}, \left[ r^{(1)}, \frac{1}{2} (J_- J_-) + \alpha_0^2 \partial J_- \right] \right] = 0
\end{multline}
whose unique local solution is
\begin{equation}
r^{(3)} = -\frac{1}{12} (J_-(J_-(J_- J_-)))_0 + \frac{\alpha_0^2(1+2\alpha_0^2)}{12} (\partial J_- \partial J_-)_0
\end{equation}
up to addition of central terms. Writing this in terms of the creation-annihilation normal order (using expressions in appendix \ref{instapmodes}), we find
\begin{equation}
r^{(3)} = -\frac{1}{12} :J_-^4:_0 + \frac{\alpha_0^2}{12} :J_-^2:_0 - \frac{\alpha_0^4}{144} + \frac{\alpha_0^2(1+2\alpha_0^2)}{12} :\partial J_- \partial J_-:_0 - \frac{\alpha_0^2(1+2\alpha_0^2)}{12} \frac{\alpha_0^2}{60}
\end{equation}
in agreement with mode expansions (3.44) of \cite{Prochazka:2019dvu}\footnote{This agrees with (3.47) if the parentheses in \cite{Prochazka:2019dvu} are understood as the creation-annihilation normal ordering.}.

Employing the identity \eqref{bosonicradexp}, we can find perturbative solution for $r^{(j)}$ much more efficiently than by manipulating the mode expansions as was done by hand in \cite{Zhu:2015nha,Prochazka:2019dvu}. All we need to do is to calculate multiple commutators of a zero mode of a local operator with another local operator. As discussed in appendix \ref{apmodes}, this in turn amounts to taking the first order pole of OPE between two local operators which can be done very quickly using the package {\tt OPEdefs} by Thielemans \cite{Thielemans:1991uw}. In this way, the $r^{(j)}$ can be found to rather high order ($j \sim 20$) in few seconds. The first few of these expressions are given in appendix \ref{bosonicrlist}. Note that since in terms of Fourier modes on cylinder the zero mode of total derivative vanishes, there is an ambiguity in writing $r^{(j)}$ as zero modes of local operators -- although the zero mode is well-defined and uniquely determined, the local operator itself is not.

The most important property of $\mathcal{R}$ whose construction we just reviewed is the fact that it satisfies the Yang-Baxter equation
\begin{equation}
\label{instantonybe}
\mathcal{R}_{12}(u_1-u_2) \mathcal{R}_{13}(u_1-u_3) \mathcal{R}_{23}(u_2-u_3) = \mathcal{R}_{23}(u_2-u_3) \mathcal{R}_{13}(u_1-u_3) \mathcal{R}_{12}(u_1-u_2)
\end{equation}
In order to understand the origin of this equation, we need to consider three Fock spaces and two ways how to change the order $123 \to 321$. The operator written as the exponential of a zero mode of local field is unique up to central terms. Therefore, \eqref{instantonybe} has to hold up to possible difference in normalization.

As usual in the context of quantum integrability, once we have a solution of Yang-Baxter equation \eqref{instantonybe}, the procedure of algebraic Bethe ansatz can immediately be applied to construct commuting Hamiltonians and study their spectrum. This is what we will do in the following sections.

\section{Universal R-matrix}

In this section we would like to find an expression for $\mathcal{R}$ for representations not restricted to Yangian level $1$ (i.e. we are interested in algebras with generating fields of spin greater than $1$). In order to do that, we will take the elementary level one $\mathcal{R}$-matrices defined in the previous section and fuse them $N$ times in the left factor and $\bar{N}$ times in the right factor. Having an expression for general $N$ and $\bar{N}$, this is equivalent to knowing a universal $\mathcal{R}$-matrix of $\mathcal{W}_{1+\infty}$. This $\mathcal{R}$-matrix will satisfy Drinfe\v{l}d's relations
\begin{align}
\label{drinfelda}
\mathbf{R} \Delta(x) \mathbf{R}^{-1} & = \Delta^{op}(x) \\
\label{drinfeldb}
(\Delta \otimes \mathbbm{1})(\mathbf{R}) & = \mathbf{R}_{13} \mathbf{R}_{23} \\
\label{drinfeldc}
(\mathbbm{1} \otimes \Delta)(\mathbf{R}) & = \mathbf{R}_{13} \mathbf{R}_{12}
\end{align}
as we will verify perturbatively in large central charge expansion. Here $x$ is any local field (or its Fourier mode) and $\Delta(x)$ is the coproduct \cite{Prochazka:2014gqa}.

\subsection{Universal $\mathcal{R}$-matrix from fusion}
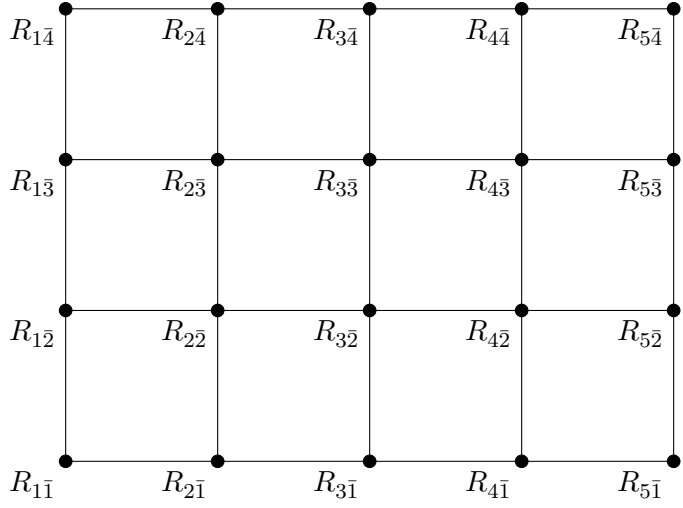
\begin{figure}
\begin{center}
\begin{tikzpicture}
\draw[step=2cm] (0,0) grid (8,6);
\foreach \i in {0,...,4}
  \foreach \j in {0,...,3}
	  \draw[black,fill=black] (2*\i,2*\j) circle (.5ex);
\node at (0,0) [below left] {$R_{1\bar{1}}$};
\node at (2,0) [below left] {$R_{2\bar{1}}$};
\node at (4,0) [below left] {$R_{3\bar{1}}$};
\node at (6,0) [below left] {$R_{4\bar{1}}$};
\node at (8,0) [below left] {$R_{5\bar{1}}$};
\node at (0,2) [below left] {$R_{1\bar{2}}$};
\node at (2,2) [below left] {$R_{2\bar{2}}$};
\node at (4,2) [below left] {$R_{3\bar{2}}$};
\node at (6,2) [below left] {$R_{4\bar{2}}$};
\node at (8,2) [below left] {$R_{5\bar{2}}$};
\node at (0,4) [below left] {$R_{1\bar{3}}$};
\node at (2,4) [below left] {$R_{2\bar{3}}$};
\node at (4,4) [below left] {$R_{3\bar{3}}$};
\node at (6,4) [below left] {$R_{4\bar{3}}$};
\node at (8,4) [below left] {$R_{5\bar{3}}$};
\node at (0,6) [below left] {$R_{1\bar{4}}$};
\node at (2,6) [below left] {$R_{2\bar{4}}$};
\node at (4,6) [below left] {$R_{3\bar{4}}$};
\node at (6,6) [below left] {$R_{4\bar{4}}$};
\node at (8,6) [below left] {$R_{5\bar{4}}$};
\end{tikzpicture}
\end{center}
\caption{Illustration of the fusion of elementary $R$-matrices for $(N,\bar{N})=(5,4)$. We should multiply $R_{j\bar{k}}$ starting from the upper left corner and proceeding all the way to the lower right corner. Since the elementary factors that are neither on same row nor on the same column commute, the multiplication column-by-column or row-by-row gives the same answer.}
\label{fusiondiagram}
\end{figure}

We start with $N + \bar{N}$ copies of $\hat{\mathfrak{gl}}(1)$ Fock space, labeled by $\mathcal{H}_j$, $j=1,\ldots,N$ and $\mathcal{H}_{\bar{k}}$, $k=1,\ldots,\bar{N}$. We will denote the elementary $\mathcal{R}$-matrix acting on $\mathcal{H}_j \otimes \mathcal{H}_{\bar{k}}$ by $\mathcal{R}_{j\bar{k}}$. The fused $\mathcal{R}$-matrix is then given by a product of $N \times \bar{N}$ terms
\begin{align}
\label{rfused}
\nonumber
\mathbf{R} & \equiv R_{1\bar{N}} R_{2\bar{N}} \cdots R_{N\bar{N}} R_{1\overline{N-1}} \cdots R_{N\overline{N-1}} \cdots R_{1\bar{2}} \cdots R_{N\bar{2}} R_{1\bar{1}} R_{2\bar{1}} \cdots R_{N\bar{1}} \\
& = \prod_{\overrightarrow{j=1}}^N \prod_{\overleftarrow{k=1}}^{\bar{N}} R_{j\bar{k}}.
\end{align}
The factors corresponding to $\mathcal{H}_j$ are multiplied from left to right while the factors corresponding to $\mathcal{H}_{\bar{k}}$ are multiplied in the opposite direction. This is necessary in order for the result to be expressible in terms of $\mathcal{W}_N \times \mathcal{W}_{\bar{N}}$ generators (this is also consistent with the difference between Drinfe\v{l}d's relations \eqref{drinfeldb} and \eqref{drinfeldc}). This is illustrated in Figure \ref{fusiondiagram}. We multiply the individual factors starting from the upper left corner and proceeding all the way to the lower right corner, with every step going either right or down. All the factors appearing to the left or above any given elementary $\mathcal{R}$-matrix must appear in the product on the left. This is the only restriction on the order of factors in the product and any two orders satisfying this restriction are equivalent due to commutativity of $R_{j\bar{k}}$ which sit on different rows and columns.

Since every elementary $\mathcal{R}$-matrix is of the form of an exponential of a zero mode of local fields, and since the commutator of two zero modes of local fields is again a zero mode of a local field, Baker-Campbell-Hausdorff formula guarantees that the fused $\mathcal{R}$-matrix is of this form as well. Furthermore, we can use the Baker-Campbell-Hausdorff formula to find the large spectral parameter expansion of the fused $\mathcal{R}$-matrix. We will denote the fused $\mathcal{R}$-matrix by $\mathbf{R}(u)$ (in contrast to $R(u)$ which denotes the elementary $\mathcal{R}$-matrix). The large spectral charge expansion of $\mathbf{R}(u)$ is of the form
\begin{equation}
\label{largeuexpunir}
\mathbf{R}(u) = \exp \left[ \frac{\mathbf{r}^{(1)}}{u} + \frac{\mathbf{r}^{(2)}}{u^2} + \frac{\mathbf{r}^{(3)}}{u^3} + \ldots \right] \equiv \exp \mathbf{r}(u)
\end{equation}
Unlike in the case of \eqref{largeuexpelr} here we absorb the powers of $\alpha_0$ into coefficients $\mathbf{r}^{(j)}$ (the former was kept to be compatible with conventions of \cite{Prochazka:2019dvu}). As before, each $\mathbf{r}^{(j)}$ will be a zero mode of a local field of engineering dimension $j+1$. Furthermore, this local field is determined uniquely up to total derivatives.

In order to determine the coefficients $\mathbf{r}^{(j)}$ in the large spectral charge expansion of $\mathbf{R}(u)$, we use the Baker-Campbell-Hausdorff formula generalized to an arbitrary number of factors in the product, i.e. the expansion
\begin{align}
\nonumber
\log (e^{X_1} e^{X_2} \cdots e^{X_N}) & = \sum_j X_j + \frac{1}{2} \sum_{j<k} \left[ X_j, X_k \right] \\
& + \frac{1}{12} \sum_{j<k} \left( \left[ X_j, \left[ X_j, X_k \right] \right] + \left[ \left[ X_j, X_k \right], X_k \right] \right) \\
\nonumber
& + \frac{1}{6} \sum_{j<k<l} \left( \left[ X_j, \left[ X_k, X_l \right] \right] + \left[ \left[ X_j, X_k \right], X_l \right] \right) + \ldots
\end{align}
and similar but more complicated expressions at higher orders. It is important that everything is expressed in terms of repeated commutators, so in particular assuming that the elementary $\mathcal{R}$-matrix was of the form of exponential of zero mode of a local field, the same will be automatically true for $\mathbf{R}(u)$. For the leading coefficient, we find
\begin{align}
\nonumber
\mathbf{r}^{(1)} & = -\frac{\alpha_0}{2} \sum_{(j,\bar{j})} ((J_j-J_{\bar{j}})(J_j-J_{\bar{j}}))_0 \\
& = -\frac{\alpha_0 \bar{N}}{2} \sum_j (J_j J_j)_0 -\frac{\alpha_0 N}{2} \sum_{\bar{j}} (J_{\bar{j}} J_{\bar{j}})_0 + \alpha_0 \sum_{j,\bar{j}} (J_j J_{\bar{j}})_0
\end{align}
where $j$ and $\bar{j}$ are considered as independent summation variables. Using the expressions for free field representations of $\mathcal{W}_N$ generators (for example section 3.1 in \cite{Prochazka:2014gqa}), we can identify this with
\begin{align}
\label{universalr1}
\nonumber
\mathbf{r}^{(1)} & = -\alpha_0 \bar{N} \left(-U_2+\tfrac{1}{2}(U_1 U_1)\right)_0 -\alpha_0 N \left(-\bar{U}_2+\tfrac{1}{2}(\bar{U}_1 \bar{U}_1)\right)_0 + \alpha_0 (U_1 \bar{U}_1)_0 \\
& = -\alpha_0 \bar{N} T_0 -\alpha_0 N \bar{T}_0 + \alpha_0 (U_1 \bar{U}_1)_0
\end{align}
where
\begin{align}
\nonumber
T & = -U_2 + \frac{1}{2} (U_1 U_1) + \frac{(N-1)\alpha_0}{2} \partial U_1 \\
& = \frac{1}{2} \sum_j (J_j J_j) + \frac{\alpha_0}{2} \sum_j (N+1-2j) \partial J_j
\end{align}
is the total stress-energy tensor of the left copy of $\hat{\mathfrak{gl}}(1) \times \mathcal{W}_N$ and similarly for $\bar{T}$. Here and in the following we use bars for all quantities associated to the right factor, i.e. the local fields $U_j$ satisfy the OPEs of $\hat{\mathfrak{gl}}(1) \times \mathcal{W}_N$ with parameter $\alpha_0$ \cite{Prochazka:2014gqa} while the fields $\bar{U}_j$ satisfy OPEs of an independent $\hat{\mathfrak{gl}}(1) \times W_{\bar{N}}$ with parameter $\alpha_0$. Note that by construction both algebras have the same value of the parameter $\alpha_0$. This is equivalent to having same Kapustin-Witten parameter $\psi$ in the notation of \cite{Gaiotto:2017euk,Prochazka:2017qum} or in the language of \cite{Tsymbaliuk:2014fvq,Prochazka:2015deb} having the same Nekrasov-like parameters $(h_1,h_2,h_3)$ up to an overall scale. This constraint reflects the fact that the fusion operation using $\mathcal{W}_{1+\infty}$ coproduct exists at fixed value of $\alpha_0$ while acting additively in $N$ \cite{Prochazka:2014gqa}.

Although the fusion construction that we are discussing is assuming $N$ and $\bar{N}$ to be positive integers, the resulting expressions are written as expressions in terms of local fields with coefficients that are rational functions of $N, \bar{N}$ and $\alpha_0$. We can therefore interpret them as expressions in $\mathcal{W}_{\infty}$, without any restrictions on $N$ and $\bar{N}$ \cite{Prochazka:2014gqa}. In the following, we will freely transition between $\mathcal{W}_N$ and $\mathcal{W}_\infty$ picture, depending on what is more convenient.

At the next order, it follows from the Baker-Campbell-Hausdorff formula that
\begin{equation}
\mathbf{r}^{(2)} = \frac{1}{\alpha_0^2} \sum_{j,\bar{j}} r^{(2)}_{(j,\bar{j})} + \frac{1}{2\alpha_0^2} \sum_{(j,\bar{j})<(k,\bar{k})} \left[ r^{(1)}_{(j,\bar{j})}, r^{(1)}_{(k,\bar{k})} \right].
\end{equation}
Here the notation $(j,\bar{j}) < (k,\bar{k})$ is related to our fixed order of the vertices as in Figure \ref{fusiondiagram}, e.g. we can sum over all pairs that have either ($j<k$) or ($j=k$ and $\bar{k}<\bar{j}$) -- remember the opposite order of the barred indices. Plugging in the expressions for $r^{(j)}$, this equals
\begin{align}
\nonumber
\mathbf{r}^{(2)} & = \frac{\alpha_0}{6} \sum_{j,\bar{j}} \left( (J_j (J_j J_j)) - 3 ((J_j J_j) J_{\bar{j}}) + 3 (J_j (J_{\bar{j}} J_{\bar{j}})) - (J_{\bar{j}} (J_{\bar{j}}J_{\bar{j}})) \right)_0 \\
\nonumber
& +\frac{\alpha_0^2}{8} \sum_{(j,\bar{j})<(k,\bar{k})} \left[ (J_j J_j) - 2(J_j J_{\bar{j}}) + (J_{\bar{j}} J_{\bar{j}})_0, (J_k J_k) - 2(J_k J_{\bar{k}}) + (J_{\bar{k}} J_{\bar{k}})_0 \right] \\
& = \frac{\alpha_0 \bar{N}}{6} \sum_{j} (J_j (J_j J_j))_0 +\frac{\alpha_0^2\bar{N}}{2} \sum_{j<k} (\partial J_j J_k)_0 - \frac{\alpha_0 N}{6} \sum_{\bar{j}} (J_{\bar{j}} (J_{\bar{j}}J_{\bar{j}}))_0 \\
\nonumber
& -\frac{\alpha_0^2 N}{2} \sum_{\bar{j}<\bar{k}} (\partial J_{\bar{j}} J_{\bar{k}})_0 - \frac{\alpha_0}{2} \sum_{j,\bar{j}} ((J_j J_j) J_{\bar{j}})_0 + \frac{\alpha_0}{2} \sum_{j,\bar{j}} (J_j (J_{\bar{j}} J_{\bar{j}}))_0 \\
\nonumber
& -\frac{\alpha_0^2}{2} \sum_j \sum_{\bar{k}} (\bar{N}+1-2\bar{k}) (\partial J_j J_{\bar{k}})_0 + \frac{\alpha_0^2}{2} \sum_{j} (N+1-2j) \sum_{\bar{k}} (J_j \partial J_{\bar{k}})_0.
\end{align}
Defining a local spin $3$ field
\begin{align}
\nonumber
\phi_3 & = U_3 - (U_1 U_2) + \frac{1}{3} (U_1(U_1 U_1)) - \frac{(N-2)\alpha_0}{2} U_2^\prime \\
\nonumber
& + \frac{(N-1)\alpha_0}{2} (U_1^\prime U_1) + \frac{N^2\alpha_0^2-N\alpha_0^2+4\alpha_0^2+2}{12} U_1^{\prime\prime} \\
& = \frac{1}{3} \sum_j (J_j (J_j J_j)) + \frac{\alpha_0}{2} \sum_{j<k} (\partial J_j J_k - J_j \partial J_k) \\
\nonumber
& + \frac{\alpha_0}{2} \sum_j (N+1-2j) (\partial J_j J_j) + \frac{1}{6} \sum_j \partial^2 J_j \\
\nonumber
& + \frac{\alpha_0^2}{12} \sum_j (3j^2+3(N+1-j)^2-(N+1)(2N-1)) \partial^2 J_j
\end{align}
we finally find
\begin{equation}
\label{universalr2}
\mathbf{r}^{(2)} = \frac{\alpha_0 \bar{N}}{2} \phi_{3,0} - \alpha_0 (T \bar{U}_1)_0 + \alpha_0 (U_1 \bar{T})_0 - \frac{\alpha_0 N}{2} \bar{\phi}_{3,0}.
\end{equation}
Note that the last two lines in the free field representation of $\phi_3$ are total derivatives so they do not contribute to the zero mode and therefore to $\mathbf{r}^{(2)}$, but this choice of $\phi_3$ will slightly simplify the expression for $\mathbf{r}^{(3)}$. If we require $\mathbf{r}^{(j)}$ to be a zero mode of a field of engineering dimension $j+1$, the only ambiguity is in total derivatives, which do not contribute to the zero mode. We also see that if we did not choose the order of the barred indices in \eqref{rfused} to be the opposite of the order of the unbarred indices, we would not be able to write $\mathbf{r}^{(2)}$ in terms of $\hat{\mathfrak{gl}}(1) \times W_{\bar{N}}$ generators -- the sign of
\begin{equation}
\sum_{\bar{j}<\bar{k}} \partial J_{\bar{j}} J_{\bar{k}}
\end{equation}
would be wrong, and it is the sign of this term that reflects the way we ordered our free bosons $J_{\bar{j}}$.

The calculation could be alternatively done entirely in terms of oscillators, but such a calculation is computationally harder (while leading to the same result). We can use the free field representation and fusion as just described to proceed to higher orders. In this way we calculated $\mathbf{r}^{(3)}$ and $\mathbf{r}^{(4)}$. The expressions for these are given in the appendix \ref{universalrlist}.

\subsection{Drinfe\v{l}d's relations I.}

It is easy to verify Drinfe\v{l}d's relations \eqref{drinfeldb} and \eqref{drinfeldc} perturbatively in large $u$. Since the coproduct is a homomorphism of algebras, we can take the logarithm of \eqref{drinfeldb},
\begin{equation}
(\Delta \otimes \mathbbm{1}) \mathbf{r}(u) = \log \left[ \exp \left(\mathbf{r}_{13}(u)\right) \exp\left(\mathbf{r}_{23}(u)\right) \right]
\end{equation}
and apply the BCH formula on the right-hand side. Expanding at large $u$, we find for the first few orders
\begin{align}
(\Delta \otimes \mathbbm{1}) (\mathbf{r}^{(1)}(u)) & = \mathbf{r}^{(1)}_{13}(u) + \mathbf{r}^{(1)}_{23}(u) \\
(\Delta \otimes \mathbbm{1}) (\mathbf{r}^{(2)}(u)) & = \mathbf{r}^{(2)}_{13}(u) + \mathbf{r}^{(2)}_{23}(u) + \frac{1}{2} \left[\mathbf{r}^{(1)}_{13}(u), \mathbf{r}^{(1)}_{23}(u) \right] \\
\nonumber
(\Delta \otimes \mathbbm{1}) (\mathbf{r}^{(3)}(u)) & = \mathbf{r}^{(3)}_{13}(u) + \mathbf{r}^{(3)}_{23}(u) + \frac{1}{2} \left[\mathbf{r}^{(1)}_{13}(u), \mathbf{r}^{(2)}_{23}(u) \right] + \frac{1}{2} \left[\mathbf{r}^{(2)}_{13}(u), \mathbf{r}^{(1)}_{23}(u) \right] \\
& + \frac{1}{12} \left[\mathbf{r}^{(1)}_{13}(u),\left[\mathbf{r}^{(1)}_{13}(u), \mathbf{r}^{(1)}_{23}(u) \right] \right] + \frac{1}{12} \left[\left[\mathbf{r}^{(1)}_{13}(u), \mathbf{r}^{(1)}_{23}(u) \right], \mathbf{r}^{(1)}_{23}(u) \right]
\end{align}
etc. Unlike in the previous section, here we only need the two-variable BCH formula to evaluate the right-hand side, but for the left-hand side we need to know the coproduct of $\hat{\mathfrak{gl}}(1) \times \mathcal{W}_N$ generators \cite{Prochazka:2014gqa}.

Let us explicitly verify the first of these relations. We need to calculate the coproduct of \eqref{universalr1} in the first factor. We have (see section 3.8 in \cite{Prochazka:2014gqa})
\begin{align}
\Delta(N) & = N_{(1)} + N_{(2)} \\
\Delta(U_1) & = U_{(1)1} + U_{(2)1} \\
\Delta(U_2) & = U_{(1)2} + U_{(2)2} + (U_{(1)1} U_{(2)1}) + \alpha_0 N_{(1)} \partial U_{(2)1}
\end{align}
so
\begin{equation}
\Delta(T) = T_{(1)} + T_{(2)} + \frac{\alpha_0 N_{(2)}}{2} \partial U_{(1)1} - \frac{\alpha_0 N_{(1)}}{2} \partial U_{(2)1}
\end{equation}
and
\begin{align}
\nonumber
(\Delta \otimes \mathbbm{1})(\mathbf{r}^{(1)}(u)) & = -\alpha_0 N_{(3)} \Delta(T_0) - \alpha_0 \Delta(N) (T_{(3)})_0 + \alpha_0 (\Delta(U_1) U_{(3)1})_0 \\
\nonumber
& = -\alpha_0 N_{(3)} \left((T_{(1)})_0 + (T_{(2)})_0 + \frac{\alpha_0 N_{(2)}}{2} (\partial U_{(1)1})_0 - \frac{\alpha_0 N_{(1)}}{2} (\partial U_{(2)1})_0\right) \\
& - \alpha_0 (N_{(1)} + N_{(2)}) (T_{(3)})_0 + \alpha_0 (U_{(1)1} U_{(3)1})_0 + \alpha_0 (U_{(2)1} U_{(3)1})_0
\end{align}
while the right-hand side is
\begin{align}
\nonumber
\mathbf{r}^{(1)}_{13}(u) + \mathbf{r}^{(1)}_{23}(u) & = -\alpha_0 N_{(3)} (T_{(1)})_0 - \alpha_0 N_{(1)} (T_{(3)})_0 + \alpha_0 (U_{(1)1} U_{(3)1})_0 \\
& -\alpha_0 N_{(3)} (T_{(2)})_0 - \alpha_0 N_{(2)} (T_{(3)})_0 + \alpha_0 (U_{(2)1} U_{(3)1})_0.
\end{align}
Since the zero mode of a total derivative vanishes, we see that both of these expressions agree.

We may similarly verify these two Drinfe\v{l}d's relations up to order $u^{-4}$ involving $\mathbf{r}^{(4)}$. We calculated these coefficients explicitly by fusing the elementary $R$-matrices (and explicit expressions for these are given in appendix \ref{universalrlist}).
Since the formula for comultiplication of generators of $\mathcal{W}_{1+\infty}$ is known \cite{Prochazka:2014gqa}, we may as well use Drinfe\v{l}d's relations to determine higher $\mathbf{r}^{(j)}$ coefficients. In this way we determined $\mathbf{r}^{(5)}$ and $\mathbf{r}^{(6)}$, but these are too long to list here (the expression for $\mathbf{r}^{(6)}$ has 22 pages in Mathematica). The calculation using Drinfe\v{l}d's relations is more efficient way to determine $\mathbf{r}^{(j)}$ than using the free field representations as was done in the previous section, especially because the BCH formula for large number of variables is complicated.

\subsection{Spectral shift and normalization of $\mathbf{R}(u)$}
The algebras $\hat{\mathfrak{gl}}(1) \times \mathcal{W}_N$ have continuous automorphisms shifting the (central) zero mode of $\hat{\mathfrak{gl}}(1)$ generator by a constant and leaving the $\mathcal{W}_N$ part intact. In the quadratic basis of generators of the algebra that we are using, the $\hat{\mathfrak{gl}}(1)$ and $\mathcal{W}_N$ factors are mixed, so in particular all $U_j$ fields transform under this transformation linearly as
\begin{equation}
\label{spectralu}
U_j \to \mathcal{S}_\gamma(U_j) \equiv \sum_{k=0}^j (N-j+1)_k \frac{\gamma^k}{k!} U_{j-k}
\end{equation}
where $(N)_k$ is the raising factorial (Pochhammer symbol) and $\gamma$ is a parameter of the symmetry transformation. Under such transformation of the left sector, the coefficients $\mathbf{r}^{(j)}$ transform as
\begin{equation}
\label{spectralr}
\mathbf{r}^{(j)} \to \sum_{k=0}^{j-1} {j-1 \choose k} (-\gamma)^k \mathbf{r}^{(j-k)} + \frac{(-\gamma)^j}{j} \alpha_0 (\bar{N} U_1 - N \bar{U}_1)_0 - \frac{(-\gamma)^{j+1}}{j(j+1)} \alpha_0 N\bar{N} \mathbbm{1}
\end{equation}
This is almost equivalent to a shift of the spectral parameter $u$:
\begin{equation}
\mathbf{r}(u) \to \sum_{j=1}^\infty \frac{\mathbf{r}^{(j)}}{(u+\gamma)^j} + \log \left( \frac{u}{u+\gamma} \right) \alpha_0 (\bar{N} U_1 - N \bar{U}_1)_0 + \left[ (u+\gamma) \log \left( \frac{u}{u+\gamma} \right) + \gamma \right] \alpha_0 N \bar{N}
\end{equation}
The reason why this is not simply translation of $u$ parameter is that in \eqref{largeuexpunir} and in similar expansion of the elementary $\mathcal{R}$-matrix we chose a specific (unnatural) normalization of $\mathbf{R}(u)$. We can always rescale $\mathbf{R}(u)$ (or shift $\mathbf{r}(u)$) by any central function, i.e. function of $N, \bar{N}, \alpha_0$ and of zero modes of $U_1$ and $\bar{U}_1$. If we define
\begin{equation}
\label{rtildedef}
\mathbf{\tilde{r}}(u) \equiv \mathbf{r}(u) - \log u \, \alpha_0 (\bar{N} U_1 - N \bar{U}_1)_0 - (u \log u - u) \alpha_0 N \bar{N},
\end{equation}
this object transforms simply under the spectral transformation of the left sector:
\begin{equation}
\label{spectraltranslunir}
\mathbf{\tilde{r}}(u) \to \mathbf{\tilde{r}}(u+\gamma).
\end{equation}
In the following, intermediate calculations with $\mathbf{R}(u)$ are simpler, but using $\mathbf{\tilde{R}}(u)$ leads to some simpler results. The translation between them is easy since they are related by an overall normalization factor.

\subsection{Drinfe\v{l}d's relations II.}

We still have not verified Drinfe\v{l}d's relation \eqref{drinfelda}. It is slightly more subtle because of large spectral parameter expansion that we are using. In particular, we calculated the asymptotic expansion of $\mathbf{R}(u)$ as in \eqref{largeuexpunir}. If we take equation \eqref{drinfelda} literally, the right-hand side as well as the coproduct on the left-hand side would be $u$-independent, so the term of order $u^0$ would immediately tell us that the coproduct and its opposite agree which does not make sense. In order to make sense of \eqref{drinfelda} order by order in $u$, we need to introduce the spectral parameter $u$ also in the coproduct, analogously to \eqref{instspecparam}. Therefore, we will consider the relation in the form
\begin{equation}
\label{drinfelda2}
\mathbf{R}(u) \mathcal{S}_u(\Delta(x)) \mathbf{R}(u)^{-1} \stackrel{!}{=} \mathcal{S}_{u}(\Delta^{op}(x))
\end{equation}
where $\mathcal{S}_u$ is the spectral shift by $\gamma = u$ in the first factor. With no loss of generality we do not need to do any spectral shift in the second factor, because of the intertwining property
\begin{equation}
\Delta(\mathcal{S}_\gamma(x)) = \mathcal{S}_\gamma(\bar{\mathcal{S}}_{\gamma}(\Delta(x))).
\end{equation}
Returning to \eqref{drinfelda2}, let us see what these equations explicitly imply for fields of lower dimension. The simplest choice is taking $x = U_1(z)$. For this choice, both coproduct and the opposite coproduct agree,
\begin{equation}
\Delta(U_1) = U_1 + \bar{U}_1 = \Delta^{op}(U_1)
\end{equation}
and
\begin{equation}
\mathcal{S}_u(\Delta(U_1)) = U_1 + \bar{U}_1 + u N = \mathcal{S}_u(\Delta^{op}(U_1))
\end{equation}
so \eqref{drinfelda2} is in this case equivalent to commutativity of $\mathbf{R}(u)$ with the total spin $1$ generator $U_1 + \bar{U}_1$. The first non-trivial choice is $x = U_2(z)$. Here we have
\begin{align}
\Delta(U_2) & = U_2 + \bar{U}_2 + (U_1 \bar{U}_1) + \alpha_0 N \partial \bar{U}_1 \\
\Delta^{op}(U_2) & = U_2 + \bar{U}_2 + (U_1 \bar{U}_1) + \alpha_0 \bar{N} \partial U_1
\end{align}
and therefore
\begin{align}
\nonumber
\mathcal{S}_u(\Delta(U_2)) & = \frac{u^2}{2} N(N-1) + u (N-1) U_1 + u N \bar{U}_1 + U_2 + \bar{U}_2 + (U_1 \bar{U}_1) + \alpha_0 N \partial \bar{U}_1 \\
\mathcal{S}_u(\Delta^{op}(U_2)) & = \frac{u^2}{2} N(N-1) + u (N-1) U_1 + u N \bar{U}_1 + U_2 + \bar{U}_2 + (U_1 \bar{U}_1) + \alpha_0 \bar{N} \partial U_1.
\end{align}
The identity \eqref{drinfelda2} with $x = U_2(z)$ is therefore equivalent to
\begin{multline}
\mathbf{R}(u) \left[ -u U_1 + U_2 + \bar{U}_2 + (U_1 \bar{U}_1) + \alpha_0 N \partial \bar{U}_1 \right] = \\
= \left[ -u U_1 + U_2 + \bar{U}_2 + (U_1 \bar{U}_1) + \alpha_0 \bar{N} \partial U_1 \right] \mathbf{R}(u).
\end{multline}
Using explicit expressions for $\mathbf{r}^{(j)}$ given in appendix \ref{universalrlist}, we can verify this identity is indeed satisfied (to the order in large $u$ expansion that we considered). We can similarly verify for $x = U_3(z), \ldots$ that to order that we calculated $\mathbf{R}(u)$ Drinfe\v{l}d's relation \eqref{drinfelda} is satisfied.

\section{ILW Hamiltonians}

We will now use the universal $\mathcal{R}$-matrix to construct commuting Hamiltonians of ILW hierarchy \cite{Litvinov:2013zda,Litvinov:2020zeq}. Having constructed an $\mathcal{R}$-matrix satisfying the Yang-Baxter equation with spectral parameter, we can use the standard technology from quantum integrability to construct an infinite collection of commuting Hamiltonians \cite{Faddeev:1996iy,Nepomechie:1998jf}. For $\mathcal{R}$-matrices acting in finite dimensional vector spaces it is enough to take a trace of $\mathcal{R}$-matrix over one of the factors (usually called the auxiliary space) that it is acting on. We are left with one-parametric family (parametrized by spectral parameter $u$) of operators that act on the remaining factor. These operators commute for different values of $u$ as follows from the Yang-Baxter equation and therefore we obtain a collection of commuting Hamiltonians.

In our setup there is a slight twist to this procedure, because the vector spaces involved are infinite dimensional. For this reason the trace is not well-defined and the construction of the commuting Hamiltonians needs to be modified. One possibility used in \cite{Zhu:2015nha,Prochazka:2019dvu} is to consider the vacuum-to-vacuum matrix element instead of the trace. In general it is not true that taking an arbitrary matrix element would lead to commuting quantities, but for our choice of $\mathcal{R}$-matrix and the vacuum matrix element this is true (essentially due to the fact that $\mathcal{R}$ preserves the highest weight states). The resulting Hamiltonians, although non-local from CFT point of view, are related to Yangian generators of the algebra \cite{Tsymbaliuk:2014fvq,Prochazka:2019dvu} and have spectrum that is easy to describe explicitly. In particular, their spectrum can be written in terms of combinatorics of partitions or plane partitions (3d Young diagrams).

There is another possibility of producing commuting Hamiltonians: instead of taking the trace of $R$ over the auxiliary space, we can take the regularized trace \cite{Litvinov:2013zda,Litvinov:2020zeq}, schematically of the form
\begin{equation}
\overline{\Tr} \left[ q^{\bar{L}_0} \mathbf{R}(u) \right].
\end{equation}
Here $\bar{L}_0$ is the `energy' acting on the auxiliary space and $|q|<1$ is a complex regulator. In the $q \to 1$ limit we expect to reduce to the usual (ill-defined) trace. On the other hand, in the limit $q \to 0$ (after an appropriate rescaling) only the vacuum-to-vacuum matrix element contributes and we expect to obtain the Yangian Hamiltonians. It turns out that the Hamiltonians constructed using this regularized trace mutually commute for an arbitrary fixed value of $q$ (and for varying value of spectral parameter) and can be identified with Hamiltonians of quantum ILW hierarchy \cite{Litvinov:2013zda}. From the CFT point of view we are studying the situation where the right part of $\mathcal{R}$-matrix is inserted along a circle on a torus whose shape (complex structure) is controlled by $q$ (see Figure \ref{figcyltor}). As we will see later, the parameter $q$ enters the Bethe ansatz equations in the same way as the twist parameter in quantum spin chains, i.e. controls the twisting that the we can do when we are closing an open spin chain into a closed one. For general $q$ the Hamiltonians that we obtain are non-local, but they become local in the (singular) $q \to 1$ limit and as we will see later they agree with the local Hamiltonians considered in \cite{Bazhanov:1994ft,Bazhanov:2001xm,Dorey:2007zx}.

We will now find explicit expressions for the first few ILW Hamiltonians. As usual, the expressions simplify when we take the logarithm of the generating function of Hamiltonians. In this way, although they are non-local, their non-locality takes a very specific form. From now on we will restrict to $\bar{N}=1$ so
\begin{equation}
\bar{U}_1 \equiv \bar{J}, \qquad \bar{U}_j = 0, \; j \geq 2
\end{equation}
and the vacuum representation on the right, i.e. the lowest weight vector in the right factor will satisfy
\begin{equation}
\bar{J}_m \ket{0} = 0, \qquad m \geq 0.
\end{equation}
Different choices of the lowest weight vector (in particular $\bar{J}_0$ eigenvalue) can be related to our calculation by a shift of the spectral parameter.

For any operator $\mathcal{O}$ we will define the \emph{normalized} expectation value (regularized trace over the auxiliary space) by
\begin{equation}
\langle \mathcal{O} \rangle_q \equiv \frac{\overline{\Tr} q^{\bar{T}_0} \mathcal{O}}{\overline{\Tr} q^{\bar{T}_0}}.
\end{equation}
Note that since the zero mode of the stress-energy tensor $\bar{L}_0$ on the plane and on the cylinder $\bar{T}_0$ differ only by a constant, such a constant would cancel between the numerator and the denominator so we can equivalently replace $\bar{T}_0$ by $\bar{L}_0$. Since we are restricting to $\bar{N}=1$, the Fock space over which we are taking the trace has states labeled by Young diagrams and $\bar{L}_0$ simply counts the number of boxes in the corresponding Young diagram. Therefore, the expectation value can also be written as
\begin{equation}
\langle \mathcal{O} \rangle_q = \frac{\sum_{\lambda} q^{|\lambda|} \bra{\bar{\lambda}} \mathcal{O} \ket{\bar{\lambda}}}{\sum_{\lambda} q^{|\lambda|}}.
\end{equation}
The sum is over all Young diagrams $\lambda$ and $|\lambda|$ denotes the number of boxes. The Hilbert space vectors $\ket{\bar{\lambda}}$ form a basis of the barred (auxiliary) Hilbert space. By definition, the expectation value of any constant (or of any operator that acts proportional to the identity operator in the barred Hilbert space) is the operator itself. For more interesting example, let us consider the expectation value of $\bar{L}_0$:
\begin{align}
\nonumber
\langle \bar{L}_0 \rangle_q & = \frac{\sum_{\lambda} |\lambda| q^{|\lambda|}}{\sum_{\lambda} q^{|\lambda|}} = q \partial_q \log \sum_{\lambda} q^{|\lambda|} = q \partial_q \log \left( \prod_{k=1}^\infty \frac{1}{1-q^k} \right) \\
& = \sum_{k=1}^\infty \frac{k q^k}{1-q^k} = \frac{1-E_2(q)}{24}.
\end{align}
where $E_2(q)$ is the holomorphic Eisenstein series (weight 2 quasi-modular form), see appendix \ref{apilw}. The result is simpler when expressed in terms of the cylinder zero mode $\bar{T}_0$:
\begin{equation}
\label{texpval}
\langle \bar{T}_0 \rangle_q = \langle \bar{L}_0 - \tfrac{1}{24} \rangle_q = -\frac{E_2(q)}{24}.
\end{equation}
We are now ready to evaluate the first ILW Hamiltonians: we define
\begin{equation}
\mathcal{H}_q(u) \equiv \langle \mathbf{R}(u) \rangle_q
\end{equation}
as well as the large $u$ expansion of its logarithm
\begin{equation}
\log \mathcal{H}_q(u) \equiv \sum_{j=1}^\infty \frac{(\log \mathcal{H}_q)_j}{u^j}.
\end{equation}
Since $\mathbf{R}(u)$ is itself an exponential of integral of local fields, we are taking here a logarithm of an expectation value of an exponential, i.e. the quantities $(\log \mathcal{H}_q)_j$ will be (non-commutative) connected expectation values of $\mathbf{r}^{(j)}$. For the first few of these quantities we find
\begin{align}
(\log \mathcal{H}_q)_1 & = \langle \mathbf{r}^{(1)} \rangle \\
\label{logh2gen}
(\log \mathcal{H}_q)_2 & = \langle \mathbf{r}^{(2)} \rangle + \frac{1}{2} \left( \langle \mathbf{r}^{(1)2} \rangle - \langle \mathbf{r}^{(1)} \rangle^2 \right) \\
\nonumber
(\log \mathcal{H}_q)_3 & = \langle \mathbf{r}^{(3)} \rangle + \frac{1}{2} \left( \langle \mathbf{r}^{(1)} \mathbf{r}^{(2)} \rangle - \langle \mathbf{r}^{(1)} \rangle \langle \mathbf{r}^{(2)} \rangle + \langle \mathbf{r}^{(2)} \mathbf{r}^{(1)} \rangle - \langle \mathbf{r}^{(2)} \rangle \langle \mathbf{r}^{(1)} \rangle\right) \\
& + \left( \frac{1}{6} \langle \mathbf{r}^{(1)3} \rangle - \frac{1}{4} \langle \mathbf{r}^{(1)2} \rangle \langle \mathbf{r}^{(1)} \rangle - \frac{1}{4} \langle \mathbf{r}^{(1)} \rangle \langle \mathbf{r}^{(1)2} \rangle + \frac{1}{3} \langle \mathbf{r}^{(1)} \rangle^3 \right) \\
\nonumber
(\log \mathcal{H}_q)_4 & = \langle \mathbf{r}^{(4)} \rangle + \frac{1}{2} \left( \langle \mathbf{r}^{(1)} \mathbf{r}^{(3)} \rangle - \langle \mathbf{r}^{(1)} \rangle \langle \mathbf{r}^{(3)} \rangle + \langle \mathbf{r}^{(2)2} \rangle - \langle \mathbf{r}^{(2)} \rangle^2 + \langle \mathbf{r}^{(3)} \mathbf{r}^{(1)} \rangle - \langle \mathbf{r}^{(3)} \rangle \langle \mathbf{r}^{(1)} \rangle\right) \\
\nonumber
& + \left( \frac{1}{6} \langle \mathbf{r}^{(1)2} \mathbf{r}^{(2)} \rangle - \frac{1}{4} \langle \mathbf{r}^{(1)2} \rangle \langle \mathbf{r}^{(2)} \rangle - \frac{1}{4} \langle \mathbf{r}^{(1)} \rangle \langle \mathbf{r}^{(1)} \mathbf{r}^{(2)} \rangle + \frac{1}{3} \langle \mathbf{r}^{(1)} \rangle^2\langle \mathbf{r}^{(2)} \rangle \right) \\
\nonumber
& + \left( \frac{1}{6} \langle \mathbf{r}^{(1)} \mathbf{r}^{(2)} \mathbf{r}^{(1)} \rangle - \frac{1}{4} \langle \mathbf{r}^{(1)} \mathbf{r}^{(2)} \rangle \langle \mathbf{r}^{(1)} \rangle - \frac{1}{4} \langle \mathbf{r}^{(1)} \rangle \langle \mathbf{r}^{(2)} \mathbf{r}^{(1)} \rangle + \frac{1}{3} \langle \mathbf{r}^{(1)} \rangle \langle \mathbf{r}^{(2)} \rangle \langle \mathbf{r}^{(1)} \rangle \right) \\
& + \left( \frac{1}{6} \langle \mathbf{r}^{(2)} \mathbf{r}^{(1)2} \rangle - \frac{1}{4} \langle \mathbf{r}^{(2)} \mathbf{r}^{(1)} \rangle \langle \mathbf{r}^{(1)} \rangle - \frac{1}{4} \langle \mathbf{r}^{(2)} \rangle \langle \mathbf{r}^{(1)2} \rangle + \frac{1}{3} \langle \mathbf{r}^{(2)} \rangle \langle \mathbf{r}^{(1)} \rangle^2 \right) \\
\nonumber
& + \Big( \frac{1}{24} \langle \mathbf{r}^{(1)4} \rangle - \frac{1}{12} \langle \mathbf{r}^{(1)3} \rangle \langle \mathbf{r}^{(1)} \rangle - \frac{1}{12} \langle \mathbf{r}^{(1)} \rangle \langle \mathbf{r}^{(1)3} \rangle - \frac{1}{8} \langle \mathbf{r}^{(1)2} \rangle^2 \\
\nonumber
& + \frac{1}{6} \langle \mathbf{r}^{(1)} \rangle^2 \langle \mathbf{r}^{(1)2} \rangle + \frac{1}{6} \langle \mathbf{r}^{(1)} \rangle \langle \mathbf{r}^{(1)2} \rangle \langle \mathbf{r}^{(1)} \rangle + \frac{1}{6} \langle \mathbf{r}^{(1)2} \rangle \langle \mathbf{r}^{(1)} \rangle^2 - \frac{1}{4} \langle \mathbf{r}^{(1)} \rangle^4 \Big)
\end{align}
where we write expectation values $\langle \ldots \rangle_q$ without the subscript $q$ when this is clear from the context. As already mentioned, taking these non-commutative connected expectation values will make the non-localities in $(\log \mathcal{H})_j$ milder in the sense that will become clearer as we proceed.

\paragraph{First ILW Hamiltonian}
We can now turn to the first Hamiltonian $(\log \mathcal{H})_1$. All we need to do is to take the expectation value of \eqref{universalr1},
\begin{align}
(\log \mathcal{H}_q)_1 & = \langle -\alpha_0 T_0 -\alpha_0 N \bar{T}_0 + \alpha_0 (U_1 \bar{U}_1)_0 \rangle_q \\
& = -\alpha_0 T_0 - \alpha_0 N \langle \bar{T}_0 \rangle_q + \alpha_0 \langle (U_1 \bar{U}_1)_0 \rangle_q.
\end{align}
In \eqref{texpval} we already calculated the expectation value of $\bar{T}_0$ so it only remains to evaluate the expectation value of the last term, but this clearly vanishes (because $\bar{J}_0$ acts as zero and non-zero mode operators have vanishing expectation values). We therefore find
\begin{equation}
(\log \mathcal{H}_q)_1 = -\alpha_0 T_0 + \frac{\alpha_0 N E_2}{24}.
\end{equation}
We see that the first Hamiltonian is up to a constant equal to $L_0$, i.e. the standard (holomorphic part of the) Hamiltonian associated to radial quantization in CFT, independently of $q$. Only the higher Hamiltonians will have non-trivial $q$-dependence, the Hamiltonian $L_0$ is shared by the BLZ, Yangian or the general ILW family of Hamiltonians. The typical representation spaces are infinite dimensional, but since the higher commuting quantities commute with $L_0$, the diagonalization of these is reduced to (finite dimensional) eigensubspaces of $L_0$ and therefore the diagonalization is effectively a finite dimensional problem.

\paragraph{Second ILW Hamiltonian}
Let us now turn to the second ILW Hamiltonian. From \eqref{logh2gen} we need to evaluate $\langle \mathbf{r}^{(2)} \rangle$ as well as the connected expectation value of $\mathbf{r}^{(1)2}$. We have
\begin{align}
\nonumber
\langle \mathbf{r}^{(2)} \rangle_q & = \left\langle \frac{\alpha_0 \bar{N}}{2} \phi_{3,0} - \alpha_0 (T \bar{U}_1)_0 + \alpha_0 (U_1 \bar{T})_0 - \frac{\alpha_0 N}{2} \bar{\phi}_{3,0} \right\rangle_q \\
& = \frac{\alpha_0}{2} \phi_{3,0} + \alpha_0 U_{1,0} \langle \bar{T}_0 \rangle_q - \frac{\alpha_0 N}{6} \langle (\bar{J}(\bar{J}\bar{J}))_0 \rangle_q \\
\nonumber
& = \frac{\alpha_0}{2} \phi_{3,0} - \frac{\alpha_0 E_2}{24} U_{1,0}
\end{align}
where we used the fact that the expectation values of objects involving odd powers of $\bar{J}$ vanish by $\mathbbm{Z}_2$ symmetry (this would not be the case if we did not restrict to auxiliary representation with $\bar{J}_0 \sim 0$). For the connected expectation value of $\mathbf{r}^{(1)2}$ we have
\begin{align}
\nonumber
\langle \mathbf{r}^{(1)2} \rangle_q - \langle \mathbf{r}^{(1)} \rangle_q^2 & = \left\langle \left(-\alpha_0 T_0 -\alpha_0 N \bar{T}_0 + \alpha_0 (U_1 \bar{U}_1)_0\right)^2 \right\rangle_q - \left\langle -\alpha_0 T_0 -\alpha_0 N \bar{T}_0 + \alpha_0 (U_1 \bar{U}_1)_0 \right\rangle_q^2 \\
& = +\alpha_0^2 N^2 \langle \bar{T}_0^2 \rangle_c -\alpha_0^2 \langle T_0 (U_1 \bar{U}_1)_0 \rangle_c -\alpha_0^2 \langle (U_1 \bar{U}_1)_0 T_0 \rangle_c \\
\nonumber
& -\alpha_0^2 N \langle \bar{T}_0 (U_1 \bar{U}_1)_0 \rangle_c -\alpha_0^2 N \langle (U_1 \bar{U}_1)_0 \bar{T}_0 \rangle_c +\alpha_0^2 \langle (U_1 \bar{U}_1)_0^2 \rangle_c
\end{align}
Here we introduced a notation for connected expectation values,
\begin{equation}
\langle \mathcal{A} \mathcal{B} \rangle_c \equiv \langle \mathcal{A} \mathcal{B} \rangle_q - \langle \mathcal{A} \rangle_q \langle \mathcal{B} \rangle_q
\end{equation}
Using the calculations and formulas for expectation values in appendix \ref{apilw}, we can simplify this to
\begin{align}
\langle \mathbf{r}^{(1)2} \rangle_q - \langle \mathbf{r}^{(1)} \rangle_q^2 & = \alpha_0^2 N^2 \frac{E_4 - E_2^2}{288} +\alpha_0^2 \left( \sum_{m>0} m\frac{1+q^m}{1-q^m} U_{1,-m} U_{1,m} + N \sum_{m>0} \frac{m^2 q^m}{1-q^m} \right)
\end{align}
Notice that most terms disappear when we take the connected expectation values. Adding this result together with the expression for the expectation value of $\mathbf{r}^{(2)}$, we finally find
\begin{align}
\nonumber
(\log \mathcal{H}_q)_2 & = \frac{\alpha_0}{2} \left( U_3 - (U_1 U_2) + \frac{1}{3} (U_1 (U_1 U_1)) \right)_0 + \frac{\alpha_0^2}{2} \sum_{m>0} m\frac{1+q^m}{1-q^m} U_{1,-m} U_{1,m} \\
& - \frac{\alpha_0 E_2}{24} U_{1,0} + \alpha_0^2 N^2 \frac{E_4 - E_2^2}{576} + \frac{\alpha_0^2 N}{2} \sum_{m>0} \frac{m^2 q^m}{1-q^m}
\end{align}
On the first line we see the first non-trivial ILW Hamiltonian. It is a sum of two terms, first being local while the second one involving
\begin{equation}
\label{nonlocalterm}
\sum_{m>0} m \frac{1+q^m}{1-q^m} U_{1,-m} U_{1,m} \equiv \sum_{m>0} (\mathfrak{d}_m + \mathfrak{d}_{-m}) U_{1,-m} U_{1,m}
\end{equation}
which cannot be written as a zero mode of a local operator. Here we use the modified derivative
\begin{equation}
\mathfrak{d}_m \equiv \frac{m}{1-q^m}.
\end{equation}
The non-locality of the product of two $U_1(z)$ fields is controlled by $q$-dependent
\begin{equation}
(\mathfrak{d}_m + \mathfrak{d}_{-m}) = m \frac{1+q^m}{1-q^m} = m \coth \left( \frac{m \log q}{2} \right)
\end{equation}
which at fixed $q$ is an even function of $m$. In the limit $q \to 1$ we have
\begin{equation}
m \frac{1+q^m}{1-q^m} \sim \frac{2}{1-q} - 1 + \frac{m^2-1}{6} (1-q)^1 + \mathcal{O}((1-q)^2).
\end{equation}
We see that every coefficient of $(1-q)^j$ is a polynomial in $m^2$, so all the coefficients in $1-q$ expansion of $(\log \mathcal{H}_q)_2$ are zero modes of local operators. On the other hand, in the limit $q \to 0$ we have instead
\begin{equation}
m \frac{1+q^m}{1-q^m} \to |m|
\end{equation}
so in this limit $(\log \mathcal{H}_q)_2$ cannot be written as a zero mode of a local operator (one would need to use the Hilbert transform to produce the absolute value of the Fourier mode $|m|$). We find a non-locality characteristic of Benjamin-Ono hierarchy which also appears in the affine Yangian context \cite{nazarovskl,Prochazka:2015deb}. In the limit $q \to \infty$ we find the same absolute value of the mode number (which corresponds to charge conjugate Yangian Hamiltonians). It is very interesting that the parameter $q$ which in our construction appears geometrically as a shape (complex structure) of the auxiliary torus naturally controls the non-locality of the higher Hamiltonians and in particular interpolates between the local limit at $q=1$ (which is expected to be related to BLZ-like integrals of motion) and non-local limit of the Benjamin-Ono / Yangian type (at $q \to 0$ and $q \to \infty$). For $q$ equal to a root of unity we get local twisted Hamiltonians as will be illustrated later. For other values of $q$, we get generically the intermediate long wave Hamiltonians. There exists also an elliptic generalization of this type of non-local interaction terms, see for instance \cite{Bonelli:2014iza}.

In a similar manner, we can find explicit expressions for higher order ILW Hamiltonians. Since the calculations get very long very quickly, we only determine $(\log \mathcal{H}_q)_3$, $(\log \mathcal{H}_q)_4$ and $(\log \mathcal{H}_q)_5$. Explicit expressions for first two of these as well as some details of the calculation are given in the appendix \ref{apilw}.

\paragraph{Spectral translation}
It is useful to verify how the ILW Hamiltonians that we found transform under the shift of the spectral parameter. The transformation of $(\log \mathcal{H})_j$ under \eqref{spectralu} is inherited from that of $\mathcal{R}$-matrix \eqref{spectralr},
\begin{equation}
(\log\mathcal{H}_q)_j \to \sum_{k=0}^{j-1} {j-1 \choose k} (-\gamma)^k (\log\mathcal{H}_q)_{j-k} + \frac{(-\gamma)^j}{j} \alpha_0 U_{1,0} - \frac{(-\gamma)^{j+1}}{j(j+1)} N\alpha_0.
\end{equation}
More explicitly, it is easy to check using the explicit forms of the Hamiltonians that
\begin{align}
(\log \mathcal{H}_q)_1 & \to (\log \mathcal{H}_q)_1 - \gamma \alpha_0 U_1 - \frac{\gamma^2}{2} N\alpha_0 \\
(\log \mathcal{H}_q)_2 & \to (\log \mathcal{H}_q)_2 - \gamma (\log \mathcal{H}_q)_1 + \frac{\gamma^2}{2} \alpha_0 U_{1,0} + \frac{\gamma^3}{6} N\alpha_0 \\
(\log \mathcal{H}_q)_3 & \to (\log \mathcal{H}_q)_3 -2\gamma (\log \mathcal{H}_q)_2 +\gamma^2 (\log \mathcal{H}_q)_1 - \frac{\gamma^3}{3} \alpha_0 U_{1,0} - \frac{\gamma^4}{12} \alpha_0 N \\
(\log \mathcal{H}_q)_4 & \to (\log \mathcal{H}_q)_4 -3\gamma (\log \mathcal{H}_q)_3 + 3\gamma^2 (\log \mathcal{H}_q)_2 -\gamma^3 (\log \mathcal{H}_q)_1 \\
& +\frac{\gamma^4}{4} \alpha_0 U_{1,0} +\frac{\gamma^5}{20} \alpha_0 N 
\end{align}
where we used the transformation properties
\begin{align}
U_{1,0} & \to U_{1,0} + \gamma N \\
T_m & \to T_m + \gamma U_{1,m} + \gamma^2 \frac{N}{2} \delta_{m,0} \\
\phi_{3,m} & \to \phi_{3,m} + 2\gamma T_m + \gamma^2 U_{1,m} + \gamma^3 \frac{N}{3} \delta_{m,0} \\
\phi_{4,0} & \to \phi_{4,0} - 3\gamma \phi_{3,0} - 3\gamma^2 T_0 - \gamma^3 U_{1,0} - \gamma^4 \frac{N}{4} \\
\phi_{5,0} & \to \phi_{5,0} - 4\gamma \phi_{4,0} + 6\gamma^2 \phi_{3,0} + 4\gamma^3 T_0 + \gamma^4 U_{1,0} + \gamma^5 \frac{N}{5}.
\end{align}

\section{Spectrum and Bethe equations}
After having constructed the ILW Hamiltonians, we can now study their spectra. We will proceed in few steps. First of all, since we are normalizing the $R$-matrices as in \eqref{largeuexpelr} and \eqref{largeuexpunir}, the lowest weight vector (vacuum) eigenvalue of $R$ on the cylinder is non-trivial, capturing the higher spin Casimir charges of the zero modes of the local fields, so we first determine these. Afterwards, we determine the spectrum of operators obtained by taking the vacuum matrix element on the left or on the right side. The resulting operators are Yangian Hamiltonians, so they can be diagonalized explicitly in terms of combinatorics of partitions (or plane partitions). Their spectrum will give us on one hand the ground state eigenvalue of ILW Hamiltonians and on the other hand the $q \to 0$ limit of their spectrum. Finally, we diagonalize the general $q$-dependent ILW Hamiltonians that we determined explicitly in the previous section. We will parametrize their eigenvalues in terms of solutions of Bethe ansatz equations of Litvinov \cite{Litvinov:2013zda} and compare the result with Yangian version of a formula derived by Feigin-Jimbo-Miwa-Mukhin in the context of quantum toroidal algebra \cite{Feigin:2015raa,Feigin:2016pld}.

\subsection{Normalization and higher Casimir charges}
Using the formula \eqref{cylindernormalorder}, it is easy to evaluate the vacuum expectation values of $r^{(j)}$. The first few values are
\begin{align}
\bra{a} r^{(1)} \ket{a} & = \frac{\alpha_0^2}{12} - \frac{\alpha_0^2 a^2}{2} \\
\bra{a} r^{(2)} \ket{a} & = -\frac{\alpha_0^3 a}{12} + \frac{\alpha_0^3 a^3}{6} \\
\bra{a} r^{(3)} \ket{a} & = -\frac{\alpha_0^4(3+\alpha_0^2)}{360} + \frac{\alpha_0^4 a^2}{12} - \frac{\alpha_0^4 a^4}{12} \\
\bra{a} r^{(4)} \ket{a} & = \frac{\alpha_0^5(3+\alpha_0^2) a}{120} - \frac{\alpha_0^5 a^3}{12} + \frac{\alpha_0^5 a^5}{20} \\
\bra{a} r^{(5)} \ket{a} & = \frac{\alpha_0^6(5+5\alpha_0^2+\alpha_0^4)}{1260} - \frac{\alpha_0^6(3+\alpha_0^2) a^2}{60} + \frac{\alpha_0^6 a^4}{12} - \frac{\alpha_0^6 a^6}{30}
\end{align}
where $\ket{a}$ is the lowest weight vector with respect to $J_-$ with zero mode eigenvalue $\alpha_0 a$ (so eigenvalue $a$ of $J_1$, see \eqref{jminus}). The $a$-dependence is completely determined from the spectral translation \eqref{spectraltranslunir} so we may focus on the case of $a=0$. In this case, the coefficients of even powers of $u$ vanish and for the remaining ones we have
\begin{equation}
\bra{0} r^{(2l-1)} \ket{0} = \sum_{k=0}^{l-1} \frac{B_{2l} \times (2l-k-2)! \alpha_0^{4l-2k-2}}{2l \times k!(2l-2k-1)!}.
\end{equation}
This means that the vacuum eigenvalue of $r(u)$ has an asymptotic expansion
\begin{equation}
\label{casimirasymptotic}
\bra{0} r(u) \ket{0} = \sum_{l=1}^\infty \sum_{k=0}^{l-1} \frac{B_{2l} \times (2l-k-2)! \alpha_0^{4l-2k-2}}{2l \times k!(2l-2k-1)!} \frac{1}{(\alpha_0 u)^{2l-1}}
\end{equation}
The internal sum can be evaluated if we introduce the Nekrasov-like parameters \cite{Tsymbaliuk:2014fvq,Prochazka:2015deb}
\begin{equation}
\label{epsilonparameters}
\epsilon_3 = \alpha_0, \qquad \epsilon_1+\epsilon_2+\epsilon_3 = 0, \qquad \epsilon_1 \epsilon_2 = -1.
\end{equation}
We have
\begin{equation}
\sum_{k=0}^{l-1} \frac{(2l-1)(2l-k-2)! \alpha_0^{2l-1-2k}}{k!(2l-2k-1)!} = -\epsilon_1^{2l-1} - \epsilon_2^{2l-1}
\end{equation}
so as asymptotic series at large $u$,
\begin{equation}
\bra{0} r(u) \ket{0} = -\sum_{l=1}^\infty \frac{B_{2l}}{2l(2l-1)} \left[ \left(\frac{\epsilon_1}{u} \right)^{2l-1} + \left(\frac{\epsilon_2}{u} \right)^{2l-1} \right].
\end{equation}
This asympotic expansion captures all the higher Casimir charges of the vacuum state. To proceed, it is useful to consider $\tilde{r}$ defined in \eqref{rtildedef} which includes zero mode of spin $0$ and spin $1$ operators and transforms nicely under spectral translation. We have
\begin{align}
\bra{0} \tilde{r}(u) \ket{0} = -\alpha_0 (u \log u-u) -\sum_{l=1}^\infty \frac{B_{2l}}{2l(2l-1)} \left[ \left(\frac{\epsilon_1}{u} \right)^{2l-1} + \left(\frac{\epsilon_2}{u} \right)^{2l-1} \right].
\end{align}
This is very reminiscent of the asymptotic expansion of $\Gamma$ function, the Stirling formula
\begin{equation}
\log \Gamma(z) \sim \left(z-\frac{1}{2}\right)\log z - z + \frac{1}{2} \log(2\pi) + \sum_{j=1}^\infty \frac{B_{2j}}{2j(2j-1)z^{2j-1}}
\end{equation}
so
\begin{align}
\label{rtildevacuumexp}
\nonumber
\bra{0} \tilde{r}(u) \ket{0} & \sim -\log \Gamma\left(\frac{u}{\epsilon_1}\right) -\log \Gamma\left(\frac{u}{\epsilon_2}\right) +\left(\frac{u}{\epsilon_1}\right)\log \frac{1}{\epsilon_1} +\left(\frac{u}{\epsilon_2}\right)\log \frac{1}{\epsilon_2} \\
& -\frac{1}{2} \log \frac{u}{\epsilon_1} -\frac{1}{2} \log \frac{u}{\epsilon_2} + \log(2\pi).
\end{align}
Since we calculated the $\mathcal{R}$-matrix perturbatively in large $u$ expansion and the asymptotic series for gamma function at infinity is not convergent (due to infinite sequence of poles at negative integers), strictly speaking the normalization of $R$ that we used is not well-defined. But this is easy to fix: we can just choose different normalization of $R$, for example the non-perturbative completion in terms of gamma functions as in \eqref{rtildevacuumexp} or alternatively we can normalize $R$ such that the vacuum expectation value is unity, as was done in \cite{Zhu:2015nha,Prochazka:2019dvu}. For our purposes these issues are not relevant since they affect only the overall normalization factor, the ratios of matrix elements of $R$ between any two states (of definite $L_0$ level) are given by rational functions of the spectral parameter so there is no issue of convergence here (see the next paragraph or \cite{Prochazka:2019dvu} where some of these matrix elements are evaluated).

\subsection{Spectrum of ILW Hamiltonians at $q \to 0$}
In the limit $q \to 0$ the ILW Hamiltonians reduce to Yangian Hamiltonians \cite{Prochazka:2019dvu,Litvinov:2020zeq}. The spectrum of these has nice combinatorial description in terms of Young diagrams or their generalizations ($N$-tuples of these or plane partitions with various asymptotics). At lower levels, the eigenvalues can be calculated as exact functions of the spectral parameter $u$ directly from the definition of the $\mathcal{R}$-matrix. Let us illustrate this on an example. The elementary $\mathcal{R}$-matrix at level $1$ acts as
\begin{align}
R(u) J_{-1} \ket{a,\bar{a}} & = \frac{u+a-\bar{a}}{u+a-\bar{a}+\epsilon_3} \mathcal{E}_{Cas}(u) J_{-1} \ket{a,\bar{a}} + \frac{\epsilon_3}{u+a-\bar{a}+\epsilon_3} \mathcal{E}_{Cas}(u) \bar{J}_{-1} \ket{a,\bar{a}} \\
R(u) \bar{J}_{-1} \ket{a,\bar{a}} & = \frac{\epsilon_3}{u+a-\bar{a}+\epsilon_3} \mathcal{E}_{Cas}(u) J_{-1} \ket{a,\bar{a}} + \frac{u+a-\bar{a}}{u+a-\bar{a}+\epsilon_3} \mathcal{E}_{Cas}(u) \bar{J}_{-1} \ket{a,\bar{a}}
\end{align}
as immediately follows from \eqref{rmatrixcomm} \cite{Prochazka:2019dvu}. Here $a$ is the lowest weight eigenvalue of $J_0$ and $\bar{a}$ is the lowest weight eigenvalue of $\bar{J}_0$. $\mathcal{E}_{Cas}(u)$ is the eigenvalue of $R(u)$ acting on the lowest weight state $\ket{a,\bar{a}}$ (the normalization factor). In the limit $q \to 0$ instead of taking the normalized trace over the right Hilbert space we simply take the vacuum expectation value on the right. Therefore, on level $1$ in the case of $N = 1 = \bar{N}$ (and $\bar{a}=0$) we have
\begin{equation}
\mathcal{H}_{q=0}(u) J_{-1} \ket{a} = \frac{u+a}{u+a+\epsilon_3} \mathcal{E}_{Cas}(u) J_{-1} \ket{a}.
\end{equation}
A similar calculation at level $2$ shows that the eigenvalues of $\mathcal{H}_{q=0}(u)$ at level $2$ are
\begin{align}
\frac{(u+a)(u+a-\epsilon_1)}{(u+a+\epsilon_3)(u+a-\epsilon_1+\epsilon_3)} \mathcal{E}_{Cas}(u), \qquad \frac{(u+a)(u+a-\epsilon_2)}{(u+a+\epsilon_3)(u+a-\epsilon_2+\epsilon_3)} \mathcal{E}_{Cas}(u).
\end{align}
This calculation can be performed also for general values of integers $N$ and $\bar{N}$. At level $1$ we have $N$ eigenvalues
\begin{equation}
\mathcal{E}_{Cas}(u) \times \prod_{l=1}^{\bar{N}} \frac{u+a_k-\bar{a}_l}{u+a_k-\bar{a}_l+\epsilon_3}, \qquad k=1,\ldots,N.
\end{equation}
(corresponding to a single box inserted at $u = -a_k$, $k = 1, \ldots N$) and similarly at level $2$ we have $N$ eigenvalues
\begin{equation}
\mathcal{E}_{Cas}(u) \times \prod_{l=1}^{\bar{N}} \frac{(u+a_k-\bar{a}_l)(u+a_k-\bar{a}_l-\epsilon_1)}{(u+a_k-\bar{a}_l+\epsilon_3)(u+a_k-\bar{a}_l-\epsilon_1+\epsilon_3)}, \qquad k=1,\ldots N,
\end{equation}
$N$ eigenvalues
\begin{equation}
\mathcal{E}_{Cas}(u) \times \prod_{l=1}^{\bar{N}} \frac{(u+a_k-\bar{a}_l)(u+a_k-\bar{a}_l-\epsilon_2)}{(u+a_k-\bar{a}_l+\epsilon_3)(u+a_k-\bar{a}_l-\epsilon_2+\epsilon_3)}, \qquad k=1,\ldots N,
\end{equation}
and $N(N-1)/2$ eigenvalues
\begin{equation}
\mathcal{E}_{Cas}(u) \times \prod_{l=1}^{\bar{N}} \frac{(u+a_{k_1}-\bar{a}_l)(u+a_{k_2}-\bar{a}_l)}{(u+a_{k_1}-\bar{a}_l+\epsilon_3)(u+a_{k_2}-\bar{a}_l+\epsilon_3)}, \qquad 1 \leq k_1 < k_2 \leq N.
\end{equation}

The general form of the eigenvalues should be clear: a general state in the Fock space of $N$ free bosons is labeled by $N$-tuple of Young diagrams $\lambda^{(k)}, k=1,\ldots,N$. Each box of each Young diagram has an associated Bethe root
\begin{equation}
\label{betherootq0}
x = -a_k + \epsilon_1 c_1 (\Box) + \epsilon_2 c_2 (\Box) \equiv -a_k + \epsilon_{\Box}
\end{equation}
where $c_j(\Box)$ are the cartesian coordinates of the box of the Young diagram $\lambda^{(k)}$, starting with the first box at coordinates $(0,0)$.\footnote{A better and more symmetric way of writing this would be to consider the Young diagram lying in $1-2$ plane of three-dimensional space and write the corresponding Bethe root as $-a_k + \epsilon_1 c_1(\Box) + \epsilon_2 c_2(\Box) + \epsilon_3 c_3(\Box)$. Here $c_3(\Box)$ is the same for all boxes of the Young diagram (since it lies in $1-2$ plane) and it is natural to choose the coordinates of the first box to be either $(0,0,0)$ or $(1,1,1)$ which does not affect the associated Bethe roots due to the identity $\epsilon_1+\epsilon_2+\epsilon_3=0$.} The eigenvalues of $\mathcal{H}_{q=0}(u)$ acting on a state $\ket{\lambda^{(1)},\ldots,\lambda^{(N)}}$ which has associated Bethe roots $x_\ell, \, \ell=1,\ldots,M$ where $M = |\lambda^{(1)}| + \ldots + |\lambda^{(N)}|$ is the total number of boxes is
\begin{equation}
\label{spectrumHq0}
\mathcal{H}_{q=0}(u) \ket{\lambda^{(1)},\ldots,\lambda^{(N)}} = \mathcal{E}_{Cas}(u) \prod_{\ell=1}^M \prod_{k=1}^{\bar{N}} \frac{u-x_\ell-\bar{a}_k}{u-x_\ell-\bar{a}_k+\epsilon_3} \ket{\lambda^{(1)},\ldots,\lambda^{(N)}}.
\end{equation}
In the previous example at level $2$, we had $N$ states with Young diagram $\ydiagram{2}$ in the $k$-th layer, $k=1,\ldots,N$, another $N$ states with Young diagram $\ydiagram{1,1}$ in the $k$-th layer, $k=1,\ldots,N$ and finally $N(N-1)/2$ states with two Young diagrams $\ydiagram{1}$, one in the $k_1$-th layer and another in the $k_2$-th layer.

For later purposes, it will be convenient to write the spectrum of $(\log \mathcal{H}_{q=0})_j$ in terms of symmetric polynomials in Bethe roots (specializing again to $\bar{N}=1$ and $\bar{a}=0$). In order to do that, we just need to take the logarithm of \eqref{spectrumHq0} and extract the corresponding coefficients. We find
\begin{align}
(\log \mathcal{H}_{q=0}) & \rightarrow \log \mathcal{E}_{Cas}(u) + \sum_{\ell=1}^M \log \left( \frac{u-x_\ell}{u-x_\ell+\epsilon_3} \right) \\
& \simeq \log \mathcal{E}_{Cas}(u) + \sum_{j=1}^\infty \frac{1}{j u^j} \left(\sum_{k=0}^{j-1} (-1)^{j-k} {j \choose k} p_k \epsilon_3^{j-k} \right)
\end{align}
where
\begin{equation}
p_j = \sum_{\ell=1}^M x_{\ell}^j
\end{equation}
are the Newton power sum polynomials of the Bethe roots and $p_0 \equiv M = \sum_j |\lambda^{(j)}|$, the total number of Bethe roots (which is the level of the corresponding state). Explicitly,
\begin{align}
(\log \mathcal{H}_{q=0})_1 & \rightarrow (\log \mathcal{H}_{q=0})_{1,hw} - \epsilon_3 p_0 \\
(\log \mathcal{H}_{q=0})_2 & \rightarrow (\log \mathcal{H}_{q=0})_{2,hw} - \epsilon_3 p_1 + \frac{\epsilon_3^2}{2} p_0 \\
(\log \mathcal{H}_{q=0})_3 & \rightarrow (\log \mathcal{H}_{q=0})_{3,hw} - \epsilon_3 p_2 + \epsilon_3^2 p_1 - \frac{\epsilon_3^3}{3} p_0 \\
(\log \mathcal{H}_{q=0})_4 & \rightarrow (\log \mathcal{H}_{q=0})_{4,hw} - \epsilon_3 p_3 + \frac{3}{2} \epsilon_3^2 p_2 - \epsilon_3^3 p_1 + \frac{\epsilon_3^4}{4} p_0
\end{align}
In particular, the Hamiltonians extracted from the logarithm of $\mathcal{H}_{q=0}$ are linear functions of the Newton sums $p_k$ with coefficients that are independent of $N$ and of lowest weight charges. The dependence on $N$ and on the lowest weight charges comes both from the lowest weight state contribution as well as from the explicit dependence of Bethe roots on $N$ and on the lowest weight charges. The linearity of these charges in $p_j$ is special to $q = 0$ limit of the ILW Hamiltonians and will not survive the deformation to $q \to 0$.

With the choice of normalization of $R$ that we are using here, the contribution from the Casimir charges can be obtained from \eqref{casimirasymptotic} and we find
\begin{align}
(\log \mathcal{H}_{q=0})_{1,hw} & = \alpha_0 \Big(u_2 - \frac{u_1^2}{2} + \frac{N}{12} \Big) \\
(\log \mathcal{H}_{q=0})_{2,hw} & = \frac{\alpha_0}{2} \Big( u_3 - u_1 u_2 + \frac{u_1^3}{3} - \frac{u_1}{6} \Big) \\
(\log \mathcal{H}_{q=0})_{3,hw} & = \frac{\alpha_0}{3} \Big( u_4 - u_1 u_3 - \frac{u_2^2}{2} + u_1^2 u_2 - \frac{u_1^4}{4} - \frac{u_2}{2} + \frac{u_1^2}{4} - \frac{N(3+\alpha_0^2)}{120} \Big) \\
\nonumber
(\log \mathcal{H}_{q=0})_{4,hw} & = \frac{\alpha_0}{4} \Big( u_5 - u_1 u_4 - u_2 u_3 + u_1^2 u_3 + u_1 u_2^2 - u_1^3 u_2 + \frac{u_1^5}{5} \\
& - u_3 + u_1 u_2 - \frac{u_1^3}{3} + \frac{(3+\alpha_0^2)u_1}{30} \Big).
\end{align}

\subsection{Ground state eigenvalue of ILW Hamiltonians}
\label{highestweightILWspectrum}

We can next use the knowledge of the spectrum of Yangian Hamiltonians to determine the lowest weight state eigenvalue of $\mathcal{H}_q$ for $q \neq 0$. Similarly to the discussion in the previous section, if we take the vacuum-to-vacuum matrix element on the left and allow for an arbitrary $\bar{N}=1$ state on the right, we find eigenvalues
\begin{equation}
\label{spectrumHbq0}
\bar{\mathcal{H}}_{\bar{q}=0}(u) \ket{\bar{\lambda}} = \mathcal{E}_{Cas}(u) \prod_{\Box \in \lambda} \prod_{j=1}^{N} \frac{u+a_j-\bar{a}-\epsilon_{\Box}}{u+a_j-\bar{a}-\epsilon_{\Box}+\epsilon_3} \ket{\bar{\lambda}}
\end{equation}
were $\lambda$ runs over all Young diagrams (states of the right Fock space). Therefore, the lowest weight state eigenvalue of $\mathcal{H}_q$ is
\begin{equation}
\bra{\{a_j\}} \mathcal{H}_q(u) \ket{\{a_j\}} = \frac{\mathcal{E}_{Cas}(u)}{\sum_{\lambda} q^{|\lambda|}} \times \sum_{\lambda} q^{|\lambda|} \prod_{\Box \in \lambda} \prod_{j=1}^{N} \frac{u+a_j-\epsilon_{\Box}}{u+a_j-\epsilon_{\Box}+\epsilon_3}
\end{equation}
where we put $\bar{a}=0$. Writing this in the form
\begin{equation}
\bra{\{a_j\}} \log \mathcal{H}_q(u) \ket{\{a_j\}} = \log \mathcal{E}_{Cas}(u) + \log \Big\langle \exp \sum_{\Box\in\lambda} \sum_{j=1}^N \log \frac{u+a_j-\epsilon_{\Box}}{u+a_j-\epsilon_{\Box}+\epsilon_3} \Big\rangle
\end{equation}
The second term on the right-hand side can be Taylor expanded at large $u$ as
\begin{equation}
\log \Big\langle \exp \sum_{k=1}^\infty \frac{1}{k u^k} \sum_{l=1}^k \sum_{m=0}^{k-l} \frac{(-1)^{l+m} k! \epsilon_3^l \left( \sum_{j=1}^N a_j^m \right) \left( \sum_{\Box\in\lambda}\epsilon_\Box^{k-l-m} \right)}{l!m!(k-l-m)!} \Big\rangle
\end{equation}
This is of the form of the logarithm of the exponential of the expectation value of an expression bilinear in power sums of Bethe roots and power sums of zero modes $a_j$. As such, its large $u$ expansion coefficients are determined in terms of connected expectation values of these Newton polynomials. Denoting the sum of $k$-th powers of Bethe roots by
\begin{equation}
\label{yangrootpowersum}
\mathcal{O}_k = \sum_{\Box\in\lambda} \epsilon_{\Box}^k,
\end{equation}
we find the large $u$ expansion of the vacuum eigenvalue of ILW Hamiltonians to be
\begin{equation}
\log \mathcal{E}_{Cas}(u) + \log \Big\langle \exp \sum_{k=1}^\infty \frac{1}{k u^k} \sum_{l=1}^k \sum_{m=0}^{k-l} \frac{(-1)^{l+m} k! \epsilon_3^l \left( \sum_{j=1}^N a_j^m \right)}{l!m!(k-l-m)!} \mathcal{O}_{k-l-m} \Big\rangle.
\end{equation}
For first few Hamiltonians, this is explicitly
\begin{align}
(\log \mathcal{H}_q)_{1,hw}^\prime & = -\alpha_0 N \langle \mathcal{O}_0 \rangle \\
(\log \mathcal{H}_q)_{2,hw}^\prime & = -\frac{\alpha_0}{2} \Big[ 2N \langle \mathcal{O}_1 \rangle - \alpha_0 N^2 \langle \mathcal{O}_0^2 \rangle_c - (\alpha_0 N + 2u_1) \langle \mathcal{O}_0 \rangle \Big] \\
\nonumber
(\log \mathcal{H}_q)_{3,hw}^\prime & = -\frac{\alpha_0}{6} \Big[ 6N \langle \mathcal{O}_2 \rangle - 6\alpha_0 N^2 \langle \mathcal{O}_0 \mathcal{O}_1 \rangle_c + \alpha_0^2 N^3 \langle \mathcal{O}_0^3 \rangle_c \\
& - 6 (2u_1 + \alpha_0 N) \langle \mathcal{O}_1 \rangle + 3\alpha_0 N(2u_1+\alpha_0 N) \langle \mathcal{O}_0^2 \rangle_c \\
\nonumber
& + 2 (3u_1^2 - 6u_2 + 3\alpha_0 u_1 + \alpha_0^2 N) \langle \mathcal{O}_0 \rangle \Big] \\
\nonumber
(\log \mathcal{H}_q)_{4,hw}^\prime & = -\frac{\alpha_0}{24} \Big[ 24N \langle \mathcal{O}_3 \rangle - 24\alpha_0 N^2 \langle \mathcal{O}_0 \mathcal{O}_2 \rangle_c - 12\alpha_0 N^2 \langle \mathcal{O}_1^2 \rangle_c \\
\nonumber
& + 12\alpha_0^2 N^3 \langle \mathcal{O}_0^2 \mathcal{O}_1 \rangle_c - \alpha_0^3 N^4 \langle \mathcal{O}_0^4 \rangle_c +36\alpha_0 N(2u_1+\alpha_0 N) \langle \mathcal{O}_0 \mathcal{O}_1 \rangle_c \\
\nonumber
& - 6\alpha_0^2 N^2(2u_1+\alpha_0 N) \langle \mathcal{O}_0^3 \rangle_c - 36(2u_1+\alpha_0 N) \langle \mathcal{O}_2 \rangle \\
& + 24(3u_1^2-6u_2+3\alpha_0 u_1+\alpha_0^2 N) \langle \mathcal{O}_1 \rangle \\
\nonumber
& - \alpha_0(12u_1^2+24N u_1^2-48N u_2+36\alpha_0 N u_1 + 11\alpha_0^2 N^2) \langle \mathcal{O}_0^2 \rangle_c \\
\nonumber
& -6(4u_1^3-12u_1 u_2+12u_3+6\alpha_0 u_1^2-12\alpha_0 u_2+4\alpha_0^2 u_1+N\alpha_0^3) \langle \mathcal{O}_0 \rangle \Big]
\end{align}
where we did not yet add the Casimir contribution (that is the reason for putting a prime on the left-hand side). In order to compare with the explicit expressions for $(\log \mathcal{H}_q)_j$, we need to evaluate the expectation values such as
\begin{align}
\langle \mathcal{O}_0 \rangle & = \sum_{m>0} \mathfrak{d}_{-m} = \frac{1-E_2}{24} \\
\langle \mathcal{O}_1 \rangle & = \frac{\alpha_0}{2} \sum_{m>0} (1-\mathfrak{d}_m+\mathfrak{d}_{-m}) \mathfrak{d}_{-m} = \frac{\alpha_0(1-E_2)}{48} - \frac{\alpha_0}{2} \sum_{k>0} \frac{k^2 q^k}{1-q^k} \\
\langle \mathcal{O}_0^2 \rangle_c & = \sum_{m>0} \mathfrak{d}_m \mathfrak{d}_{-m} = \frac{E_4-E_2^2}{288}
\end{align}
as well as higher ones whose expressions are longer so they are collected in appendix \ref{secilwexplw}. Using these, we finally find the prediction for eigenvalues of ILW Hamiltonian acting on the lowest weight state. The first two are
\begin{align}
(\log \mathcal{H}_q)_{1,hw} & = -\alpha_0 \left(-u_2 + \frac{u_1^2}{2}-\frac{N}{24} \right) + \frac{\alpha_0 N E_2}{24} \\
\nonumber
(\log \mathcal{H}_q)_{2,hw} & = \frac{\alpha_0}{2} \left( u_3 - u_1 u_2 + \frac{1}{3} u_1^3 - \frac{u_1}{12} \right) - \frac{\alpha_0 E_2}{24} u_1 \\
& + \alpha_0^2 N^2 \frac{E_4 - E_2^2}{576} + \frac{\alpha_0^2 N}{2} \sum_{m>0} \frac{m^2 q^m}{1-q^m}
\end{align}
and the expressions for $(\log \mathcal{H}_q)_{3,hw}$ and $(\log \mathcal{H}_q)_{4,hw}$ are given in appendix \ref{secilwexplw}. It is easy to see that both ways of calculating these expectation values lead to the same result.

\subsection{Spectrum of $(\log\mathcal{H}_q)_j$ and Bethe ansatz}
According to conjecture of Litvinov, the ILW Hamiltonians that we have constructed can be diagonalized by solving the Bethe ansatz equations \cite{Litvinov:2013zda}. More concretely, to describe the joint eigenspectrum at Virasoro level $M$, consider the following system of $M$ algebraic equations for $M$ unknown Bethe roots $x_j$, $j=1,\ldots,M$
\begin{equation}
\label{betheequations}
q \prod_{l=1}^N \frac{x_j+a_l-\epsilon_3}{x_j+a_l} \prod_{k \neq j} \frac{(x_j-x_k+\epsilon_1)(x_j-x_k+\epsilon_2)(x_j-x_k+\epsilon_3)}{(x_j-x_k-\epsilon_1)(x_j-x_k-\epsilon_2)(x_j-x_k-\epsilon_3)} = 1, \qquad j=1,\ldots,M.
\end{equation}
or more concisely
\begin{equation}
q A(x_j) \prod_{k \neq j} \varphi(x_j-x_k) = 1, \qquad j=1,\ldots,M
\end{equation}
where
\begin{equation}
\label{structurefunction}
\varphi(u) = \frac{(u+\epsilon_1)(u+\epsilon_2)(u+\epsilon_3)}{(u-\epsilon_1)(u-\epsilon_2)(u-\epsilon_3)}
\end{equation}
is the structure function of $\mathcal{W}_{1+\infty}$ and
\begin{equation}
A(u) = \prod_{l=1}^N \frac{u+a_l-\epsilon_3}{u+a_l}
\end{equation}
captures the information about the higher spin charges of the lowest weight vector of the representation (primary). \footnote{For concreteness $A(u)$ was written with $Y_{00N}$ truncation of $\mathcal{W}_{1+\infty}$ in mind in terms of the corresponding Coulomb parametres $a_j$ (these are the zero modes of free fields in the usual free field representation of $Y_{00N}$). For such representations and generic $a_j$ the representation spaces are $N$-tuples of Young diagrams stacked in the 3rd direction. We can replace $A(u)$ by a product of such functions to parametrize the lowest weight charges of $Y_{N_1 N_2 N_3}$ algebra (see \cite{Prochazka:2018tlo} for more details) or consider more general rational function with value $1$ at $u \to \infty$ to get more general quasi-finite representations of the algebra \cite{Prochazka:2015deb}. Representations of this kind have the property that there are only finite number of states at every Virasoro level.}

From the point of view of $\mathcal{W}_{1+\infty}$, these Bethe ansatz equations are very natural: the product over roots describes interaction between pairs of Bethe roots and is controlled by the unique structure constant of $\mathcal{W}_{1+\infty}$ \cite{Prochazka:2015deb} \footnote{The trigonometric and elliptic versions of these Bethe equations appear for example in \cite{Feigin:2015raa,Feigin:2016pld,Koroteev:2015dja,Koroteev:2016znb,Benini:2018ywd}.}. The first product encoding the information about higher spin charges of the lowest weight state of the representation corresponds from the point of view of Bethe ansatz equations to an interaction of Bethe roots with external fields. Each (physical) solution of BAE is associated to a eigenstate of ILW Hamiltonians. The solution is given by $M$ complex Bethe roots $x_j$, $j=1,\ldots,M$ and eigenvalues of all $(\log\mathcal{H}_q)_k$ are given by symmetric polynomials in $x_j$.

At $q=0$, we can explicitly solve these Bethe ansatz equations for any given $M$ and the Bethe roots can be combinatorially associated to boxes of Young diagrams as in \eqref{betherootq0}. Similar situation happens for $q \to \infty$ but apart from these special values of $q$, we can solve the equations only numerically. In terms of these Bethe roots, the eigenvalues of the first ILW Hamiltonians are given by
\begin{align}
\label{ilwspectra1}
(\log\mathcal{H}_q)_1 & \to -\alpha_0 p_0 + (\log\mathcal{H}_q)_{1,hw} \\
\label{ilwspectra2}
(\log\mathcal{H}_q)_2 & \to -\alpha_0 p_1 + \frac{\alpha_0^2}{2} p_0 + (\log\mathcal{H}_q)_{2,hw} \\
\label{ilwspectra3}
(\log\mathcal{H}_q)_3 & \to -\alpha_0 p_2 + \alpha_0^2 p_1 - \frac{\alpha_0(1+4\alpha_0^2)}{12} p_0 + \frac{\alpha_0 E_2}{12} p_0 + (\log\mathcal{H}_q)_{3,hw} \\
\label{ilwspectra4}
(\log\mathcal{H}_q)_4 & \to -\alpha_0 p_3 +\frac{3\alpha_0^2}{2} p_2 - \frac{\alpha_0(1+4\alpha_0^2)}{4} p_1 + \frac{\alpha_0^2(1+2\alpha_0^2)}{8} p_0 \\
\nonumber
& +\frac{\alpha_0 E_2}{4} p_1 - \frac{\alpha_0^2 E_2}{8} p_0 + \alpha_0^2 N \frac{E_4-E_2^2}{144} p_0 + 3\alpha_0^2 \sum_{m>0} \frac{m^2 q^m}{1-q^m} p_0 + (\log\mathcal{H}_q)_{4,hw} \\
\label{ilwspectra5}
\nonumber
(\log\mathcal{H}_q)_5 & \to -\alpha_0 p_4 + 2\alpha_0^2 p_3 - 2\alpha_0^3 p_2 + \alpha_0^4 p_1 - \frac{\alpha_0^5}{5} p_0 -12 \alpha_0 \langle \mathcal{O}_0 \rangle p_2 - 24 \alpha_0 \langle \mathcal{O}_1 \rangle p_1 \\
& + 24 \alpha_0^2 \langle \mathcal{O}_0 \rangle p_1 - 12\alpha_0 \langle \mathcal{O}_2 \rangle p_0 + 24\alpha_0^2 \langle \mathcal{O}_1 \rangle p_0 - 2\alpha_0 (1+7\alpha_0^2) \langle \mathcal{O}_0 \rangle p_0 \\
\nonumber
& + 6N\alpha_0^2 \langle \mathcal{O}_0^2 \rangle_c p_1 - 2\alpha_0^2 u_1 \langle \mathcal{O}_0^2 \rangle_c p_0 + 8N\alpha_0^2 \langle \mathcal{O}_0 \mathcal{O}_1 \rangle_c p_0 - 7N\alpha_0^3 \langle \mathcal{O}_0^2 \rangle_c p_0 \\
\nonumber
& - N^2\alpha_0^3 (\langle \mathcal{O}_0^3 \rangle - 3 \langle \mathcal{O}_0^2 \rangle \langle \mathcal{O}_0 \rangle +2 \langle \mathcal{O}_0 \rangle^3) \, p_0 + (\log\mathcal{H}_q)_{5,hw}
\end{align}
These expressions were obtained by explicit numerical diagonalization at lower levels as well as by using the $q \to 0$ limit where the spectrum is known. The formulas are also constrained by requiring compatibility with the spectral shift. Note that although all these expressions are at most linear in $p_j$, at order $u^{-6}$ terms such as $\langle\mathcal{O}_0^2\rangle_c \, p_0^2$ are expected to appear. The coefficients of the symmetric polynomials in Bethe roots have explicit dependence on the twist parameter $q$ via transcendental functions such as the Eisenstein series $E_j$. From the explicit form of ILW Hamiltonians we see that the coefficients of these transcendental functions multiply operators that are expressible in terms of lower degree ILW Hamiltonians. In other words, by taking a suitable triangular linear combination of ILW Hamiltonians we can eliminate these transcendental terms and we end up with ILW Hamiltonians which involve only rational functions of $q$ and whose spectrum is given by symmetric polynomials in Bethe roots without any additional explicit $q$-dependence (all $q$-dependence comes from $q$-dependence of Bethe roots themselves).

\subsection{Feigin-Jimbo-Miwa-Mukhin formula}
Based on the result of these calculations, we will now conjecture a formula for eigenvalues of the generating function $\mathcal{H}_q(u)$. The result is in the form of a rational limit of an analogous result proven by Feigin, Jimbo, Miwa and Mukhin in \cite{Feigin:2015raa,Feigin:2016pld} in the related $q$-deformed context of quantum toroidal $\mathfrak{gl}(1)$.

We look for the eigenvalues of the generating function $\mathcal{H}_q(u)$ in the form
\begin{equation}
\label{ilwhgenansatz}
\mathcal{H}_q(u) \to \mathcal{N} \sum_\lambda q^{|\lambda|} \prod_{i=1}^M g(u,x_i) \prod_{\Box\in\lambda} h(u,\epsilon_\Box) \prod_{i=1}^M \prod_{\Box\in\lambda} f(u,x_i,\epsilon_\Box).
\end{equation}
where $\mathcal{N}$ is an overall $q$- and $x$-independent normalization prefactor. We can now use the two Yangian limits that we studied previously to put restrictions on functions $g$ and $h$: first of all, if there are no Bethe roots (i.e. the Virasoro level $M=0$), we should reproduce the lowest weight charges that we calculated in section \ref{highestweightILWspectrum}. In particular, we require
\begin{equation}
\frac{\mathcal{E}_{Cas}(u)}{\sum_\lambda q^{|\lambda|}} \times \sum_\lambda q^{|\lambda|} \prod_{\Box\in\lambda} \prod_{j=1}^N \frac{u+a_j-\epsilon_\Box}{u+a_j-\epsilon_\Box+\epsilon_3} \stackrel{!}{=} \mathcal{N} \sum_{\lambda} q^{|\lambda|} \prod_{\Box\in\lambda} h(u,\epsilon_\Box)
\end{equation}
and we see that we need to identify
\begin{equation}
\mathcal{N} = \frac{\mathcal{E}_{Cas}(u)}{\sum_\lambda q^{|\lambda|}}
\end{equation}
as well as
\begin{equation}
h(u,\epsilon_\Box) = \prod_{i=1}^N \frac{u+a_i-\epsilon_\Box}{u+a_i-\epsilon_\Box+\epsilon_3}.
\end{equation}
The second restriction comes from the $q \to 0$ limit. In this limit, only the auxilliary vacuum state $\lambda = \emptyset$ contributes to the right-hand side of \eqref{ilwhgenansatz} while the left-hand side is given by \eqref{spectrumHbq0}. Bethe roots are in this case given as in \eqref{betherootq0} so we need
\begin{equation}
\prod_{i=1}^M g(u,x_i) \stackrel{!}{=} \prod_{i=1}^M \frac{u-x_i}{u-x_i+\epsilon_3}
\end{equation}
or
\begin{equation}
g(u,x_i) = \frac{u-x_i}{u-x_i+\epsilon_3}.
\end{equation}
Finally, it remains to identify the function $f(u,x_i,\epsilon_\Box)$. Let us use the ansatz
\begin{equation}
f(u,x_i,\epsilon_\Box) = \varphi(u-\alpha x_i-\beta \epsilon_\Box - \gamma)
\end{equation}
where $\varphi(u)$ is the structure constant of $\mathcal{W}_{1+\infty}$ as in \eqref{structurefunction}. Due to restriction $\epsilon_1+\epsilon_2+\epsilon_3 = 0$, the Taylor series of $\log \varphi(u)$ at $u \to \infty$ starts at order $u^{-4}$, so the constants $\alpha$, $\beta$ and $\gamma$ can be fixed by comparing the spectrum with that of $(\log\mathcal{H}_q)_4$. This determines these to be
\begin{equation}
\alpha = 1, \qquad \beta = 1, \qquad \gamma = -\alpha_0.
\end{equation}
Therefore, the spectrum of the generating function of ILW Hamiltonians $\mathcal{H}_q(u)$ is given by
\begin{equation}
\label{spectrumilwhamiltonians}
\mathcal{H}_q(u) \to \frac{\mathcal{E}_{Cas}(u)}{\sum_\lambda q^{|\lambda|}} \prod_{i=1}^M \frac{u-x_i}{u-x_i+\epsilon_3} \sum_\lambda q^{|\lambda|} \prod_{\Box\in\lambda} \prod_{i=1}^N \frac{u+a_i-\epsilon_\Box}{u+a_i-\epsilon_\Box+\epsilon_3}  \prod_{i=1}^M \prod_{\Box\in\lambda} \varphi(u-x_i-\epsilon_\Box+\epsilon_3).
\end{equation}
which is a rational analogue of a formula by Feigin, Jimbo, Miwa and Mukhin. To summarize, the elements of this formula are
\begin{itemize}
\item $\mathcal{E}_{Cas}(u)$ is the normalization factor of the $\mathcal{R}$-matrix
\item $\epsilon_j$ are the Nekrasov-like parameters of $\mathcal{W}_\infty$; these are determined by the rank $N$ of $\mathcal{W}_{N}$ and the central charge (i.e by the physical model such as Lee-Yang or Ising model)
\item the rational function $\varphi(u)$ is the structure function of $\mathcal{W}_{1+\infty}$ which governs the representation theory of the algebra; it can also be viewed as a scattering phase of the particles that are described by these Bethe ansatz equations
\item $q$ is the complex number parametrizing which family of ILW Hamiltonians we are diagonalizing, different values of $q$ correspond to different integrable structures that the algebra has; from $\mathcal{R}$-matrix point of view it encodes the shape of the auxiliary torus, from spin chain point of view it is the twist parameter; it controls the form of the non-locality and its inversion $q \to q^{-1}$ corresponds to charge conjugation
\item the sum over $\lambda$ represents the trace over the auxiliary space (with states parametrized by Young diagrams since we focused on $\bar{N}=1$)
\item $\epsilon_\Box$ are the Yangian Bethe roots associated to the auxiliary sector; the explicit expressions for these can be read off combinatorially from the associated Young diagram $\lambda$
\item $a_i$ are the parameters parametrizing the lowest weights in the quantum space (Coulomb parameters in the context of Nekrasov partition functions, external fields from the point of view of Bethe equations, eigenvalues of zero modes of scalar fields in free field representations of $\mathcal{W}_{1+\infty}$)
\item $x_i$ are the Bethe roots associated to the quantum space which are determined by solution of Bethe equations; these parameters (and $M$ which is the number of these Bethe roots) are the only parameters in this formula that depend on the particular eigenstate whose eigenvalues we are evaluating
\end{itemize}
Since we used all the Hamiltonians up to $(\log \mathcal{H}_q)_4$ to fix the parameters in this formula, we independently verified that also the spectrum of $(\log \mathcal{H}_q)_5$ agrees with the prediction coming from this formula.

In the Yangian limit a convenient generating function of higher spin charges of a state $\Lambda$ with associated Bethe roots $x_j$ is \cite{Tsymbaliuk:2014fvq,Prochazka:2015deb}
\begin{equation}
\psi_\Lambda(u) = A(u) \prod_j \varphi(u-x_j).
\end{equation}
If we use the same definition for $\psi_\Lambda(u)$ in terms of Bethe roots even for $q \neq 0$, we can interpret the FJMM generating function as a dressed or averaged version of $\psi_\Lambda(u)$,
\begin{align}
\frac{\mathcal{H}_q(u)}{\mathcal{H}_{q=0}(u)} & \to \frac{1}{\sum_\lambda q^{|\lambda|}} \sum_\lambda q^{|\lambda|} \prod_{\Box\in\lambda} \psi_\Lambda(u-\epsilon_\Box+\epsilon_3) \\
& = \Bigg\langle \prod_{\Box\in\lambda} \psi_\Lambda(u-\epsilon_\Box+\epsilon_3) \Bigg\rangle_q
\end{align}
where the average is physically the canonical Gibbs ensemble over the auxiliary space. Note that correct interpretation of the factor $\mathcal{H}_{q=0}(u)$ on the left-hand side is that we put $q \to 0$ in the averaging over states in the auxilliary space (i.e. restrict to the ground state contribution) but we do not put $q \to 0$ in the $q$-dependence of the Bethe roots $x_j(q)$. In other words,
\begin{equation}
\mathcal{H}_{q=0}(u) \to \mathcal{E}_{Cas}(u) \prod_{i=1}^M \frac{u-x_i}{u-x_i+\epsilon_3}
\end{equation}
with $x_i$ the $q$-dependent Bethe roots.

\section{Local limit $q \to 1$}
\label{seclocal}
It is interesting to understand what happens in the local limit ($q \to 1$) where the ILW Hamiltonians reduce to local BLZ/KP Hamiltonians. This limit is singular: in the construction of ILW Hamiltonians we used the parameter $q$ to regularize the trace over the infinite dimensional auxiliary space which would otherwise not be well-defined. The local limit corresponds to removal of this regulator. Therefore it is not surprising that the ILW Hamiltonians explicitly depend on $q$ in such a way that they become singular as $q \to 1$. But this is not the only singular behavior that we encounter - clearing the denominators in the Bethe equations \eqref{betheequations} to make them polynomial shows that in the limit $q \to 1$ the leading coefficients of the polynomials cancel, i.e. the degree of Bethe equations drops. As a consequence of that, some of the Bethe roots blow up in the $q \to 1$ limit.

As we will see, in the $q \to 1$ limit the $\mathcal{W}_\infty$ and Heisenberg factors of $\mathcal{W}_{1+\infty}$ algebra decouple and the divergences of Bethe roots as $q \to 1$ make this decoupling possible. It turns out that the Bethe roots that go to infinity as $q \to 1$ are precisely those that describe the excitations in the Heisenberg subsector of $\mathcal{W}_{1+\infty}$ while the roots that remain finite in the limit correspond to Bethe roots associated to $\mathcal{W}_\infty$ subalgebra. In particular, the divergence of Heisenberg Bethe roots guarantees that they decouple from Bethe equations for the remaining $\mathcal{W}_\infty$ Bethe roots which is an essential ingredient in the decoupling mechanism. From the point of view of ILW Hamiltonians, for generic $q$ we have one independent conserved quantity of every spin. On the other hand, if the Heisenberg and $\mathcal{W}_\infty$ subalgebras of $\mathcal{W}_{1+\infty}$ decouple in $q \to 1$ limit, one can expect two independent conserved quantities of every spin greater than or equal to $2$, one conserved quantity being associated to Heisenberg subalgebra and the other coming from $\mathcal{W}_\infty$. Again, it is the singular behaviour of ILW integrals of motion as $q \to 1$ that makes it possible to diagonalize all of these quantities, although naively there seem to be twice as many of them as for generic values of $q$.

\paragraph{$q \to 1$ limit of $(\log \mathcal{H}_q)_2$}

Numerical as well as analytic solution of Bethe ansatz equations at lower levels (for examples see the following section) shows the following pattern: first of all, given a solution of BAE, the individual Bethe roots are either regular in the $q \to 1$ limit or they diverge as
\begin{equation}
\label{bethesing}
x = \frac{C}{1-q} + o((1-q)^{-1})
\end{equation}
where $C$ is a universal constant, same for all Bethe roots (and independent of the specific solution of Bethe equations), but depending on the $\epsilon_j$ parameters as well as the function $A(u)$ parametrizing the lowest weights. Furthermore, the number of singular Bethe roots exactly corresponds to the number of excitations associated to the $\hat{\mathfrak{gl}}(1)$ subalgebra of $\mathcal{W}_{1+\infty}$ while the remaining Bethe roots that are finite in the limit correspond to $\mathcal{W}_\infty$ factor of the algebra. Therefore, all the Bethe roots split into two groups as $q \to 1$, those which remain finite $(\mathcal{W}_\infty$ roots) and those that go to infinity (Heisenberg roots).

This behaviour is consistent with the explicit form of ILW Hamiltonians: consider the first non-trivial ILW Hamiltonian
\begin{equation}
\mathcal{A}_2 \equiv -\frac{1}{2} \phi_{3,0} - \frac{\alpha_0}{2} \sum_{m>0} m \frac{1+q^m}{1-q^m} U_{1,-m} U_{1,m}
\end{equation}
which is obtained from $(\log \mathcal{H}_q)_2$ after subtracting the transcendental terms (which are in this case central). It has spectrum given by
\begin{equation}
\label{a2spectrum}
p_1 - \frac{\alpha_0}{2} p_0 - \frac{1}{2} u_3 + \frac{1}{2} u_1 u_2 - \frac{u_1^3}{6} + \frac{u_1}{24} \equiv p_1 - \frac{\alpha_0}{2} p_0 + hw.
\end{equation}
where $hw$ denotes a constant (central) contribution that can be evaluated by acting with $\mathcal{A}_2$ on the lowest weight vector. Since $\phi_{3,0}$ is a $q$-independent operator with finite matrix elements in representations spaces, the only singular part in the $q \to 1$ limit comes from the second term,
\begin{equation}
-\frac{\alpha_0}{2} \sum_{m>0} m \frac{1+q^m}{1-q^m} U_{1,-m} U_{1,m} \sim -\frac{\alpha_0}{1-q} \sum_{m>0} U_{1,-m} U_{1,m} + \mathcal{O}((1-q)^0).
\end{equation}
The coefficient of $(1-q)^{-1}$ term is up to normalization the (Sugawara) stress-energy tensor of the Heisenberg subalgebra. Inspecting the expression for eigenvalues of $\mathcal{A}_2$, we see that the only possible source of such singular behavior is from the term
\begin{equation}
p_1 = \sum_{j=1}^M x_j.
\end{equation}
Therefore, if $M_\infty$ out of $M$ Bethe roots are regular while remaining $M_1$ roots have leading singular behavior of the form
\begin{equation}
\frac{C}{1-q}
\end{equation}
with universal constant $C$, we are lead to identify
\begin{equation}
C \stackrel{!}{=} -\alpha_0 N = \psi_0 \epsilon_1 \epsilon_2 \epsilon_3.
\end{equation}
For easier comparison, we wrote the value of $C$ in terms of triality and scaling-covariant parameters \cite{Prochazka:2015deb}. Here $\psi_0$ is the central element of the Yangian algebra which can be extracted from the generating function of higher spin charges of the lowest weight state via \cite{Prochazka:2015deb}
\begin{equation}
\label{atopsi}
A(u) = 1 + \epsilon_1 \epsilon_2 \epsilon_3 \sum_{j=0}^\infty \frac{\psi_j}{u^{j+1}}.
\end{equation}
This numerical value of $C$ is confirmed by the results of numeric as well as analytic calculations at lower levels.

Since in the $q \to 1$ limit we are diagonalizing both $(\log\mathcal{H}_q)_1$ (which is the zero mode of the total $\mathcal{W}_{1+\infty}$ stress-energy tensor) as well as the zero mode of the stress-energy tensor of $\hat{\mathfrak{gl}}(1)$ (which as we saw is the dominant term in $(\log\mathcal{H}_q)_2$ in the $q \to 1$ limit), we see that among the commuting conserved quantities in the $q \to 1$ limit we have also the zero mode $T^{(\infty)}_0$ of $\mathcal{W}_{\infty}$. This suggests factorization of ILW conserved quantities in the local limit into Heisenberg conserved quantities and $\mathcal{W}_\infty$ conserved quantities. This is further confirmed by studying Bethe equations \eqref{betheequations}: since Bethe roots associated to $\mathcal{W}_\infty$ are regular in the local limit while those of $\hat{\mathfrak{gl}}(1)$ are singular, we see that the contribution of $\hat{\mathfrak{gl}}(1)$ roots decouples from the equations determining $\mathcal{W}_\infty$ roots. This is again confirmed by the numerics \cite{kudprochvelk}. In particular, to each $\mathcal{W}_\infty$ solution of BAE (which has all Bethe roots finite) we can associate an infinite set of BAE solutions at higher levels which have the same collection of finite Bethe roots in the $q \to 1$ limit. All of these states correspond to the same state in $\mathcal{W}_\infty$ subsector but different state in the $\hat{\mathfrak{gl}}(1)$ sector.

\paragraph{$q \to 1$ limit of $(\log \mathcal{H}_q)_3$}

In order to extract the next local Hamiltonian, we need to study the singular behavior of $(\log\mathcal{H}_q)_3$. Note that as we see from the lower level examples and numerics \cite{kudprochvelk}, in general the behavior of individual Bethe roots as $q \to 1$ is not simply a Laurent series in integer powers of $(1-q)$, i.e. there are non-trivial monodromies of Bethe roots around $q=1$ (as well as around other points in the $q$-plane) and we need to use the Puiseux series (Laurent series in fractional powers of $1-q$). But given any solution of Bethe equations, the examples indicate that analytic continuation around $q=1$ only permutes Bethe roots of that solution, i.e. the power sum polynomials of Bethe roots of a solution have no monodromy. This is unlike the general situation around other special values of parameter $q$ where even different solutions mix under the monodromy, so the analytic continuation around all special points in the $q$-plane connects different eigenstates of the ILW Hamiltonians, analogously to the situation in other integrable models \cite{Dorey:1996re}. Actually, as we will see in examples in the next section, the monodromy seems to connect all eigenstates at given level, i.e. there is a single multivalued function of $x(q)$ representing all the Bethe roots associated to physical solutions of Bethe equations at a given Virasoro level. The fact that monodromy around $q = 1$ does not mix different solutions is also compatible with $q \to 1$ expansion of ILW Hamiltonians which after subtraction of transcendental terms have Laurent expansion in integer powers of $(1-q)$ (since the matrix elements are rational functions of $q$ with poles only at roots of unity).

Subtracting the term in $(\log \mathcal{H}_q)_3$ proportional to $E_2 T_0$ as well as the central terms, we see that the operator
\begin{align}
\mathcal{A}_3 & \equiv -\frac{1}{3} \phi_{4,0} +\frac{\alpha_0}{2} \sum_{m>0} \frac{m(1+q^m)}{1-q^m} (U_{1,-m} T_m + T_{-m} U_{1,m}) \\
& -\frac{\alpha_0^2 (N+2)}{12} \sum_{m>0} m^2 U_{1,-m} U_{1,m} +\frac{\alpha_0^2 N}{4} \sum_{m>0} \frac{m^2 (1+q^m)^2}{(1-q^m)^2} U_{1,-m} U_{1,m}
\end{align}
has spectrum given by
\begin{equation}
p_2 - \alpha_0 p_1 + \frac{1+4\alpha_0^2}{12} p_0 + hw.
\end{equation}
The leading singular behavior of $\mathcal{A}_3$ is the quadratic pole $(1-q)^{-2}$ with coefficient proportional to the $\widehat{\mathfrak{gl}}(1)$ Sugawara Hamiltonian which we already encountered in $(\log\mathcal{H}_q)_2$. Therefore, in order to isolate an interesting higher Hamiltonian from $(\log\mathcal{H}_q)_3$, we need to cancel this leading order term by a similar term from $\mathcal{A}_2$. Therefore, we consider the limit
\begin{align}
\mathcal{B}_3 & \equiv \lim_{q \to 1} \Big[ (1-q) \mathcal{A}_3 + \alpha_0 N \mathcal{A}_2 \Big] \\
& = -\frac{\alpha_0 N}{2} \phi_{3,0} + \alpha_0 \sum_{m>0} (U_{1,-m} T_m + T_{-m} U_{1,m}) - \frac{\alpha_0^2 N}{2} \sum_{m>0} U_{1,-m} U_{1,m}
\end{align}
with spectrum given by
\begin{equation}
\label{spectruma3reg}
\lim_{q \to 1} \left[ (1-q) p_2 - (1-q) \alpha_0 p_1 + \alpha_0 N p_1 - \frac{\alpha_0^2 N}{2} p_0 \right] + hw.
\end{equation}
By construction, this operator commutes with both $T^{(1)}_0$ and $T^{(\infty)}_0$ so in particular should be a sum of a zero mode of spin $3$ operator in $\mathcal{W}_\infty$ and in $\hat{\mathfrak{gl}}(1)$. This is indeed the case: in terms of Heisenberg and $\mathcal{W}_\infty$ modes, this operator can be written as
\begin{equation}
\mathcal{B}_3 = -\frac{\alpha_0 N}{2} W_{3,0} - \alpha_0 U_{1,0} T_0 + \frac{\alpha_0}{3N} (U_1(U_1 U_1))_0 - \frac{\alpha_0^2 N}{4} (U_1 U_1)_0 + hw.
\end{equation}
We used the fact that
\begin{equation}
\sum_{m>0} (U_{1,-m} T_m + T_{-m} U_{1,m}) = (U_1 T)_0 - U_{1,0} T_0 + \frac{1}{12} U_{1,0}
\end{equation}
and introduced the unique (up to normalization) spin $3$ primary field in $\mathcal{W}_\infty$
\begin{align}
W_3 & = U_3 - \frac{N-2}{N} (U_1 U_2) + \frac{(N-1)(N-2)}{3N^2} (U_1 (U_1 U_1)) - \frac{(N-2)\alpha_0}{2} U_2^\prime \\
& + \frac{(N-1)(N-2)\alpha_0}{2N} (U_1^\prime U_1) + \frac{(N-1)(N-2)\alpha_0^2}{12} U_1^{\prime\prime}.
\end{align}
Let us Taylor expand the power sum polynomials of Bethe roots as
\begin{align}
p_1 & = \frac{p_{1,-1}}{1-q} + p_{1,0} + p_{1,1} (1-q) + \ldots \\
p_2 & = \frac{p_{2,-2}}{(1-q)^2} + \frac{p_{2,-1}}{1-q} + p_{2,0} + p_{2,1} (1-q) + \ldots
\end{align}
The monodromy properties of Bethe roots around $1-q$ as discussed above imply that the series expansion has only integer powers of $1-q$, i.e. is a Laurent expansion. Furthermore, the leading singular behaviour of power sums is determined by \eqref{bethesing}. Plugging this form of $p_j$ into \eqref{spectruma3reg}, we find the expansion
\begin{equation}
\frac{p_{2,-2} + \alpha_0 N p_{1,-1}}{1-q} + p_{2,-1} + \alpha_0 N p_{1,0} + \frac{\alpha_0^2 N}{2} M_1 - \frac{\alpha_0^2 N}{2} M_\infty + hw.
\end{equation}
where we used that $p_0 \equiv M = M_1+M_\infty$ and
\begin{equation}
p_{1,-1} = -\alpha_0 N M_1
\end{equation}
as follows from \eqref{bethesing}. We see that in order for this quantity to be non-singular as $q \to 1$, we need to have
\begin{equation}
p_{2,-2} = - \alpha_0 N p_{1,-1} = (-\alpha_0 N)^2 M_1.
\end{equation}
This condition together with analogous conditions coming from higher conserved quantities actually imply \eqref{bethesing} (with universal value of $C$), so we find an a posteriori representation theoretic jutification for the validity of \eqref{bethesing}. In order to disentangle the spin $3$ quantities associated to $\hat{\mathfrak{gl}}(1)$ and $\mathcal{W}_{\infty}$, we proceed to the next ILW Hamiltonian.

\paragraph{$q \to 1$ limit of $(\log \mathcal{H}_q)_4$}

By looking at the local expansion of $\mathcal{A}_4$ in appendix \ref{applocal}, we see that the following combination is finite as $q \to 1$:
\begin{align}
\label{locopb4}
\mathcal{B}_4 & \equiv \lim_{q \to 1} \left[ (1-q)^2 \mathcal{A}_4 + \alpha_0 N (1-q) \mathcal{A}_3 \right] \\
& = \frac{N^2 \alpha_0^3}{2} \sum_{m>0} U_{1,-m} U_{1,m} - \alpha_0^2 U_{1,0} \sum_{m>0} U_{1,-m} U_{1,m} \\
& - \frac{3}{2} \alpha_0^2 \sum_{m_1,m_2>0} (U_{1,-m_1} U_{1,-m_2} U_{1,m_1+m_2} + U_{1,-m_1-m_2} U_{1,m_1} U_{1,m_2}) \\
& = - \frac{\alpha_0^2}{2} (U_1 (U_1 U_1))_0 + \alpha_0^2 U_{1,0} (U_1 U_1)_0 + \frac{N^2 \alpha_0^3}{4} (U_1 U_1)_0 + hw.
\end{align}
Its spectrum is given in terms of Bethe roots by
\begin{equation}
\label{locopb4sp}
\mathcal{B}_4 \leadsto p_{3,-2} + N \alpha_0 p_{2,-1} + \frac{N \alpha_0^2}{2} p_{1,-1} + hw.
\end{equation}
and the finiteness of the limit in \eqref{locopb4} requires in addition that
\begin{equation}
p_{3,-3} = -N\alpha_0 p_{2,-2} = (-N\alpha_0)^3 M_1.
\end{equation}

\paragraph{$q \to 1$ limit of $(\log \mathcal{H}_q)_5$}
Finally, we can have a look at the local limit of $(\log \mathcal{H}_q)_5$. The quantity that is finite in the local limit is
\begin{equation}
\label{locopb5}
\mathcal{B}_5 \equiv (1-q)^2 \mathcal{A}_5 + \frac{7N\alpha_0}{3} (1-q) \mathcal{A}_4 + \frac{4N^2\alpha_0^2}{3} \mathcal{A}_3 + \frac{N^3\alpha_0^3}{6} \mathcal{A}_2 - \frac{N\alpha_0^2 u_1}{3} \mathcal{A}_2
\end{equation}
corresponding to operator
\begin{align}
\label{b5local}
\nonumber
\mathcal{B}_5 & = -\frac{4\alpha_0^2 N^2}{9} \left( W_{4,0} - \frac{3(N-3)(\alpha_0^2 N^2+3\alpha_0^2 N-9)}{2N(5\alpha_0^2 N^3-5\alpha_0^2 N-5N-17)} (T_{\infty} T_{\infty})_0 \right) \\
\nonumber
& + \frac{\alpha_0^2}{2N} \left( (U_1(U_1(U_1 U_1)))_0 - N (U_1^\prime U_1^\prime)_0 \right) \\
\nonumber
& + \frac{2\alpha_0^2 N}{3} T_0^2 - \alpha_0^2 T_0 (U_1 U_1)_0 \\
& + \frac{3\alpha_0^2 N}{2} U_{1,0} W_{3,0} - \frac{\alpha_0^3 N^3}{12} W_{3,0} \\
\nonumber
& - \frac{\alpha_0^2}{N} U_{1,0} (U_1 (U_1 U_1))_0 - \frac{19\alpha_0^3 N}{36} (U_1 (U_1 U_1))_0 \\
\nonumber
& + 2\alpha_0^2 U_{1,0}^2 T_0 - \frac{\alpha_0^3 N^2}{6} U_{1,0} T_0 - \frac{\alpha_0^2 N}{4} T_0 \\
\nonumber
& + \frac{4\alpha_0^3 N}{3} U_{1,0} (U_1 U_1)_0 + \frac{\alpha_0^2 (36+7\alpha_0^2 N^3)}{72} (U_1 U_1)_0 + hw.
\end{align}
It is written manifestly in terms of commuting conserved quantities of $\mathcal{W}_\infty$ and of the Heisenberg algebra. In fact, the operator
\begin{equation}
\label{winfspin4current}
\mathcal{S}_4 \equiv W_{4,0} - \frac{3(N-3)(\alpha_0^2 N^2+3\alpha_0^2 N-9)}{2N(5\alpha_0^2 N^3-5\alpha_0^2 N-5N-17)} (T_{\infty} T_{\infty})_0
\end{equation}
is the unique zero mode of spin $4$ local field in $\mathcal{W}_\infty$ commuting in $W_{3,0}$. Here $W_4$ is the unique spin $4$ primary field in $\mathcal{W}_\infty$ normalized such that
\begin{equation}
W_4 = U_4 + \ldots
\end{equation}
In $\mathcal{W}_\infty$ we have $4$ fields of dimension $4$, $W_4, (T_\infty T_\infty), \partial W_3$ and $\partial^2 T_\infty$. The latter two fields are total derivatives so have vanishing zero modes. The requirement of commutativity with $W_{3,0}$ is equivalent to the simple pole of OPE of $W_3$ with our candidate for spin $4$ conserved quantity being a total derivative. We have
\begin{align}
W_3(z) (T_\infty T_\infty)(w) & \sim \ldots + \frac{4(T_\infty \partial W_3)(w)}{z-w} + \ldots \\
W_3(z) W_4(w) & \sim \ldots + \frac{6(N-3)(\alpha_0^2 N^2+3\alpha_0^2 N-9)}{N(5\alpha_0^2N^3-5\alpha_0^2N-5N-17)} \frac{(T_\infty \partial W_3)(w)}{z-w} + \ldots
\end{align}
where we did not write explicitly the regular terms, poles of order higher than one and operators that are total derivatives (which are in this case fields $\partial W_5$, $\partial^3 W_3$ and $\partial (T_\infty W_3)$). We see that indeed the combination \eqref{winfspin4current} is uniquely determined by the commutativity of its zero mode with $W_{3,0}$.

A similar situation happens with spin $4$ quantity in the Heisenberg subalgebra. There are two dimension $4$ local fields modulo total derivatives, $(U_1(U_1(U_1 U_1)))$ and $(\partial U_1 \partial U_1)$. Their OPE with $(U_1(U_1 U_1))$ is
\begin{align}
\nonumber
(U_1(U_1 U_1))(z) (U_1(U_1(U_1 U_1)))(w) & \sim \ldots + \frac{24N}{5(z-w)} \partial (U_1(U_1(U_1(U_1 U_1))))(w) \\
& + \frac{6N^2}{z-w} (\partial^3 U_1 (U_1 U_1))(w) + \ldots \\
(U_1(U_1 U_1))(z) (\partial U_1 \partial U_1)(w) & \sim \ldots + \frac{12N}{z-w} (\partial U_1 (\partial U_1 \partial U_1))(w) \\
\nonumber
& + \frac{12N}{z-w} (\partial^2 U_1(\partial U_1 U_1))(w) + \frac{N^2}{5(z-w)} \partial^5 U_1(w) + \ldots
\end{align}
where we listed all the simple poles. The first and the last term are obviously total derivatives. Furthermore, the identity
\begin{equation}
(\partial U_1 (\partial U_1 \partial U_1)) + (\partial^2 U_1(\partial U_1 U_1)) = \frac{1}{2}(\partial^3 U_1 (U_1 U_1)) - \frac{1}{2}\partial (\partial^2 U_1(U_1 U_1)) + \partial(\partial U_1(\partial U_1 U_1))
\end{equation}
implies that the combination
\begin{equation}
(U_1(U_1(U_1 U_1)))_0 - N (\partial U_1 \partial U_1)_0
\end{equation}
commutes with $(U_1(U_1 U_1))_0$ and this is indeed the combination that appears in \eqref{b5local}. To summarize, by taking correctly the local limit $q \to 1$ we extracted from $(\log \mathcal{H}_q)_5$ and lower order ILW Hamiltonians the quantity $\mathcal{B}_5$ in \eqref{b5local}. It is a function of conserved quantities of spin up to $4$ both in $\mathcal{W}_\infty$ and in the Heisenberg subalgebra. Furthermore, its spectrum (as follows from \eqref{ilwspectra1}-\eqref{ilwspectra5}) is given by
\begin{align}
\mathcal{B}_5 & \leadsto p_{4,-2} + \frac{7\alpha_0 N}{3} p_{3,-1} + \frac{4\alpha_0^2 N^2}{3} p_{2,0} - 2\alpha_0 p_{3,-2} - \frac{7\alpha_0^2 N}{2} p_{2,-1} \\
\nonumber
& + \frac{\alpha_0^3 N^2(N-8)}{6} p_{1,0} - \frac{\alpha_0^2 N}{3} u_1 p_{1,0} + \frac{\alpha_0 N(1+4\alpha_0^2)}{12} p_{1,-1} \\
& + \frac{\alpha_0^2 N^2(4+16\alpha_0^2-3\alpha_0^2 N)}{36} p_{0,0} + \frac{\alpha_0^3 N}{6} u_1 p_{0,0}.
\end{align}
The finiteness of the limit in \eqref{locopb5} furthermore requires additional two conditions to be satisfied by $p_{j,k}$,
\begin{align}
p_{4,-4} & = (-\alpha_0 N)^3 p_{1,-1} = (-\alpha_0 N)^4 M_1 \\
p_{4,-3} & = -\frac{7\alpha_0 N}{3} p_{3,-2} - \frac{4\alpha_0^2 N^2}{3} p_{2,-1} - \frac{\alpha_0^2 N}{6}(N(N+1)\alpha_0-2u_1) p_{1,-1}.
\end{align}
In order to isolate the contribution of $\mathcal{W}_\infty$ conserved charge and Heisenberg conserved charge, we would need to consider the next quantity, $(\log\mathcal{H}_q)_6$, but such calculation would be technically very difficult.

\paragraph{$\mathcal{W}_\infty$ conserved charges and Bethe ansatz equations}
We can now focus on states that have no Heisenberg excitations, i.e. states in $\mathcal{W}_\infty$ subrepresentations. In this case, all Bethe roots have finite limit as $q \to 1$ so $p_{j,k} = 0$ for $k<0$. From the expression for eigenvalues of $\mathcal{B}_3$ we deduce that the operator $W_{3,0}$ has spectrum
\begin{equation}
\label{w3spectrum}
W_{3,0} \leadsto -2 \left( p_{1,0} + \frac{u_1}{N} M_\infty \right) + \alpha_0 M_\infty + hw.
\end{equation}
Let's emphasize again that this equation only holds if there are no Heisenberg excitations, i.e. if all Bethe roots are finite in the local limit. It might seem strange at first that the formula for the spectrum of $W_{3,0}$ depends on $u_1$ which is the zero mode of the Heisenberg current. But this is actually needed for consistency: Bethe equations are invariant under rigid translations of Bethe roots if we simultaneously translate the Coulomb parameters $a_l$ (this symmetry is just the spectral shift symmetry), i.e.
\begin{align}
x_j & \longrightarrow x_j - \gamma \\
a_l & \longrightarrow a_l + \gamma \\
u_1 & \longrightarrow u_1 + N\gamma
\end{align}
is the symmetry of the equations which simply shifts the zero mode of the Heisenberg field. The combination in parentheses in \eqref{w3spectrum} is invariant under these translations as it must be since the left-hand side does not involve the Heisenberg algebra. If we are only interested in $\mathcal{W}_\infty$, we can always choose the Coulomb parameters in such a way that $u_1 = 0$ and the formula simplifies further.

Proceeding similarly with spin $4$ conserved quantity, we find the eigenvalues
\begin{align}
\label{s4spectrum}
\nonumber
\mathcal{S}_4 & \leadsto -3 \left( p_{2,0} + \frac{2u_1}{N} p_{1,0} + \frac{u_1^2}{N^2} M_{\infty} \right) + 3\alpha_0 \left( p_{1,0} + \frac{u_1}{N} M_{\infty} \right) + \frac{3}{2N} (M_{\infty}^2+2\Delta M_{\infty}) \\
& + \frac{N^3\alpha_0^2-9N\alpha_0^2-3N-3}{8N} M_{\infty} + hw.
\end{align}

\paragraph{Heisenberg conserved charges and Bethe ansatz equations}
Let us now turn to the spectrum of Heisenberg conserved quantities. From \eqref{locopb4sp} we see that the spectrum of $(U_1(U_1 U_1))_0$ is given in terms of Bethe roots by
\begin{equation}
(U_1(U_1 U_1))_0 \leadsto - \frac{2}{\alpha_0^2} p_{3,-2} - \frac{2N}{\alpha_0} p_{2,-1} -\frac{4 u_1}{\alpha_0} p_{1,-1} - N(N+1) p_{1,-1} + hw.
\end{equation}
In terms of mode expansion, we have
\begin{align}
\nonumber
(U_1(U_1 U_1))_0 & = U_{1,0}^3 - \frac{N}{4} U_{1,0} + 6 U_{1,0} \sum_{m>0} U_{1,-m} U_{1,m} \\
& + 3 \sum_{m_1,m_2>0} (U_{1,-m_1} U_{1,-m_2} U_{1,m_1+m_2} + U_{1,-m_1-m_2} U_{1,m_1} U_{1,m_2} ).
\end{align}
The operator on the second line is the cut-and-join operator whose spectrum is known to be
\begin{equation}
6N^{\frac{3}{2}} \sum_{\Box\in\lambda} (x_{\Box}-y_{\Box})
\end{equation}
where $\lambda$ is a Young diagram parametrizing an excitation of Heisenberg algebra \footnote{This formula only applies to Heisenberg algebra and should not be confused with the parametrization of states in $\mathcal{W}_{1+\infty}$ in the Yangian limit $q \to 0$ or $q \to \infty$.}. The prefactor $N^{\frac{3}{2}}$ is due to the fact that our normalization of $U_1$ is such that the two-point function is proportional to $N$. In order for these two to match, the physical Bethe roots need to satisfy a non-trivial constraint,
\begin{equation}
6N^{\frac{3}{2}} \sum_{\Box\in\lambda} (x_{\Box}-y_{\Box}) = -\frac{2}{\alpha_0^2} p_{3,-2} - \frac{2N}{\alpha_0} p_{2,-1} + \frac{2u_1}{\alpha_0} p_{1,-1} - N(N+1) p_{1,-1}.
\end{equation}
Numerical evidence shows that there is in fact an additional relation
\begin{equation}
p_{3,-2} = -2\alpha_0 N p_{2,-1} - \alpha_0 u_1 p_{1,-1} + \frac{N(N+1) \alpha_0^2}{2} p_{1,-1}
\end{equation}
so assuming this is true in general, we find a constraint
\begin{equation}
6N^{\frac{3}{2}} \sum_{\Box\in\lambda} (x_{\Box}-y_{\Box}) = \frac{2N}{\alpha_0} \left( p_{2,-1} +\frac{2 u_1}{N} p_{1,-1} \right) - 2N(N+1) p_{1,-1}.
\end{equation}
Writing this purely in terms of Bethe ansatz equations in manifestly triality and scaling invariant way (in particular not restricting to $\epsilon_1 \epsilon_2 = -1$), we find a representation-theoretic restriction on Bethe roots
\begin{equation}
3 \sum_{\Box\in\lambda} (x_\Box-y_\Box) = -\frac{1}{\epsilon_1\epsilon_2\epsilon_3 \sqrt{\psi_0}} \left( p_{2,-1} - \frac{2\psi_1}{\psi_0} p_{1,-1}  + 2\psi_0 \epsilon_1 \epsilon_2 \epsilon_3 p_{1,-1} \right)
\end{equation}
The quantities $\psi_0$ and $\psi_1$ can be read off from the generating function of lowest weight charges $A(u)$ via \eqref{atopsi}. In terms of $\alpha_0, N$ and $u_1$ we have
\begin{align}
\psi_0 & = -\frac{N}{\epsilon_1 \epsilon_2} \\
\psi_1 & = \frac{u_1}{\epsilon_1 \epsilon_2} + \frac{N(N-1)}{2} \frac{\epsilon_3}{\epsilon_1 \epsilon_2}.
\end{align}
It is quite interesting how the local behaviour of the Bethe roots around $q = 1$ encodes discrete (quantized) quantities such as the eigenvalues of the cut-and-join operator. In particular, although the Bethe ansatz equations are non-linear algebraic equations, there still exist infinitely many constraints on the coefficients of Puiseux expansion around $q \to 1$ which encode the shape of Young diagram of the associated state of Heisenberg algebra.

\paragraph{Specialization to Virasoro algebra}
We can specialize the previous discussion to $N=2$ (Virasoro algebra) and compare to results of BLZ. The central charge is conveniently parametrized by
\begin{equation}
c = 1 - 6 \frac{(p^\prime-p)^2}{p^\prime p} = 1 - 6\alpha_0^2
\end{equation}
and
\begin{equation}
\epsilon_1 = \sqrt{\frac{p^\prime}{p}}, \qquad \epsilon_2 = -\sqrt{\frac{p}{p^\prime}}, \qquad \epsilon_3 = -\sqrt{\frac{p^\prime}{p}} + \sqrt{\frac{p}{p^\prime}} = \alpha_0.
\end{equation}
We choose Coulomb parameters $a_1$ and $a_2$ such that $u_1 = 0$ and express them in terms of conformal dimension $\Delta$ of the primary,
\begin{equation}
a_1 + a_2 = 0, \qquad \Delta = \left( a_1 + \frac{\epsilon_3}{2} \right) \left( a_1 - \frac{\epsilon_3}{2} \right).
\end{equation}
The function encoding the higher spin charges of the primary state is
\begin{equation}
\frac{(u + a_1 - \epsilon_3)(u + a_2 - \epsilon_3)}{(u + a_1)(u + a_2)} \longrightarrow \frac{(u-\epsilon_3)^2 - \Delta - \frac{\epsilon_3^3}{4}}{u^2 - \Delta - \frac{\epsilon_3^3}{4}}.
\end{equation}
It is convenient to shift the Bethe roots according to
\begin{equation}
x_j = y_j + \frac{\epsilon_3}{2}
\end{equation}
so the Bethe equations \eqref{betheequations} become
\begin{equation}
q \frac{\left(y_j - \frac{\epsilon_3}{2}\right)^2 - \Delta - \frac{\epsilon_3^2}{4}}{\left(y_j + \frac{\epsilon_3}{2}\right)^2 - \Delta - \frac{\epsilon_3^2}{4}} \prod_{k \neq j} \frac{(y_j-y_k+\epsilon_1)(y_j-y_k+\epsilon_2)(y_j-y_k+\epsilon_3)}{(y_j-y_k-\epsilon_1)(y_j-y_k-\epsilon_2)(y_j-y_k-\epsilon_3)} = 1, \qquad j=1,\ldots,M
\end{equation}
or
\begin{equation}
q \frac{\left(y_j - \frac{\epsilon_3}{2}\right)^2 - \Delta - \frac{\epsilon_3^2}{4}}{\left(y_j + \frac{\epsilon_3}{2}\right)^2 - \Delta - \frac{\epsilon_3^2}{4}} \prod_{k \neq j} \frac{\left[(y_j-y_k)^2-\epsilon_3(y_j-y_k)-1\right] (y_j-y_k+\epsilon_3)}{\left[(y_j-y_k)^2+\epsilon_3(y_j-y_k)-1\right](y_j-y_k-\epsilon_3)} = 1, \qquad j=1,\ldots,M.
\end{equation}
where in the second equation we eliminated $\epsilon_1$ and $\epsilon_2$ and expressed everything in terms of $\epsilon_3 = \alpha_0$. These equations have the obvious charge conjugation symmetry (exchanging the numerators and the denominators)
\begin{equation}
q \to q^{-1}, \qquad y_j \to -y_j.
\end{equation}
Specializing to quantum KdV charges at $q=1$, we find that the spin $3$ charge is
\begin{equation}
W_{3,0} \leadsto -2\sum_{j=1}^M y_j
\end{equation}
which vanishes for Virasoro physical states while for spin $4$ charge we find
\begin{equation}
\label{s4virasoro}
\mathcal{S}_4 \leadsto -3\sum_{j=1}^M y_j^2 + \frac{3}{4}(M^2+2M\Delta) + \frac{\alpha_0^2}{8} M - \frac{9}{16}M + hw.
\end{equation}
This should be compared to formula for spectrum of $I_3$ in \eqref{blzintegralsev2} where for $N=2$ we have
\begin{equation}
\mathcal{S}_4 = \frac{1}{4} I_3.
\end{equation}
The charge conjugation symmetry flips sign of odd spin generators and keeps the even spin generators invariant. In the case of Virasoro algebra there are no odd spin generators so Virasoro representations are self-conjugate. The solutions of Bethe equations in this case have to be self-conjugate, i.e. given any solution, the Bethe roots can be either zero or come in pairs of roots differing by a sign. This is consistent with the form of eigenvalues of $W_{3,0}$ and $\mathcal{S}_4$. As an illustration, at level $1$, the unique solution of Bethe equations is
\begin{equation}
y_1 = 0
\end{equation}
with corresponding $\mathcal{S}_4$ eigenvalue
\begin{equation}
\frac{3}{2} \Delta + \frac{\alpha_0^2}{8} + \frac{3}{16} + hw.
\end{equation}
which agrees with formula \eqref{s4virasoro}. At level $2$ there are two states, $L_{-2} \ket{\Delta}$ and $L_{-1}^2 \ket{\Delta}$ and from explicit diagonalization we find that $\mathcal{S}_4$ has eigenvalues
\begin{equation}
\label{virasoros4level2}
3\Delta - \frac{\alpha_0^2}{2} + \frac{3}{2} \pm \frac{3}{8} \sqrt{32\Delta+4\alpha_0^4+4\alpha_0^2+1} + hw.
\end{equation}
The solutions of Bethe ansatz equations are given by pairs of roots $\{y,-y\}$ with $y$ satisfying
\begin{equation}
8y^4 - (1+2\alpha_0^2) y^2 - \Delta = 0.
\end{equation}
Solving this biquadratic equation and plugging in \eqref{s4virasoro}, we reproduce exactly \eqref{virasoros4level2}.

\subsection{Roots of unity and twisted subalgebras}
From the explicit expressions for ILW Hamiltonians we see that $q = 1$ is not the only singular point in the $q$-plane. In fact, from the term \eqref{nonlocalterm} we see that all the roots of unity are singular at sufficiently high level. Therefore, in the limit of highly excited states, the interior $|q|<1$ of the unit circle and its exterior $|q|>1$ are separated by a dense set of singular points.

Let us choose $q_0$ to be a primitive $K$-th root of unity (i.e. for $q_0=1$ we have $K=1$, for $q_0=-1$ we have $K=2$, for $q_0=\pm i$ we have $K=4$ etc.). The ILW Hamiltonian $\mathcal{A}_2$ has simple pole at $q = q_0$ with residue
\begin{equation}
\label{b2twisted}
\lim_{q \to q_0} (q-q_0) \mathcal{A}_2(q) = \alpha_0 q_0 \sum_{m>0} U_{1,-Km} U_{1,Km}.
\end{equation}
Some of the Bethe roots again diverge in the $q \to q_0$ limit with leading singular behaviour
\begin{equation}
x \sim \frac{\tilde{C}}{q-q_0} + o((q-q_0)^{-1})
\end{equation}
where the coefficient $\tilde{C}$ is a common value for all of the singular Bethe roots analogously to the situation at $q \to 1$. Let us assume analogously to $q \to 1$ case that there is no monodromy around $q \sim q_0$ after we take power sums of Bethe roots. The spectrum of \eqref{b2twisted} is given by
\begin{equation}
\alpha_0 q_0 NK |\lambda|
\end{equation}
where $\lambda$ is the Young diagram labeling a state in the Hilbert space associated to $K$-twisted Heisenberg algebra (i.e. Heisenberg algebra generated by Fourier modes $U_{1,Km}, \; m \in \mathbbm{Z}$). In our conventions we label the state $U_{1,-K} \ket{0}$ by $\ydiagram{1}$ with single box. Comparing this spectrum with the spectrum of $\mathcal{A}_2$ \eqref{a2spectrum} which is
\begin{equation}
\tilde{C} \tilde{M}_1,
\end{equation}
where $\tilde{M}_1$ is the number of singular Bethe roots, we see that the constant $\tilde{C}$ has to be equal to
\begin{equation}
\tilde{C} = \alpha_0 q_0 N \frac{K|\lambda|}{\tilde{M}_1}.
\end{equation}
The numerical data at lower levels indicate that every box of $\lambda$ corresponds to $K$ singular roots, i.e. the fraction in equation above is equal to unity and that the constant $\tilde{C}$ is
\begin{equation}
\tilde{C} = q_0 \alpha_0 N
\end{equation}
which is up to a phase the same result as we found in the local limit $q \to 1$. This value is confirmed by the numerical experiments. It would be very interesting to see which other Hamiltonians can be extracted from higher $\mathcal{A}_j$ and whether twisting by $K$ happens only in the Heisenberg subsector or affects also the rest of the algebra.

\section{Examples of solutions}
Let us discuss some concrete examples of solutions of Bethe ansatz equations \eqref{betheequations}.

\subsection{Lee-Yang model}

\subsubsection{Argyres-Douglas minimal models}

We will start with the Lee-Yang model with $c=-\frac{22}{5}$. This model is special in several ways. It is at the same time a minimal model of Virasoro algebra as well as of $\mathcal{W}_3$ algebra. It is also the simplest non-trivial minimal model, having the smallest number of states in the vacuum representation. Unlike the more famous Ising model it is non-unitary. Via the correspondence between 4d $\mathcal{N}=2$ superconformal field theories and 2d vertex operator algebras \cite{Beem:2013sza} it corresponds to the first model in the Argyres-Douglas series, labeled by a pair of Dynkin diagrams $(A_1,A_2)$ \cite{Argyres:1995jj,Cecotti:2010fi} and is associated to a plane algebraic curve $p^2 + x^3 = 0$.

\begin{figure}
\centering
\begin{minipage}{0.45\textwidth}
\centering
\begin{tikzpicture}[scale=0.4]
\definecolor{mbr}{RGB}{128,71,0}
\definecolor{mgr}{RGB}{148,166,0}
\definecolor{mor}{RGB}{242,198,0}
\filldraw[fill=mbr,draw=black] (330:0) ++(90:0) -- ++(330:1) -- ++(90:1) -- ++(150:1) -- ++(270:1);
\filldraw[fill=mbr,draw=black] (330:1) ++(90:0) -- ++(330:1) -- ++(90:1) -- ++(150:1) -- ++(270:1);
\filldraw[fill=mbr,draw=black] (330:2) ++(90:0) -- ++(330:1) -- ++(90:1) -- ++(150:1) -- ++(270:1);
\filldraw[fill=mbr,draw=black] (330:3) ++(90:0) -- ++(330:1) -- ++(90:1) -- ++(150:1) -- ++(270:1);
\filldraw[fill=mbr,draw=black] (330:4) ++(90:0) -- ++(330:1) -- ++(90:1) -- ++(150:1) -- ++(270:1);
\filldraw[fill=mbr,draw=black] (330:5) ++(90:0) -- ++(330:1) -- ++(90:1) -- ++(150:1) -- ++(270:1);
\filldraw[fill=mbr,draw=black] (330:6) ++(90:0) -- ++(330:1) -- ++(90:1) -- ++(150:1) -- ++(270:1);
\filldraw[fill=mbr,draw=black] (330:7) ++(90:0) -- ++(330:1) -- ++(90:1) -- ++(150:1) -- ++(270:1);
\filldraw[fill=mbr,draw=black] (330:8) ++(90:0) -- ++(330:1) -- ++(90:1) -- ++(150:1) -- ++(270:1);
\filldraw[fill=mbr,draw=black] (330:0) ++(90:1) -- ++(330:1) -- ++(90:1) -- ++(150:1) -- ++(270:1);
\filldraw[fill=mbr,draw=black] (330:1) ++(90:1) -- ++(330:1) -- ++(90:1) -- ++(150:1) -- ++(270:1);
\filldraw[fill=mbr,draw=black] (330:2) ++(90:1) -- ++(330:1) -- ++(90:1) -- ++(150:1) -- ++(270:1);
\filldraw[fill=mbr,draw=black] (330:3) ++(90:1) -- ++(330:1) -- ++(90:1) -- ++(150:1) -- ++(270:1);
\filldraw[fill=mbr,draw=black] (330:4) ++(90:1) -- ++(330:1) -- ++(90:1) -- ++(150:1) -- ++(270:1);
\filldraw[fill=mbr,draw=black] (330:5) ++(90:1) -- ++(330:1) -- ++(90:1) -- ++(150:1) -- ++(270:1);
\filldraw[fill=mbr,draw=black] (330:6) ++(90:1) -- ++(330:1) -- ++(90:1) -- ++(150:1) -- ++(270:1);
\filldraw[fill=mbr,draw=black] (330:7) ++(90:1) -- ++(330:1) -- ++(90:1) -- ++(150:1) -- ++(270:1);
\filldraw[fill=mbr,draw=black] (330:8) ++(90:1) -- ++(330:1) -- ++(90:1) -- ++(150:1) -- ++(270:1);
\filldraw[fill=mor,draw=black] (330:0) ++(210:1) ++(30:1) -- ++(210:1) -- ++(330:1) -- ++(30:1) -- ++(150:1);
\filldraw[fill=mor,draw=black] (330:0) ++(210:2) ++(30:1) -- ++(210:1) -- ++(330:1) -- ++(30:1) -- ++(150:1);
\filldraw[fill=mor,draw=black] (330:0) ++(210:3) ++(30:1) -- ++(210:1) -- ++(330:1) -- ++(30:1) -- ++(150:1);
\filldraw[fill=mor,draw=black] (330:0) ++(210:4) ++(30:1) -- ++(210:1) -- ++(330:1) -- ++(30:1) -- ++(150:1);
\filldraw[fill=mor,draw=black] (330:0) ++(210:5) ++(30:1) -- ++(210:1) -- ++(330:1) -- ++(30:1) -- ++(150:1);
\filldraw[fill=mor,draw=black] (330:0) ++(210:6) ++(30:1) -- ++(210:1) -- ++(330:1) -- ++(30:1) -- ++(150:1);
\filldraw[fill=mor,draw=black] (330:0) ++(210:7) ++(30:1) -- ++(210:1) -- ++(330:1) -- ++(30:1) -- ++(150:1);
\filldraw[fill=mor,draw=black] (330:0) ++(210:8) ++(30:1) -- ++(210:1) -- ++(330:1) -- ++(30:1) -- ++(150:1);
\filldraw[fill=mor,draw=black] (330:0) ++(210:9) ++(30:1) -- ++(210:1) -- ++(330:1) -- ++(30:1) -- ++(150:1);
\filldraw[fill=mor,draw=black] (330:1) ++(210:1) ++(30:1) -- ++(210:1) -- ++(330:1) -- ++(30:1) -- ++(150:1);
\filldraw[fill=mor,draw=black] (330:1) ++(210:2) ++(30:1) -- ++(210:1) -- ++(330:1) -- ++(30:1) -- ++(150:1);
\filldraw[fill=mor,draw=black] (330:1) ++(210:3) ++(30:1) -- ++(210:1) -- ++(330:1) -- ++(30:1) -- ++(150:1);
\filldraw[fill=mor,draw=black] (330:1) ++(210:4) ++(30:1) -- ++(210:1) -- ++(330:1) -- ++(30:1) -- ++(150:1);
\filldraw[fill=mor,draw=black] (330:1) ++(210:5) ++(30:1) -- ++(210:1) -- ++(330:1) -- ++(30:1) -- ++(150:1);
\filldraw[fill=mor,draw=black] (330:1) ++(210:6) ++(30:1) -- ++(210:1) -- ++(330:1) -- ++(30:1) -- ++(150:1);
\filldraw[fill=mor,draw=black] (330:1) ++(210:7) ++(30:1) -- ++(210:1) -- ++(330:1) -- ++(30:1) -- ++(150:1);
\filldraw[fill=mor,draw=black] (330:1) ++(210:8) ++(30:1) -- ++(210:1) -- ++(330:1) -- ++(30:1) -- ++(150:1);
\filldraw[fill=mor,draw=black] (330:1) ++(210:9) ++(30:1) -- ++(210:1) -- ++(330:1) -- ++(30:1) -- ++(150:1);
\filldraw[fill=mor,draw=black] (330:2) ++(210:1) ++(30:1) -- ++(210:1) -- ++(330:1) -- ++(30:1) -- ++(150:1);
\filldraw[fill=mor,draw=black] (330:2) ++(210:2) ++(30:1) -- ++(210:1) -- ++(330:1) -- ++(30:1) -- ++(150:1);
\filldraw[fill=mor,draw=black] (330:2) ++(210:3) ++(30:1) -- ++(210:1) -- ++(330:1) -- ++(30:1) -- ++(150:1);
\filldraw[fill=mor,draw=black] (330:2) ++(210:4) ++(30:1) -- ++(210:1) -- ++(330:1) -- ++(30:1) -- ++(150:1);
\filldraw[fill=mor,draw=black] (330:2) ++(210:5) ++(30:1) -- ++(210:1) -- ++(330:1) -- ++(30:1) -- ++(150:1);
\filldraw[fill=mor,draw=black] (330:2) ++(210:6) ++(30:1) -- ++(210:1) -- ++(330:1) -- ++(30:1) -- ++(150:1);
\filldraw[fill=mor,draw=black] (330:2) ++(210:7) ++(30:1) -- ++(210:1) -- ++(330:1) -- ++(30:1) -- ++(150:1);
\filldraw[fill=mor,draw=black] (330:2) ++(210:8) ++(30:1) -- ++(210:1) -- ++(330:1) -- ++(30:1) -- ++(150:1);
\filldraw[fill=mor,draw=black] (330:2) ++(210:9) ++(30:1) -- ++(210:1) -- ++(330:1) -- ++(30:1) -- ++(150:1);
\filldraw[fill=mgr,draw=black] (210:1) ++(330:4) -- ++(90:1) -- ++(210:1) -- ++(270:1) -- ++(30:1);
\filldraw[fill=mgr,draw=black] (210:2) ++(330:4) -- ++(90:1) -- ++(210:1) -- ++(270:1) -- ++(30:1);
\filldraw[fill=mgr,draw=black] (210:3) ++(330:4) -- ++(90:1) -- ++(210:1) -- ++(270:1) -- ++(30:1);
\filldraw[fill=mgr,draw=black] (210:4) ++(330:4) -- ++(90:1) -- ++(210:1) -- ++(270:1) -- ++(30:1);
\filldraw[fill=mgr,draw=black] (210:5) ++(330:4) -- ++(90:1) -- ++(210:1) -- ++(270:1) -- ++(30:1);
\filldraw[fill=mgr,draw=black] (210:6) ++(330:4) -- ++(90:1) -- ++(210:1) -- ++(270:1) -- ++(30:1);
\filldraw[fill=mgr,draw=black] (210:7) ++(330:4) -- ++(90:1) -- ++(210:1) -- ++(270:1) -- ++(30:1);
\filldraw[fill=mgr,draw=black] (210:8) ++(330:4) -- ++(90:1) -- ++(210:1) -- ++(270:1) -- ++(30:1);
\filldraw[fill=mgr,draw=black] (210:9) ++(330:4) -- ++(90:1) -- ++(210:1) -- ++(270:1) -- ++(30:1);
\filldraw[fill=mgr,draw=black] (210:2) ++(330:5) -- ++(90:1) -- ++(210:1) -- ++(270:1) -- ++(30:1);
\filldraw[fill=mgr,draw=black] (210:3) ++(330:5) -- ++(90:1) -- ++(210:1) -- ++(270:1) -- ++(30:1);
\filldraw[fill=mgr,draw=black] (210:4) ++(330:5) -- ++(90:1) -- ++(210:1) -- ++(270:1) -- ++(30:1);
\filldraw[fill=mgr,draw=black] (210:5) ++(330:5) -- ++(90:1) -- ++(210:1) -- ++(270:1) -- ++(30:1);
\filldraw[fill=mgr,draw=black] (210:6) ++(330:5) -- ++(90:1) -- ++(210:1) -- ++(270:1) -- ++(30:1);
\filldraw[fill=mgr,draw=black] (210:7) ++(330:5) -- ++(90:1) -- ++(210:1) -- ++(270:1) -- ++(30:1);
\filldraw[fill=mgr,draw=black] (210:8) ++(330:5) -- ++(90:1) -- ++(210:1) -- ++(270:1) -- ++(30:1);
\filldraw[fill=mgr,draw=black] (210:9) ++(330:5) -- ++(90:1) -- ++(210:1) -- ++(270:1) -- ++(30:1);
\filldraw[fill=mgr,draw=black] (210:10) ++(330:5) -- ++(90:1) -- ++(210:1) -- ++(270:1) -- ++(30:1);
\filldraw[fill=mbr,draw=black] (330:3) ++(270:1) -- ++(330:1) -- ++(90:1) -- ++(150:1) -- ++(270:1);
\filldraw[fill=mbr,draw=black] (330:4) ++(270:1) -- ++(330:1) -- ++(90:1) -- ++(150:1) -- ++(270:1);
\filldraw[fill=mbr,draw=black] (330:5) ++(270:1) -- ++(330:1) -- ++(90:1) -- ++(150:1) -- ++(270:1);
\filldraw[fill=mbr,draw=black] (330:6) ++(270:1) -- ++(330:1) -- ++(90:1) -- ++(150:1) -- ++(270:1);
\filldraw[fill=mbr,draw=black] (330:7) ++(270:1) -- ++(330:1) -- ++(90:1) -- ++(150:1) -- ++(270:1);
\filldraw[fill=mbr,draw=black] (330:8) ++(270:1) -- ++(330:1) -- ++(90:1) -- ++(150:1) -- ++(270:1);
\filldraw[fill=mbr,draw=black] (330:3) ++(270:2) -- ++(330:1) -- ++(90:1) -- ++(150:1) -- ++(270:1);
\filldraw[fill=mbr,draw=black] (330:4) ++(270:2) -- ++(330:1) -- ++(90:1) -- ++(150:1) -- ++(270:1);
\filldraw[fill=mbr,draw=black] (330:5) ++(270:2) -- ++(330:1) -- ++(90:1) -- ++(150:1) -- ++(270:1);
\filldraw[fill=mbr,draw=black] (330:6) ++(270:2) -- ++(330:1) -- ++(90:1) -- ++(150:1) -- ++(270:1);
\filldraw[fill=mbr,draw=black] (330:7) ++(270:2) -- ++(330:1) -- ++(90:1) -- ++(150:1) -- ++(270:1);
\filldraw[fill=mbr,draw=black] (330:8) ++(270:2) -- ++(330:1) -- ++(90:1) -- ++(150:1) -- ++(270:1);
\filldraw[fill=mor,draw=black] (330:3) ++(270:2) ++(210:1) ++(30:1) -- ++(210:1) -- ++(330:1) -- ++(30:1) -- ++(150:1);
\filldraw[fill=mor,draw=black] (330:3) ++(270:2) ++(210:2) ++(30:1) -- ++(210:1) -- ++(330:1) -- ++(30:1) -- ++(150:1);
\filldraw[fill=mor,draw=black] (330:3) ++(270:2) ++(210:3) ++(30:1) -- ++(210:1) -- ++(330:1) -- ++(30:1) -- ++(150:1);
\filldraw[fill=mor,draw=black] (330:3) ++(270:2) ++(210:4) ++(30:1) -- ++(210:1) -- ++(330:1) -- ++(30:1) -- ++(150:1);
\filldraw[fill=mor,draw=black] (330:3) ++(270:2) ++(210:5) ++(30:1) -- ++(210:1) -- ++(330:1) -- ++(30:1) -- ++(150:1);
\filldraw[fill=mor,draw=black] (330:3) ++(270:2) ++(210:6) ++(30:1) -- ++(210:1) -- ++(330:1) -- ++(30:1) -- ++(150:1);
\filldraw[fill=mor,draw=black] (330:3) ++(270:2) ++(210:7) ++(30:1) -- ++(210:1) -- ++(330:1) -- ++(30:1) -- ++(150:1);
\filldraw[fill=mor,draw=black] (330:3) ++(270:2) ++(210:8) ++(30:1) -- ++(210:1) -- ++(330:1) -- ++(30:1) -- ++(150:1);
\filldraw[fill=mor,draw=black] (330:3) ++(270:2) ++(210:9) ++(30:1) -- ++(210:1) -- ++(330:1) -- ++(30:1) -- ++(150:1);
\filldraw[fill=mor,draw=black] (330:4) ++(270:2) ++(210:1) ++(30:1) -- ++(210:1) -- ++(330:1) -- ++(30:1) -- ++(150:1);
\filldraw[fill=mor,draw=black] (330:4) ++(270:2) ++(210:2) ++(30:1) -- ++(210:1) -- ++(330:1) -- ++(30:1) -- ++(150:1);
\filldraw[fill=mor,draw=black] (330:4) ++(270:2) ++(210:3) ++(30:1) -- ++(210:1) -- ++(330:1) -- ++(30:1) -- ++(150:1);
\filldraw[fill=mor,draw=black] (330:4) ++(270:2) ++(210:4) ++(30:1) -- ++(210:1) -- ++(330:1) -- ++(30:1) -- ++(150:1);
\filldraw[fill=mor,draw=black] (330:4) ++(270:2) ++(210:5) ++(30:1) -- ++(210:1) -- ++(330:1) -- ++(30:1) -- ++(150:1);
\filldraw[fill=mor,draw=black] (330:4) ++(270:2) ++(210:6) ++(30:1) -- ++(210:1) -- ++(330:1) -- ++(30:1) -- ++(150:1);
\filldraw[fill=mor,draw=black] (330:4) ++(270:2) ++(210:7) ++(30:1) -- ++(210:1) -- ++(330:1) -- ++(30:1) -- ++(150:1);
\filldraw[fill=mor,draw=black] (330:4) ++(270:2) ++(210:8) ++(30:1) -- ++(210:1) -- ++(330:1) -- ++(30:1) -- ++(150:1);
\filldraw[fill=mor,draw=black] (330:4) ++(270:2) ++(210:9) ++(30:1) -- ++(210:1) -- ++(330:1) -- ++(30:1) -- ++(150:1);
\filldraw[fill=mor,draw=black] (330:5) ++(270:2) ++(210:1) ++(30:1) -- ++(210:1) -- ++(330:1) -- ++(30:1) -- ++(150:1);
\filldraw[fill=mor,draw=black] (330:5) ++(270:2) ++(210:2) ++(30:1) -- ++(210:1) -- ++(330:1) -- ++(30:1) -- ++(150:1);
\filldraw[fill=mor,draw=black] (330:5) ++(270:2) ++(210:3) ++(30:1) -- ++(210:1) -- ++(330:1) -- ++(30:1) -- ++(150:1);
\filldraw[fill=mor,draw=black] (330:5) ++(270:2) ++(210:4) ++(30:1) -- ++(210:1) -- ++(330:1) -- ++(30:1) -- ++(150:1);
\filldraw[fill=mor,draw=black] (330:5) ++(270:2) ++(210:5) ++(30:1) -- ++(210:1) -- ++(330:1) -- ++(30:1) -- ++(150:1);
\filldraw[fill=mor,draw=black] (330:5) ++(270:2) ++(210:6) ++(30:1) -- ++(210:1) -- ++(330:1) -- ++(30:1) -- ++(150:1);
\filldraw[fill=mor,draw=black] (330:5) ++(270:2) ++(210:7) ++(30:1) -- ++(210:1) -- ++(330:1) -- ++(30:1) -- ++(150:1);
\filldraw[fill=mor,draw=black] (330:5) ++(270:2) ++(210:8) ++(30:1) -- ++(210:1) -- ++(330:1) -- ++(30:1) -- ++(150:1);
\filldraw[fill=mor,draw=black] (330:5) ++(270:2) ++(210:9) ++(30:1) -- ++(210:1) -- ++(330:1) -- ++(30:1) -- ++(150:1);
\filldraw[fill=mgr,draw=black] (210:1) ++(270:2) ++(330:7) -- ++(90:1) -- ++(210:1) -- ++(270:1) -- ++(30:1);
\filldraw[fill=mgr,draw=black] (210:2) ++(270:2) ++(330:7) -- ++(90:1) -- ++(210:1) -- ++(270:1) -- ++(30:1);
\filldraw[fill=mgr,draw=black] (210:3) ++(270:2) ++(330:7) -- ++(90:1) -- ++(210:1) -- ++(270:1) -- ++(30:1);
\filldraw[fill=mgr,draw=black] (210:4) ++(270:2) ++(330:7) -- ++(90:1) -- ++(210:1) -- ++(270:1) -- ++(30:1);
\filldraw[fill=mgr,draw=black] (210:5) ++(270:2) ++(330:7) -- ++(90:1) -- ++(210:1) -- ++(270:1) -- ++(30:1);
\filldraw[fill=mgr,draw=black] (210:6) ++(270:2) ++(330:7) -- ++(90:1) -- ++(210:1) -- ++(270:1) -- ++(30:1);
\filldraw[fill=mgr,draw=black] (210:7) ++(270:2) ++(330:7) -- ++(90:1) -- ++(210:1) -- ++(270:1) -- ++(30:1);
\filldraw[fill=mgr,draw=black] (210:8) ++(270:2) ++(330:7) -- ++(90:1) -- ++(210:1) -- ++(270:1) -- ++(30:1);
\filldraw[fill=mgr,draw=black] (210:9) ++(270:2) ++(330:7) -- ++(90:1) -- ++(210:1) -- ++(270:1) -- ++(30:1);
\filldraw[fill=mgr,draw=black] (210:2) ++(270:2) ++(330:8) -- ++(90:1) -- ++(210:1) -- ++(270:1) -- ++(30:1);
\filldraw[fill=mgr,draw=black] (210:3) ++(270:2) ++(330:8) -- ++(90:1) -- ++(210:1) -- ++(270:1) -- ++(30:1);
\filldraw[fill=mgr,draw=black] (210:4) ++(270:2) ++(330:8) -- ++(90:1) -- ++(210:1) -- ++(270:1) -- ++(30:1);
\filldraw[fill=mgr,draw=black] (210:5) ++(270:2) ++(330:8) -- ++(90:1) -- ++(210:1) -- ++(270:1) -- ++(30:1);
\filldraw[fill=mgr,draw=black] (210:6) ++(270:2) ++(330:8) -- ++(90:1) -- ++(210:1) -- ++(270:1) -- ++(30:1);
\filldraw[fill=mgr,draw=black] (210:7) ++(270:2) ++(330:8) -- ++(90:1) -- ++(210:1) -- ++(270:1) -- ++(30:1);
\filldraw[fill=mgr,draw=black] (210:8) ++(270:2) ++(330:8) -- ++(90:1) -- ++(210:1) -- ++(270:1) -- ++(30:1);
\filldraw[fill=mgr,draw=black] (210:9) ++(270:2) ++(330:8) -- ++(90:1) -- ++(210:1) -- ++(270:1) -- ++(30:1);
\filldraw[fill=mgr,draw=black] (210:10) ++(270:2) ++(330:8) -- ++(90:1) -- ++(210:1) -- ++(270:1) -- ++(30:1);
\filldraw[fill=mbr,draw=black] (330:6) ++(270:3) -- ++(330:1) -- ++(90:1) -- ++(150:1) -- ++(270:1);
\filldraw[fill=mbr,draw=black] (330:7) ++(270:3) -- ++(330:1) -- ++(90:1) -- ++(150:1) -- ++(270:1);
\filldraw[fill=mbr,draw=black] (330:8) ++(270:3) -- ++(330:1) -- ++(90:1) -- ++(150:1) -- ++(270:1);
\filldraw[fill=mbr,draw=black] (330:6) ++(270:4) -- ++(330:1) -- ++(90:1) -- ++(150:1) -- ++(270:1);
\filldraw[fill=mbr,draw=black] (330:7) ++(270:4) -- ++(330:1) -- ++(90:1) -- ++(150:1) -- ++(270:1);
\filldraw[fill=mbr,draw=black] (330:8) ++(270:4) -- ++(330:1) -- ++(90:1) -- ++(150:1) -- ++(270:1);
\filldraw[fill=mor,draw=black] (330:3) ++(270:7) ++(210:1) ++(30:4) -- ++(210:1) -- ++(330:1) -- ++(30:1) -- ++(150:1);
\filldraw[fill=mor,draw=black] (330:3) ++(270:7) ++(210:2) ++(30:4) -- ++(210:1) -- ++(330:1) -- ++(30:1) -- ++(150:1);
\filldraw[fill=mor,draw=black] (330:3) ++(270:7) ++(210:3) ++(30:4) -- ++(210:1) -- ++(330:1) -- ++(30:1) -- ++(150:1);
\filldraw[fill=mor,draw=black] (330:3) ++(270:7) ++(210:4) ++(30:4) -- ++(210:1) -- ++(330:1) -- ++(30:1) -- ++(150:1);
\filldraw[fill=mor,draw=black] (330:3) ++(270:7) ++(210:5) ++(30:4) -- ++(210:1) -- ++(330:1) -- ++(30:1) -- ++(150:1);
\filldraw[fill=mor,draw=black] (330:3) ++(270:7) ++(210:6) ++(30:4) -- ++(210:1) -- ++(330:1) -- ++(30:1) -- ++(150:1);
\filldraw[fill=mor,draw=black] (330:3) ++(270:7) ++(210:7) ++(30:4) -- ++(210:1) -- ++(330:1) -- ++(30:1) -- ++(150:1);
\filldraw[fill=mor,draw=black] (330:3) ++(270:7) ++(210:8) ++(30:4) -- ++(210:1) -- ++(330:1) -- ++(30:1) -- ++(150:1);
\filldraw[fill=mor,draw=black] (330:3) ++(270:7) ++(210:9) ++(30:4) -- ++(210:1) -- ++(330:1) -- ++(30:1) -- ++(150:1);
\filldraw[fill=mor,draw=black] (330:4) ++(270:7) ++(210:1) ++(30:4) -- ++(210:1) -- ++(330:1) -- ++(30:1) -- ++(150:1);
\filldraw[fill=mor,draw=black] (330:4) ++(270:7) ++(210:2) ++(30:4) -- ++(210:1) -- ++(330:1) -- ++(30:1) -- ++(150:1);
\filldraw[fill=mor,draw=black] (330:4) ++(270:7) ++(210:3) ++(30:4) -- ++(210:1) -- ++(330:1) -- ++(30:1) -- ++(150:1);
\filldraw[fill=mor,draw=black] (330:4) ++(270:7) ++(210:4) ++(30:4) -- ++(210:1) -- ++(330:1) -- ++(30:1) -- ++(150:1);
\filldraw[fill=mor,draw=black] (330:4) ++(270:7) ++(210:5) ++(30:4) -- ++(210:1) -- ++(330:1) -- ++(30:1) -- ++(150:1);
\filldraw[fill=mor,draw=black] (330:4) ++(270:7) ++(210:6) ++(30:4) -- ++(210:1) -- ++(330:1) -- ++(30:1) -- ++(150:1);
\filldraw[fill=mor,draw=black] (330:4) ++(270:7) ++(210:7) ++(30:4) -- ++(210:1) -- ++(330:1) -- ++(30:1) -- ++(150:1);
\filldraw[fill=mor,draw=black] (330:4) ++(270:7) ++(210:8) ++(30:4) -- ++(210:1) -- ++(330:1) -- ++(30:1) -- ++(150:1);
\filldraw[fill=mor,draw=black] (330:4) ++(270:7) ++(210:9) ++(30:4) -- ++(210:1) -- ++(330:1) -- ++(30:1) -- ++(150:1);
\filldraw[fill=mor,draw=black] (330:5) ++(270:7) ++(210:1) ++(30:4) -- ++(210:1) -- ++(330:1) -- ++(30:1) -- ++(150:1);
\filldraw[fill=mor,draw=black] (330:5) ++(270:7) ++(210:2) ++(30:4) -- ++(210:1) -- ++(330:1) -- ++(30:1) -- ++(150:1);
\filldraw[fill=mor,draw=black] (330:5) ++(270:7) ++(210:3) ++(30:4) -- ++(210:1) -- ++(330:1) -- ++(30:1) -- ++(150:1);
\filldraw[fill=mor,draw=black] (330:5) ++(270:7) ++(210:4) ++(30:4) -- ++(210:1) -- ++(330:1) -- ++(30:1) -- ++(150:1);
\filldraw[fill=mor,draw=black] (330:5) ++(270:7) ++(210:5) ++(30:4) -- ++(210:1) -- ++(330:1) -- ++(30:1) -- ++(150:1);
\filldraw[fill=mor,draw=black] (330:5) ++(270:7) ++(210:6) ++(30:4) -- ++(210:1) -- ++(330:1) -- ++(30:1) -- ++(150:1);
\filldraw[fill=mor,draw=black] (330:5) ++(270:7) ++(210:7) ++(30:4) -- ++(210:1) -- ++(330:1) -- ++(30:1) -- ++(150:1);
\filldraw[fill=mor,draw=black] (330:5) ++(270:7) ++(210:8) ++(30:4) -- ++(210:1) -- ++(330:1) -- ++(30:1) -- ++(150:1);
\filldraw[fill=mor,draw=black] (330:5) ++(270:7) ++(210:9) ++(30:4) -- ++(210:1) -- ++(330:1) -- ++(30:1) -- ++(150:1);
\end{tikzpicture}
\caption{Periodic plane partition corresponding to $\Delta=0$ primary of Lee-Yang model}
\label{leeyangdelta0}
\end{minipage}\hfill
\begin{minipage}{0.45\textwidth}
\centering
\begin{tikzpicture}[scale=0.4]
\definecolor{mbr}{RGB}{128,71,0}
\definecolor{mgr}{RGB}{148,166,0}
\definecolor{mor}{RGB}{242,198,0}
\filldraw[fill=mor,draw=black] (30:1) -- ++(210:1) -- ++(330:1) -- ++(30:1) -- ++(150:1);
\filldraw[fill=mor,draw=black] (330:1) ++(30:1) -- ++(210:1) -- ++(330:1) -- ++(30:1) -- ++(150:1);
\filldraw[fill=mor,draw=black] (330:2) ++(30:1) -- ++(210:1) -- ++(330:1) -- ++(30:1) -- ++(150:1);
\filldraw[fill=mor,draw=black] (330:3) ++(30:1) -- ++(210:1) -- ++(330:1) -- ++(30:1) -- ++(150:1);
\filldraw[fill=mor,draw=black] (330:4) ++(30:1) -- ++(210:1) -- ++(330:1) -- ++(30:1) -- ++(150:1);
\filldraw[fill=mor,draw=black] (330:5) ++(30:1) -- ++(210:1) -- ++(330:1) -- ++(30:1) -- ++(150:1);
\filldraw[fill=mor,draw=black] (330:6) ++(30:1) -- ++(210:1) -- ++(330:1) -- ++(30:1) -- ++(150:1);
\filldraw[fill=mor,draw=black] (330:7) ++(30:1) -- ++(210:1) -- ++(330:1) -- ++(30:1) -- ++(150:1);
\filldraw[fill=mor,draw=black] (330:8) ++(30:1) -- ++(210:1) -- ++(330:1) -- ++(30:1) -- ++(150:1);
\filldraw[fill=mor,draw=black] (330:9) ++(30:1) -- ++(210:1) -- ++(330:1) -- ++(30:1) -- ++(150:1);
\filldraw[fill=mor,draw=black] (210:1) ++(30:1) -- ++(210:1) -- ++(330:1) -- ++(30:1) -- ++(150:1);
\filldraw[fill=mor,draw=black] (210:2) ++(30:1) -- ++(210:1) -- ++(330:1) -- ++(30:1) -- ++(150:1);
\filldraw[fill=mor,draw=black] (210:3) ++(30:1) -- ++(210:1) -- ++(330:1) -- ++(30:1) -- ++(150:1);
\filldraw[fill=mor,draw=black] (210:4) ++(30:1) -- ++(210:1) -- ++(330:1) -- ++(30:1) -- ++(150:1);
\filldraw[fill=mor,draw=black] (210:5) ++(30:1) -- ++(210:1) -- ++(330:1) -- ++(30:1) -- ++(150:1);
\filldraw[fill=mor,draw=black] (210:6) ++(30:1) -- ++(210:1) -- ++(330:1) -- ++(30:1) -- ++(150:1);
\filldraw[fill=mor,draw=black] (210:7) ++(30:1) -- ++(210:1) -- ++(330:1) -- ++(30:1) -- ++(150:1);
\filldraw[fill=mor,draw=black] (210:8) ++(30:1) -- ++(210:1) -- ++(330:1) -- ++(30:1) -- ++(150:1);
\filldraw[fill=mor,draw=black] (210:9) ++(30:1) -- ++(210:1) -- ++(330:1) -- ++(30:1) -- ++(150:1);
\filldraw[fill=mor,draw=black] (330:1) ++(210:1) ++(30:1) -- ++(210:1) -- ++(330:1) -- ++(30:1) -- ++(150:1);
\filldraw[fill=mor,draw=black] (330:1) ++(210:2) ++(30:1) -- ++(210:1) -- ++(330:1) -- ++(30:1) -- ++(150:1);
\filldraw[fill=mor,draw=black] (330:1) ++(210:3) ++(30:1) -- ++(210:1) -- ++(330:1) -- ++(30:1) -- ++(150:1);
\filldraw[fill=mor,draw=black] (330:1) ++(210:4) ++(30:1) -- ++(210:1) -- ++(330:1) -- ++(30:1) -- ++(150:1);
\filldraw[fill=mor,draw=black] (330:1) ++(210:5) ++(30:1) -- ++(210:1) -- ++(330:1) -- ++(30:1) -- ++(150:1);
\filldraw[fill=mor,draw=black] (330:1) ++(210:6) ++(30:1) -- ++(210:1) -- ++(330:1) -- ++(30:1) -- ++(150:1);
\filldraw[fill=mor,draw=black] (330:1) ++(210:7) ++(30:1) -- ++(210:1) -- ++(330:1) -- ++(30:1) -- ++(150:1);
\filldraw[fill=mor,draw=black] (330:1) ++(210:8) ++(30:1) -- ++(210:1) -- ++(330:1) -- ++(30:1) -- ++(150:1);
\filldraw[fill=mor,draw=black] (330:1) ++(210:9) ++(30:1) -- ++(210:1) -- ++(330:1) -- ++(30:1) -- ++(150:1);
\filldraw[fill=mgr,draw=black] (210:1) ++(330:3) -- ++(90:1) -- ++(210:1) -- ++(270:1) -- ++(30:1);
\filldraw[fill=mgr,draw=black] (210:2) ++(330:3) -- ++(90:1) -- ++(210:1) -- ++(270:1) -- ++(30:1);
\filldraw[fill=mgr,draw=black] (210:3) ++(330:3) -- ++(90:1) -- ++(210:1) -- ++(270:1) -- ++(30:1);
\filldraw[fill=mgr,draw=black] (210:4) ++(330:3) -- ++(90:1) -- ++(210:1) -- ++(270:1) -- ++(30:1);
\filldraw[fill=mgr,draw=black] (210:5) ++(330:3) -- ++(90:1) -- ++(210:1) -- ++(270:1) -- ++(30:1);
\filldraw[fill=mgr,draw=black] (210:6) ++(330:3) -- ++(90:1) -- ++(210:1) -- ++(270:1) -- ++(30:1);
\filldraw[fill=mgr,draw=black] (210:7) ++(330:3) -- ++(90:1) -- ++(210:1) -- ++(270:1) -- ++(30:1);
\filldraw[fill=mgr,draw=black] (210:8) ++(330:3) -- ++(90:1) -- ++(210:1) -- ++(270:1) -- ++(30:1);
\filldraw[fill=mgr,draw=black] (210:9) ++(330:3) -- ++(90:1) -- ++(210:1) -- ++(270:1) -- ++(30:1);
\filldraw[fill=mgr,draw=black] (210:2) ++(330:4) -- ++(90:1) -- ++(210:1) -- ++(270:1) -- ++(30:1);
\filldraw[fill=mgr,draw=black] (210:3) ++(330:4) -- ++(90:1) -- ++(210:1) -- ++(270:1) -- ++(30:1);
\filldraw[fill=mgr,draw=black] (210:4) ++(330:4) -- ++(90:1) -- ++(210:1) -- ++(270:1) -- ++(30:1);
\filldraw[fill=mgr,draw=black] (210:5) ++(330:4) -- ++(90:1) -- ++(210:1) -- ++(270:1) -- ++(30:1);
\filldraw[fill=mgr,draw=black] (210:6) ++(330:4) -- ++(90:1) -- ++(210:1) -- ++(270:1) -- ++(30:1);
\filldraw[fill=mgr,draw=black] (210:7) ++(330:4) -- ++(90:1) -- ++(210:1) -- ++(270:1) -- ++(30:1);
\filldraw[fill=mgr,draw=black] (210:8) ++(330:4) -- ++(90:1) -- ++(210:1) -- ++(270:1) -- ++(30:1);
\filldraw[fill=mgr,draw=black] (210:9) ++(330:4) -- ++(90:1) -- ++(210:1) -- ++(270:1) -- ++(30:1);
\filldraw[fill=mgr,draw=black] (210:10) ++(330:4) -- ++(90:1) -- ++(210:1) -- ++(270:1) -- ++(30:1);
\filldraw[fill=mbr,draw=black] (330:2) ++(270:1) -- ++(330:1) -- ++(90:1) -- ++(150:1) -- ++(270:1);
\filldraw[fill=mbr,draw=black] (330:3) ++(270:1) -- ++(330:1) -- ++(90:1) -- ++(150:1) -- ++(270:1);
\filldraw[fill=mbr,draw=black] (330:4) ++(270:1) -- ++(330:1) -- ++(90:1) -- ++(150:1) -- ++(270:1);
\filldraw[fill=mbr,draw=black] (330:5) ++(270:1) -- ++(330:1) -- ++(90:1) -- ++(150:1) -- ++(270:1);
\filldraw[fill=mbr,draw=black] (330:6) ++(270:1) -- ++(330:1) -- ++(90:1) -- ++(150:1) -- ++(270:1);
\filldraw[fill=mbr,draw=black] (330:7) ++(270:1) -- ++(330:1) -- ++(90:1) -- ++(150:1) -- ++(270:1);
\filldraw[fill=mbr,draw=black] (330:8) ++(270:1) -- ++(330:1) -- ++(90:1) -- ++(150:1) -- ++(270:1);
\filldraw[fill=mbr,draw=black] (330:9) ++(270:1) -- ++(330:1) -- ++(90:1) -- ++(150:1) -- ++(270:1);
\filldraw[fill=mbr,draw=black] (330:2) ++(270:2) -- ++(330:1) -- ++(90:1) -- ++(150:1) -- ++(270:1);
\filldraw[fill=mbr,draw=black] (330:3) ++(270:2) -- ++(330:1) -- ++(90:1) -- ++(150:1) -- ++(270:1);
\filldraw[fill=mbr,draw=black] (330:4) ++(270:2) -- ++(330:1) -- ++(90:1) -- ++(150:1) -- ++(270:1);
\filldraw[fill=mbr,draw=black] (330:5) ++(270:2) -- ++(330:1) -- ++(90:1) -- ++(150:1) -- ++(270:1);
\filldraw[fill=mbr,draw=black] (330:6) ++(270:2) -- ++(330:1) -- ++(90:1) -- ++(150:1) -- ++(270:1);
\filldraw[fill=mbr,draw=black] (330:7) ++(270:2) -- ++(330:1) -- ++(90:1) -- ++(150:1) -- ++(270:1);
\filldraw[fill=mbr,draw=black] (330:8) ++(270:2) -- ++(330:1) -- ++(90:1) -- ++(150:1) -- ++(270:1);
\filldraw[fill=mbr,draw=black] (330:9) ++(270:2) -- ++(330:1) -- ++(90:1) -- ++(150:1) -- ++(270:1);
\filldraw[fill=mor,draw=black] (270:4) ++ (30:2) -- ++(210:1) -- ++(330:1) -- ++(30:1) -- ++(150:1);
\filldraw[fill=mor,draw=black] (270:4) ++ (30:2) ++ (330:1) -- ++(210:1) -- ++(330:1) -- ++(30:1) -- ++(150:1);
\filldraw[fill=mor,draw=black] (270:4) ++ (30:2) ++ (330:2) -- ++(210:1) -- ++(330:1) -- ++(30:1) -- ++(150:1);
\filldraw[fill=mor,draw=black] (270:4) ++ (30:2) ++ (330:3) -- ++(210:1) -- ++(330:1) -- ++(30:1) -- ++(150:1);
\filldraw[fill=mor,draw=black] (270:4) ++ (30:2) ++ (330:4) -- ++(210:1) -- ++(330:1) -- ++(30:1) -- ++(150:1);
\filldraw[fill=mor,draw=black] (270:4) ++ (30:2) ++ (330:5) -- ++(210:1) -- ++(330:1) -- ++(30:1) -- ++(150:1);
\filldraw[fill=mor,draw=black] (270:4) ++ (30:2) ++ (330:6) -- ++(210:1) -- ++(330:1) -- ++(30:1) -- ++(150:1);
\filldraw[fill=mor,draw=black] (270:4) ++ (30:2) ++ (330:7) -- ++(210:1) -- ++(330:1) -- ++(30:1) -- ++(150:1);
\filldraw[fill=mor,draw=black] (270:4) ++ (30:2) ++ (210:1) -- ++(210:1) -- ++(330:1) -- ++(30:1) -- ++(150:1);
\filldraw[fill=mor,draw=black] (270:4) ++ (30:2) ++ (210:2) -- ++(210:1) -- ++(330:1) -- ++(30:1) -- ++(150:1);
\filldraw[fill=mor,draw=black] (270:4) ++ (30:2) ++ (210:3) -- ++(210:1) -- ++(330:1) -- ++(30:1) -- ++(150:1);
\filldraw[fill=mor,draw=black] (270:4) ++ (30:2) ++ (210:4) -- ++(210:1) -- ++(330:1) -- ++(30:1) -- ++(150:1);
\filldraw[fill=mor,draw=black] (270:4) ++ (30:2) ++ (210:5) -- ++(210:1) -- ++(330:1) -- ++(30:1) -- ++(150:1);
\filldraw[fill=mor,draw=black] (270:4) ++ (30:2) ++ (210:6) -- ++(210:1) -- ++(330:1) -- ++(30:1) -- ++(150:1);
\filldraw[fill=mor,draw=black] (270:4) ++ (30:2) ++ (210:7) -- ++(210:1) -- ++(330:1) -- ++(30:1) -- ++(150:1);
\filldraw[fill=mor,draw=black] (270:4) ++ (30:2) ++ (210:8) -- ++(210:1) -- ++(330:1) -- ++(30:1) -- ++(150:1);
\filldraw[fill=mor,draw=black] (270:4) ++ (330:1) ++(210:-1) -- ++(210:1) -- ++(330:1) -- ++(30:1) -- ++(150:1);
\filldraw[fill=mor,draw=black] (270:4) ++ (330:1) ++(210:0) -- ++(210:1) -- ++(330:1) -- ++(30:1) -- ++(150:1);
\filldraw[fill=mor,draw=black] (270:4) ++ (330:1) ++(210:1) -- ++(210:1) -- ++(330:1) -- ++(30:1) -- ++(150:1);
\filldraw[fill=mor,draw=black] (270:4) ++ (330:1) ++(210:2) -- ++(210:1) -- ++(330:1) -- ++(30:1) -- ++(150:1);
\filldraw[fill=mor,draw=black] (270:4) ++ (330:1) ++(210:3) -- ++(210:1) -- ++(330:1) -- ++(30:1) -- ++(150:1);
\filldraw[fill=mor,draw=black] (270:4) ++ (330:1) ++(210:4) -- ++(210:1) -- ++(330:1) -- ++(30:1) -- ++(150:1);
\filldraw[fill=mor,draw=black] (270:4) ++ (330:1) ++(210:5) -- ++(210:1) -- ++(330:1) -- ++(30:1) -- ++(150:1);
\filldraw[fill=mor,draw=black] (270:4) ++ (330:1) ++(210:6) -- ++(210:1) -- ++(330:1) -- ++(30:1) -- ++(150:1);
\filldraw[fill=mgr,draw=black] (210:4) ++(330:7) -- ++(90:1) -- ++(210:1) -- ++(270:1) -- ++(30:1);
\filldraw[fill=mgr,draw=black] (210:5) ++(330:7) -- ++(90:1) -- ++(210:1) -- ++(270:1) -- ++(30:1);
\filldraw[fill=mgr,draw=black] (210:6) ++(330:7) -- ++(90:1) -- ++(210:1) -- ++(270:1) -- ++(30:1);
\filldraw[fill=mgr,draw=black] (210:7) ++(330:7) -- ++(90:1) -- ++(210:1) -- ++(270:1) -- ++(30:1);
\filldraw[fill=mgr,draw=black] (210:8) ++(330:7) -- ++(90:1) -- ++(210:1) -- ++(270:1) -- ++(30:1);
\filldraw[fill=mgr,draw=black] (210:9) ++(330:7) -- ++(90:1) -- ++(210:1) -- ++(270:1) -- ++(30:1);
\filldraw[fill=mgr,draw=black] (210:10) ++(330:7) -- ++(90:1) -- ++(210:1) -- ++(270:1) -- ++(30:1);
\filldraw[fill=mgr,draw=black] (210:11) ++(330:7) -- ++(90:1) -- ++(210:1) -- ++(270:1) -- ++(30:1);
\filldraw[fill=mgr,draw=black] (210:5) ++(330:8) -- ++(90:1) -- ++(210:1) -- ++(270:1) -- ++(30:1);
\filldraw[fill=mgr,draw=black] (210:6) ++(330:8) -- ++(90:1) -- ++(210:1) -- ++(270:1) -- ++(30:1);
\filldraw[fill=mgr,draw=black] (210:7) ++(330:8) -- ++(90:1) -- ++(210:1) -- ++(270:1) -- ++(30:1);
\filldraw[fill=mgr,draw=black] (210:8) ++(330:8) -- ++(90:1) -- ++(210:1) -- ++(270:1) -- ++(30:1);
\filldraw[fill=mgr,draw=black] (210:9) ++(330:8) -- ++(90:1) -- ++(210:1) -- ++(270:1) -- ++(30:1);
\filldraw[fill=mgr,draw=black] (210:10) ++(330:8) -- ++(90:1) -- ++(210:1) -- ++(270:1) -- ++(30:1);
\filldraw[fill=mgr,draw=black] (210:11) ++(330:8) -- ++(90:1) -- ++(210:1) -- ++(270:1) -- ++(30:1);
\filldraw[fill=mgr,draw=black] (210:12) ++(330:8) -- ++(90:1) -- ++(210:1) -- ++(270:1) -- ++(30:1);
\filldraw[fill=mbr,draw=black] (330:3) ++(270:4) -- ++(330:1) -- ++(90:1) -- ++(150:1) -- ++(270:1);
\filldraw[fill=mbr,draw=black] (330:4) ++(270:4) -- ++(330:1) -- ++(90:1) -- ++(150:1) -- ++(270:1);
\filldraw[fill=mbr,draw=black] (330:5) ++(270:4) -- ++(330:1) -- ++(90:1) -- ++(150:1) -- ++(270:1);
\filldraw[fill=mbr,draw=black] (330:6) ++(270:4) -- ++(330:1) -- ++(90:1) -- ++(150:1) -- ++(270:1);
\filldraw[fill=mbr,draw=black] (330:7) ++(270:4) -- ++(330:1) -- ++(90:1) -- ++(150:1) -- ++(270:1);
\filldraw[fill=mbr,draw=black] (330:8) ++(270:4) -- ++(330:1) -- ++(90:1) -- ++(150:1) -- ++(270:1);
\filldraw[fill=mbr,draw=black] (330:3) ++(270:5) -- ++(330:1) -- ++(90:1) -- ++(150:1) -- ++(270:1);
\filldraw[fill=mbr,draw=black] (330:4) ++(270:5) -- ++(330:1) -- ++(90:1) -- ++(150:1) -- ++(270:1);
\filldraw[fill=mbr,draw=black] (330:5) ++(270:5) -- ++(330:1) -- ++(90:1) -- ++(150:1) -- ++(270:1);
\filldraw[fill=mbr,draw=black] (330:6) ++(270:5) -- ++(330:1) -- ++(90:1) -- ++(150:1) -- ++(270:1);
\filldraw[fill=mbr,draw=black] (330:7) ++(270:5) -- ++(330:1) -- ++(90:1) -- ++(150:1) -- ++(270:1);
\filldraw[fill=mbr,draw=black] (330:8) ++(270:5) -- ++(330:1) -- ++(90:1) -- ++(150:1) -- ++(270:1);
\filldraw[fill=mor,draw=black] (270:8) ++ (30:3) -- ++(210:1) -- ++(330:1) -- ++(30:1) -- ++(150:1);
\filldraw[fill=mor,draw=black] (270:8) ++ (30:3) ++ (330:1) -- ++(210:1) -- ++(330:1) -- ++(30:1) -- ++(150:1);
\filldraw[fill=mor,draw=black] (270:8) ++ (30:3) ++ (330:2) -- ++(210:1) -- ++(330:1) -- ++(30:1) -- ++(150:1);
\filldraw[fill=mor,draw=black] (270:8) ++ (30:3) ++ (330:3) -- ++(210:1) -- ++(330:1) -- ++(30:1) -- ++(150:1);
\filldraw[fill=mor,draw=black] (270:8) ++ (30:3) ++ (330:4) -- ++(210:1) -- ++(330:1) -- ++(30:1) -- ++(150:1);
\filldraw[fill=mor,draw=black] (270:8) ++ (30:3) ++ (330:5) -- ++(210:1) -- ++(330:1) -- ++(30:1) -- ++(150:1);
\filldraw[fill=mor,draw=black] (270:8) ++ (30:3) ++ (210:1) -- ++(210:1) -- ++(330:1) -- ++(30:1) -- ++(150:1);
\filldraw[fill=mor,draw=black] (270:8) ++ (30:3) ++ (210:2) -- ++(210:1) -- ++(330:1) -- ++(30:1) -- ++(150:1);
\filldraw[fill=mor,draw=black] (270:8) ++ (30:3) ++ (210:3) -- ++(210:1) -- ++(330:1) -- ++(30:1) -- ++(150:1);
\filldraw[fill=mor,draw=black] (270:8) ++ (30:3) ++ (210:4) -- ++(210:1) -- ++(330:1) -- ++(30:1) -- ++(150:1);
\filldraw[fill=mor,draw=black] (270:8) ++ (30:3) ++ (210:5) -- ++(210:1) -- ++(330:1) -- ++(30:1) -- ++(150:1);
\filldraw[fill=mor,draw=black] (270:8) ++ (30:3) ++ (210:6) -- ++(210:1) -- ++(330:1) -- ++(30:1) -- ++(150:1);
\filldraw[fill=mor,draw=black] (270:8) ++ (30:3) ++ (210:7) -- ++(210:1) -- ++(330:1) -- ++(30:1) -- ++(150:1);
\filldraw[fill=mor,draw=black] (270:8) ++ (330:1) ++(210:-2) -- ++(210:1) -- ++(330:1) -- ++(30:1) -- ++(150:1);
\filldraw[fill=mor,draw=black] (270:8) ++ (330:1) ++(210:-1) -- ++(210:1) -- ++(330:1) -- ++(30:1) -- ++(150:1);
\filldraw[fill=mor,draw=black] (270:8) ++ (330:1) ++(210:0) -- ++(210:1) -- ++(330:1) -- ++(30:1) -- ++(150:1);
\filldraw[fill=mor,draw=black] (270:8) ++ (330:1) ++(210:1) -- ++(210:1) -- ++(330:1) -- ++(30:1) -- ++(150:1);
\filldraw[fill=mor,draw=black] (270:8) ++ (330:1) ++(210:2) -- ++(210:1) -- ++(330:1) -- ++(30:1) -- ++(150:1);
\filldraw[fill=mor,draw=black] (270:8) ++ (330:1) ++(210:3) -- ++(210:1) -- ++(330:1) -- ++(30:1) -- ++(150:1);
\filldraw[fill=mor,draw=black] (270:8) ++ (330:1) ++(210:4) -- ++(210:1) -- ++(330:1) -- ++(30:1) -- ++(150:1);
\end{tikzpicture}
\caption{Periodic plane partition corresponding to $\Delta=0$ primary of Ising model}
\label{isingdelta0}
\end{minipage}
\end{figure}

\begin{figure}
\centering
\begin{minipage}{0.45\textwidth}
\centering
\begin{tikzpicture}[scale=0.4]
\definecolor{mbr}{RGB}{128,71,0}
\definecolor{mgr}{RGB}{148,166,0}
\definecolor{mor}{RGB}{242,198,0}
\definecolor{mrd}{RGB}{255,0,0}
\definecolor{mbl}{RGB}{0,0,255}
\filldraw[fill=mbr,draw=black] (330:1) ++(90:0) -- ++(330:1) -- ++(90:1) -- ++(150:1) -- ++(270:1);
\filldraw[fill=mbr,draw=black] (330:2) ++(90:0) -- ++(330:1) -- ++(90:1) -- ++(150:1) -- ++(270:1);
\filldraw[fill=mbr,draw=black] (330:3) ++(90:0) -- ++(330:1) -- ++(90:1) -- ++(150:1) -- ++(270:1);
\filldraw[fill=mbr,draw=black] (330:4) ++(90:0) -- ++(330:1) -- ++(90:1) -- ++(150:1) -- ++(270:1);
\filldraw[fill=mbr,draw=black] (330:5) ++(90:0) -- ++(330:1) -- ++(90:1) -- ++(150:1) -- ++(270:1);
\filldraw[fill=mbr,draw=black] (330:6) ++(90:0) -- ++(330:1) -- ++(90:1) -- ++(150:1) -- ++(270:1);
\filldraw[fill=mbr,draw=black] (330:7) ++(90:0) -- ++(330:1) -- ++(90:1) -- ++(150:1) -- ++(270:1);
\filldraw[fill=mbr,draw=black] (330:8) ++(90:0) -- ++(330:1) -- ++(90:1) -- ++(150:1) -- ++(270:1);
\filldraw[fill=mbr,draw=black] (330:0) ++(90:1) -- ++(330:1) -- ++(90:1) -- ++(150:1) -- ++(270:1);
\filldraw[fill=mbr,draw=black] (330:1) ++(90:1) -- ++(330:1) -- ++(90:1) -- ++(150:1) -- ++(270:1);
\filldraw[fill=mbr,draw=black] (330:2) ++(90:1) -- ++(330:1) -- ++(90:1) -- ++(150:1) -- ++(270:1);
\filldraw[fill=mbr,draw=black] (330:3) ++(90:1) -- ++(330:1) -- ++(90:1) -- ++(150:1) -- ++(270:1);
\filldraw[fill=mbr,draw=black] (330:4) ++(90:1) -- ++(330:1) -- ++(90:1) -- ++(150:1) -- ++(270:1);
\filldraw[fill=mbr,draw=black] (330:5) ++(90:1) -- ++(330:1) -- ++(90:1) -- ++(150:1) -- ++(270:1);
\filldraw[fill=mbr,draw=black] (330:6) ++(90:1) -- ++(330:1) -- ++(90:1) -- ++(150:1) -- ++(270:1);
\filldraw[fill=mbr,draw=black] (330:7) ++(90:1) -- ++(330:1) -- ++(90:1) -- ++(150:1) -- ++(270:1);
\filldraw[fill=mbr,draw=black] (330:8) ++(90:1) -- ++(330:1) -- ++(90:1) -- ++(150:1) -- ++(270:1);
\filldraw[fill=mor,draw=black] (330:1) ++(210:1) ++(30:1) -- ++(210:1) -- ++(330:1) -- ++(30:1) -- ++(150:1);
\filldraw[fill=mor,draw=black] (330:1) ++(210:2) ++(30:1) -- ++(210:1) -- ++(330:1) -- ++(30:1) -- ++(150:1);
\filldraw[fill=mor,draw=black] (330:1) ++(210:3) ++(30:1) -- ++(210:1) -- ++(330:1) -- ++(30:1) -- ++(150:1);
\filldraw[fill=mor,draw=black] (330:1) ++(210:4) ++(30:1) -- ++(210:1) -- ++(330:1) -- ++(30:1) -- ++(150:1);
\filldraw[fill=mor,draw=black] (330:1) ++(210:5) ++(30:1) -- ++(210:1) -- ++(330:1) -- ++(30:1) -- ++(150:1);
\filldraw[fill=mor,draw=black] (330:1) ++(210:6) ++(30:1) -- ++(210:1) -- ++(330:1) -- ++(30:1) -- ++(150:1);
\filldraw[fill=mor,draw=black] (330:1) ++(210:7) ++(30:1) -- ++(210:1) -- ++(330:1) -- ++(30:1) -- ++(150:1);
\filldraw[fill=mor,draw=black] (330:1) ++(210:8) ++(30:1) -- ++(210:1) -- ++(330:1) -- ++(30:1) -- ++(150:1);
\filldraw[fill=mor,draw=black] (330:1) ++(210:9) ++(30:1) -- ++(210:1) -- ++(330:1) -- ++(30:1) -- ++(150:1);
\filldraw[fill=mor,draw=black] (330:2) ++(210:1) ++(30:1) -- ++(210:1) -- ++(330:1) -- ++(30:1) -- ++(150:1);
\filldraw[fill=mor,draw=black] (330:2) ++(210:2) ++(30:1) -- ++(210:1) -- ++(330:1) -- ++(30:1) -- ++(150:1);
\filldraw[fill=mor,draw=black] (330:2) ++(210:3) ++(30:1) -- ++(210:1) -- ++(330:1) -- ++(30:1) -- ++(150:1);
\filldraw[fill=mor,draw=black] (330:2) ++(210:4) ++(30:1) -- ++(210:1) -- ++(330:1) -- ++(30:1) -- ++(150:1);
\filldraw[fill=mor,draw=black] (330:2) ++(210:5) ++(30:1) -- ++(210:1) -- ++(330:1) -- ++(30:1) -- ++(150:1);
\filldraw[fill=mor,draw=black] (330:2) ++(210:6) ++(30:1) -- ++(210:1) -- ++(330:1) -- ++(30:1) -- ++(150:1);
\filldraw[fill=mor,draw=black] (330:2) ++(210:7) ++(30:1) -- ++(210:1) -- ++(330:1) -- ++(30:1) -- ++(150:1);
\filldraw[fill=mor,draw=black] (330:2) ++(210:8) ++(30:1) -- ++(210:1) -- ++(330:1) -- ++(30:1) -- ++(150:1);
\filldraw[fill=mor,draw=black] (330:2) ++(210:9) ++(30:1) -- ++(210:1) -- ++(330:1) -- ++(30:1) -- ++(150:1);
\filldraw[fill=mgr,draw=black] (210:1) ++(330:4) -- ++(90:1) -- ++(210:1) -- ++(270:1) -- ++(30:1);
\filldraw[fill=mgr,draw=black] (210:2) ++(330:4) -- ++(90:1) -- ++(210:1) -- ++(270:1) -- ++(30:1);
\filldraw[fill=mgr,draw=black] (210:3) ++(330:4) -- ++(90:1) -- ++(210:1) -- ++(270:1) -- ++(30:1);
\filldraw[fill=mgr,draw=black] (210:4) ++(330:4) -- ++(90:1) -- ++(210:1) -- ++(270:1) -- ++(30:1);
\filldraw[fill=mgr,draw=black] (210:5) ++(330:4) -- ++(90:1) -- ++(210:1) -- ++(270:1) -- ++(30:1);
\filldraw[fill=mgr,draw=black] (210:6) ++(330:4) -- ++(90:1) -- ++(210:1) -- ++(270:1) -- ++(30:1);
\filldraw[fill=mgr,draw=black] (210:7) ++(330:4) -- ++(90:1) -- ++(210:1) -- ++(270:1) -- ++(30:1);
\filldraw[fill=mgr,draw=black] (210:8) ++(330:4) -- ++(90:1) -- ++(210:1) -- ++(270:1) -- ++(30:1);
\filldraw[fill=mgr,draw=black] (210:9) ++(330:4) -- ++(90:1) -- ++(210:1) -- ++(270:1) -- ++(30:1);
\filldraw[fill=mbr,draw=black] (330:3) ++(270:1) -- ++(330:1) -- ++(90:1) -- ++(150:1) -- ++(270:1);
\filldraw[fill=mbr,draw=black] (330:4) ++(270:1) -- ++(330:1) -- ++(90:1) -- ++(150:1) -- ++(270:1);
\filldraw[fill=mbr,draw=black] (330:5) ++(270:1) -- ++(330:1) -- ++(90:1) -- ++(150:1) -- ++(270:1);
\filldraw[fill=mbr,draw=black] (330:6) ++(270:1) -- ++(330:1) -- ++(90:1) -- ++(150:1) -- ++(270:1);
\filldraw[fill=mbr,draw=black] (330:7) ++(270:1) -- ++(330:1) -- ++(90:1) -- ++(150:1) -- ++(270:1);
\filldraw[fill=mbr,draw=black] (330:8) ++(270:1) -- ++(330:1) -- ++(90:1) -- ++(150:1) -- ++(270:1);
\filldraw[fill=mbr,draw=black] (330:4) ++(270:2) -- ++(330:1) -- ++(90:1) -- ++(150:1) -- ++(270:1);
\filldraw[fill=mbr,draw=black] (330:5) ++(270:2) -- ++(330:1) -- ++(90:1) -- ++(150:1) -- ++(270:1);
\filldraw[fill=mbr,draw=black] (330:6) ++(270:2) -- ++(330:1) -- ++(90:1) -- ++(150:1) -- ++(270:1);
\filldraw[fill=mbr,draw=black] (330:7) ++(270:2) -- ++(330:1) -- ++(90:1) -- ++(150:1) -- ++(270:1);
\filldraw[fill=mbr,draw=black] (330:8) ++(270:2) -- ++(330:1) -- ++(90:1) -- ++(150:1) -- ++(270:1);
\filldraw[fill=mor,draw=black] (330:4) ++(270:2) ++(210:1) ++(30:1) -- ++(210:1) -- ++(330:1) -- ++(30:1) -- ++(150:1);
\filldraw[fill=mor,draw=black] (330:4) ++(270:2) ++(210:2) ++(30:1) -- ++(210:1) -- ++(330:1) -- ++(30:1) -- ++(150:1);
\filldraw[fill=mor,draw=black] (330:4) ++(270:2) ++(210:3) ++(30:1) -- ++(210:1) -- ++(330:1) -- ++(30:1) -- ++(150:1);
\filldraw[fill=mor,draw=black] (330:4) ++(270:2) ++(210:4) ++(30:1) -- ++(210:1) -- ++(330:1) -- ++(30:1) -- ++(150:1);
\filldraw[fill=mor,draw=black] (330:4) ++(270:2) ++(210:5) ++(30:1) -- ++(210:1) -- ++(330:1) -- ++(30:1) -- ++(150:1);
\filldraw[fill=mor,draw=black] (330:4) ++(270:2) ++(210:6) ++(30:1) -- ++(210:1) -- ++(330:1) -- ++(30:1) -- ++(150:1);
\filldraw[fill=mor,draw=black] (330:4) ++(270:2) ++(210:7) ++(30:1) -- ++(210:1) -- ++(330:1) -- ++(30:1) -- ++(150:1);
\filldraw[fill=mor,draw=black] (330:4) ++(270:2) ++(210:8) ++(30:1) -- ++(210:1) -- ++(330:1) -- ++(30:1) -- ++(150:1);
\filldraw[fill=mor,draw=black] (330:4) ++(270:2) ++(210:9) ++(30:1) -- ++(210:1) -- ++(330:1) -- ++(30:1) -- ++(150:1);
\filldraw[fill=mor,draw=black] (330:5) ++(270:2) ++(210:1) ++(30:1) -- ++(210:1) -- ++(330:1) -- ++(30:1) -- ++(150:1);
\filldraw[fill=mor,draw=black] (330:5) ++(270:2) ++(210:2) ++(30:1) -- ++(210:1) -- ++(330:1) -- ++(30:1) -- ++(150:1);
\filldraw[fill=mor,draw=black] (330:5) ++(270:2) ++(210:3) ++(30:1) -- ++(210:1) -- ++(330:1) -- ++(30:1) -- ++(150:1);
\filldraw[fill=mor,draw=black] (330:5) ++(270:2) ++(210:4) ++(30:1) -- ++(210:1) -- ++(330:1) -- ++(30:1) -- ++(150:1);
\filldraw[fill=mor,draw=black] (330:5) ++(270:2) ++(210:5) ++(30:1) -- ++(210:1) -- ++(330:1) -- ++(30:1) -- ++(150:1);
\filldraw[fill=mor,draw=black] (330:5) ++(270:2) ++(210:6) ++(30:1) -- ++(210:1) -- ++(330:1) -- ++(30:1) -- ++(150:1);
\filldraw[fill=mor,draw=black] (330:5) ++(270:2) ++(210:7) ++(30:1) -- ++(210:1) -- ++(330:1) -- ++(30:1) -- ++(150:1);
\filldraw[fill=mor,draw=black] (330:5) ++(270:2) ++(210:8) ++(30:1) -- ++(210:1) -- ++(330:1) -- ++(30:1) -- ++(150:1);
\filldraw[fill=mor,draw=black] (330:5) ++(270:2) ++(210:9) ++(30:1) -- ++(210:1) -- ++(330:1) -- ++(30:1) -- ++(150:1);
\filldraw[fill=mgr,draw=black] (210:1) ++(270:2) ++(330:7) -- ++(90:1) -- ++(210:1) -- ++(270:1) -- ++(30:1);
\filldraw[fill=mgr,draw=black] (210:2) ++(270:2) ++(330:7) -- ++(90:1) -- ++(210:1) -- ++(270:1) -- ++(30:1);
\filldraw[fill=mgr,draw=black] (210:3) ++(270:2) ++(330:7) -- ++(90:1) -- ++(210:1) -- ++(270:1) -- ++(30:1);
\filldraw[fill=mgr,draw=black] (210:4) ++(270:2) ++(330:7) -- ++(90:1) -- ++(210:1) -- ++(270:1) -- ++(30:1);
\filldraw[fill=mgr,draw=black] (210:5) ++(270:2) ++(330:7) -- ++(90:1) -- ++(210:1) -- ++(270:1) -- ++(30:1);
\filldraw[fill=mgr,draw=black] (210:6) ++(270:2) ++(330:7) -- ++(90:1) -- ++(210:1) -- ++(270:1) -- ++(30:1);
\filldraw[fill=mgr,draw=black] (210:7) ++(270:2) ++(330:7) -- ++(90:1) -- ++(210:1) -- ++(270:1) -- ++(30:1);
\filldraw[fill=mgr,draw=black] (210:8) ++(270:2) ++(330:7) -- ++(90:1) -- ++(210:1) -- ++(270:1) -- ++(30:1);
\filldraw[fill=mgr,draw=black] (210:9) ++(270:2) ++(330:7) -- ++(90:1) -- ++(210:1) -- ++(270:1) -- ++(30:1);
\filldraw[fill=mbr,draw=black] (330:6) ++(270:3) -- ++(330:1) -- ++(90:1) -- ++(150:1) -- ++(270:1);
\filldraw[fill=mbr,draw=black] (330:7) ++(270:3) -- ++(330:1) -- ++(90:1) -- ++(150:1) -- ++(270:1);
\filldraw[fill=mbr,draw=black] (330:8) ++(270:3) -- ++(330:1) -- ++(90:1) -- ++(150:1) -- ++(270:1);
\filldraw[fill=mbr,draw=black] (330:7) ++(270:4) -- ++(330:1) -- ++(90:1) -- ++(150:1) -- ++(270:1);
\filldraw[fill=mbr,draw=black] (330:8) ++(270:4) -- ++(330:1) -- ++(90:1) -- ++(150:1) -- ++(270:1);
\filldraw[fill=mgr,draw=black] (210:2) ++(330:6) -- ++(90:1) -- ++(210:1) -- ++(270:1) -- ++(30:1);
\filldraw[fill=mgr,draw=black] (210:3) ++(330:6) -- ++(90:1) -- ++(210:1) -- ++(270:1) -- ++(30:1);
\filldraw[fill=mgr,draw=black] (210:4) ++(330:6) -- ++(90:1) -- ++(210:1) -- ++(270:1) -- ++(30:1);
\filldraw[fill=mgr,draw=black] (210:5) ++(330:6) -- ++(90:1) -- ++(210:1) -- ++(270:1) -- ++(30:1);
\filldraw[fill=mgr,draw=black] (210:6) ++(330:6) -- ++(90:1) -- ++(210:1) -- ++(270:1) -- ++(30:1);
\filldraw[fill=mgr,draw=black] (210:7) ++(330:6) -- ++(90:1) -- ++(210:1) -- ++(270:1) -- ++(30:1);
\filldraw[fill=mgr,draw=black] (210:8) ++(330:6) -- ++(90:1) -- ++(210:1) -- ++(270:1) -- ++(30:1);
\filldraw[fill=mgr,draw=black] (210:9) ++(330:6) -- ++(90:1) -- ++(210:1) -- ++(270:1) -- ++(30:1);
\filldraw[fill=mgr,draw=black] (210:10) ++(330:6) -- ++(90:1) -- ++(210:1) -- ++(270:1) -- ++(30:1);
\filldraw[fill=mor,draw=black] (330:4) ++(210:2) ++(30:1) -- ++(210:1) -- ++(330:1) -- ++(30:1) -- ++(150:1);
\filldraw[fill=mor,draw=black] (330:4) ++(210:3) ++(30:1) -- ++(210:1) -- ++(330:1) -- ++(30:1) -- ++(150:1);
\filldraw[fill=mor,draw=black] (330:4) ++(210:4) ++(30:1) -- ++(210:1) -- ++(330:1) -- ++(30:1) -- ++(150:1);
\filldraw[fill=mor,draw=black] (330:4) ++(210:5) ++(30:1) -- ++(210:1) -- ++(330:1) -- ++(30:1) -- ++(150:1);
\filldraw[fill=mor,draw=black] (330:4) ++(210:6) ++(30:1) -- ++(210:1) -- ++(330:1) -- ++(30:1) -- ++(150:1);
\filldraw[fill=mor,draw=black] (330:4) ++(210:7) ++(30:1) -- ++(210:1) -- ++(330:1) -- ++(30:1) -- ++(150:1);
\filldraw[fill=mor,draw=black] (330:4) ++(210:8) ++(30:1) -- ++(210:1) -- ++(330:1) -- ++(30:1) -- ++(150:1);
\filldraw[fill=mor,draw=black] (330:4) ++(210:9) ++(30:1) -- ++(210:1) -- ++(330:1) -- ++(30:1) -- ++(150:1);
\filldraw[fill=mor,draw=black] (330:4) ++(210:10) ++(30:1) -- ++(210:1) -- ++(330:1) -- ++(30:1) -- ++(150:1);
\filldraw[fill=mgr,draw=black] (210:0) ++(330:1) -- ++(90:1) -- ++(210:1) -- ++(270:1) -- ++(30:1);
\filldraw[fill=mgr,draw=black] (210:1) ++(330:1) -- ++(90:1) -- ++(210:1) -- ++(270:1) -- ++(30:1);
\filldraw[fill=mgr,draw=black] (210:2) ++(330:1) -- ++(90:1) -- ++(210:1) -- ++(270:1) -- ++(30:1);
\filldraw[fill=mgr,draw=black] (210:3) ++(330:1) -- ++(90:1) -- ++(210:1) -- ++(270:1) -- ++(30:1);
\filldraw[fill=mgr,draw=black] (210:4) ++(330:1) -- ++(90:1) -- ++(210:1) -- ++(270:1) -- ++(30:1);
\filldraw[fill=mgr,draw=black] (210:5) ++(330:1) -- ++(90:1) -- ++(210:1) -- ++(270:1) -- ++(30:1);
\filldraw[fill=mgr,draw=black] (210:6) ++(330:1) -- ++(90:1) -- ++(210:1) -- ++(270:1) -- ++(30:1);
\filldraw[fill=mgr,draw=black] (210:7) ++(330:1) -- ++(90:1) -- ++(210:1) -- ++(270:1) -- ++(30:1);
\filldraw[fill=mgr,draw=black] (210:8) ++(330:1) -- ++(90:1) -- ++(210:1) -- ++(270:1) -- ++(30:1);
\filldraw[fill=mor,draw=black] (150:1) ++(210:0) ++(30:1) -- ++(210:1) -- ++(330:1) -- ++(30:1) -- ++(150:1);
\filldraw[fill=mor,draw=black] (150:1) ++(210:1) ++(30:1) -- ++(210:1) -- ++(330:1) -- ++(30:1) -- ++(150:1);
\filldraw[fill=mor,draw=black] (150:1) ++(210:2) ++(30:1) -- ++(210:1) -- ++(330:1) -- ++(30:1) -- ++(150:1);
\filldraw[fill=mor,draw=black] (150:1) ++(210:3) ++(30:1) -- ++(210:1) -- ++(330:1) -- ++(30:1) -- ++(150:1);
\filldraw[fill=mor,draw=black] (150:1) ++(210:4) ++(30:1) -- ++(210:1) -- ++(330:1) -- ++(30:1) -- ++(150:1);
\filldraw[fill=mor,draw=black] (150:1) ++(210:5) ++(30:1) -- ++(210:1) -- ++(330:1) -- ++(30:1) -- ++(150:1);
\filldraw[fill=mor,draw=black] (150:1) ++(210:6) ++(30:1) -- ++(210:1) -- ++(330:1) -- ++(30:1) -- ++(150:1);
\filldraw[fill=mor,draw=black] (150:1) ++(210:7) ++(30:1) -- ++(210:1) -- ++(330:1) -- ++(30:1) -- ++(150:1);
\filldraw[fill=mor,draw=black] (150:1) ++(210:8) ++(30:1) -- ++(210:1) -- ++(330:1) -- ++(30:1) -- ++(150:1);
\filldraw[fill=mor,draw=black] (330:6) ++(270:3) ++(210:0) -- ++(210:1) -- ++(330:1) -- ++(30:1) -- ++(150:1);
\filldraw[fill=mor,draw=black] (330:6) ++(270:3) ++(210:1) -- ++(210:1) -- ++(330:1) -- ++(30:1) -- ++(150:1);
\filldraw[fill=mor,draw=black] (330:6) ++(270:3) ++(210:2) -- ++(210:1) -- ++(330:1) -- ++(30:1) -- ++(150:1);
\filldraw[fill=mor,draw=black] (330:6) ++(270:3) ++(210:3) -- ++(210:1) -- ++(330:1) -- ++(30:1) -- ++(150:1);
\filldraw[fill=mor,draw=black] (330:6) ++(270:3) ++(210:4) -- ++(210:1) -- ++(330:1) -- ++(30:1) -- ++(150:1);
\filldraw[fill=mor,draw=black] (330:6) ++(270:3) ++(210:5) -- ++(210:1) -- ++(330:1) -- ++(30:1) -- ++(150:1);
\filldraw[fill=mor,draw=black] (330:6) ++(270:3) ++(210:6) -- ++(210:1) -- ++(330:1) -- ++(30:1) -- ++(150:1);
\filldraw[fill=mor,draw=black] (330:6) ++(270:3) ++(210:7) -- ++(210:1) -- ++(330:1) -- ++(30:1) -- ++(150:1);
\filldraw[fill=mor,draw=black] (330:6) ++(270:3) ++(210:8) -- ++(210:1) -- ++(330:1) -- ++(30:1) -- ++(150:1);
\filldraw[fill=mgr,draw=black] (210:0) ++(330:7) ++(270:3) -- ++(210:1) -- ++(270:1) -- ++(30:1);
\filldraw[fill=mgr,draw=black] (210:1) ++(330:7) ++(270:3) -- ++(210:1) -- ++(270:1) -- ++(30:1);
\filldraw[fill=mgr,draw=black] (210:2) ++(330:7) ++(270:3) -- ++(210:1) -- ++(270:1) -- ++(30:1);
\filldraw[fill=mgr,draw=black] (210:3) ++(330:7) ++(270:3) -- ++(210:1) -- ++(270:1) -- ++(30:1);
\filldraw[fill=mgr,draw=black] (210:4) ++(330:7) ++(270:3) -- ++(210:1) -- ++(270:1) -- ++(30:1);
\filldraw[fill=mgr,draw=black] (210:5) ++(330:7) ++(270:3) -- ++(210:1) -- ++(270:1) -- ++(30:1);
\filldraw[fill=mgr,draw=black] (210:6) ++(330:7) ++(270:3) -- ++(210:1) -- ++(270:1) -- ++(30:1);
\filldraw[fill=mgr,draw=black] (210:7) ++(330:7) ++(270:3) -- ++(210:1) -- ++(270:1) -- ++(30:1);
\filldraw[fill=mgr,draw=black] (210:8) ++(330:7) ++(270:3) -- ++(210:1) -- ++(270:1) -- ++(30:1);
\filldraw[fill=mor,draw=black] (330:7) ++(270:4) ++(210:0) -- ++(210:1) -- ++(330:1) -- ++(30:1) -- ++(150:1);
\filldraw[fill=mor,draw=black] (330:7) ++(270:4) ++(210:1) -- ++(210:1) -- ++(330:1) -- ++(30:1) -- ++(150:1);
\filldraw[fill=mor,draw=black] (330:7) ++(270:4) ++(210:2) -- ++(210:1) -- ++(330:1) -- ++(30:1) -- ++(150:1);
\filldraw[fill=mor,draw=black] (330:7) ++(270:4) ++(210:3) -- ++(210:1) -- ++(330:1) -- ++(30:1) -- ++(150:1);
\filldraw[fill=mor,draw=black] (330:7) ++(270:4) ++(210:4) -- ++(210:1) -- ++(330:1) -- ++(30:1) -- ++(150:1);
\filldraw[fill=mor,draw=black] (330:7) ++(270:4) ++(210:5) -- ++(210:1) -- ++(330:1) -- ++(30:1) -- ++(150:1);
\filldraw[fill=mor,draw=black] (330:7) ++(270:4) ++(210:6) -- ++(210:1) -- ++(330:1) -- ++(30:1) -- ++(150:1);
\filldraw[fill=mor,draw=black] (330:7) ++(270:4) ++(210:7) -- ++(210:1) -- ++(330:1) -- ++(30:1) -- ++(150:1);
\filldraw[fill=mor,draw=black] (330:7) ++(270:4) ++(210:8) -- ++(210:1) -- ++(330:1) -- ++(30:1) -- ++(150:1);
\filldraw[fill=mor,draw=black] (330:8) ++(270:4) ++(210:0) -- ++(210:1) -- ++(330:1) -- ++(30:1) -- ++(150:1);
\filldraw[fill=mor,draw=black] (330:8) ++(270:4) ++(210:1) -- ++(210:1) -- ++(330:1) -- ++(30:1) -- ++(150:1);
\filldraw[fill=mor,draw=black] (330:8) ++(270:4) ++(210:2) -- ++(210:1) -- ++(330:1) -- ++(30:1) -- ++(150:1);
\filldraw[fill=mor,draw=black] (330:8) ++(270:4) ++(210:3) -- ++(210:1) -- ++(330:1) -- ++(30:1) -- ++(150:1);
\filldraw[fill=mor,draw=black] (330:8) ++(270:4) ++(210:4) -- ++(210:1) -- ++(330:1) -- ++(30:1) -- ++(150:1);
\filldraw[fill=mor,draw=black] (330:8) ++(270:4) ++(210:5) -- ++(210:1) -- ++(330:1) -- ++(30:1) -- ++(150:1);
\filldraw[fill=mor,draw=black] (330:8) ++(270:4) ++(210:6) -- ++(210:1) -- ++(330:1) -- ++(30:1) -- ++(150:1);
\filldraw[fill=mor,draw=black] (330:8) ++(270:4) ++(210:7) -- ++(210:1) -- ++(330:1) -- ++(30:1) -- ++(150:1);
\filldraw[fill=mor,draw=black] (330:8) ++(270:4) ++(210:8) -- ++(210:1) -- ++(330:1) -- ++(30:1) -- ++(150:1);
\end{tikzpicture}
\caption{Periodic plane partition corresponding to $\Delta=-\frac{1}{5}$ primary of Lee-Yang model}
\label{leeyangdelta15}
\end{minipage}\hfill
\begin{minipage}{0.45\textwidth}
\centering
\begin{tikzpicture}[scale=0.4]
\definecolor{mbr}{RGB}{128,71,0}
\definecolor{mgr}{RGB}{148,166,0}
\definecolor{mor}{RGB}{242,198,0}
\definecolor{mrd}{RGB}{255,0,0}
\definecolor{mbl}{RGB}{0,0,255}
\filldraw[fill=mbr,draw=black] (330:4) ++(90:0) -- ++(330:1) -- ++(90:1) -- ++(150:1) -- ++(270:1);
\filldraw[fill=mbr,draw=black] (330:5) ++(90:0) -- ++(330:1) -- ++(90:1) -- ++(150:1) -- ++(270:1);
\filldraw[fill=mbr,draw=black] (330:6) ++(90:0) -- ++(330:1) -- ++(90:1) -- ++(150:1) -- ++(270:1);
\filldraw[fill=mbr,draw=black] (330:7) ++(90:0) -- ++(330:1) -- ++(90:1) -- ++(150:1) -- ++(270:1);
\filldraw[fill=mbr,draw=black] (330:8) ++(90:0) -- ++(330:1) -- ++(90:1) -- ++(150:1) -- ++(270:1);
\filldraw[fill=mbr,draw=black] (330:0) ++(90:1) -- ++(330:1) -- ++(90:1) -- ++(150:1) -- ++(270:1);
\filldraw[fill=mbr,draw=black] (330:1) ++(90:1) -- ++(330:1) -- ++(90:1) -- ++(150:1) -- ++(270:1);
\filldraw[fill=mbr,draw=black] (330:2) ++(90:1) -- ++(330:1) -- ++(90:1) -- ++(150:1) -- ++(270:1);
\filldraw[fill=mbr,draw=black] (330:3) ++(90:1) -- ++(330:1) -- ++(90:1) -- ++(150:1) -- ++(270:1);
\filldraw[fill=mbr,draw=black] (330:4) ++(90:1) -- ++(330:1) -- ++(90:1) -- ++(150:1) -- ++(270:1);
\filldraw[fill=mbr,draw=black] (330:5) ++(90:1) -- ++(330:1) -- ++(90:1) -- ++(150:1) -- ++(270:1);
\filldraw[fill=mbr,draw=black] (330:6) ++(90:1) -- ++(330:1) -- ++(90:1) -- ++(150:1) -- ++(270:1);
\filldraw[fill=mbr,draw=black] (330:7) ++(90:1) -- ++(330:1) -- ++(90:1) -- ++(150:1) -- ++(270:1);
\filldraw[fill=mbr,draw=black] (330:8) ++(90:1) -- ++(330:1) -- ++(90:1) -- ++(150:1) -- ++(270:1);
\filldraw[fill=mor,draw=black] (330:0) ++(210:2) ++(30:1) -- ++(210:1) -- ++(330:1) -- ++(30:1) -- ++(150:1);
\filldraw[fill=mor,draw=black] (330:0) ++(210:3) ++(30:1) -- ++(210:1) -- ++(330:1) -- ++(30:1) -- ++(150:1);
\filldraw[fill=mor,draw=black] (330:0) ++(210:4) ++(30:1) -- ++(210:1) -- ++(330:1) -- ++(30:1) -- ++(150:1);
\filldraw[fill=mor,draw=black] (330:0) ++(210:5) ++(30:1) -- ++(210:1) -- ++(330:1) -- ++(30:1) -- ++(150:1);
\filldraw[fill=mor,draw=black] (330:0) ++(210:6) ++(30:1) -- ++(210:1) -- ++(330:1) -- ++(30:1) -- ++(150:1);
\filldraw[fill=mor,draw=black] (330:0) ++(210:7) ++(30:1) -- ++(210:1) -- ++(330:1) -- ++(30:1) -- ++(150:1);
\filldraw[fill=mor,draw=black] (330:0) ++(210:8) ++(30:1) -- ++(210:1) -- ++(330:1) -- ++(30:1) -- ++(150:1);
\filldraw[fill=mor,draw=black] (330:0) ++(210:9) ++(30:1) -- ++(210:1) -- ++(330:1) -- ++(30:1) -- ++(150:1);
\filldraw[fill=mor,draw=black] (330:1) ++(210:2) ++(30:1) -- ++(210:1) -- ++(330:1) -- ++(30:1) -- ++(150:1);
\filldraw[fill=mor,draw=black] (330:1) ++(210:3) ++(30:1) -- ++(210:1) -- ++(330:1) -- ++(30:1) -- ++(150:1);
\filldraw[fill=mor,draw=black] (330:1) ++(210:4) ++(30:1) -- ++(210:1) -- ++(330:1) -- ++(30:1) -- ++(150:1);
\filldraw[fill=mor,draw=black] (330:1) ++(210:5) ++(30:1) -- ++(210:1) -- ++(330:1) -- ++(30:1) -- ++(150:1);
\filldraw[fill=mor,draw=black] (330:1) ++(210:6) ++(30:1) -- ++(210:1) -- ++(330:1) -- ++(30:1) -- ++(150:1);
\filldraw[fill=mor,draw=black] (330:1) ++(210:7) ++(30:1) -- ++(210:1) -- ++(330:1) -- ++(30:1) -- ++(150:1);
\filldraw[fill=mor,draw=black] (330:1) ++(210:8) ++(30:1) -- ++(210:1) -- ++(330:1) -- ++(30:1) -- ++(150:1);
\filldraw[fill=mor,draw=black] (330:1) ++(210:9) ++(30:1) -- ++(210:1) -- ++(330:1) -- ++(30:1) -- ++(150:1);
\filldraw[fill=mor,draw=black] (330:2) ++(210:2) ++(30:1) -- ++(210:1) -- ++(330:1) -- ++(30:1) -- ++(150:1);
\filldraw[fill=mor,draw=black] (330:2) ++(210:3) ++(30:1) -- ++(210:1) -- ++(330:1) -- ++(30:1) -- ++(150:1);
\filldraw[fill=mor,draw=black] (330:2) ++(210:4) ++(30:1) -- ++(210:1) -- ++(330:1) -- ++(30:1) -- ++(150:1);
\filldraw[fill=mor,draw=black] (330:2) ++(210:5) ++(30:1) -- ++(210:1) -- ++(330:1) -- ++(30:1) -- ++(150:1);
\filldraw[fill=mor,draw=black] (330:2) ++(210:6) ++(30:1) -- ++(210:1) -- ++(330:1) -- ++(30:1) -- ++(150:1);
\filldraw[fill=mor,draw=black] (330:2) ++(210:7) ++(30:1) -- ++(210:1) -- ++(330:1) -- ++(30:1) -- ++(150:1);
\filldraw[fill=mor,draw=black] (330:2) ++(210:8) ++(30:1) -- ++(210:1) -- ++(330:1) -- ++(30:1) -- ++(150:1);
\filldraw[fill=mor,draw=black] (330:2) ++(210:9) ++(30:1) -- ++(210:1) -- ++(330:1) -- ++(30:1) -- ++(150:1);
\filldraw[fill=mgr,draw=black] (210:2) ++(330:4) -- ++(90:1) -- ++(210:1) -- ++(270:1) -- ++(30:1);
\filldraw[fill=mgr,draw=black] (210:3) ++(330:4) -- ++(90:1) -- ++(210:1) -- ++(270:1) -- ++(30:1);
\filldraw[fill=mgr,draw=black] (210:4) ++(330:4) -- ++(90:1) -- ++(210:1) -- ++(270:1) -- ++(30:1);
\filldraw[fill=mgr,draw=black] (210:5) ++(330:4) -- ++(90:1) -- ++(210:1) -- ++(270:1) -- ++(30:1);
\filldraw[fill=mgr,draw=black] (210:6) ++(330:4) -- ++(90:1) -- ++(210:1) -- ++(270:1) -- ++(30:1);
\filldraw[fill=mgr,draw=black] (210:7) ++(330:4) -- ++(90:1) -- ++(210:1) -- ++(270:1) -- ++(30:1);
\filldraw[fill=mgr,draw=black] (210:8) ++(330:4) -- ++(90:1) -- ++(210:1) -- ++(270:1) -- ++(30:1);
\filldraw[fill=mgr,draw=black] (210:9) ++(330:4) -- ++(90:1) -- ++(210:1) -- ++(270:1) -- ++(30:1);
\filldraw[fill=mgr,draw=black] (210:3) ++(330:5) -- ++(90:1) -- ++(210:1) -- ++(270:1) -- ++(30:1);
\filldraw[fill=mgr,draw=black] (210:4) ++(330:5) -- ++(90:1) -- ++(210:1) -- ++(270:1) -- ++(30:1);
\filldraw[fill=mgr,draw=black] (210:5) ++(330:5) -- ++(90:1) -- ++(210:1) -- ++(270:1) -- ++(30:1);
\filldraw[fill=mgr,draw=black] (210:6) ++(330:5) -- ++(90:1) -- ++(210:1) -- ++(270:1) -- ++(30:1);
\filldraw[fill=mgr,draw=black] (210:7) ++(330:5) -- ++(90:1) -- ++(210:1) -- ++(270:1) -- ++(30:1);
\filldraw[fill=mgr,draw=black] (210:8) ++(330:5) -- ++(90:1) -- ++(210:1) -- ++(270:1) -- ++(30:1);
\filldraw[fill=mgr,draw=black] (210:9) ++(330:5) -- ++(90:1) -- ++(210:1) -- ++(270:1) -- ++(30:1);
\filldraw[fill=mgr,draw=black] (210:10) ++(330:5) -- ++(90:1) -- ++(210:1) -- ++(270:1) -- ++(30:1);
\filldraw[fill=mbr,draw=black] (330:4) ++(270:1) -- ++(330:1) -- ++(90:1) -- ++(150:1) -- ++(270:1);
\filldraw[fill=mbr,draw=black] (330:5) ++(270:1) -- ++(330:1) -- ++(90:1) -- ++(150:1) -- ++(270:1);
\filldraw[fill=mbr,draw=black] (330:6) ++(270:1) -- ++(330:1) -- ++(90:1) -- ++(150:1) -- ++(270:1);
\filldraw[fill=mbr,draw=black] (330:7) ++(270:1) -- ++(330:1) -- ++(90:1) -- ++(150:1) -- ++(270:1);
\filldraw[fill=mbr,draw=black] (330:8) ++(270:1) -- ++(330:1) -- ++(90:1) -- ++(150:1) -- ++(270:1);
\filldraw[fill=mbr,draw=black] (330:7) ++(270:2) -- ++(330:1) -- ++(90:1) -- ++(150:1) -- ++(270:1);
\filldraw[fill=mbr,draw=black] (330:8) ++(270:2) -- ++(330:1) -- ++(90:1) -- ++(150:1) -- ++(270:1);
\filldraw[fill=mor,draw=black] (330:3) ++(270:2) ++(210:2) ++(30:1) -- ++(210:1) -- ++(330:1) -- ++(30:1) -- ++(150:1);
\filldraw[fill=mor,draw=black] (330:3) ++(270:2) ++(210:3) ++(30:1) -- ++(210:1) -- ++(330:1) -- ++(30:1) -- ++(150:1);
\filldraw[fill=mor,draw=black] (330:3) ++(270:2) ++(210:4) ++(30:1) -- ++(210:1) -- ++(330:1) -- ++(30:1) -- ++(150:1);
\filldraw[fill=mor,draw=black] (330:3) ++(270:2) ++(210:5) ++(30:1) -- ++(210:1) -- ++(330:1) -- ++(30:1) -- ++(150:1);
\filldraw[fill=mor,draw=black] (330:3) ++(270:2) ++(210:6) ++(30:1) -- ++(210:1) -- ++(330:1) -- ++(30:1) -- ++(150:1);
\filldraw[fill=mor,draw=black] (330:3) ++(270:2) ++(210:7) ++(30:1) -- ++(210:1) -- ++(330:1) -- ++(30:1) -- ++(150:1);
\filldraw[fill=mor,draw=black] (330:3) ++(270:2) ++(210:8) ++(30:1) -- ++(210:1) -- ++(330:1) -- ++(30:1) -- ++(150:1);
\filldraw[fill=mor,draw=black] (330:3) ++(270:2) ++(210:9) ++(30:1) -- ++(210:1) -- ++(330:1) -- ++(30:1) -- ++(150:1);
\filldraw[fill=mor,draw=black] (330:4) ++(270:2) ++(210:2) ++(30:1) -- ++(210:1) -- ++(330:1) -- ++(30:1) -- ++(150:1);
\filldraw[fill=mor,draw=black] (330:4) ++(270:2) ++(210:3) ++(30:1) -- ++(210:1) -- ++(330:1) -- ++(30:1) -- ++(150:1);
\filldraw[fill=mor,draw=black] (330:4) ++(270:2) ++(210:4) ++(30:1) -- ++(210:1) -- ++(330:1) -- ++(30:1) -- ++(150:1);
\filldraw[fill=mor,draw=black] (330:4) ++(270:2) ++(210:5) ++(30:1) -- ++(210:1) -- ++(330:1) -- ++(30:1) -- ++(150:1);
\filldraw[fill=mor,draw=black] (330:4) ++(270:2) ++(210:6) ++(30:1) -- ++(210:1) -- ++(330:1) -- ++(30:1) -- ++(150:1);
\filldraw[fill=mor,draw=black] (330:4) ++(270:2) ++(210:7) ++(30:1) -- ++(210:1) -- ++(330:1) -- ++(30:1) -- ++(150:1);
\filldraw[fill=mor,draw=black] (330:4) ++(270:2) ++(210:8) ++(30:1) -- ++(210:1) -- ++(330:1) -- ++(30:1) -- ++(150:1);
\filldraw[fill=mor,draw=black] (330:4) ++(270:2) ++(210:9) ++(30:1) -- ++(210:1) -- ++(330:1) -- ++(30:1) -- ++(150:1);
\filldraw[fill=mor,draw=black] (330:5) ++(270:2) ++(210:2) ++(30:1) -- ++(210:1) -- ++(330:1) -- ++(30:1) -- ++(150:1);
\filldraw[fill=mor,draw=black] (330:5) ++(270:2) ++(210:3) ++(30:1) -- ++(210:1) -- ++(330:1) -- ++(30:1) -- ++(150:1);
\filldraw[fill=mor,draw=black] (330:5) ++(270:2) ++(210:4) ++(30:1) -- ++(210:1) -- ++(330:1) -- ++(30:1) -- ++(150:1);
\filldraw[fill=mor,draw=black] (330:5) ++(270:2) ++(210:5) ++(30:1) -- ++(210:1) -- ++(330:1) -- ++(30:1) -- ++(150:1);
\filldraw[fill=mor,draw=black] (330:5) ++(270:2) ++(210:6) ++(30:1) -- ++(210:1) -- ++(330:1) -- ++(30:1) -- ++(150:1);
\filldraw[fill=mor,draw=black] (330:5) ++(270:2) ++(210:7) ++(30:1) -- ++(210:1) -- ++(330:1) -- ++(30:1) -- ++(150:1);
\filldraw[fill=mor,draw=black] (330:5) ++(270:2) ++(210:8) ++(30:1) -- ++(210:1) -- ++(330:1) -- ++(30:1) -- ++(150:1);
\filldraw[fill=mor,draw=black] (330:5) ++(270:2) ++(210:9) ++(30:1) -- ++(210:1) -- ++(330:1) -- ++(30:1) -- ++(150:1);
\filldraw[fill=mgr,draw=black] (210:2) ++(270:2) ++(330:7) -- ++(90:1) -- ++(210:1) -- ++(270:1) -- ++(30:1);
\filldraw[fill=mgr,draw=black] (210:3) ++(270:2) ++(330:7) -- ++(90:1) -- ++(210:1) -- ++(270:1) -- ++(30:1);
\filldraw[fill=mgr,draw=black] (210:4) ++(270:2) ++(330:7) -- ++(90:1) -- ++(210:1) -- ++(270:1) -- ++(30:1);
\filldraw[fill=mgr,draw=black] (210:5) ++(270:2) ++(330:7) -- ++(90:1) -- ++(210:1) -- ++(270:1) -- ++(30:1);
\filldraw[fill=mgr,draw=black] (210:6) ++(270:2) ++(330:7) -- ++(90:1) -- ++(210:1) -- ++(270:1) -- ++(30:1);
\filldraw[fill=mgr,draw=black] (210:7) ++(270:2) ++(330:7) -- ++(90:1) -- ++(210:1) -- ++(270:1) -- ++(30:1);
\filldraw[fill=mgr,draw=black] (210:8) ++(270:2) ++(330:7) -- ++(90:1) -- ++(210:1) -- ++(270:1) -- ++(30:1);
\filldraw[fill=mgr,draw=black] (210:9) ++(270:2) ++(330:7) -- ++(90:1) -- ++(210:1) -- ++(270:1) -- ++(30:1);
\filldraw[fill=mgr,draw=black] (210:3) ++(270:2) ++(330:8) -- ++(90:1) -- ++(210:1) -- ++(270:1) -- ++(30:1);
\filldraw[fill=mgr,draw=black] (210:4) ++(270:2) ++(330:8) -- ++(90:1) -- ++(210:1) -- ++(270:1) -- ++(30:1);
\filldraw[fill=mgr,draw=black] (210:5) ++(270:2) ++(330:8) -- ++(90:1) -- ++(210:1) -- ++(270:1) -- ++(30:1);
\filldraw[fill=mgr,draw=black] (210:6) ++(270:2) ++(330:8) -- ++(90:1) -- ++(210:1) -- ++(270:1) -- ++(30:1);
\filldraw[fill=mgr,draw=black] (210:7) ++(270:2) ++(330:8) -- ++(90:1) -- ++(210:1) -- ++(270:1) -- ++(30:1);
\filldraw[fill=mgr,draw=black] (210:8) ++(270:2) ++(330:8) -- ++(90:1) -- ++(210:1) -- ++(270:1) -- ++(30:1);
\filldraw[fill=mgr,draw=black] (210:9) ++(270:2) ++(330:8) -- ++(90:1) -- ++(210:1) -- ++(270:1) -- ++(30:1);
\filldraw[fill=mgr,draw=black] (210:10) ++(270:2) ++(330:8) -- ++(90:1) -- ++(210:1) -- ++(270:1) -- ++(30:1);
\filldraw[fill=mbr,draw=black] (330:7) ++(270:3) -- ++(330:1) -- ++(90:1) -- ++(150:1) -- ++(270:1);
\filldraw[fill=mbr,draw=black] (330:8) ++(270:3) -- ++(330:1) -- ++(90:1) -- ++(150:1) -- ++(270:1);
\filldraw[fill=mor,draw=black] (330:3) ++(270:7) ++(210:2) ++(30:4) -- ++(210:1) -- ++(330:1) -- ++(30:1) -- ++(150:1);
\filldraw[fill=mor,draw=black] (330:3) ++(270:7) ++(210:3) ++(30:4) -- ++(210:1) -- ++(330:1) -- ++(30:1) -- ++(150:1);
\filldraw[fill=mor,draw=black] (330:3) ++(270:7) ++(210:4) ++(30:4) -- ++(210:1) -- ++(330:1) -- ++(30:1) -- ++(150:1);
\filldraw[fill=mor,draw=black] (330:3) ++(270:7) ++(210:5) ++(30:4) -- ++(210:1) -- ++(330:1) -- ++(30:1) -- ++(150:1);
\filldraw[fill=mor,draw=black] (330:3) ++(270:7) ++(210:6) ++(30:4) -- ++(210:1) -- ++(330:1) -- ++(30:1) -- ++(150:1);
\filldraw[fill=mor,draw=black] (330:3) ++(270:7) ++(210:7) ++(30:4) -- ++(210:1) -- ++(330:1) -- ++(30:1) -- ++(150:1);
\filldraw[fill=mor,draw=black] (330:3) ++(270:7) ++(210:8) ++(30:4) -- ++(210:1) -- ++(330:1) -- ++(30:1) -- ++(150:1);
\filldraw[fill=mor,draw=black] (330:3) ++(270:7) ++(210:9) ++(30:4) -- ++(210:1) -- ++(330:1) -- ++(30:1) -- ++(150:1);
\filldraw[fill=mor,draw=black] (330:4) ++(270:7) ++(210:2) ++(30:4) -- ++(210:1) -- ++(330:1) -- ++(30:1) -- ++(150:1);
\filldraw[fill=mor,draw=black] (330:4) ++(270:7) ++(210:3) ++(30:4) -- ++(210:1) -- ++(330:1) -- ++(30:1) -- ++(150:1);
\filldraw[fill=mor,draw=black] (330:4) ++(270:7) ++(210:4) ++(30:4) -- ++(210:1) -- ++(330:1) -- ++(30:1) -- ++(150:1);
\filldraw[fill=mor,draw=black] (330:4) ++(270:7) ++(210:5) ++(30:4) -- ++(210:1) -- ++(330:1) -- ++(30:1) -- ++(150:1);
\filldraw[fill=mor,draw=black] (330:4) ++(270:7) ++(210:6) ++(30:4) -- ++(210:1) -- ++(330:1) -- ++(30:1) -- ++(150:1);
\filldraw[fill=mor,draw=black] (330:4) ++(270:7) ++(210:7) ++(30:4) -- ++(210:1) -- ++(330:1) -- ++(30:1) -- ++(150:1);
\filldraw[fill=mor,draw=black] (330:4) ++(270:7) ++(210:8) ++(30:4) -- ++(210:1) -- ++(330:1) -- ++(30:1) -- ++(150:1);
\filldraw[fill=mor,draw=black] (330:4) ++(270:7) ++(210:9) ++(30:4) -- ++(210:1) -- ++(330:1) -- ++(30:1) -- ++(150:1);
\filldraw[fill=mor,draw=black] (330:5) ++(270:7) ++(210:2) ++(30:4) -- ++(210:1) -- ++(330:1) -- ++(30:1) -- ++(150:1);
\filldraw[fill=mor,draw=black] (330:5) ++(270:7) ++(210:3) ++(30:4) -- ++(210:1) -- ++(330:1) -- ++(30:1) -- ++(150:1);
\filldraw[fill=mor,draw=black] (330:5) ++(270:7) ++(210:4) ++(30:4) -- ++(210:1) -- ++(330:1) -- ++(30:1) -- ++(150:1);
\filldraw[fill=mor,draw=black] (330:5) ++(270:7) ++(210:5) ++(30:4) -- ++(210:1) -- ++(330:1) -- ++(30:1) -- ++(150:1);
\filldraw[fill=mor,draw=black] (330:5) ++(270:7) ++(210:6) ++(30:4) -- ++(210:1) -- ++(330:1) -- ++(30:1) -- ++(150:1);
\filldraw[fill=mor,draw=black] (330:5) ++(270:7) ++(210:7) ++(30:4) -- ++(210:1) -- ++(330:1) -- ++(30:1) -- ++(150:1);
\filldraw[fill=mor,draw=black] (330:5) ++(270:7) ++(210:8) ++(30:4) -- ++(210:1) -- ++(330:1) -- ++(30:1) -- ++(150:1);
\filldraw[fill=mor,draw=black] (330:5) ++(270:7) ++(210:9) ++(30:4) -- ++(210:1) -- ++(330:1) -- ++(30:1) -- ++(150:1);
\filldraw[fill=mor,draw=black] (330:1) ++(90:2) ++(210:1) -- ++(210:1) -- ++(330:1) -- ++(30:1) -- ++(150:1);
\filldraw[fill=mor,draw=black] (330:2) ++(90:2) ++(210:1) -- ++(210:1) -- ++(330:1) -- ++(30:1) -- ++(150:1);
\filldraw[fill=mor,draw=black] (330:3) ++(90:2) ++(210:1) -- ++(210:1) -- ++(330:1) -- ++(30:1) -- ++(150:1);
\filldraw[fill=mor,draw=black] (330:4) ++(90:2) ++(210:1) -- ++(210:1) -- ++(330:1) -- ++(30:1) -- ++(150:1);
\filldraw[fill=mor,draw=black] (330:5) ++(90:0) ++(210:1) -- ++(210:1) -- ++(330:1) -- ++(30:1) -- ++(150:1);
\filldraw[fill=mor,draw=black] (330:6) ++(90:0) ++(210:1) -- ++(210:1) -- ++(330:1) -- ++(30:1) -- ++(150:1);
\filldraw[fill=mor,draw=black] (330:7) ++(90:0) ++(210:1) -- ++(210:1) -- ++(330:1) -- ++(30:1) -- ++(150:1);
\filldraw[fill=mor,draw=black] (330:8) ++(270:2) ++(210:1) -- ++(210:1) -- ++(330:1) -- ++(30:1) -- ++(150:1);
\filldraw[fill=mor,draw=black] (330:9) ++(270:2) ++(210:1) -- ++(210:1) -- ++(330:1) -- ++(30:1) -- ++(150:1);
\filldraw[fill=mbr,draw=black] (150:1) ++(270:1) -- ++(330:1) -- ++(90:1) -- ++(150:1) -- ++(270:1);
\filldraw[fill=mbr,draw=black] (270:1) -- ++(330:1) -- ++(90:1) -- ++(150:1) -- ++(270:1);
\filldraw[fill=mbr,draw=black] (330:1) ++(270:1) -- ++(330:1) -- ++(90:1) -- ++(150:1) -- ++(270:1);
\filldraw[fill=mbr,draw=black] (330:2) ++(270:1) -- ++(330:1) -- ++(90:1) -- ++(150:1) -- ++(270:1);
\filldraw[fill=mbr,draw=black] (330:2) ++(270:2) -- ++(330:1) -- ++(90:1) -- ++(150:1) -- ++(270:1);
\filldraw[fill=mbr,draw=black] (330:2) ++(270:3) -- ++(330:1) -- ++(90:1) -- ++(150:1) -- ++(270:1);
\filldraw[fill=mbr,draw=black] (330:3) ++(270:3) -- ++(330:1) -- ++(90:1) -- ++(150:1) -- ++(270:1);
\filldraw[fill=mbr,draw=black] (330:4) ++(270:3) -- ++(330:1) -- ++(90:1) -- ++(150:1) -- ++(270:1);
\filldraw[fill=mbr,draw=black] (330:5) ++(270:3) -- ++(330:1) -- ++(90:1) -- ++(150:1) -- ++(270:1);
\filldraw[fill=mbr,draw=black] (330:5) ++(270:4) -- ++(330:1) -- ++(90:1) -- ++(150:1) -- ++(270:1);
\filldraw[fill=mbr,draw=black] (330:5) ++(270:5) -- ++(330:1) -- ++(90:1) -- ++(150:1) -- ++(270:1);
\filldraw[fill=mbr,draw=black] (330:6) ++(270:5) -- ++(330:1) -- ++(90:1) -- ++(150:1) -- ++(270:1);
\filldraw[fill=mbr,draw=black] (330:7) ++(270:5) -- ++(330:1) -- ++(90:1) -- ++(150:1) -- ++(270:1);
\filldraw[fill=mgr,draw=black] (210:1) ++(90:1) ++(330:5) -- ++(90:1) -- ++(210:1) -- ++(270:1) -- ++(30:1);
\filldraw[fill=mgr,draw=black] (210:1) ++(90:0) ++(330:5) -- ++(90:1) -- ++(210:1) -- ++(270:1) -- ++(30:1);
\filldraw[fill=mgr,draw=black] (210:1) ++(270:1) ++(330:8) -- ++(90:1) -- ++(210:1) -- ++(270:1) -- ++(30:1);
\filldraw[fill=mgr,draw=black] (210:1) ++(270:2) ++(330:8) -- ++(90:1) -- ++(210:1) -- ++(270:1) -- ++(30:1);
\filldraw[fill=mgr,draw=black] (210:1) ++(270:3) ++(330:10) -- ++(90:1) -- ++(210:1) -- ++(270:1) -- ++(30:1);
\end{tikzpicture}
\caption{Periodic plane partition corresponding to level $5$ excited state in vacuum representation of Lee-Yang model }
\label{leeyangdelta0lvl5}
\end{minipage}
\end{figure}

This minimal model has two primaries, one with $\Delta=0$ and one with $\Delta=-\frac{1}{5}$. Recall that if we specialize parameters of $\mathcal{W}_\infty$ by imposing a condition of the form $\lambda_3 = N$, corresponding to truncation to $Y_{00N}$ algebra in the notation of \cite{Gaiotto:2017euk,Prochazka:2018tlo}, the plane partitions corresponding to states in the lowest weight representations are truncated in the direction associated to $\lambda_3$ so that there are at most $N$ layers of boxes \cite{Prochazka:2015deb}. On the other hand, if we impose two such conditions in two directions (as we should do in $(A_{N_1-1},A_{N_2-1})$ Argyres-Douglas minimal model), the irreducible representations have more states than if we naively imposed both truncation conditions at the same time. Actually, what happens in situations like this is that we effectively develop a periodic direction in the space of plane partitions, i.e. the plane partitions should be drawn on a cylinder. The periodic plane partition corresponding to Lee-Yang $\Delta=0$ lowest weight state is illustrated in Figure \ref{leeyangdelta0}. Counting of states in $\Delta=0$ irreducible representation is equivalent to counting of periodic configurations of boxes that can be stacked on top of such ground state. We can notice several features of this box counting problem. First of all, starting from the corner, we can stack horizontally at most three boxes along the brown wall. The fourth box in this direction would not be supported from below. This corresponds to $\mathcal{W}_3$ algebra truncation with null state at level four (corresponding to absent spin $4$ generator). Analogously, starting from an empty configuration, we can only put two boxes on top of each other vertically. The third box would not be supported by anything from behind so it does not satisfy the plane partition rule. This corresponds to truncation of $\mathcal{W}_\infty$ to Virasoro algebra (and absence of spin $3$ generator of the algebra). But imposing simultaneously two conditions of this type leads to periodicity and in turn to a relaxation of these rules. We see that it we consider an L-shaped configuration of $4$ boxes with three boxes stacked horizontally and 2 vertically, this configuration admits the addition of fourth box in the horizontal direction (or which is the same by periodicity, third box in the vertical direction). Therefore, there is an additional state at level $5$ that would not be allowed if we would naively impose both truncation conditions, which would effectively restrict the growth of the plane partitions to $2 \times 3$ rectangle. The configuration of boxes corresponding to this level $5$ state in the vacuum representation of Lee-Yang is illustrated in Figure \ref{leeyangdelta0lvl5}.

The fact that the characters of irreducible representations count exactly such periodic plane partitions applies to all $\mathcal{W}_N$ minimal models and all their allowed primaries. The choice of a specific minimal model controls the periodicity of the corresponding plane partitions. On the other hand, the choice of a primary is related to the choice of non-trivial asymptotics along non-periodic directions (there are generically two such directions for $\mathcal{W}_N$ minimal models but only one such direction in the special case of Argyres-Douglas minimal models). In Figure \ref{leeyangdelta15} we show the configuration corresponding to the primary field $\Delta = -\frac{1}{5}$ of the Lee-Yang model. It can be obtained from the configuration of figure \ref{leeyangdelta0} by adding an infinite column of boxes. The periodicity conditions are not affected. Figure \ref{isingdelta0} on the other hand illustrates the periodic configuration corresponding to the vacuum representation of Virasoro $c=\frac{1}{2}$ model, the Ising model. We see that the periodic stairs are offset relative to Lee-Yang model. The fact that this model is a Virasoro minimal model can be seen from the fact that the stairs have height $2$. On the other hand, Ising model just like all the first unitary minimal models of $\mathcal{W}_N$ series is also a representation of $Y_{012}$ truncation of $\mathcal{W}_\infty$. Such truncations are characterized by having a null vector at level $6$ in the vacuum representation and we can clearly see this null vector by presence of a \emph{pit} at coordinates $(2,3)$ in the horizontal plane \cite{bershtein2018plane,Prochazka:2018tlo}.

The characters of irreducible representations have particularly simple form in the case of minimal models of Argyres-Douglas type $(A_{N_1-1},A_{N_2-1})$ corresponding to minimal models of simultaneously $\mathcal{W}_{N_1}$ and $\mathcal{W}_{N_2}$ algebras \cite{Fukuda:2015ura,Prochazka:2015deb} (we always assume that $\gcd(N_1,N_2)=1$). In this case, a non-trivial asymptotic is possible only in one of the directions and the allowed primaries are thus labeled by Young diagrams that fit in asymptotic $N_1 \times N_2$ rectangle. Since there are
\begin{equation}
{N_1 + N_2 \choose N_1} = {N_1 + N_2 \choose N_2}
\end{equation}
such Young diagrams, we would expect the same number of primaries. But not all of these give inequivalent representations of $\mathcal{W}_\infty$. Whenever there is a full row or column (of length $N_1$ or $N_2$), we can remove it by a shift of the spectral parameter (which affects only the zero mode of the Heisenberg algebra). There are $N_1+N_2$ such identifications of primaries so in total there are
\begin{equation}
\frac{(N_1+N_2-1)!}{N_1! N_2!}
\end{equation}
inequivalent primaries. The irreducible characters of the corresponding $\mathcal{W}_{1+\infty}$ representation can be read off from the asymptotic Young diagram $\lambda$ as follows: the character is given by the hook length formula
\begin{equation}
\frac{q^{\Delta-\frac{c}{24}} (q;q)_\infty}{(q^{N_1+N_2};q^{N_1+N_2})_\infty} \prod_{\Box\in\lambda} \frac{1}{(q^{\hook(\Box)};q^{N_1+N_2})_\infty}.
\end{equation}
where $\Delta$ is the conformal dimension of the primary which can be combinatorially determined as explained for example in \cite{Prochazka:2015deb}. Furthermore
\begin{equation}
(a;q)_\infty = \prod_{j=0}^\infty (1-a q^j)
\end{equation}
is the $q$-Pochhammer symbol and $\hook(\Box)$ is the (periodic) hook length corresponding to the box $\Box$ as illustrated in the following diagram for $(A_1,A_2)$ minimal model:
\begin{center}
\begin{tikzpicture}[scale=0.50]
\filldraw[fill=yellow!60] (0,1) rectangle (1,2);
\filldraw[fill=yellow!60] (4,1) rectangle (5,2);
\filldraw[fill=yellow!60] (5,1) rectangle (6,2);
\filldraw[fill=yellow!60] (4,0) rectangle (5,1);
\filldraw[fill=yellow!60] (8,1) rectangle (9,2);
\filldraw[fill=yellow!60] (9,1) rectangle (10,2);
\filldraw[fill=yellow!60] (10,1) rectangle (11,2);
\filldraw[fill=yellow!60] (8,0) rectangle (9,1);
\filldraw[fill=yellow!60] (9,0) rectangle (10,1);
\filldraw[fill=yellow!60] (12,1) rectangle (13,2);
\filldraw[fill=yellow!60] (13,1) rectangle (14,2);
\filldraw[fill=yellow!60] (16,1) rectangle (17,2);
\filldraw[fill=yellow!60] (17,1) rectangle (18,2);
\filldraw[fill=yellow!60] (18,1) rectangle (19,2);
\filldraw[fill=yellow!60] (16,0) rectangle (17,1);
\filldraw[fill=yellow!60] (4,5) rectangle (5,6);
\filldraw[fill=yellow!60] (4,4) rectangle (5,5);
\filldraw[fill=yellow!60] (8,5) rectangle (9,6);
\filldraw[fill=yellow!60] (9,5) rectangle (10,6);
\filldraw[fill=yellow!60] (8,4) rectangle (9,5);
\filldraw[fill=yellow!60] (9,4) rectangle (10,5);
\filldraw[fill=yellow!60] (12,5) rectangle (13,6);
\filldraw[fill=yellow!60] (13,5) rectangle (14,6);
\filldraw[fill=yellow!60] (14,5) rectangle (15,6);
\filldraw[fill=yellow!60] (12,4) rectangle (13,5);
\filldraw[fill=yellow!60] (13,4) rectangle (14,5);
\filldraw[fill=yellow!60] (14,4) rectangle (15,5);
\filldraw[fill=yellow!60] (16,5) rectangle (17,6);
\filldraw[fill=yellow!60] (17,5) rectangle (18,6);
\filldraw[fill=yellow!60] (18,5) rectangle (19,6);
\draw[help lines] (0,4) grid (3,6);
\draw[thick,red] (3,6) -- (0,6) -- (0,4);
\draw[help lines] (4,4) grid (7,6);
\draw[thick,red] (7,6) -- (5,6) -- (5,4) -- (4,4);
\draw[help lines] (8,4) grid (11,6);
\draw[thick,red] (11,6) -- (10,6) -- (10,4) -- (8,4);
\draw[help lines] (12,4) grid (15,6);
\draw[thick,red] (15,6) -- (15,4) -- (12,4);
\draw[help lines] (16,4) grid (19,6);
\draw[thick,red] (19,6) -- (19,5) -- (16,5) -- (16,4);
\draw[help lines] (0,0) grid (3,2);
\draw[thick,red] (3,2) -- (1,2) -- (1,1) -- (0,1) -- (0,0);
\draw[help lines] (4,0) grid (7,2);
\draw[thick,red] (7,2) -- (6,2) -- (6,1) -- (5,1) -- (5,0) -- (4,0);
\draw[help lines] (8,0) grid (11,2);
\draw[thick,red] (11,2) -- (11,1) -- (10,1) -- (10,0) -- (8,0);
\draw[help lines] (12,0) grid (15,2);
\draw[thick,red] (15,2) -- (14,2) -- (14,1) -- (12,1) -- (12,0);
\draw[help lines] (16,0) grid (19,2);
\draw[thick,red] (19,2) -- (19,1) -- (17,1) -- (17,0) -- (16,0);
\node at (0.5,0.5) {$1$};
\node at (1.5,0.5) {$3$};
\node at (2.5,0.5) {$4$};
\node at (0.5,1.5) {$4$};
\node at (1.5,1.5) {$1$};
\node at (2.5,1.5) {$2$};
\node at (4.5,0.5) {$4$};
\node at (5.5,0.5) {$1$};
\node at (6.5,0.5) {$3$};
\node at (4.5,1.5) {$2$};
\node at (5.5,1.5) {$4$};
\node at (6.5,1.5) {$1$};
\node at (8.5,0.5) {$3$};
\node at (9.5,0.5) {$4$};
\node at (10.5,0.5) {$1$};
\node at (8.5,1.5) {$1$};
\node at (9.5,1.5) {$2$};
\node at (10.5,1.5) {$4$};
\node at (12.5,0.5) {$1$};
\node at (13.5,0.5) {$2$};
\node at (14.5,0.5) {$4$};
\node at (12.5,1.5) {$3$};
\node at (13.5,1.5) {$4$};
\node at (14.5,1.5) {$1$};
\node at (16.5,0.5) {$4$};
\node at (17.5,0.5) {$1$};
\node at (18.5,0.5) {$2$};
\node at (16.5,1.5) {$1$};
\node at (17.5,1.5) {$3$};
\node at (18.5,1.5) {$4$};
\node at (0.5,4.5) {$2$};
\node at (1.5,4.5) {$3$};
\node at (2.5,4.5) {$4$};
\node at (0.5,5.5) {$1$};
\node at (1.5,5.5) {$2$};
\node at (2.5,5.5) {$3$};
\node at (4.5,4.5) {$4$};
\node at (5.5,4.5) {$2$};
\node at (6.5,4.5) {$3$};
\node at (4.5,5.5) {$3$};
\node at (5.5,5.5) {$1$};
\node at (6.5,5.5) {$2$};
\node at (8.5,4.5) {$3$};
\node at (9.5,4.5) {$4$};
\node at (10.5,4.5) {$2$};
\node at (8.5,5.5) {$2$};
\node at (9.5,5.5) {$3$};
\node at (10.5,5.5) {$1$};
\node at (12.5,4.5) {$2$};
\node at (13.5,4.5) {$3$};
\node at (14.5,4.5) {$4$};
\node at (12.5,5.5) {$1$};
\node at (13.5,5.5) {$2$};
\node at (14.5,5.5) {$3$};
\node at (16.5,4.5) {$1$};
\node at (17.5,4.5) {$2$};
\node at (18.5,4.5) {$3$};
\node at (16.5,5.5) {$2$};
\node at (17.5,5.5) {$3$};
\node at (18.5,5.5) {$4$};
\node at (3.5,1) {$\simeq$};
\node at (7.5,1) {$\simeq$};
\node at (11.5,1) {$\simeq$};
\node at (15.5,1) {$\simeq$};
\node at (3.5,5) {$\simeq$};
\node at (7.5,5) {$\simeq$};
\node at (11.5,5) {$\simeq$};
\node at (15.5,5) {$\simeq$};
\node at (-2,1) {$\Delta=-\frac{1}{5}:$};
\node at (-2,5) {$\Delta=0:$};
\end{tikzpicture}
\end{center}
The yellow boxes form the corresponding asymptotic Young diagram $\lambda$ and they are separated from white boxes by the red line which is the projection of the asymptotic staircase. Given a box, its hook length is given by the number of boxes to the left (the arm length) plus the number of boxes above (the leg length) plus one (for the box itself). We count the boxes until we reach the red line, if needed periodically extending from the left to the right and from the top to the bottom. We see that all the Young diagrams related by $N_1+N_2$ identifications have the same collection of hook lengths. In the case of $(A_1,A_2)$ minimal model we see that
\begin{align}
\chi_{\Delta=0} & = \frac{q^{11/60}}{(q^2;q^5)_\infty (q^3;q^5)_\infty} \simeq q^{11/60}(1+q^2+q^3+q^4+q^5+2q^6+\ldots) \\
\chi_{\Delta=-\frac{1}{5}} & = \frac{q^{-1/60}}{(q;q^5)_\infty (q^4;q^5)_\infty} \simeq q^{-1/60} (1+q+q^2+q^3+2q^4+2q^5+3q^6+\ldots).
\end{align}
These functions count the states in the corresponding irreducible $\mathcal{W}_\infty$ representations. To get the counting of states in $\mathcal{W}_{1+\infty}$, we should multiply these by the Heisenberg contribution $(q;q)_\infty^{-1}$ obtaining expansions
\begin{align}
\frac{1}{\prod_{n=1}^\infty (1-q^n)} \chi_{\Delta=0} & \sim 1+q+3q^2+5q^3+9q^4+14q^5+\ldots \\
\frac{1}{\prod_{n=1}^\infty (1-q^n)} \chi_{\Delta=-\frac{1}{5}} & \sim 1+2q+4q^2+7q^3+13q^4+21q^5+\ldots
\end{align}
which will be useful in the following. As another example, for $W_3 \cap W_4$ minimal model (corresponding to $(A_2,A_3)$ Argyres-Douglas with $c_{\infty}=-\frac{114}{7}$) the primary $\Delta=-\frac{4}{7}$ with asymptotic $\ydiagram{2}$ has hook lengths
\begin{center}
\begin{tikzpicture}[scale=0.50]
\filldraw[fill=yellow!60] (0,2) rectangle (1,3);
\filldraw[fill=yellow!60] (1,2) rectangle (2,3);
\draw[help lines] (0,0) grid (4,3);
\draw[thick,red] (0,0) -- (0,2) -- (2,2) -- (2,3) -- (4,3);
\node at (0.5,0.5) {$2$};
\node at (1.5,0.5) {$3$};
\node at (2.5,0.5) {$5$};
\node at (3.5,0.5) {$6$};
\node at (0.5,1.5) {$1$};
\node at (1.5,1.5) {$2$};
\node at (2.5,1.5) {$4$};
\node at (3.5,1.5) {$5$};
\node at (0.5,2.5) {$5$};
\node at (1.5,2.5) {$6$};
\node at (2.5,2.5) {$1$};
\node at (3.5,2.5) {$2$};
\end{tikzpicture}
\end{center}
corresponding to $\mathcal{W}_\infty$ character
\begin{equation}
\frac{q^{3/28}}{(q;q^7)_\infty (q^2;q^7)_\infty^2 (q^5;q^7)_\infty^2 (q^6;q^7)_\infty}.
\end{equation}

\subsubsection{Solutions of Bethe ansatz equations for $\Delta=-\frac{1}{5}$}
Now that we know enough information about the representations and their characters, we can have a look at the corresponding solutions of Bethe ansatz equations \eqref{betheequations}. First of all we make a choice of the Nekrasov-like parameters of the algebra. Let us choose
\begin{equation}
\epsilon_1 = 2, \qquad \epsilon_2 = 3, \qquad \epsilon_3 = -5
\end{equation}
with $\psi_0 = \frac{1}{5}$. The corresponding $\lambda$-parameters are
\begin{equation}
\lambda_1 = 3, \qquad \lambda_2 = 2, \qquad \lambda_3 = -\frac{6}{5}
\end{equation}
and the $\mathcal{W}_\infty$ central charge is
\begin{equation}
c_\infty = (\lambda_1-1)(\lambda_2-1)(\lambda_3-1) = -\frac{22}{5}
\end{equation}
which is exactly the central charge of the non-unitary Lee-Yang model. As for the function parametrizing the lowest weights, we choose
\begin{equation}
A(u) = \frac{(u-5)(u-6)}{(u-2)(u-3)}.
\end{equation}
As discussed in \cite{Prochazka:2018tlo}, this can be done from the point of view of second asymptotic direction, i.e. Virasoro algebra ($N \leftrightarrow \lambda_2$) where we would write
\begin{equation}
\label{leeyangfact1}
\frac{(u-2-\epsilon_2)(u-3-\epsilon_2)}{(u-2)(u-3)} = \frac{(u-5)(u-6)}{(u-2)(u-3)}
\end{equation}
or from the point of view of first direction, i.e. $\mathcal{W}_3$ algebra ($N \leftrightarrow \lambda_1$) where the factorization is instead
\begin{equation}
\label{leeyangfact2}
\frac{(u-2-\epsilon_1)(u-3-\epsilon_1)(u-4-\epsilon_1)}{(u-2)(u-3)(u-4)} = \frac{(u-5)(u-6)}{(u-2)(u-3)}.
\end{equation}
The requirement of factorizability both as \eqref{leeyangfact1} and \eqref{leeyangfact2} (i.e. corresponding to both truncations) determines all allowed primaries \cite{Prochazka:2018tlo}. In our case there are only two of them up to constant shifts of $u$-parameter (which does not change the $\mathcal{W}_\infty$ representation but changes the lowest weight with respect to Heisenberg algebra).

\paragraph{Level 1}
At level $1$, there is a single Bethe ansatz equation
\begin{equation}
q \frac{(x_1-5)(x_1-6)}{(x_1-2)(x_1-3)} = 1
\end{equation}
with two solutions
\begin{equation}
\frac{5-11q +\sqrt{1+34q+q^2}}{2(1-q)} \simeq 3 + 6q - 66q^2 + 1158q^3 + \ldots
\end{equation}
and
\begin{equation}
\frac{5-11q -\sqrt{1+34q+q^2}}{2(1-q)} \simeq 2 - 12q + 60q^2 - 1164q^3 + \ldots.
\end{equation}
We see that in the Yangian limit $q \to 0$ the solutions are analytic and the Bethe roots approach the weighted coordinates of positions where we can put the boxes: choosing the corner of the vacuum periodic plane partition to be at coordinates $(0,0,0)$, in the $\Delta=-\frac{1}{5}$ we can put the first box either at coordinates $(1,0,0)$ or $(0,1,0)$. The weighted coordinates of these are $\epsilon_1 = 2$ and $\epsilon_2 = 3$ which are exactly the limiting values of the Bethe roots as $q \to 0$. We could similarly study the behaviour as $q \to \infty$ where we would find the conjugate Yangian prediction for these limiting values.

In the local limit $q \to 1$, we have
\begin{equation}
\frac{5-11q +\sqrt{1+34q+q^2}}{2(1-q)} \simeq 4 - \frac{1}{3}(1-q) - \frac{1}{6} (1-q)^2 + \ldots
\end{equation}
and
\begin{equation}
\frac{5-11q -\sqrt{1+34q+q^2}}{2(1-q)} \simeq -\frac{6}{1-q} + 7 + \frac{1}{3}(1-q) + \frac{1}{6} (1-q)^2 + \ldots.
\end{equation}
We see that the first of these is regular as $q \to 1$ and so it corresponds to $\mathcal{W}_\infty$ excitation. On the other hand, the second solution has a simple pole as $q \to 1$ with residue given by $-\psi_0 \epsilon_1 \epsilon_2 \epsilon_3$ so it corresponds to a Heisenberg excitation in line with the general discussion in section \ref{seclocal}.

Finally, we have in addition two branch points in the $q$-plane at points
\begin{equation}
q = -17 \pm 12 \sqrt{2}
\end{equation}
(note that these are inverse of one another as required by charge conjugation). These two points both lie on the negative real axis and as we continue the solutions of Bethe equations around them, the solutions get exchanged. This shows that the two sheets corresponding to two different states at level $1$ actually connect as we analytically continue in $q$-parameter space, so there is a single connected Riemann surface covering the $q$-plane which captures both Bethe roots at this level. We also see that the excitation purely in Heisenberg subalgebra at $q = 1$ can be transformed to purely $\mathcal{W}_\infty$ excitation by analytic continuation in the twist parameter $q$ (and vice versa), so for generic values of $q$ we cannot expect decoupling of these degrees of freedom.

Note also that for $q$ equal to one of these two branch points the corresponding eigenstates of ILW Hamiltonians have identical spectrum of all the conserved charges. In the case of level $1$ states, the fact that by tuning one complex parameter $q$ we can make two Bethe roots agree is not so surprising, but as we will see such phenomenon persists also at higher excited levels although generically we would expect these to be of a higher codimension.

\paragraph{Level 2}
At the second level, each solution of BAE is given by a pair of Bethe roots $(x_1,x_2)$ solving a system of equations
\begin{align}
1 & = q \frac{(x_1-5)(x_1-6)}{(x_1-2)(x_1-3)} \frac{(x_1-x_2+\epsilon_1)(x_1-x_2+\epsilon_2)(x_1-x_2+\epsilon_3)}{(x_1-x_2-\epsilon_1)(x_1-x_2-\epsilon_2)(x_1-x_2-\epsilon_3)} \\
1 & = q \frac{(x_2-5)(x_2-6)}{(x_2-2)(x_2-3)} \frac{(x_2-x_1+\epsilon_1)(x_2-x_1+\epsilon_2)(x_2-x_1+\epsilon_3)}{(x_2-x_1-\epsilon_1)(x_2-x_1-\epsilon_2)(x_2-x_1-\epsilon_3)}.
\end{align}
We see that we have a system of two equations of fifth degree for two unknowns which is much more complicated than a single quadratic equation that we were solving previously. There are various possible approaches to solve these equations.

The most naive one is to use Newton's method to find numerically an approximate solution starting from a random initial seed. In this way we find generically $16$ solutions. Since the equations themselves are invariant under the exchange $x_1 \leftrightarrow x_2$, the solutions will also form either pairs of solutions related by this exchange or solutions with $x_1 = x_2$. From the numerics we find three groups of solutions that do not mix among each other as we analytically continue in $q$. The first group are three solutions
\begin{equation}
(2,5), \qquad (3,5) \qquad (3,6)
\end{equation}
which are $q$-independent (and can be associated to null states). Next there are two (non-physical) degenerate solutions with $x_1 = x_2$ which are given by
\begin{equation}
\label{leeyanglevel2unphysical}
x_1 = x_2 = \frac{5+11q \pm \sqrt{1-34q+q^2}}{2(1+q)}.
\end{equation}
Finally we have four solutions which are $q$-dependent and which have $x_1 \neq x_2$ for generic $q$. We have no explicit formula for these eight Bethe roots, but it is precisely these solutions that correspond to four states in the irreducible representation at level $2$ and this is why in the following we will mainly focus on these. The Bethe roots corresponding to states in irreducible representations of $\mathcal{W}_{1+\infty}$ will be called \emph{physical} in the following.

We can also try to learn something about the solutions analytically. We can eliminate one of the variables using the resultant of the two Bethe equations. The resultant of BAE with respect to $x_2$ is a polynomial of degree $16$ in $x_1$ (with coefficients still depending on $q$). This polynomial has linear factors corresponding to the integer roots $x_1 \in (2,3,5,6)$ with multiplicity $1$ or $2$. Next we have a quadratic factor
\begin{equation}
(1+q)x_1^2 - (5+11q)x_1 + 6+30q
\end{equation}
whose solutions are the degenerate roots \eqref{leeyanglevel2unphysical}. Finally, there is an irreducible degree $8$ polynomial
\begin{multline}
\label{leeyanglevel2degree8}
(q-1)^5 (q+1)^2 x_1^8 -(q-1)^4 (q+1) \left(53 q^2+18 q-11\right) x_1^7 \\
+(q-1)^3 \left(1207 q^4+828 q^3-446 q^2-180 q+31\right) x_1^6 \\
-(q-1)^2 \left(15443 q^5+571 q^4-8062 q^3+226 q^2+395 q+67\right) x_1^5 \\
+4 (q-1) \left(30388 q^6-18560 q^5-10357 q^4+4678 q^3+53 q^2+400 q-122\right) x_1^4 \\
-4 \left(150836 q^7-190341 q^6+25286 q^5+24859 q^4-4261 q^3+2758 q^2-1509 q+148\right) x_1^3 \\
+48 \left(38460 q^7-35489 q^6+3469 q^5+1592 q^4-674 q^3+479 q^2-41 q-20\right) x_1^2 \\
-144 \left(22125 q^7-13310 q^6+583 q^5+1020 q^4-756 q^3+119 q^2+66 q-19\right) x_1 \\
+1728 \left(1375 q^7-425 q^6-55 q^5+132 q^4-60 q^3+q^2+5 q-1\right)
\end{multline}
whose roots are the $4$ pairs of non-degenerate $q$-dependent Bethe roots that we identify with the physical states. Solving this degree $8$ equation for $x_1$ determines $x_1$ as a function of $q$. Finally by calculating the greatest common divisor with respect to $x_2$ of Bethe ansatz equations reduces them to a linear equation for $x_2$ which allows us to uniquely determine $x_2$ in terms of $x_1$ and $q$. The formula is
\begin{align}
\label{leeyangnumsecondroot}
\nonumber
x_2 & = \frac{N}{D_1 D_2} \\
\nonumber
N & = (q-1)^3 (q+1)^2 x_1^7 -(q-1)^2 (q+1) \left(25 q^2+6 q-7\right) x_1^6 \\
\nonumber
& +3 (q-1) \left(43 q^4+36 q^3-2 q^2-12 q-5\right) x_1^5 \\
\nonumber
& +\left(1993 q^5-2059 q^4-2186 q^3+1958 q^2+373 q-295\right) x_1^4 \\
\nonumber
& -2 \left(15839 q^5-10325 q^4-10612 q^3+5956 q^2+1037 q-599\right) x_1^3 \\
\nonumber
& +12 \left(15315 q^5-7279 q^4-6216 q^3+2648 q^2+333 q-193\right) x_1^2 \\
\nonumber
& -72 \left(6975 q^5-2405 q^4-1536 q^3+560 q^2+37 q-31\right) x_1 \\
& +864 (5 q+1) \left(125 q^4-55 q^3-2 q^2+5 q-1\right) \\
\nonumber
D_1 & = (q-1) x_1^2+(5-11 q) x_1+6 (5 q-1) \\
\nonumber
D_2 & = \left(1-q^2\right)^2 x_1^4+\left(1-q^2\right) \left(25 q^2-7\right) x_1^3+4 \left(58 q^4-41 q^2+4\right) x_1^2 \\
\nonumber
& -12 \left(79 q^4-24 q^2+1\right) x_1+144 q^2 \left(10 q^2-1\right).
\end{align}
We see that it is a complicated yet very explicit map which we can use to find the second Bethe root once we know the first one (for example from the resultant), i.e. we can use this map to pair solutions of \eqref{leeyanglevel2degree8} to form solutions of BAE. By construction the corresponding transformation of Bethe roots is involutive as it must be since the Bethe roots appear in pairs.

In the Yangian limit $q \to 0$, the zeros of \eqref{leeyanglevel2degree8} can found using ordinary power series: the eight solutions are
\begin{align}
\nonumber
& \begin{cases} -3 - \frac{18}{7}q - \frac{55458}{1715}q^2 + \frac{6113358}{420175}q^3 + \mathcal{O}(q^4) \\
+2 - \frac{144}{5}q^2 + \frac{1692}{175}q^3 + \mathcal{O}(q^4)
\end{cases} \\
\nonumber
& \begin{cases} -2 - 3q + \frac{249}{20}q^2 - \frac{4839}{200}q^3 + \mathcal{O}(q^4) \\
+3 + \frac{84}{5}q^2 - \frac{459}{50}q^3 + \mathcal{O}(q^4)
\end{cases} \\
& \begin{cases} +2 + 3q + \frac{555}{4}q^2 + \frac{22623}{8}q^3 + \mathcal{O}(q^4) \\
+3 - 24q + 24q^2 - \frac{15789}{2}q^3 + \mathcal{O}(q^4)
\end{cases} \\
\nonumber
& \begin{cases} +2 - 12q^2 + \frac{2160}{7}q^3 + \mathcal{O}(q^4) \\
+4 + \frac{60}{7}q - \frac{55176}{343}q^2 + \frac{79819980}{16807}q^3 + \mathcal{O}(q^4)
\end{cases}
\end{align}
The eight roots of the resultant were paired into $4$ solutions of Bethe ansatz equations using equation \eqref{leeyangnumsecondroot} which gives us the second Bethe root once we know the first one. We can identify the Yangian $q \to 0$ limit of these solutions with configurations of boxes in periodic plane partitions. The positions in which we can put two boxes and the corresponding Yangian Bethe roots are
\begin{align}
(\epsilon_1,2\epsilon_1) = (2,4) \\
(\epsilon_1,\epsilon_1+\epsilon_3) = (2,-3) \\
(\epsilon_2,\epsilon_2+\epsilon_3) = (3,-2) \\
(\epsilon_1,\epsilon_2) = (2,3)
\end{align}
which are exactly the $q \to 0$ limits of the perturbative solutions that we found. A similar analysis would apply to $q \to \infty$ where we would find a very similar picture corresponding to a conjugate representation (which in this case is the same one) up to a spectral shift.

\begin{figure}
\centering
\begin{subfigure}{0.45\textwidth}
\includegraphics[width=\textwidth]{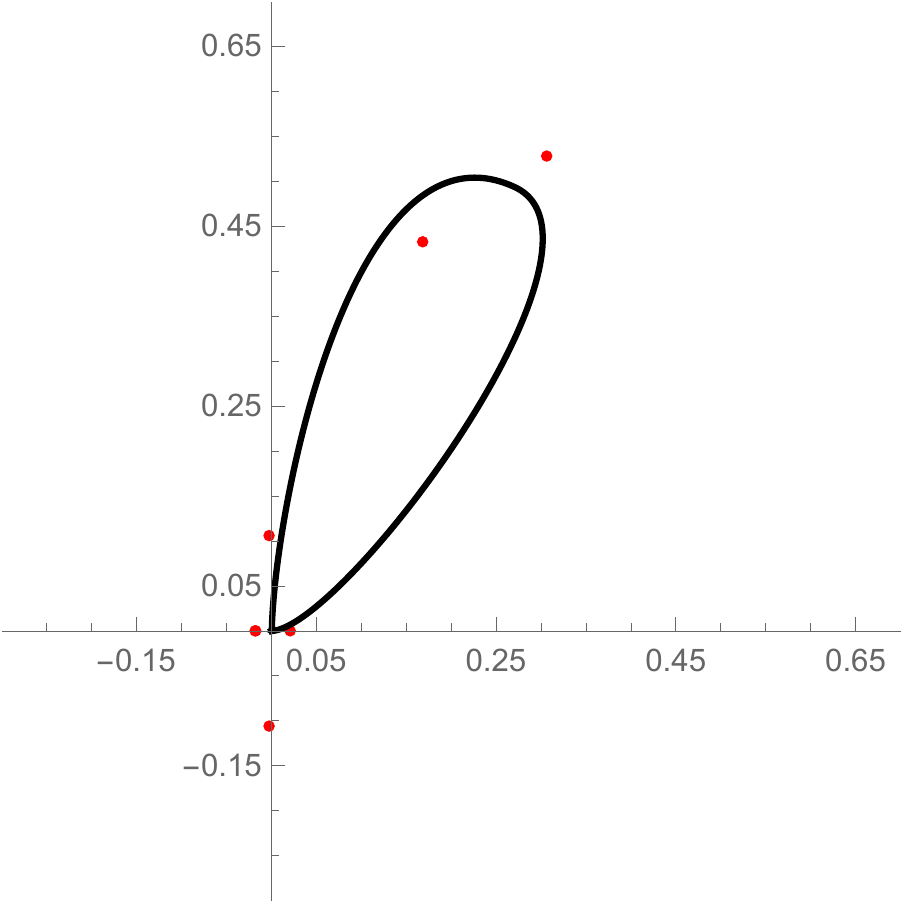}
\end{subfigure}
\qquad
\begin{subfigure}{0.45\textwidth}
\includegraphics[width=\textwidth]{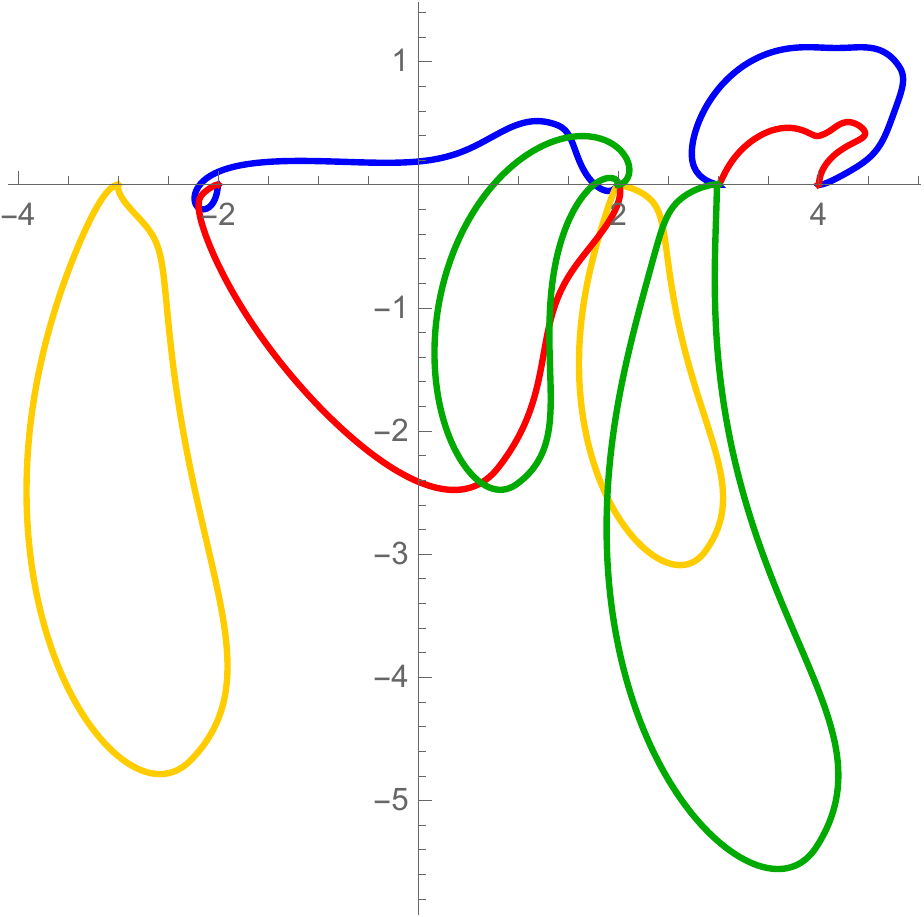}
\end{subfigure}
\caption{An illustration of non-trivial homotopy of solutions of Bethe ansatz equations around $q = q_* \sim 0.168469+0.432772i$. The left figure shows the curve in the $q$-plane. The red dots are zeros of the discriminant so around these points the monodromy is potentially non-trivial. The curve on the left encloses only one such singular point. The figure on the right shows how four pairs of Bethe roots evolve in the complex plane. We start at $q=0$ with Yangian solutions blue $\rightarrow (2,4)$, red $\rightarrow (-2,3)$, yellow $\rightarrow (2,-3)$ and green $\rightarrow (2,3)$ and go around $q = q_*$ and return back to $q=0$. We see that while the yellow and green Bethe roots go back to their original values, the red and green Bethe roots are instead pairwise exchanged.}
\label{fighomotopy}
\end{figure}

We can also use the analytic techniques to find points in $q$-plane around which the solutions have non-trivial monodromy, i.e. points in $q$-plane such that analytically continuing around them allows us to go from one solution of Bethe ansatz equations to another. First of all, the solutions of the three groups of solutions cannot mix under the monodromy because this would not be compatible with analytic continuation in $q$ (i.e. analytic continuation around a point in $q$-plane of a solution with non-trivial $q$-dependence cannot produce $q$-independent solution and analytic continuation of solution satisfying $x_1 = x_2$ will never result in solution that does not satisfy this constraint). The most interesting solutions which are those that solve \eqref{leeyanglevel2degree8}, i.e. the physical solutions. The discriminant of this equation is $q$-dependent polynomial which tells us where in $q$-plane there can be a non-trivial monodromy. The discriminant factorizes such that the factors are of the following form:
\begin{align}
\nonumber
D & \sim F_1^{30} F_2^{10} F_3^4 F_4 F_5^2 F_6^2 \\
\nonumber
F_1 & = q \\
\nonumber
F_2 & = q-1 \\
\nonumber
F_3 & = q+1 \\
\nonumber
\label{leeyanglevel2discr}
F_4 & = 25q^4-1180q^3-282q^2-1180q+25 \\
F_5 & = q^8+59q^7+55q^6-119q^5+494q^4-119q^3 \\
\nonumber
& +55q^2+59q+1 \\
\nonumber
F_6 & = 49q^{12}+2753q^{11}+6044q^{10}+250824q^9+67633q^8 \\
\nonumber
& +669175q^7+1739524q^6+669175q^5+67633q^4 \\
\nonumber
& +250824q^3+6044q^2+2753q+49
\end{align}
All these factors are palindromic polynomials, i.e. $q$ is a root of these if and only if $q^{-1}$ is a root which is compatible with the charge conjugation symmetry of the algebra. Since $\epsilon_j$ parameters as well as the parameters characterizing charges of the lowest weight vector are real, the roots of these polynomials also appear in complex conjugate pairs.

We can use the homotopy method to study the monodromy of the solutions. Since at the Yangian points $q \to 0$ or $q \to \infty$ we can solve the Bethe ansatz equations exactly, we can start from there, slowly deforming the homotopy parameter $q$ in the complex plane and at every step using Newton's method to update the solution. Since the Newton's method converges very rapidly away from the special points in the $q$-plane, this is a very convenient way to numerically solve BAE at any given value of $q$. Starting with four physical solutions around $q=0$, we can study what happens as we go around zeros of the discriminant \eqref{leeyanglevel2discr}. It turns out that $q=0$ (zero of $F_1$) is a regular point where our solutions have no monodromy and actually we can find a power series solutions around this point (converging up to nearest singular point in $q$-plane). The local point $q = 1$ has monodromy associated to it, but this monodromy is such that only two Bethe roots of a given solution of BAE get exchanged. The solutions of BAE corresponding to different physical states do not get mixed as we go around $q=1$. Also, as discussed in detail in the previous section, some solutions of BAE blow up as we approach $q=1$. These Bethe roots correspond to excitations of the Heisenberg subalgebra. We have a similar situation around $q=-1$, but at this level there is no monodromy (since there can be only one excitation of the twist $2$ Heisenberg subalgebra). For values of $q$ that are roots of $F_4$ the monodromy is similarly only exchanging two Bethe roots in a given solution. There is no monodromy at all associated to roots of $F_5$. Finally, the most interesting case are the roots of $F_6$. Around all these points the solutions of Bethe ansatz equations are exchanged in a transitive manner, i.e. starting with any solution around $q = 0$, by a suitable path around roots of $F_6$ we can get any other of the four physical solutions. This is somewhat analogous to the situation in other integrable models \cite{Dorey:1996re} where various eigenstates of the set of commuting operators are connected as we analytically continue in the parameters. One example of such monodromy around $q \sim 0.168469+0.432772i$ is shown in Figure \ref{fighomotopy}. To summarize, the monodromy of solutions of \eqref{leeyanglevel2degree8} as we vary $q$ must preserve the pairs of Bethe roots in a given solution, so the monodromy group is reduced from the generic $S_8$ to $S_2^4 \rtimes S_4$.

Finally we want to study the solutions around $q = \pm 1$. At local point $q = 1$, we expect to find two solutions whose both Bethe roots blow up (corresponding to $\ydiagram{2}$ and $\ydiagram{1,1}$ excitations of Heisenberg algebra), one solution where only one Bethe root blows up while the other one corresponds to level $1$ state of $\mathcal{W}_\infty$ and finally one finite solution corresponding to level $2$ state in $\mathcal{W}_\infty$. This is exactly the case as we can see from the solutions
\begin{align}
\nonumber
& \begin{cases}
-\frac{6}{1-q} + \sqrt{3} \sqrt[4]{5} (1-q)^{-\frac{1}{2}} + \left( 7 - \frac{\sqrt{5}}{2} \right) - \frac{1}{4} \sqrt{-12+\frac{27}{\sqrt{5}}} (1-q)^{\frac{1}{2}} + \mathcal{O}((1-q)^1) \\
-\frac{6}{1-q} - \sqrt{3} \sqrt[4]{5} (1-q)^{-\frac{1}{2}} + \left( 7 - \frac{\sqrt{5}}{2} \right) + \frac{1}{4} \sqrt{-12+\frac{27}{\sqrt{5}}} (1-q)^{\frac{1}{2}} + \mathcal{O}((1-q)^1) \\
\end{cases} \\
\nonumber
& \begin{cases}
-\frac{6}{1-q} + i \sqrt{3} \sqrt[4]{5} (1-q)^{-\frac{1}{2}} + \left( 7 + \frac{\sqrt{5}}{2} \right) - \frac{i}{4} \sqrt{-12+\frac{27}{\sqrt{5}}} (1-q)^{\frac{1}{2}} + \mathcal{O}((1-q)^1) \\
-\frac{6}{1-q} - i \sqrt{3} \sqrt[4]{5} (1-q)^{-\frac{1}{2}} + \left( 7 + \frac{\sqrt{5}}{2} \right) + \frac{i}{4} \sqrt{-12+\frac{27}{\sqrt{5}}} (1-q)^{\frac{1}{2}} + \mathcal{O}((1-q)^1) \\
\end{cases} \\
\nonumber
& \begin{cases}
-\frac{6}{1-q} + 7 - \frac{4}{3} (1-q) - \frac{2}{3} (1-q)^2 + \mathcal{O}((1-q)^3) \\
4 - \frac{1}{3}(1-q) - \frac{1}{6}(1-q)^2 + \mathcal{O}((1-q)^3)
\end{cases} \\
& \begin{cases}
4 - \sqrt{5}{2} - \frac{3}{4}(1-q) - \frac{3}{64}(8-\sqrt{10})(1-q)^2 + \mathcal{O}((1-q)^3) \\
4 + \sqrt{5}{2} - \frac{3}{4}(1-q) - \frac{3}{64}(8+\sqrt{10})(1-q)^2 + \mathcal{O}((1-q)^3).
\end{cases}
\end{align}
As we go around $q=1$, the monodromy exchanges the Bethe roots in a given solution of Bethe ansatz equations, i.e. symmetric polynomials of Bethe roots do not undergo any monodromy and we expect this to be the general behaviour at $q = 1$.

By using the homotopy to continue four solutions corresponding to four level $2$ states from $q=0$ to $q=-1$, we see that three of the solutions have usual Taylor series expansion around $q=-1$ and one solution that blows up as $q \to -1$ (corresponding to twisted Heisenberg algebra eigenstate),
\begin{align}
\nonumber
& \begin{cases}
4+\sqrt{3-\sqrt{5}} + \frac{1}{24} \left(9 \sqrt{5}-53\right) (q+1) + \mathcal{O}((1+q)^2) \\
4-\sqrt{3+\sqrt{5}} + \frac{1}{24} \left(-9 \sqrt{5}-53\right) (q+1) + \mathcal{O}((1+q)^2)
\end{cases} \\
\nonumber
& \begin{cases}
4-\sqrt{3-\sqrt{5}} + \frac{1}{24} \left(9 \sqrt{5}-53\right) (q+1) + \mathcal{O}((1+q)^2) \\
4+\sqrt{3+\sqrt{5}} + \frac{1}{24} \left(-9 \sqrt{5}-53\right) (q+1) + \mathcal{O}((1+q)^2)
\end{cases} \\
& \begin{cases}
4-\frac{\sqrt{7}}{2} + \frac{9}{8}(q+1) + \mathcal{O}((1+q)^2) \\
4+\frac{\sqrt{7}}{2} + \frac{9}{8}(q+1) + \mathcal{O}((1+q)^2)
\end{cases} \\
\nonumber
& \begin{cases}
-\frac{6}{1+q} + 7-\frac{\sqrt{19}}{2} - \frac{11}{24}(1+q) - \left( \frac{11}{48} + \frac{5}{16\sqrt{19}} \right) (1+q)^2 + \mathcal{O}((1+q)^3) \\
-\frac{6}{1+q} + 7+\frac{\sqrt{19}}{2} - \frac{11}{24}(1+q) - \left( \frac{11}{48} - \frac{5}{16\sqrt{19}} \right) (1+q)^2 + \mathcal{O}((1+q)^3).
\end{cases}
\end{align}
We see that the twisted level $2$ Heisenberg excitation has the same singular behaviour (with the same residue) as $q \to -1$ as the usual Heisenberg excitations have for $q \to 1$. Although it corresponds to a lowest excitation of the twisted Heisenberg algebra of even modes, at the level of Bethe ansatz equations it is represented by a pair of singular Bethe roots.

\subsubsection{Solutions of Bethe ansatz equations for $\Delta=0$}
Let us solve the Bethe ansatz equations also for the other primary, the translation-invariant vacuum state $\Delta=0$ (which is \emph{not} the lowest energy or ground state of the system). In this case, the generating function of Yangian charges of the lowest weight state is simply
\begin{equation}
\frac{u+\psi_0 \epsilon_1 \epsilon_2 \epsilon_3}{u} = \frac{u-6}{u}
\end{equation}
so Bethe ansatz equations are of one lower degree which makes the analysis simpler.

\paragraph{Level 1}
The Bethe equation is simply
\begin{equation}
1 = q \frac{x_1-6}{x_1}
\end{equation}
so we have only one solution with Bethe root
\begin{equation}
x_1 = -\frac{6q}{1-q}
\end{equation}
whose only singularity is at the local point $q=1$. It corresponds to first Heisenberg excitation. There are no $\mathcal{W}_\infty$ excitations in the translationally invariant vacuum.

\paragraph{Level 2}
At level $2$ Bethe equations are
\begin{align}
1 & = q \, \frac{x_1-6}{x_1} \, \frac{(x_1-x_2+\epsilon_1)(x_1-x_2+\epsilon_2)(x_1-x_2+\epsilon_3)}{(x_1-x_2-\epsilon_1)(x_1-x_2-\epsilon_2)(x_1-x_2-\epsilon_3)} \\
1 & = q \, \frac{x_2-6}{x_2} \, \frac{(x_2-x_1+\epsilon_1)(x_2-x_1+\epsilon_2)(x_2-x_1+\epsilon_3)}{(x_2-x_1-\epsilon_1)(x_2-x_1-\epsilon_2)(x_2-x_1-\epsilon_3)}.
\end{align}
Numerical solution indicates that there are three pairs of non-trivial $q$-dependent solutions corresponding to three physical states at level $2$ in the vacuum representation and one degenerate solution with $x_1 = x_2$. The factor of resultant corresponding to physical states is degree $6$ polynomial
\begin{multline}
\label{leeyangvac2resultant}
(q-1)^4 (q+1)^2 x_1^6 -12 (q-1)^3 (q+1) (3q+1) q x_1^5 \\
+(q-1)^2 \left(521 q^4+360 q^3+2 q^2-19\right) x_1^4 -6 (q-1) \left(649 q^5+43 q^4-40 q^3-48 q^2-33 q+5\right) x_1^3 \\
+36 \left(441 q^4-246 q^3+29 q^2-76 q-4\right) q^2 x_1^2 -216 \left(155 q^2-34 q+23\right) q^4 x_1 +28512 q^6
\end{multline}
whose $6$ roots correspond to three pairs of Bethe roots. The identification between the pairs is obtained as before by taking the greatest common divisor of Bethe equations and we find a fractional linear map
\begin{equation}
x_2 = \frac{6q^2(x_1-6)}{(q^2-1)x_1-6q^2}
\end{equation}
which can be checked to be involutive on the set of solutions of \eqref{leeyangvac2resultant}. The discriminant of the relevant factor of the resultant (corresponding to Bethe roots that are not identically equal) is
\begin{multline}
D \sim q^{12}(1-q)^{18}(1+q)^4(5-66q+5q^2) \times \\
\times (196+8923q+12800q^2+59842q^3+12800q^4+8923q^5+196q^6)^2
\end{multline}
and the points in the $q$-plane where there are non-trivial monodromies are different than the ones of $\Delta=-\frac{1}{5}$ representation.

\paragraph{Level 3}
The vacuum representation $\Delta=0$ is slightly simpler than the primary $\Delta=-\frac{1}{5}$ due to the fact that the generating function of Yangian lowest weight charges
\begin{equation}
A(u) = \frac{u-6}{u}
\end{equation}
has only single zero and a single pole. The number of zeros and poles of this function is the quantity that (along with the level $M$) has the main influence on the degrees of Bethe ansatz equations and the associated difficulty of solving it. The vacuum representation of $\mathcal{W}_{1+\infty}$ would have six states at level $3$, but since we have a singular vector at level $3$ (due to Virasoro condition $\lambda_2 = 2$), we will only have five states.

The naive numerical solution of the system of level $3$ Bethe ansatz equations fails due to the fact that at this level there appear continuous families of solutions. In fact, consider a solutions of the form
\begin{align}
\begin{cases} x_1 & = \quad f(q) \\ x_2 & = \quad f(q) + \epsilon_1 = f(q) + 2 \\ x_3 & = \quad f(q) + \epsilon_1 + \epsilon_2 = f(q) + 5 \end{cases} \\
\begin{cases} x_1 & = \quad f(q) \\ x_2 & = \quad f(q) + \epsilon_2 = f(q) + 3 \\ x_3 & = \quad f(q) + \epsilon_1 + \epsilon_2 = f(q) + 5 \end{cases}
\end{align}
i.e. triples of Bethe roots satisfying a non-trivial resonance condition (related to the wheel condition in shuffle algebra \cite{Negut01012014}). It is clear that these are solutions of Bethe ansatz equations for any function $f(q)$ if we write these equations in the polynomial form, i.e. if we move all the numerators on one side and denominators on the other side. Since there is a one-parametric family of such solutions for any value of $q$, if we apply Newton's method with random initial seed, we will almost surely end up with a solution of this kind.

There is no problem with the homotopy method. The solutions of BAE at $q=0$ are either solutions of the family of solutions written above, or one of the following families of discrete solutions:
\begin{enumerate}
\item solutions corresponding to physical states in the irreducible representation
\begin{equation}
\label{leeyangdelta0level3q0}
(-10,-5,0), \qquad (-5,0,2), \qquad (-5,0,3), \qquad (0,2,3), \qquad (0,2,4)
\end{equation}
\item null state solution (remnant of missing $6$th state in the vacuum representation)
\[(0,3,6)\]
\item solutions
\[(-3,0,2), \qquad (-2,0,3), \qquad (-5,-3,0), \qquad (-5,-2,0), \qquad (0,2,5), \qquad (0,3,5)\]
\item solutions with degenerate Bethe roots
\[ (-5,-5,0), \qquad (-5,0,0), \qquad (0,0,2), \qquad (0,0,3), \qquad (0,2,2), \qquad (0,3,3)\]
\end{enumerate}

\begin{figure}
\centering
\includegraphics[width=0.6\textwidth]{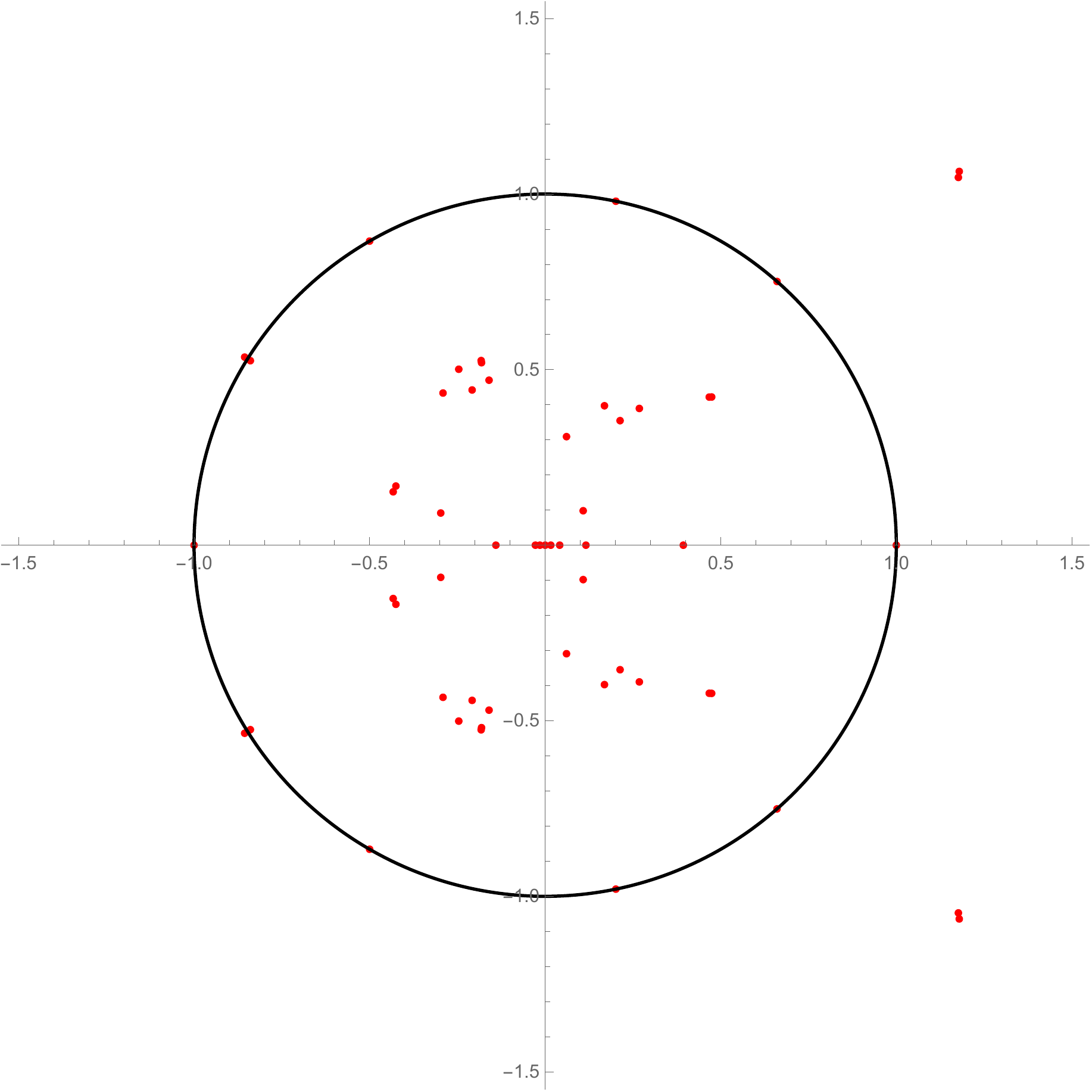}
\caption{Zeros of the discriminant of the polynomial whose zeros are exactly the $15$ Bethe roots of the Lee-Yang vacuum representation at level $3$. Only around these points in the $q$-plane the Bethe roots can have non-trivial monodromy. The zeros are symmetric under complex conjugation $q \to \bar{q}$ as well as under spherical inversion $q \to q^{-1}$ so we do not show most of the zeros outside of the unit disc. Note that apart from the special points $q=1$ and $q=-1$ which lie on the unit circle, there are also other special points with modulus $|q|=1$, in particular the cubic roots of unity $q = e^{\pm 2\pi i/3}$.}
\label{figleeyangvacl3discr}
\end{figure}

We can project the set of Bethe solutions corresponding to level $3$ states in the vacuum module (case 1 at $q=0$) to complex plane (eliminating variables $x_2$ and $x_3$ from Bethe equations) using resultants. As a first step in the reduction, we evaluate the resultant for the first and second Bethe equation with respect to variable $x_3$ and the largest irreducible factor is a polynomial in $x_2$ of degree $15$ (with coefficients which are polynomials in $x_1$ and $q$). Similarly, by taking the resultant of first and third Bethe equation in variable $x_3$ and extracting the largest irreducible factor, we get another polynomial of degree $15$ in $x_2$ with coefficients which are polynomials in $x_1$ and $q$. Note that if we did the same procedure with second and third Bethe equations, we would get a polynomial in $x_2$ of degree $24$. Now we calculate the resultant with respect to $x_2$ of the two degree $15$ polynomials in $x_2$ that we found before. The resulting polynomial is a polynomial of large degree in $x_1$ with coefficients that are polynomials in $q$. The largest irreducible factor is of degree $120$ in $x_1$, but this is not the factor we are interested in (we can for example check that the solutions obtained by homotopy from the physical solutions \eqref{leeyangdelta0level3q0} are not zeros of this factor). The factor that we are interested in is the irreducible factor of degree $15$ in $x_1$. Its explicit form is rather complicated and is given in appendix \ref{appleeyangreslvl3}. For a given $q$ its roots are exactly the five triples of Bethe roots corresponding to five solutions of BAE associated to five physical level $3$ states. Having a single polynomial whose roots are in one-to-one correspondence with the physical Bethe roots is very convenient starting point to study Puiseux series expansions of the solutions around any value of $q$, for example using the Newton's polygon. Having an irreducible polynomial with these properties is also useful in order to show that the analytic continuation in $q$ parameter does not mix these physical solutions of BAE with other solutions. Discriminant of this polynomial tells us around which points in the $q$-plane the solutions undergo a non-trivial monodromy. There are in fact $91$ such special points and their positions in the $q$-plane are illustrated in Figure \ref{figleeyangvacl3discr}.

The expansion of the solutions around $q=0$ produces is a regular Taylor series expansion with rational coefficients as shown in appendix \ref{appleeyangyanglvl3}. The expansion of physical solutions around local point $q=1$ is given in appendix \ref{appleeyangloclvl3}. The first of these solutions has limit $(0,3,6)$ as $q \to 1$ which coincides with the constant $q$-independent solution $(0,3,6)$ but for $q \neq 1$ these are different solutions of BAE. They lie on different branches (different factors of the resultant) so they cannot mix as we move around $q$-plane (but this does not prevent them from coinciding at special values of $q$). The second solution corresponds to one Heisenberg root and level $2$ excitation of $\mathcal{W}_\infty$. The third solution has three Heisenberg roots which are cyclically exchanged as we go around $q = 1$ and corresponds to state $\ydiagram{2,1}$ of Heisenberg algebra. The last two states have also three Heisenberg roots, but the monodromy around $q = 1$ only exchanges two of their roots while the third root is single-valued. These correspond to $\ydiagram{3}$ and $\ydiagram{1,1,1}$ excitations of Heisenberg algebra.

\paragraph{Level 3 - twisted local limit $q \to -1$}
The limit $q \to -1$ is also a singular limit for some Bethe roots, concretely for those which are associated to excitations of even mode subalgebra of the Heisenberg algebra. The degree $15$ factor of the resultant given in appendix \ref{appleeyangreslvl3} simplifies to
\begin{equation}
(x-3)^3(28-78x+157x^2-48x^3+4x^4)(17424-6864x+2332x^2-612x^3+141x^4-18x^5+x^6)
\end{equation}
which is of degree $13$, i.e. we lost two roots (corresponding to Bethe roots that blew up in the limit). Using the homotopy method, we see that the roots of discriminant are combined to solutions of BAE in the following manner: the six roots of the sextic factor give two complex conjugate solutions of Bethe ansatz equations. The four roots of the quartic equation together with two roots $x=3$ give other two complex conjugate solutions of BAE. The last physical solution of BAE in $q \to -1$ limit has Laurent expansion
\begin{equation}
\begin{cases}
x_1 & = \, -\frac{6}{q+1} + \frac{12+\sqrt{19}}{2} - \frac{59}{24}(q+1) + \mathcal{O}((q+1)^2) \\
x_2 & = \, -\frac{6}{q+1} + \frac{12-\sqrt{19}}{2} - \frac{59}{24}(q+1) + \mathcal{O}((q+1)^2) \\
x_3 & = \, 3 - \frac{3}{2} (q+1) - \frac{3}{4} (q+1)^2 + \mathcal{O}((q+1)^3).
\end{cases}
\end{equation}
The Bethe roots $x_1$ and $x_2$ are singular and correspond to level $2$ excitation of the Heisenberg algebra and the third root corresponds to the excitation of the algebra which is the commutant of the twisted Heisenberg algebra in $\mathcal{W}_{1+\infty}$.

\paragraph{Level 3 - other twisted local limits $q \to e^{\pm 2 \pi i/3}$}
There are other points in the $q$-plane where one of the Bethe roots becomes singular. The easiest way of seeing such points is by looking at the leading coefficient $x^{15}$ in the resultant characterizing all physical Bethe roots. Bethe roots blow up exactly when this leading coefficient vanishes. In the example studied here the number of Bethe roots that blow up is captured by this polynomial in $q$: we have
\begin{equation}
\beta_{15} = (q-1)^{10} (q+1)^2 (q^2+q+1)^3
\end{equation}
and we can see that indeed ten Bethe roots blow up as $q \to 1$ and two Bethe roots blow up as $q \to -1$ as confirmed by explicit calculations above. We also see that three Bethe roots should blow up when $q \to e^{\pm 2 \pi i / 3}$. This is exactly what is happening. The corresponding singular solution around $q = e^{2\pi i/3}$ has Bethe roots of the form
\begin{equation}
-\frac{\psi_0 \epsilon_1 \epsilon_2 \epsilon_3 e^{2\pi i/3}}{q-e^{2\pi i/3}} + \gamma_0 + \mathcal{O}((q-e^{2\pi i/3})^1)
\end{equation}
where $\gamma_0$ are the three roots of the equation
\begin{equation}
\gamma_0^3 - 18\gamma_0^2+89\gamma_0-102\sqrt{3}+10i = 0.
\end{equation}
Since the leading and subleading terms have no monodromy around $q = e^{2\pi i /3}$ and distinguish the three roots, there is no monodromy at all orders, i.e. we have Laurent expansion of the Bethe roots in $(q - e^{2\pi i /3})$.

\subsection{Perturbative expansion at $q = 0$ and residue formula}

Just as the constant coefficient in $q \to 0$ expansion can be read off directly from the combinatorics of periodic plane partitions, so can be the $\mathcal{O}(q)$ coefficient: it is simply given by the residue of the generating function of the Yangian higher spin charges. Consider for concreteness level $3$ descendants of $\Delta = 0$ primary in Lee-Yang model. We have five physical solutions as listed in \eqref{leeyangdelta0level3q0}. Their corresponding generating functions of Yangian higher spin charges are
\begin{align}
\label{psifnsleeyanglev3}
(0,\epsilon_1,2\epsilon_1) = (0,2,4) & \leadsto \psi(u) = \frac{(u-9)(u-1)(u+2)}{(u-4)(u-3)(u+5)} \\
(0,\epsilon_3,2\epsilon_3) = (0,-5,-10) & \leadsto \psi(u) = \frac{(u-6)(u-5)(u+12)(u+13)}{(u-3)(u-2)(u+10)(u+15)} \\
(0,\epsilon_1,\epsilon_2) = (0,2,3) & \leadsto \psi(u) = \frac{(u-8)(u-7)(u-1)u(u+1)}{(u-5)(u-4)(u-3)(u-2)(u+5)} \\
(0,\epsilon_1,\epsilon_3) = (0,2,-5) & \leadsto \psi(u) = \frac{(u-7)(u-6)u(u+1)(u+7)(u+8)}{(u-4)(u-3)(u-2)(u+3)(u+5)(u+10)} \\
(0,\epsilon_2,\epsilon_3) (0,3,-5) & \leadsto \psi(u) = \frac{(u-8)(u-1)u(u+7)(u+8)}{(u-3)(u-2)(u+2)(u+5)(u+10)}
\end{align}
Combinatorially, they correspond to adding three boxes to ground state configuration of Figure \ref{leeyangdelta0}. We can put pile of three boxes along first or third directions or an $L$-shaped configuration of boxes in any of the three planes.

The $\mathcal{O}(q^1)$ term in expansion of ILW Hamiltonians comes from the matrix element between $\bra{\Box}$ and $\ket{\Box}$ states in the auxiliary space. As such, these matrix elements are related to amplitude of addition and removal of a box in a Young diagram. The product of these two amplitudes (which is independent of the normalization of the vectors in representation space) is known to be expressible in terms of a residue of the generating function of Yangian charges $\psi(u)$, see \cite{Prochazka:2015deb} section 4.4. For this reason, it is not surprising that we have the following empirical formula for $\mathcal{O}(q^1)$ term in the expansion of the Bethe roots around $q = 0$:
\begin{equation}
x = x_0 - q \cdot \res_{u \to x_0} \psi(u) + \mathcal{O}(q^2)
\end{equation}
where $x_0$ is the $q \to 0$ limit of the given Bethe root. Applying this formula to functions in \eqref{psifnsleeyanglev3} exactly reproduces the first order terms in appendix \ref{appleeyangyanglvl3}. Since $\psi(u)$ only has poles at points where a box can be added or removed \cite{Prochazka:2015deb}, the Bethe roots corresponding to boxes deeper in the Young diagram start at higher orders in $q$-expansion. We can see from explicit calculations such as those given in \ref{appleeyangyanglvl3} that the first non-trivial power of $q$ (apart from the constant term) is given by (one plus) the number of boxes that need to be removed from the Young diagram before the given box becomes removable. All of this is consistent with the analysis in appendix D of \cite{Vasko:2015swp} where Bethe equations are studied to the first order in $q$-expansion.

\subsection{Numerical tests in first unitary $\mathcal{W}_4$ model}
In this section we want to numerically test the formulas \eqref{w3spectrum} and \eqref{s4spectrum}. If we want spin $4$ field $U_4(z)$ to be non-trivial, we need to have rank of $\mathcal{W}_N$ algebra to be at least $4$. For this reason and for concreteness we choose the first unitary minimal model of $\mathcal{W}_4$ algebra. The $\lambda$-parameters of $\mathcal{W}_\infty$ are determined by the constraints
\begin{equation}
\frac{1}{\lambda_1} + \frac{1}{\lambda_2} + \frac{1}{\lambda_3} = 0, \qquad \frac{2}{\lambda_1} + \frac{1}{\lambda_2} + \frac{0}{\lambda_3} = 1, \qquad \lambda_3 = 4
\end{equation}
(the middle equation corresponds to the choice of the first unitary minimal model). These equations have unique solution
\begin{equation}
\lambda_1 = -\frac{2}{3}, \qquad \lambda_2 = \frac{4}{5}, \qquad \lambda_3 = 4.
\end{equation}
The corresponding central charge is
\begin{equation}
c_{\infty} = (\lambda_1-1)(\lambda_2-1)(\lambda_3-1) = 1.
\end{equation}
In general, for the first unitary minimal model of $\mathcal{W}_N$ algebra we would have
\begin{equation}
c_{\infty} = \frac{2(N-1)}{N+2}.
\end{equation}
Next we calculate the Nekrasov-like parameters, imposing the conventional restriction $\epsilon_1 \epsilon_2 = -1$ that fixes the overall scaling symmetry up to an overall sign. We find
\begin{align}
\epsilon_1 = \sqrt{\frac{6}{5}}, \qquad \epsilon_2 = -\sqrt{\frac{5}{6}}, \qquad \epsilon_3 = \sqrt{\frac{5}{6}} - \sqrt{\frac{6}{5}}
\end{align}
and
\begin{equation}
\psi_0 = 4, \qquad \alpha_0 = \sqrt{\frac{5}{6}} - \sqrt{\frac{6}{5}} = -\frac{1}{\sqrt{30}}.
\end{equation}
Finally, we restrict to the vacuum representation so the generating function of higher spin charges of the ground state is
\begin{equation}
A(u) = \frac{u+\psi_0 \epsilon_1 \epsilon_2 \epsilon_3}{u} = \frac{u + \frac{4}{\sqrt{30}}}{u}.
\end{equation}
With this choice, all $\psi_j, j \geq 1$ vanish. Writing this in terms of Coulomb parameters $a_j$,
\begin{equation}
A(u) = \prod_{\ell=1}^N \frac{u+a_\ell-\epsilon_3}{u+a_\ell},
\end{equation}
we can determine these to be
\begin{equation}
a_1 = 0, \qquad a_2 = \frac{1}{\sqrt{30}}, \qquad a_3 = \frac{2}{\sqrt{30}}, \qquad a_4 = \frac{3}{\sqrt{30}}.
\end{equation}
Finally, we want to know how many states there are in the vacuum representation of this unitary minimal model. We could either use the general Weyl character formula \cite{Bouwknegt:1992wg}, or simply count boxes in staircase configuration as in Figure \ref{isingdelta0}, but with the height of green-brown wall being $4$ instead of $2$ (since we are considering $\mathcal{W}_4$ minimal model). The position of the pit in the horizontal yellow plane is still the same, because both models are first unitary minimal models so have a null state at level $6$. Counting the boxes, we find the character proportional to
\begin{equation}
1 + q + 3q^2 + 6q^3 + 13q^4 + 23q^5 + 45q^6 + \ldots
\end{equation}
Probably the easiest way to get this is to consider first MacMahon function counting all plane partitions. At level $5$ we have one less state because we are in $\mathcal{W}_4$. At level $6$ there are three missing states, one of them being prohibited by the existence of the pit and the other two states correspond to plane partitions with $6$ boxes whose height exceeds $4$. In order to find number of states in $\mathcal{W}_4$ vacuum representation, we need to subtract the contribution of Heisenberg algebra and we finally find the state counting function
\begin{equation}
1 + q^2 + 2q^3 + 4q^4 + 6q^5 + 12q^6 + \ldots
\end{equation}
Let us now find the eigenvalues of $W_{3,0}$ and $\mathcal{S}_4$ and match them with solutions of BAE.

\paragraph{Level 0}
Since formulas \eqref{w3spectrum} and \eqref{s4spectrum} only give eigenvalues relative to the lowest weight state, we first calculate the eigenvalue of $\mathcal{W}_{3,0}$ and $\mathcal{S}_4$ on the vacuum state. The first one vanishes by the parity symmetry and the second one has eigenvalue
\begin{equation}
-\frac{121}{115200}.
\end{equation}
In the following, we will list the eigenvalues of $\mathcal{S}_4$ relative to this value.

\paragraph{Level 1}
The analysis at level $1$ is very simple becase there are no states at this level.

\paragraph{Level 2}
On level $2$ we expect only one physical solution that is regular as $q \to 1$, since we have only one state
\begin{equation}
L_{-2} \ket{0}.
\end{equation}
$W_{3,0}$ acting on this state is zero by parity, while $\mathcal{S}_4$ eigenvalue relative to vacuum is
\begin{equation}
-\frac{121}{240} = -0.5041\bar{6}.
\end{equation}
The physical solution of Bethe equations is
\begin{equation}
x_1 = -0.807679\ldots, \qquad x_2 = 0.077382\ldots
\end{equation}
and formulas \eqref{w3spectrum} and \eqref{s4spectrum} give us eigenvalues zero and $-0.5041\bar{6}$ in agreement with the explicit calculation.

\paragraph{Level 3}
Here we expect two states. In the basis $L_{-3} \ket{0}$ and $W_{3,-3} \ket{0}$ the matrix of $W_{3,0}$ is
\begin{equation}
\begin{pmatrix} 0 & \frac{4}{5} \\ 6 & 0 \end{pmatrix}.
\end{equation}
The matrix is off-diagonal due to $\mathbbm{Z}_2$ parity. The eigenvalues are
\begin{equation}
\pm \sqrt{\frac{24}{5}}.
\end{equation}
The matrix of $\mathcal{S}_4$ relative to the lowest weight state is
\begin{equation}
\begin{pmatrix} -\frac{363}{160} & 0 \\ 0 & -\frac{363}{160} \end{pmatrix}
\end{equation}
with obvious eigenvalues. These matrices also clearly commute. The corresponding solutions of BAE are
\begin{equation}
(-1.39495, \, -0.847418, \, 0.0514826), \qquad (-0.781779, \, 0.117121, \, 0.664658)
\end{equation}
which using \eqref{w3spectrum} and \eqref{s4spectrum} give exactly the eigenvalues we want. Note that due to the fact that every iteration of Newton's method doubles the number of digits, we can find the Bethe roots with almost arbitrary precision once we have the right initial seed.

\paragraph{Level 4}
Let us finally look at solutions at level $4$. There are $4$ states, the parity even states
\begin{equation}
L_{-4} \ket{0}, \quad L_{-2}^2 \ket{0}, \quad W_{4,-4} \ket{0}
\end{equation}
and one parity odd state
\begin{equation}
W_{3,-4} \ket{0}.
\end{equation}
The matrix of $W_{3,0}$ must be block off-diagonal in this basis and it is equal to
\begin{equation}
\begin{pmatrix} 0 & 0 & 0 & \frac{64}{45} \\ 0 & 0 & 0 & \frac{256}{135} \\ 0 & 0 & 0 & -8 \\ 8 & 8 & -\frac{121}{270} & 0 \end{pmatrix}
\end{equation}
with eigenvalues
\begin{equation}
+2\sqrt{\frac{113}{15}}, \, -2\sqrt{\frac{113}{15}}, \, 0, 0.
\end{equation}
The matrix of $\mathcal{S}_4$ is block diagonal
\begin{equation}
\begin{pmatrix} -\frac{121}{24} & -\frac{121}{60} & -\frac{242}{2025} & 0 \\ -\frac{121}{90} & -\frac{1331}{360} & \frac{3388}{6075} & 0 \\
 12 & 24 & \frac{55}{72} & 0 \\ 0 & 0 & 0 & -\frac{847}{120} \end{pmatrix}
\end{equation}
with eigenvalues
\begin{equation}
-\frac{847}{120}, \, -\frac{847}{120}, \, \frac{11}{120} \left( -5 + 6 \sqrt{31} \right), \, \frac{11}{120} \left( -5 - 6 \sqrt{31} \right).
\end{equation}
It is a useful consistency check to see that the matrices representing $W_{3,0}$ and $\mathcal{S}_4$ commute. The corresponding solutions of BAE are
\begin{align}
\xi_1 & = \{ -1.69167 \pm 0.25460 i, \, -0.861612, \, 0.0396581 \} \\
\xi_2 & = \{-0.769955, \, 0.131316, \, 0.961369 \pm 0.254598 i \} \\
\xi_3 & = \{-0.756691 \pm 0.092167 i, \, 0.0263942 \pm 0.0921672 i \} \\
\xi_4 & = \{-1.41680, \, -0.817274, \, 0.0869769, \, 0.686498\}
\end{align}
and it is easy to check that these give the right values of $W_{3,0}$ and $\mathcal{S}_4$. Since the vacuum representation is self-conjugate,
the transformation
\begin{equation}
x_j(q) \mapsto \tilde{x}_j(q) = -x_j(q^{-1}) - \psi_0 \epsilon_1 \epsilon_2 \epsilon_3
\end{equation}
maps (physical) solutions of Bethe equations to solutions of Bethe equations. Since the local point $q = 1$ is also invariant under $q \to q^{-1}$, given any solution of Bethe equations $x_j$ at $q = 1$,
\begin{equation}
-x_j - \psi_0 \epsilon_1 \epsilon_2 \epsilon_3
\end{equation}
is also a solution of Bethe equations at $q = 1$. In the present situation this transformation exchanges $\xi_1 \leftrightarrow \xi_2$ since they have opposite values of $W_{3,0}$ charge and keeps $\xi_3$ and $\xi_4$ invariant. We can use this symmetry to decrease the number of independent BAE and Bethe roots to $2$ and use this to analytically study the Bethe equations.

\section{Outlook}
The results of this article are of technical nature, but they can be a starting point or ingredients in pursuing various other future directions.

\paragraph{Decoupling of $\mathcal{W}_\infty$ and combinatorics of $\mathcal{W}_\infty$ states}
It is very well-known that states in lowest weight representations of Heisenberg algebra can be enumerated by Young diagrams. The Yangian picture of $\mathcal{W}_{1+\infty}$ allows us to enumerate states in lowest weight representations of $\mathcal{W}_{1+\infty}$ in terms of tuples of Young diagrams, in terms of plane partitions (possibly with Young diagram asymptotics) or in terms of periodic plane partitions. Which of these is realized depends on the choice of the parameters of the algebra (physical model) and on the lowest weights (choice of the primary). On the other hand, without the Heisenberg factor, there is no known natural way of combinatorially enumerating the states of $\mathcal{W}_\infty$ representations. The ILW conserved quantities studied here however seem to offer one solution: given any curve in the complex $q$-plane starting at $q = 0$ and ending at $q = 1$ and avoiding points where the solutions become ambiguous, we can interpolate from eigenstates of Yangian conserved quantities at $q = 0$ to pairs of eigenstates of Heisenberg $\times$ $\mathcal{W}_\infty$ conserved quantities at $q = 1$. A natural choice of such curve would be along the interval $(0,1)$, but since we have seen that there are singular points on the positive real axis, one should go either infinitesimally above or below the real axis. At any finite level $M$ there can be no accumulation points (because all the equations can be reduced to polynomial equations of finite order) so such prescription is well-defined. It would be interesting to understand such a map in more detail.

\paragraph{Behaviour of Bethe roots around special points}
The points $q = 0$ and $q = \infty$ are associated to Yangian conserved charges. In examples studied here all Bethe roots had nice Taylor expansion around these points (and we found nice combinatorial interpretation of the first two Taylor series coefficients). It would be nice if there was a nice combinatorial way of calculating higher order terms in the expansion. One of the reasons is that at higher levels there are continuous families of unphysical solutions of BAE so one needs to go to higher orders in Taylor expansion in order to fully specify the initial conditions in order to use the monodromy method \cite{kudprochvelk}.

The local behaviour around the point $q = 1$ as well as around other roots of unity is more puzzling. The (twisted) Heisenberg subalgebras decouple at these points and the Bethe roots associated to such subalgebras of $\mathcal{W}_{1+\infty}$ blow up. But the local behaviour of Bethe roots encodes information about these Heisenberg excitations in very subtle way: the higher conserved quantities of Heisenberg algebra appear at higher orders in Laurent series expansion, but at these orders, the (twisted) Heisenberg Bethe roots interact non-linearly with the remaining Bethe roots. It is rather surprising how simple discrete information such as the Heisenberg Young diagram is captured by the non-linear Bethe equations and how it is encoded in the local behaviour of the Bethe roots around these roots of unity.

\paragraph{Proof of FJMM formula, shuffle algebra}
The main missing step in calculations in this article is the fact that we did not attempt to prove the Feigin-Jimbo-Miwa-Mukhin formula for the spectrum of $\mathcal{H}_q(u)$ (and neither did we prove the associated Bethe ansatz equations). What we did was to numerically verify that the first five ILW Hamiltonians calculated from the instanton $\mathcal{R}$-matrix have spectrum as predicted by FJMM formula. It would be very useful to have an algebraic derivation of Bethe ansatz equations along the lines of algebraic Bethe ansatz. One problem is that in the usual spin chain one has access to ladder operators that commute for any value of the spectral parameter, so it is easy to write an off-shell Bethe state parametrized by a collection of spectral parameters, in which it is symmetric (as a consequence of the commutativity). The natural ladder operator $e(u)$ that we have in $\mathcal{W}_{1+\infty}$ however does not commute with itself at different values of $u$. This seems to be an obstruction for following the standard algebraic Bethe ansatz procedure literally. On the other hand, the algebra of ladder operators in $\mathcal{W}_{1+\infty}$ has beautiful shuffle algebra description \cite{Negut01012014}, where all the ladder operators are labeled by symmetric polynomials and the associative product of the algebra gets translated to star product remotely reminiscent of the Moyal product. It is actually this shuffle algebra picture that is an important ingredient of the proof by Feigin-Jimbo-Miwa-Mukhin. The authors prove the Bethe ansatz equations and formula for spectrum of the associated Hamiltonians for $q$-deformed (trigonometric) version of $\mathcal{W}_{1+\infty}$ and at the end take the rational limit. It would be interesting to see to what extent their proof relies on additional properties of $q$-deformed algebra (such as the existence of Miki automorphism) or whether it can be translated to affine Yangian language.

\paragraph{Map between solutions of different Bethe ansatz equations}
Currently we have at our disposal two different looking sets of Bethe ansatz equations, the equations of Gaudin type, introduced by Bazhanov, Lukyanov and Zamolodchikov (for Virasoro algebra) and the equations of XXX spin chain type (but with scattering amplitude controlled by a function of higher degree) studied by Litvinov. The advantage of Litvinov's equations is that they are uniform for the whole $\mathcal{W}_{1+\infty}$, respect the symmetries of the algebra and in addition come with natural monodromy parameter that allows their solution using the homotopy method. On the other hand, the Gaudin type equations have been studied more thoroughly and there is clear connection via ODE/IM correspondence to quantized spectral curves. The KdV conserved charges are encoded in WKB expansion of the wave function \cite{Dorey:2019ngq} which is therefore a natural generating function of the higher conserved quantities.

It would be really useful to understand the map between the two pictures. At every finite level, we have a finite number of Bethe roots and an infinite set of algebraic equations connecting them, so one would hope that there should be a natural map relating these. In $\mathcal{W}_{1+\infty}$ there are two kinds of generating functions (local quantum fields, Yangian currents) that one can introduce to describe the algebra. From the first point of view, we introduce the generating function by summing over Fourier modes and the resulting generating function becomes a local field in CFT. The parameter of the generating function is simply the coordinate on the CFT worldsheet. A second approach is to sum over the spin label of the $\mathcal{W}_{1+\infty}$ generators, obtaining the Yangian currents. The associated parameter of these generating functions is the spectral parameter and it lives in the same space where the zero modes of Heisenberg fields (or Coulomb parameters) live. The differential operators such as the Miura operator act in the CFT space, but Litvinov's Bethe ansatz equations are in the spectral parameter space. On the other hand, the picture of Bazhanov, Lukyanov and Zamolodchikov involves a Schr\"odinger differential operator with monster potential and the Bethe roots live in the space where the differential operator acts, the Bethe roots being positions of singularities of the potential. It would seem natural to identify the Miura and BLZ operators, since for $\mathcal{W}_N$ algebra they are both $N$-th order ordinary differential operators. But as argued in \cite{Feigin:2007mr}, we should not do so, in particular because for simple Lie algebras which are not Langlands self-dual the two operators are related to Langlands dual algebras. If there was a natural map between Bethe roots of Litvinov and BLZ Bethe equations, one could understand the relation between the two spaces. In the related quantum toroidal setting such relation is understood thanks to so-called Miki automorphism \cite{Sasa:2019rbk} which exchanges the mode and spin spaces (which is geometrically related to fiber-base duality \cite{Mitev:2014jza}) and the fact that there can be different systems of Bethe ansatz equations associated to the same physical system is an example of more general phenomenon called spectral duality \cite{mukhin2006bispectral,mukhin2008bispectral,Mironov:2012uh,Mironov:2012ba}. There is also a close connection to (affine) Gaudin models, starting from \cite{Feigin:2007mr} and recently discussed for example in \cite{rybnikov2020proof,Vicedo:2017cge,Kotousov:2019nvt,Kotousov:2021vih,Kotousov:2022azm}.

\paragraph{Gluings and BPS algebras}
It is an important open question to describe the space of vertex operator algebras. In \cite{Prochazka:2017qum,Prochazka:2018tlo}, a large class of vertex operator algebras was constructed with $\mathcal{W}_{1+\infty}$ or its truncations $Y_{N_1 N_2 N_3}$ serving as fundamental building blocks. A parallel construction in the Yangian language was studied in \cite{Gaberdiel:2017hcn,Li:2019nna,Li:2020rij,Galakhov:2021xum,Galakhov:2021vbo}. But there are other important vertex algebras which seem to require more complicated basic elements \cite{Eberhardt:2019xmf,Eberhardt:2020zgt}. One such generalization is to consider the Grassmannian vertex operator algebra as a more general building block. Just as $\mathcal{W}_{1+\infty}$ is a two-parametric family of algebras, the unitary Grassmannian coset algebras depend on three complex parameters. But there are indications from the representation theory \cite{Eberhardt:2020zgt} that these algebras are part of four-parametric family of algebras. Performing a OPE bootstrap for Grassmannian analogously to \cite{Gaberdiel:2012ku,Prochazka:2014gqa} is much more difficult due to significantly larger number of local fields, so it would be good to find other approaches to study these algebras. Due to their simple structure, Bethe ansatz equations seem to be a natural starting point for such generalizations. See \cite{Galakhov:2022uyu} for recent discussion in the context of BPS algebras associated to quiver gauge theories.

\paragraph{Hamiltonian structures and KP hierarchy}
On the classical level, many integrable systems admit bi-Hamiltonian structure, i.e. two Poisson brackets whose linear combinations are also Poisson brackets. Kadomtsev-Petviashvili hierarchy is one of such integrable systems. $\mathcal{W}_{1+\infty}$ discussed here can be thought of as quantized KP hierarchy with respect to so-called second Hamiltonian structure. In \cite{Kozlowski:2016too}, the authors quantized KP equation directly with respect to the first Hamiltonian structure and by directly using the coordinate Bethe ansatz obtained the same Bethe ansatz equations as those considered here. It would be really interesting to understand how to relate the different Hamiltonian structures at quantum level.

\paragraph{Thermodynamics and modular properties of the higher local charges}
There is a very nice interplay between low energy and high energy properties of the two-dimensional conformal field theories. Studying CFT on a torus for different values of the modular parameter lead Cardy \cite{Cardy:1986ie} to a formula which relates the counting of highly-excited states to low energy properties of the system. On mathematics side, this manifests by the fact that characters of irreducible representations of vertex operator algebras often transform nicely under modular transformations.

We can refine this IR/UV duality by introducing various probes in our system, the simplest being an integral of local fields such as $(TT)(z)$ along a spatial circle (for a nice recent discussion see \cite{Maloney:2018hdg,Dymarsky:2022dhi}). Since $(TT)(z)$ is quasi-primary, it transforms nicely under the conformal transformations (in fact it transforms under the global conformal transformations of Riemann sphere in the same way as primary fields), so naively we would expect nice modular properties of torus partiton functions with insertions of higher spin charges such as $(TT)_0$. The problem is that in order to calculate the cylinder zero mode, we need to integrate along the spatial cycle and this is in turn mapped to a temporal cycle under the modular $S$-transformation. A clever idea of Dijkgraaf \cite{Dijkgraaf:1996iy} is to use the translation-invariance along the time direction on the torus to average the expectation value of $(TT)_0$, extending the integral along the cycle to an integral over the whole torus. After doing this, the integration region is invariant under modular transformations so the characters refined by insertions of operators such as $(TT)_0$ still transform nicely under modular transformations. Once we insert more such operators, the argument has to be refined, because the local operators can have singularities at the coincident points and a regularization is needed to resolve that, leading to modular anomalies that are correlated to operator product expansions of the local fields whose zero modes we are inserting.

The higher spin extensions have not received so much attention (but see \cite{Gaberdiel:2012yb,Iles:2014gra,Downing:2021mfw} for recent discussion), so it would be very interesting to understand the interplay of modularity with higher spin symmetries of $\mathcal{W}_{1+\infty}$ more thoroughly. We saw many (quasi-)modular forms appearing in expansions of various ILW quantities. The construction of conserved quantities from $\mathcal{R}$-matrix by algebraic Bethe ansatz introduces an auxiliary torus. If we keep the quantum space to be on a cylinder, there is no reason to expect modular invariance to be there. From the explicit calculations, we saw that the matrix elements of ILW Hamiltonians were rational functions of $q$, all of the Eisenstein series could be eliminated by redefinition of Hamiltonians. Also the Bethe equations have purely rational form. The situation changes when we put the quantum space on a torus as well. The configuration becomes symmetric and the infinite sum of matrix elements of ILW Hamiltonians have similar form to Eisenstein series and we can ask if there are any transformations acting on both tori that would be symmetries of the setup. One would need to understand the interaction between combined modular transformations acting in certain way on both tori and the insertion of $\mathcal{R}$-matrix coupling both tori.

The thermodynamic properties enter the game also if we want to understand the local limit of ILW Hamiltonians. The Yangian limit, $q \to 0$, corresponds to low temperatures in the auxiliary torus, where the lowest energy states dominante the trace. In this limit, the Bethe roots of the quantum space are frozen to their Yangian positions, which can be described using combinatorics of plane partitions. As we increase $q$, the excited states in the auxiliary torus contribute as well. In the local limit $q \to 1$ we are removing the regulator, so the highly excited states should dominate. The typical Young diagram converges to a limit shape and it is the geometry of this limit shape that should control the Hamiltonians in the local limit.

\paragraph{Elliptic Calogero model}
To simplify the calculations, we could try to choose a simpler auxiliary space. In particular, there are representations of the algebra where the generators of algebra act as quantum mechanical differential operators such in Calogero systems where finite number of particles in one direction interact with $1/r^3$ forces when particles are close to each other. The Miura operator can be thought of as a representation of universal $\mathcal{R}$-matrix with one of the spaces being the space of functions of $z$ variable on which the derivatives act \cite{Prochazka:2019dvu,Gaiotto:2020dsq}. Applying the algebraic Bethe ansatz but with Calogero space being the auxiliary space could lead to simpler expressions for ILW Hamiltonians.

The transition from Calogero systems to 2d conformal theory is the second quantization, where the collective coordinates correspond to Fourier modes of CFT fields. The total momentum of the system (which is quantized with integer steps) becomes CFT $L_0$ operator (which has the same level spacing) and the usual Calogero Hamiltonian corresponds to spin $3$ conserved quantity in CFT. It would be nice if Litvinov's Bethe ansatz equations applied also to the finite dimensional system before the second quantization. Nekrasov and Shatashvili \cite{Nekrasov:2009rc} found TBA-like equations that describe the spectrum of elliptic Calogero model, but their equations take the form of thermodynamic limit of Litvinov's equations. Taking the thermodynamic limit to go from a system of finite number of particles to a system with infinite number of particles seems to be very natural, but it is less clear what is the role of thermodynamic limit used to parametrize the solution of elliptic Calogero model which has a finite number of particles. There are also elliptic Bethe ansatz equations by Felder and Varchenko \cite{Felder:1995iv} which could again be spectral dual of rational ILW equations studied here. It would be very interesting to see how these results about elliptic Calogero model fit in the picture discussed here.

\section*{Acknowledgements}

We would like to thank to Ilka Brunner, Lorenz Eberhardt, Mat\v{e}j Kudrna, Andrew Linshaw, Alexey Litvinov, Alex Maloney, Jan Manschot, Davide Masoero, Stefano Negro, Go Noshita, Paolo Rossi, Paolo Rossi, Martin Schnabl, Alessandro Sfondrini, Bogdan Stefa\'nski, Jan \v{S}\v{t}ov\'{i}\v{c}ek, Yuji Tachikawa, Alessandro Torrielli, Petr Va\v{s}ko, Miroslav Ve\v{l}k and especially to Yutaka Matsuo for useful discussions. The research was supported by the Grant Agency of the Czech Republic under the grant EXPRO 20-25775X. The research was partially supported by Grant-in-Aid MEXT/JSPS KAKENHI 18K03610. AW also wants to thank the FMSP program, JSPS Fellowship, and JSR Fellowship for their financial support.

\appendix

\section{Fourier expansions and conformal normal ordering}

\label{apmodes}

Here we will compare the conformal normal ordering on the complex plane as well as on the cylinder in terms of the corresponding Fourier modes.

\subsection{Complex plane}
We will start with the case of complex plane which is usually discussed in the textbooks \cite{DiFrancesco:1997nk}. See also the summary in \cite{Prochazka:2014gqa}. A field $A(z)$ (for simplicity of notation we consider holomorphic fields) of conformal dimension $h$ has Fourier expansion
\begin{equation}
\label{fourierplane}
A^{(pl)}(z) = \sum_{m \in \mathbbm{Z}} z^{-m-h} A^{(pl)}_m.
\end{equation}
There are two basic operations that produce more complicated fields -- the derivative and the normal ordered product of fields.

\paragraph{Derivatives}
In terms of their Fourier modes, the derivative $\partial A$ has modes related to those of $A$ via
\begin{equation}
(\partial A^{(pl)})_n = -(h+n) A^{(pl)}_n
\end{equation}
as can be easily seen by taking the derivative of the Fourier expansion of $A$.

\paragraph{Normal ordering in terms of Fourier modes}
The normal ordered product is slightly more complicated. By definition, the (conformal) normal ordered product $(AB)$ of (bosonic) local fields $A$ and $B$ is defined by
\begin{align}
(AB)^{(pl)}(w) & = \frac{1}{2\pi i} \oint_{w} \frac{dz}{z-w} A^{(pl)}(z)B^{(pl)}(w) \\
\nonumber
& = \frac{1}{2\pi i} \int_{|z|>|w|} \frac{dz}{z-w} A^{(pl)}(z)B^{(pl)}(w) - \frac{1}{2\pi i} \int_{|z|<|w|} \frac{dz}{z-w} B^{(pl)}(w)A^{(pl)}(z)
\end{align}
where the fields are implicitly radially ordered. Using the identities
\begin{equation}
\frac{1}{2\pi i} \int_{|z|>|w|} \frac{dz}{z-w} z^{-n} = \begin{cases} 0 & n > 0 \\ w^{-n} & n \leq 0 \end{cases}
\end{equation}
and
\begin{equation}
\frac{1}{2\pi i} \int_{|z|<|w|} \frac{dz}{z-w} z^{-n} = \begin{cases} -w^{-n} & n > 0 \\ 0 & n \leq 0 \end{cases}
\end{equation}
we find the Fourier expansion of $(AB)$ to be
\begin{equation}
(AB)^{(pl)}(w) = \sum_k w^{-h_A-h_B-k} (AB)^{(pl)}_k
\end{equation}
with
\begin{equation}
(AB)^{(pl)}_n = \sum_{k \leq -h_A} A^{(pl)}_k B^{(pl)}_{n-k} + \sum_{k>-h_A} B^{(pl)}_{n-k} A^{(pl)}_k
\end{equation}
which expresses the Fourier modes of the composite field $(AB)$ in terms of those of $A$ and $B$. If $A$ and $B$ were anticommuting, there will be an additional minus sign every time the order of $A$ and $B$ is changed.

\paragraph{Commutators of Fourier modes}
It is useful to derive the relation between the commutator of Fourier modes of two local fields $A$ and $B$ and their operator product expansion. Given two local fields $A$ and $B$ of dimensions $h_A$ and $h_B$, their Fourier modes can be extracted by the contour integral
\begin{equation}
A^{(pl)}_m = \oint_0 \frac{dz}{2\pi i} A^{(pl)}(z) z^{m+h_A-1}
\end{equation}
and similarly for $B$. We can compute the commutation relations between the modes as
\begin{align}
\label{planecommutator}
\nonumber
\left[ A^{(pl)}_m, B^{(pl)}_n \right] & = \oint_0 \frac{dw}{2\pi i} \int_{|z|>|w|} \frac{dz}{2\pi i} z^{m+h_A-1} w^{n+h_B-1} A^{(pl)}(z) B^{(pl)}(w) \\
\nonumber
& - \oint_0 \frac{dw}{2\pi i} \int_{|z|<|w|} \frac{dz}{2\pi i} z^{m+h_A-1} w^{n+h_B-1} B^{(pl)}(w) A^{(pl)}(z) \\
& = \oint_0 \frac{dw}{2\pi i} \oint_w \frac{dz}{2\pi i} z^{m+h_A-1} w^{n+h_B-1} A^{(pl)}(z) B^{(pl)}(w) \\
\nonumber
& = \sum_{k>0} {m + h_A - 1 \choose k-1} \left\{AB\right\}^{(pl)}_{-k,m+n}
\end{align}
In the second equality we performed the standard contour deformation, deforming the difference of two pairs of contours into a $w$-contour around $w=0$ and $z$-contour around $z=w$. On the right-hand side the notation $\left\{AB\right\}^{(pl)}_{-k,m+n}$ means the $m+n$-th Fourier mode of the local operator that appears as $k$-th order pole in OPE of $A$ and $B$. Let us emphasize that the expression that we derived shows that all the commutation relations between Fourier modes of local operators are determined solely the singular part of the operator product expansion of the two operators.

\subsection{Cylinder}
Let us repeat the previous calculations, this time on the cylinder, i.e. in coordinate $y$ related to the plane coordinate $z$ by the exponential map,
\begin{equation}
\label{cylinderplanemap}
z = e^y.
\end{equation}
Note that for simplicity we do not introduce any signs or imaginary units so in particular the circles around origin in the complex plane are mapped to at line segments with constant real part and imaginary part running over an interval of length $2\pi$.

The Fourier expansion of the local field $A$ on the cylinder is now
\begin{equation}
A^{(cyl)}(y) = \sum_{m \in \mathbbm{Z}} e^{-my} A^{(cyl)}_m.
\end{equation}
In general these Fourier modes are not equal to those calculated in the complex plane,
\begin{equation}
A^{(cyl)}_m \neq A^{(pl)}_m. \qquad \text{(in general)}
\end{equation}

\paragraph{Conformal primary fields}
One exception is a conformal primary field $A(z)$ of conformal dimension $h$ which by definition transforms as
\begin{equation}
\label{primaryfieldtransf}
A^{(z)}(z) dz^h = A^{(w)}(w) dw^h
\end{equation}
under the conformal transformation $z=z(w)$. Specializing to the exponential map between the complex plane and cylinder \eqref{cylinderplanemap}, this gives
\begin{equation}
A^{(pl)}(z) z^h = A^{(cyl)}(y)
\end{equation}
so in this case we indeed have $A^{(cyl)}_m = A^{(pl)}_m$ (actually this relation is the reason why the Fourier modes in \eqref{fourierplane} are conventionally shifted by $h$). But already the stress-energy tensor $T(z)$ has more complicated transformation property than \eqref{primaryfieldtransf} which is responsible for the shift between the zero mode of $T$ on the complex plane and the zero mode on the cylinder (the difference after adding the holomorphic and anti-holomorphic contributions being the Casimir energy).

\paragraph{Derivatives}
The Fourier modes of field $A$ and its derivative $\partial A$ are very simply related:
\begin{equation}
(\partial A^{(cyl)})_m = -m A^{(cyl)}_m.
\end{equation}
In particular, a zero mode of a total derivative vanishes (which was not true in the case of complex plane).

\paragraph{Normal ordering in terms of the Fourier modes}
In order to express the (conformal) normal ordering in terms of the Fourier modes on cylinder, we first invert the Fourier expansion of $A$: \begin{equation}
A^{(cyl)}_m = \frac{1}{2\pi i} \int_0^{2\pi i} e^{my} A^{(cyl)}(y) dy = \frac{1}{2\pi} \int_0^{2\pi} e^{imt+mr} A^{(cyl)}(r+it) dt
\end{equation}
where in the first integral we integrate over a segment of length $2\pi$ parallel to the imaginary axis, i.e. $y = it+r, t \in (0,2\pi), r = const$. The conformal normal ordered product of fields on the cylinder is calculated as (assuming $A$ and $B$ to be bosonic)
\begin{align}
\nonumber
(A^{(cyl)}B^{(cyl)})_{(cyl)}(x) & = \frac{1}{2\pi i} \oint_x \frac{dy}{y-x} A^{(cyl)}(y) B^{(cyl)}(x) \\
\nonumber
& = \frac{1}{2\pi i} \oint_x \frac{dy}{e^{y-x}-1} A^{(cyl)}(y) B^{(cyl)}(x) \\
& - \sum_{k=1}^\infty \oint_x \frac{B_k}{k!} (y-x)^{k-1} A^{(cyl)}(y) B^{(cyl)}(x) \\
\nonumber
& = \frac{1}{2\pi i} \oint_x \frac{dy}{e^{y-x}-1} A^{(cyl)}(y) B^{(cyl)}(x) - \sum_{k>0} \frac{B_k}{k!} \left\{A^{(cyl)} B^{(cyl)}\right\}_{-k}(x)
\end{align}
In order to be able to apply the contour deformation argument, we need the integrand to be a function globally defined on the cylinder (i.e. $2\pi$-periodic in the imaginary direction). For this reason, in the second equality we replaced a non-periodic function $(y-x)^{-1}$ by its periodic version up to a correction that is expressible in terms of singular part of OPE of $A$ and $B$. Here $B_k$ are the Bernoulli numbers ($B_1 = -\frac{1}{2}, B_2 = \frac{1}{6}, B_3 = 0, B_4 = -\frac{1}{30}, \ldots$). The fields are implicitly radially ordered (which in this case means that the field $A(x)$ with $\Re x > \Re y$ appears to the left of $B(y)$ etc.).

We can now use the contour deformation argument to write the first term as
\begin{equation}
\frac{1}{2\pi i} \left( \int_{\Re y > \Re x} - \int_{\Re y < \Re x} \right) \frac{dy}{e^{y-x}-1} A^{(cyl)}(y) B^{(cyl)}(x)
\end{equation}
The integral is along the line segment of constant real part. Plugging in the mode expansions, its remains to evaluate the integrals
\begin{align}
\int_{\Re y > \Re x} \frac{e^{-my} e^{-nx}}{e^{y-x}-1} \frac{dy}{2\pi i} & = \begin{cases} 0 & m \geq 0 \\ e^{-(m+n)x} & m \leq -1 \end{cases} \\
\int_{\Re y < \Re x} \frac{e^{-my} e^{-nx}}{e^{y-x}-1} \frac{dy}{2\pi i} & = \begin{cases} - e^{-(m+n)x} & m \geq 0 \\ 0 & m \leq -1 \end{cases}
\end{align}
where we are integrating along a $y$-contour which is a segment of length $2\pi$ parallel with imaginary axis. Finally, we find the normal ordered product in terms of the Fourier modes on cylinder,
\begin{align}
\nonumber
(A^{(cyl)}B^{(cyl)})_{(cyl)}(x) & = \sum_{n \in \mathbbm{Z}} \sum_{m<0} e^{-(m+n)x} A_m B_n + \sum_{n \in \mathbbm{Z}} \sum_{m \geq 0} e^{-(m+n)x} B_n A_m \\
& - \sum_{k>0} \frac{B_k}{k!} \left\{A^{(cyl)}, B^{(cyl)}\right\}_{-k}(x)
\end{align}
or
\begin{equation}
\label{cylindernormalorder}
(AB)^{(cyl)}_m = \sum_{n<0} A^{(cyl)}_n B_{m-n} + \sum_{n \geq 0} B_{m-n} A_n - \sum_{k>0} \frac{B_k}{k!} \left\{ AB \right\}^{(cyl)}_{-k,m}.
\end{equation}
We see that unlike for the Fourier modes on the complex plane, here the normal order depends on the singular part of the OPE of $A$ and $B$. Just as for the complex plane, anticommuting fields would have an additional minus sign in the second term.

\paragraph{Commutators of Fourier modes}
Finally we derive the expression for the commutator of the Fourier modes in terms of the singular part of the operator product expansion. We have
\begin{align}
\label{cylindercommutator}
\nonumber
\left[ A^{(cyl)}_m, B^{(cyl)}_n \right] & = \frac{1}{(2\pi i)^2} \left( \iint_{\Re y > \Re x} - \iint_{\Re x > \Re y} \right) e^{nx+my} A^{(cyl)}(y) B^{(cyl)}(x) \\
\nonumber
& = \int_0^{2\pi i} \frac{dx}{2\pi i} \oint_x \frac{dy}{2\pi i} e^{nx+my} A^{(cyl)}(y) B^{(cyl)}(x) \\
& = \int_0^{2\pi i} \frac{dx}{2\pi i} \oint_x \frac{dy}{2\pi i} \sum_{k>0} \frac{e^{nx+my}}{(y-x)^k} \left\{A^{(cyl)}B^{(cyl)}\right\}_{-k}(x) \\
\nonumber
& = \int_0^{2\pi i} \frac{dx}{2\pi i} \sum_{k>0} \frac{e^{(m+n)x} m^{k-1}}{(k-1)!} \left\{A^{(cyl)} B^{(cyl)}\right\}_{-k}(x) \\
\nonumber
& = \sum_{k>0} \frac{m^{k-1}}{(k-1)!} \left\{A^{(cyl)}B^{(cyl)}\right\}_{-k,m+n}
\end{align}
In particular, for $m=0$ only the first order pole ($k=1$) contributes and we have a simple relation
\begin{equation}
\left[ A^{(cyl)}_0, B^{(cyl)}(x) \right] = \left\{A^{(cyl)}B^{(cyl)}\right\}_{-1}(x).
\end{equation}
An analogous commutator of a zero mode on the plane is more complicated and involves all of the singular terms of OPE.

\subsection{Example - Virasoro algebra}
The stress-energy tensor $T(z)$ has the following operator product expansion with respect to any choice of complex coordinates
\begin{equation}
T(z) T(w) = \frac{c/2}{(z-w)^4} + \frac{2T(w)}{(z-w)^2} + \frac{\partial T(w)}{z-w} + reg.
\end{equation}
Denoting the Fourier modes of $T(z)$ in the complex plane by $L_m$, we can use \eqref{planecommutator} to find the commutator
\begin{align}
\nonumber
\left[ L_m, L_n \right] & = {m+1 \choose 3} \frac{c}{2} \mathbbm{1}_{m+n} + {m+1 \choose 1} 2L_{m+n} + {m+1 \choose 0} (-m-n-2) L_{m+n} \\
& = \frac{c(m+1)m(m-1)}{12} \delta_{m+n,0} + (m-n) L_{m+n}
\end{align}
which is the Virasoro algebra. Denoting the corresponding Fourier modes on the cylinder by $T_m$, the commutation relations of these are instead
\begin{align}
\nonumber
\left[ T_m, T_n \right] & = \frac{m^3}{6} \frac{c}{2} \mathbbm{1}_{m+n} + 2m T_{m+n} - (m+n) T_{m+n} \\
& = \frac{cm^3}{12} \delta_{m+n,0} + (m-n) T_{m+n}
\end{align}
as follows from \eqref{cylindercommutator}. This is the Virasoro algebra on the cylinder. The two commutation relations are related by the identification
\begin{equation}
T_m = L_m - \frac{c}{24} \delta_{m,0}
\end{equation}
which is a consequence of the transformation property of stress-energy tensor under conformal transformations
\begin{equation}
\left(\frac{dz}{dw}\right)^2 T^{(z)}(z) = T^{(w)}(w) - \frac{c}{12} \left[ \left(\frac{d^3 z}{dw^3}\right) \left( \frac{dz}{dw} \right)^{-1} - \frac{3}{2} \left( \frac{d^2 z}{dw^2} \right)^2 \left( \frac{dz}{dw} \right)^{-2} \right].
\end{equation}
which in the case of exponential map \eqref{cylinderplanemap} simplifies to
\begin{equation}
z^2 T^{(cyl)}(z) = T^{(pl)}(y) + \frac{c}{24}.
\end{equation}

To check the expression for Fourier modes of composite fields, let us consider the zero mode of a local operator $(TT)(x)$ on the cylinder. Using \eqref{cylindernormalorder} we find
\begin{align}
\nonumber
(TT)^{(cyl)}_0 & = T_0^2 + 2 \sum_{m>0} T_{-m} T_m + \frac{1}{720} \frac{c}{2} - \frac{1}{12} 2T_0 + \frac{1}{2} (\partial T)_0 \\
& = T_0^2 + 2\sum_{m>0} T_{-m} T_m + \frac{c}{1440} - \frac{1}{6} T_0 \\
\nonumber
& = L_0^2 + 2\sum_{m>0} L_{-m} L_m - \frac{c+2}{12} L_0 + \frac{c(5c+22)}{2880}
\end{align}
which agrees with expressions given in appendix of \cite{Bazhanov:1996aq}.

As a side comment, observe that the constant term vanishes for $c=-\frac{22}{5}$ which is the Lee-Yang minimal model (related to $(A_1,A_2)$ Argyres-Douglas theory) and for this value of the central charge $(TT)^{(cyl)}_0$ vanishes on the primary state if and only if $h = 0$ and $h = -\frac{1}{5}$ which are exactly the two primaries of the Lee-Yang minimal model.

\section{Large $u$ expansion of R-matrix}

\subsection{Mode expansions}

\label{instapmodes}

Using the results of appendix \ref{apmodes}, we can find the cylinder Fourier modes of the composite operators constructed out of the local current $J_-$ whose OPE is
\begin{equation}
J_-(z) J_-(w) \sim \frac{2\alpha_0^2}{(z-w)^2} + reg.
\end{equation}
that enters the large central charge expansion of the instanton $\mathcal{R}$-matrix. We find using \eqref{cylindernormalorder}
\begin{align}
(J_- J_-)_m & = : J_-^2 :_m - \frac{2\alpha_0^2}{12} \delta_{m,0} \\
(J_- (J_- J_-))_m & = : J_-^3 :_m - \frac{2\alpha_0^2}{4} :J_-:_m \\
(J_-(J_- (J_- J_-)))_m & = : J_-^4 :_m - \frac{2\alpha_0^2}{2} :J_-^2:_m + \frac{4\alpha_0^4}{48} \delta_{m,0} \\
(\partial J_- \partial J_-)_m & = :\partial J_- \partial J_-:_m - \frac{2\alpha_0^2}{120} \delta_{m,0}
\end{align}
where $:J_-^k:$ denotes the creation-annihilation normal ordering and we suppress the subscript $(cyl)$ associated to Fourier modes on the left-hand side.

\subsection{Large $u$ expansion of $\mathcal{R}$ at $N=1$}
\label{bosonicrlist}
Here we collect expressions for first coefficients $r^{(j)}$ of the large $u$ expansion of the logarithm of Maulik-Okounkov $\mathcal{R}$-matrix. First five of these were given in \cite{Prochazka:2019dvu} in terms of oscillators. Here we write them as zero modes of local currents. In this appendix we write $J$ instead of $J_-$ to simplify the notation. All the zero modes are with respect to coordinates on the cylinder.
\begin{align}
r^{(1)} & = -\frac{1}{2} (JJ)_0 \\
r^{(2)} & = \frac{1}{6} (J(JJ))_0 \\
r^{(3)} & = -\frac{1}{12} (J(J(JJ)))_0 + \frac{\alpha_0^2(1+2\alpha_0^2)}{12} (\partial J \partial J)_0 \\
r^{(4)} & = \frac{1}{20} (J(J(J(JJ))))_0 - \frac{\alpha_0^2(1+2\alpha_0^2)}{4} (\partial J (\partial J J))_0 \\
\nonumber
r^{(5)} & = -\frac{1}{30} (J(J(J(J(JJ)))))_0 + \frac{\alpha_0^2(1+2\alpha_0^2)}{2} (\partial J (\partial J (JJ)))_0 \\
& - \frac{\alpha_0^4(2+9\alpha_0^2+6\alpha_0^4)}{60} (\partial^2 J \partial^2 J)_0 \\
\nonumber
r^{(6)} & = \frac{1}{42} (J(J(J(J(J(JJ))))))_0 - \frac{5\alpha_0^2(1+2\alpha_0^2)}{6} (\partial J (\partial J (J(JJ))))_0 \\
& + \frac{\alpha_0^4(2+9\alpha_0^2+6\alpha_0^4)}{12} (\partial^2 J (\partial^2 J J))_0 \\
\nonumber
r^{(7)} & = -\frac{1}{56} (J(J(J(J(J(J(JJ)))))))_0 + \frac{5\alpha_0^2(1+2\alpha_0^2)}{4} (\partial J (\partial J (J(J(JJ)))))_0 \\
\nonumber
& - \frac{\alpha_0^4(2+9\alpha_0^2+6\alpha_0^4)}{4} (\partial^2 J (\partial^2 J (JJ)))_0 + \frac{\alpha_0^4(9+41\alpha_0^2+26\alpha_0^4)}{72} (\partial J (\partial J(\partial J \partial J)))_0 \\
& + \frac{\alpha_0^6(90+671\alpha_0^2+998\alpha_0^4+360\alpha_0^6)}{5040} (\partial^3 J \partial^3 J)_0 \\
\nonumber
r^{(8)} & = \frac{1}{72} (J(J(J(J(J(J(J(JJ))))))))_0 - \frac{7\alpha_0^2(1+2\alpha_0^2)}{4} (\partial J (\partial J (J(J(J(JJ))))))_0 \\
\nonumber
& + \frac{7\alpha_0^4(2+9\alpha_0^2+6\alpha_0^4)}{12} (\partial^2 J (\partial^2 J (J(JJ))))_0 \\
& - \frac{7\alpha_0^4(9+41\alpha_0^2+26\alpha_0^4)}{72} (\partial J (\partial J(\partial J (\partial J J))))_0 \\
\nonumber
& - \frac{\alpha_0^6(90+671\alpha_0^2+998\alpha_0^4+360\alpha_0^6)}{720} (\partial^3 J (\partial^3 J J))_0 \\
\nonumber
& + \frac{\alpha_0^6(25+186\alpha_0^2+278\alpha_0^4+100\alpha_0^6)}{180} (\partial^2 J (\partial^2 J (\partial^2 J)))_0
\end{align}

\subsection{Large $u$ expansion of universal $\mathcal{R}$}
\label{universalrlist}
In this appendix we list expressions of coefficients $\mathbf{r}^{(j)}$ of the large central charge expansion of the logarithm of $\mathbf{R}(u)$, see \eqref{largeuexpunir}. We have
\begin{align}
\mathbf{r}^{(1)} & = -\alpha_0 \bar{N} T_0 -\alpha_0 N \bar{T}_0 + \alpha_0 (U_1 \bar{U}_1)_0 \\
\mathbf{r}^{(2)} & = \frac{\alpha_0 \bar{N}}{2} \phi_{3,0} - \alpha_0 (T \bar{U}_1)_0 + \alpha_0 (U_1 \bar{T})_0 - \frac{\alpha_0 N}{2} \bar{\phi}_{3,0} \\
\nonumber
\mathbf{r}^{(3)} & = \frac{\alpha_0 \bar{N}}{3} \phi_{4,0} +\alpha_0 (\phi_3 \bar{U}_1)_0 -2\alpha_0 (T\bar{T})_0 +\alpha_0 (U_1 \bar{\phi}_3)_0 +\frac{N\alpha_0}{3} \bar{\phi}_{4,0} \\
& + \frac{\bar{N}(\bar{N}-1)\alpha_0^3}{12} (\partial U_1 \partial U_1)_0 + \frac{N(N-1)\alpha_0^3}{12} (\partial \bar{U}_1 \partial \bar{U}_1)_0 \\
\nonumber
& -\frac{1}{6} \alpha_0 (\alpha_0^2 N \bar{N}-\alpha_0^2 N-\alpha_0^2 \bar{N}-\alpha_0^2-1) (\partial U_1 \partial \bar{U}_1)_0 \\
\nonumber
\mathbf{r}^{(4)} & = \frac{\alpha_0 \bar{N}}{4} \phi_{5,0} +\alpha_0 (\bar{U}_1 \phi_4)_0 +3 \alpha_0 (\phi_3 \bar{T})_0 -3 \alpha_0 (T \bar{\phi}_3)_0 -\alpha_0 (U_1 \bar{\phi}_4)_0 -\frac{\alpha_0 N}{4} \bar{\phi}_{5,0} \\
\nonumber
& - \frac{(N\bar{N}\alpha_0^2-2\bar{N}\alpha_0^2-N\alpha_0^2-2\alpha_0^2-2)\alpha_0}{4} (\bar{U}_1 \partial^2 T)_0 - \frac{(\bar{N}-1)\alpha_0^3}{4} (\bar{U}_1 (\partial^2 U_1 U_1))_0 \\
& + \frac{(N\bar{N}\alpha_0^2-2N\alpha_0^2-\bar{N}\alpha_0^2-2\alpha_0^2-2)\alpha_0}{4} (U_1 \partial^2 \bar{T})_0 + \frac{(N-1)\alpha_0^3}{4} (U_1 (\partial^2 \bar{U}_1 \bar{U}_1))_0 \\
\nonumber
& - \frac{N(N-1)\alpha_0^3}{4} (\bar{U}_1 \partial^2 \bar{T})_0 + \frac{\bar{N}(\bar{N}-1)\alpha_0^3}{4} (U_1 \partial^2 T)_0
\end{align}
and more complicated expressions for higher orders. We defined the local fields
\begin{align}
T & = -U_2 + \frac{1}{2} (U_1 U_1) + \frac{(N-1)\alpha_0}{2} \partial U_1 \\
\nonumber
\phi_3 & = U_3 - (U_1 U_2) + \frac{1}{3} (U_1(U_1 U_1)) - \frac{(N-2)\alpha_0}{2} U_2^\prime \\
& + \frac{(N-1)\alpha_0}{2} (U_1^\prime U_1) + \frac{N^2\alpha_0^2-N\alpha_0^2+4\alpha_0^2+2}{12} U_1^{\prime\prime} \\
\nonumber
\phi_4 & = U_4 - (U_1 U_3) - \frac{1}{2} (U_2 U_2) + (U_1(U_1 U_2)) - \frac{1}{4} (U_1(U_1(U_1 U_1))) \\
\nonumber
& + \frac{(N-2)\alpha_0}{2} (U_1 U_2^\prime) + \frac{(N-1)\alpha_0}{2} (U_1^\prime U_2) - \frac{(N-1)\alpha_0}{2} (U_1^\prime (U_1 U_1)) \\
& - \frac{(N-3)\alpha_0}{2} U_3^\prime - \frac{N^2\alpha_0^2+2N\alpha_0^2+3\alpha_0^2+3}{12} (U_1^\prime U_1^\prime) \\
\nonumber
& - \frac{N^2\alpha_0^2-N\alpha_0^2+12\alpha_0^2-3N+9}{12} (U_1^{\prime\prime} U_1) + \frac{N^2\alpha_0^2-3N\alpha_0^2+12\alpha_0^2+3}{12} U_2^{\prime\prime} \\
\nonumber
& - \frac{(N-1)(2N\alpha_0^2+6\alpha_0^2-2N+5)\alpha_0}{24} U_1^{\prime\prime\prime} \\
\nonumber
\phi_5 & = U_5 - (U_1 U_4) - (U_2 U_3) + (U_1 (U_1 U_3)) + (U_1 (U_2 U_2)) - (U_1 (U_1 (U_1 U_2))) \\
\nonumber
& + \frac{1}{5} (U_1(U_1(U_1(U_1 U_1)))) - \frac{(N-2)\alpha_0}{2} (U_1 (U_1 U_2^\prime)) - (N-1)\alpha_0 (U_1^\prime (U_1 U_2)) \\
\nonumber
& + \frac{(N-3)\alpha_0}{2} (U_1 U_3^\prime) + \frac{(N-1)\alpha_0}{2} (U_1^\prime U_3) + \frac{(N-1)\alpha_0}{2} (U_1^\prime (U_1 (U_1 U_1))) \\
\nonumber
& + \frac{(N-2)\alpha_0}{2} (U_2^\prime U_2) - \frac{(N-4)\alpha_0}{2} U_4^\prime - \frac{N^2\alpha_0^2-3N\alpha_0^2+22\alpha_0^2+6}{12} (U_1
 U_2^{\prime\prime}) \\
& + \frac{N^2\alpha_0^2+3N\alpha_0^2+8\alpha_0^2+6}{6} (U_1^\prime (U_1^\prime U_1)) - \frac{N^2\alpha_0^2+2N\alpha_0^2+6\alpha_0^2+6}{6} (U_1^\prime U_2^\prime) \\
\nonumber
& - \frac{N^2\alpha_0^2-N\alpha_0^2+26\alpha_0^2-6N+24}{12} (U_1^{\prime\prime} U_2) + \frac{(N-1)(3N\alpha_0^2+8\alpha_0^2+6)\alpha_0}{6} (U_1^{\prime\prime} U_1^\prime) \\
\nonumber
& + \frac{N^2\alpha_0^2-N\alpha_0^2+24\alpha_0^2-6N+18}{12} (U_1^{\prime\prime} (U_1 U_1)) + \frac{N^2\alpha_0^2-5N\alpha_0^2+24\alpha_0^2+6}{12} U_3^{\prime\prime} \\
\nonumber
& + \frac{(N-1)(N\alpha_0^2+8\alpha_0^2-N+3)\alpha_0}{12} (U_1^{\prime\prime\prime} U_1) - \frac{(N-2)(N\alpha_0^2+6\alpha_0^2+4)\alpha_0}{12} U_2^{\prime\prime\prime} \\
\nonumber
& - \frac{N^4\alpha_0^4-16N^3\alpha_0^4-19N^2\alpha_0^4+34N\alpha_0^4-144\alpha_0^4-15N^3\alpha_0^2+5N\alpha_0^2-206\alpha_0^2-48}{720} U_1^{(4)}
\end{align}
(see the conventions used in \cite{Prochazka:2014gqa}).

\section{Explicit expressions for ILW Hamiltonians}
\label{apilw}

\subsection{(Quasi-)modular forms}
First few Eisenstein series are
\begin{align}
E_2(q) & = 1 - 24 \sum_{m>0} \frac{m q^m}{1-q^m} \\
E_4(q) & = 1 + 240 \sum_{m>0} \frac{m^3 q^m}{1-q^m} \\
E_6(q) & = 1 - 504 \sum_{m>0} \frac{m^5 q^m}{1-q^m} \\
E_8(q) & = 1 + 480 \sum_{m>0} \frac{m^7 q^m}{1-q^m} = E_4(q)^2.
\end{align}
These satisfy the useful relations
\begin{align}
q \partial_q E_2 & = -\frac{E_4 - E_2^2}{12} = -24 \sum_{m>0} \frac{m^2 q^m}{(1-q^m)^2} \\
q \partial_q E_4 & = -4\frac{E_6 - E_2 E_4}{12} = 240 \sum_{m>0} \frac{m^4 q^m}{(1-q^m)^2} \\
q \partial_q E_6 & = -6\frac{E_8 - E_2 E_6}{12}.
\end{align}
All Eistenstein series $E_{2k}$ with $2k \geq 8$ are algebraically dependent on the basic ones $E_2, E_4$ and $E_6$. $E_2$ is not modular but only quasi-modular and the ring of modular forms is generated by $E_4$ and $E_6$ only. Allowing differentiation $q \partial_q$ as well, we see that everything can be expressed in terms of $E_2$ only.

\subsection{Calculation of expectation values in auxiliary space}

In order to find explicit formulas for ILW Hamiltonians from the universal $\mathcal{R}$-matrix, we need to take torus expectation values of various quantities in the bosonic Fock space. Here we list various expectation values of these. We start with two simple lemmas:

\paragraph{Lemma 1}
\begin{align}
\langle \mathcal{O} \bar{T}_0 \rangle_q & = \left( q \partial_q - \frac{E_2}{24} \right) \langle \mathcal{O} \rangle_q \\
\langle \mathcal{O} \bar{T}_0 \rangle_{qc} & = q \partial_q \langle \mathcal{O} \rangle_q
\end{align}
(here the subscript $q$ refers to torus expectation value with modular parameter $q$ while $c$ to connected correlation function). \\
Proof: for the first equation, we have
\begin{align}
\nonumber
\langle \mathcal{O} \bar{L}_0 \rangle_q & = \frac{q \partial_q \sum_{\lambda} q^{|\lambda|} \bra{\bar{\lambda}} \mathcal{O} \ket{\bar{\lambda}}}{\sum_{\lambda} q^{|\lambda|}} \\
& = q \partial_q \left( \frac{\sum_{\lambda} q^{|\lambda|} \bra{\bar{\lambda}} \mathcal{O} \ket{\bar{\lambda}}}{\sum_{\lambda} q^{|\lambda|}} \right) + \frac{\sum_{\lambda} q^{|\lambda|} \bra{\bar{\lambda}} \mathcal{O} \ket{\bar{\lambda}}}{\sum_\lambda q^{|\lambda|}} \frac{q \partial_q \sum_{\lambda} q^{|\lambda|}}{\sum_{\lambda} q^{|\lambda|}} \\
\nonumber
& = \left( q \partial_q + \frac{1-E_2}{24} \right) \langle \mathcal{O} \rangle_q.
\end{align}
Therefore,
\begin{equation}
\langle \mathcal{O} \bar{T}_0 \rangle_q = \langle \mathcal{O} \bar{L}_0 \rangle_q - \frac{1}{24} \langle \mathcal{O} \rangle_q = \left( q \partial_q - \frac{E_2}{24} \right) \langle \mathcal{O} \rangle_q
\end{equation}
as we wanted to show. For the second claim, we have simply
\begin{equation}
\langle \mathcal{O} \bar{T}_0 \rangle_{qc} \equiv \langle \mathcal{O} \bar{T}_0 \rangle_q - \langle \mathcal{O} \rangle_q \langle \bar{T}_0 \rangle_q = q \partial_q \langle \mathcal{O} \rangle_q
\end{equation}
Analogously, we have
\begin{align}
\nonumber
\langle \bar{T}_0^2 \mathcal{O} \rangle_{qc} & \equiv \langle \bar{T}_0^2 \mathcal{O} \rangle_q - \langle\bar{T}_0^2 \rangle_q \langle\mathcal{O}\rangle_q -2\langle\bar{T}_0\rangle_q \langle \bar{T}_0 \mathcal{O} \rangle_q +2 \langle\bar{T}_0\rangle_q^2 \langle\mathcal{O}\rangle_q \\
\nonumber
& = \left( q \frac{\partial}{\partial q} - \frac{E_2}{24} \right)^2 \langle \mathcal{O} \rangle_q - \frac{E_4 - E_2^2}{288} \langle \mathcal{O} \rangle_q + \frac{E_2}{12} \left( q \frac{\partial}{\partial q} - \frac{E_2}{24} \right) \langle \mathcal{O} \rangle_q + \frac{E_2^2}{576} \langle \mathcal{O} \rangle_q \\
& = \left( q \frac{\partial}{\partial q} \right)^2 \langle \mathcal{O} \rangle_q - \left( q \frac{\partial}{\partial q} \frac{E_2}{24} \right) \langle \mathcal{O} \rangle_q - \frac{E_2}{12} \left( q \frac{\partial}{\partial q} \right) \langle \mathcal{O} \rangle_q + \frac{E_2^2}{576} \langle \mathcal{O} \rangle_q \\
\nonumber
& - \frac{E_4 - E_2^2}{288} \langle \mathcal{O} \rangle_q + \frac{E_2}{12} \left( q \frac{\partial}{\partial q} - \frac{E_2}{24} \right) \langle \mathcal{O} \rangle_q + \frac{E_2^2}{576} \langle \mathcal{O} \rangle_q \\
\nonumber
& = \left( q \frac{\partial}{\partial q} \right)^2 \langle \mathcal{O} \rangle_q.
\end{align}

\paragraph{Lemma 2}
For two local fields $A$ and $B$ in the left sector (independent of $\bar{J}$), we have
\begin{align}
\langle (A \bar{J})_0 (B \bar{J})_0 \rangle_q & = \sum_{m>0} \frac{m}{1-q^m} A_{-m} B_m + \sum_{m>0} \frac{mq^m}{1-q^m} A_m B_{-m} \\
& \equiv \sum_{m>0} \mathfrak{d}_m A_{-m} B_m + \sum_{m>0} \mathfrak{d}_{-m} A_m B_{-m}
\end{align}
where we introduced the derivative
\begin{equation}
\mathfrak{d}_m \equiv \frac{m}{1-q^m}.
\end{equation}
Proof: first of all, one can show easily by calculation similar to evaluation of $\langle \bar{T}_0 \rangle_q$ that
\begin{equation}
\langle \bar{J}_{m} \bar{J}_{-m} \rangle_q = q \partial_q \log \frac{1}{1-q^m} = \frac{m}{1-q^m} = \frac{-m q^{-m}}{1-q^{-m}}, \qquad m \neq 0
\end{equation}
which holds for both positive and negative $m$. Due to the fact that $A$ and $\bar{J}$ commute (and therefore have regular OPE), their normal ordered product is just their product (and the same for $A \leftrightarrow B$) so
\begin{align}
\langle (A \bar{J})_0 (B \bar{J})_0 \rangle_q & = \sum_{m \in \mathbbm{Z}} A_{-m} B_m \langle \bar{J}_m \bar{J}_{-m} \rangle_q \\
& = \sum_{m>0} \frac{m}{1-q^m} A_{-m} B_m + \sum_{m>0} \frac{m q^m}{1-q^m} A_m B_{-m}
\end{align}
as we wanted to show. We can in principle normal order the right-hand side,
\begin{align}
\langle (A \bar{J})_0 (B \bar{J})_0 \rangle_q & = \sum_{m>0} \frac{m}{1-q^m} A_{-m} B_m + \sum_{m>0} \frac{m q^m}{1-q^m} B_{-m} A_m + \sum_{m>0} \frac{m q^m}{1-q^m} \left[ A_m, B_{-m} \right]
\end{align}
As an application of this formula, we have
\begin{equation}
\langle (U_1 \bar{J})_0 (U_1 \bar{J})_0 \rangle_q = \sum_{m>0} m\frac{1+q^m}{1-q^m} U_{1,-m} U_{1,m} + N \sum_{m>0} \frac{m^2 q^m}{1-q^m}.
\end{equation}
Similarly, we have the following correlators
\begin{align}
\langle \bar{T}_m \bar{T}_{-m} \rangle_q & = \frac{m(m^2-E_2)}{12(1-q^m)} \\
\langle \bar{J}_m (\bar{J}(\bar{J}\bar{J}))_{-m} \rangle_q & = -\frac{mE_2}{4(1-q^m)} \\
\langle (\bar{J}(\bar{J}\bar{J}))_m \bar{J}_{-m} \rangle_q & = -\frac{mE_2}{4(1-q^m)} \\
\langle \bar{J}_m \bar{J}_n \bar{T}_{-m-n} \rangle_q & = \frac{mn}{(1-q^m)(1-q^n)} \\
\langle \bar{J}_{-m} \bar{J}_m \bar{J}_{-n} \bar{J}_n \rangle_q & = \frac{mn q^m (\delta_{mn} + q^n)}{(1-q^m)(1-q^n)}, \quad m,n>0
\end{align}
which lead to analogous formulas as in lemma 2.

\paragraph{Some properties of $\mathfrak{d}_m$}
The modified derivatives $\mathfrak{d}_m$ satisfy various relations. Their derivatives are
\begin{equation}
q \partial_q \mathfrak{d}_m = \mathfrak{d}_m \mathfrak{d}_{-m}.
\end{equation}
The difference of these derivatives gives an ordinary derivative,
\begin{equation}
\mathfrak{d}_m - \mathfrak{d}_{-m} = m
\end{equation}
while their sum
\begin{equation}
\mathfrak{d}_m + \mathfrak{d}_{-m} = m \frac{1+q^m}{1-q^m}
\end{equation}
is the integral kernel appearing in ILW equation and interpolates between Benjamin-Ono or Yangian limit and Bahanov-Lukyanov-Zamolodchikov or Kadomtsev-Petviashvili limits. There are also more complicated identities involving infinite sums such as
\begin{equation}
\sum_{\substack{j \in \mathbbm{Z} \\ j \neq 0,m}} \mathfrak{d}_{j} \mathfrak{d}_{m-j} = \frac{m^2-E_2}{6} \mathfrak{d}_m.
\end{equation}

\subsection{List of expectation values in auxiliary space}
Here is a list of useful expectation values.
\begin{align}
\langle \bar{T}_0 \rangle_q & = -\frac{E_2}{24} \\
\langle \bar{T}_0^2 \rangle_{qc} & = \frac{E_4 - E_2^2}{288} \\
\langle \bar{T}_0^3 \rangle_{qc} & = \frac{-2E_6+3E_2 E_4-E_2^3}{1728} \\
\langle \bar{T}_0^4 \rangle_{qc} & = \frac{3 E_4^2 - 8 E_2 E_6 + 6 E_2^2 E_4 - E_2^4}{6912} \\
\langle (\partial \bar{J} \partial \bar{J})_0 \rangle_q & = -\frac{1}{120} - 2\sum_{m>0} \frac{m^3 q^m}{1-q^m} = -\frac{E_4}{120} \\
\langle (\bar{J} (\bar{J} (\bar{J}\bar{J})))_0 \rangle_q & = \frac{E_2^2}{48} \\
\langle (\bar{J} (\bar{J}\bar{J}))_0 (\bar{J} (\bar{J}\bar{J}))_0 \rangle_q & = \frac{-2E_6-3E_2 E_4+5E_2^3}{720} \\
\langle (\bar{J} (\bar{J} (\bar{J}\bar{J})))_0 \bar{T}_0 \rangle_{qc} & = \frac{-E_2 E_4 + E_2^3}{288} \\
\langle (\partial \bar{J} \partial \bar{J})_0 \bar{T}_0 \rangle_{qc} & = \frac{E_6-E_2 E_4}{360}
\end{align}

\subsection{Third order ILW Hamiltonian}
We need the following expectation values:
\begin{align}
\nonumber
\langle \mathbf{r}^{(3)} \rangle & = \frac{\alpha_0}{3} \phi_{4,0} -2\alpha_0 T_0 \langle \bar{T}_0 \rangle +\frac{N\alpha_0}{3} \langle \bar{\phi}_{4,0} \rangle + \frac{N(N-1)\alpha_0^3}{12} (\partial \bar{J} \partial \bar{J})_0 \\
& = \frac{\alpha_0}{3} \phi_{4,0} +\frac{\alpha_0 E_2}{12} T_0 -\frac{N\alpha_0 E_2^2}{576} -\frac{N\alpha_0(1+\alpha_0^2+N\alpha_0^2) E_4}{1440} \\
\nonumber
\langle \mathbf{r}^{(1)} \mathbf{r}^{(2)} \rangle_{sc} & \equiv \frac{1}{2} \langle \mathbf{r}^{(1)} \mathbf{r}^{(2)} \rangle + \frac{1}{2} \langle \mathbf{r}^{(2)} \mathbf{r}^{(1)} \rangle - \frac{1}{2} \langle \mathbf{r}^{(1)} \rangle \langle \mathbf{r}^{(2)} \rangle + \frac{1}{2} \langle \mathbf{r}^{(2)} \rangle \langle \mathbf{r}^{(1)} \rangle \\
\nonumber
& = -\alpha_0^2 N U_{1,0} \langle \bar{T}_0^2 \rangle_c -\frac{\alpha_0^2}{2} \langle (U_1 \bar{J})_0 (T \bar{J})_0 \rangle_c -\frac{\alpha_0^2}{2} \langle (T \bar{J})_0 (U_1 \bar{J})_0 \rangle_c \\
\nonumber
& = -\alpha_0^2 N U_{1,0} \frac{E_4-E_2^2}{288} -\frac{\alpha_0^2}{2} \sum_{m>0} \frac{m(1+q^m)}{1-q^m} (U_{1,-m} T_m + T_{-m} U_{1,m}) \\
& -\alpha_0^2 \sum_{m>0} \frac{m^2 q^m}{1-q^m} U_{1,0} \\
\nonumber
\frac{1}{6} \langle \mathbf{r}^{(1)3} \rangle_{sc} & \equiv \frac{1}{6} \langle \mathbf{r}^{(1)3} \rangle - \frac{1}{4} \langle \mathbf{r}^{(1)2} \rangle \langle \mathbf{r}^{(1)} \rangle - \frac{1}{4} \langle \mathbf{r}^{(1)} \rangle \langle \mathbf{r}^{(1)2} \rangle + \frac{1}{3} \langle \mathbf{r}^{(1)} \rangle^3 \\
\nonumber
& = -\frac{\alpha_0^3 N^3}{6} \langle \bar{T}_0 \bar{T}_0 \bar{T}_0 \rangle_c +\frac{\alpha_0^3}{12} T_0 \langle(U_1 \bar{U}_1)_0 (U_1 \bar{U}_1)_0\rangle -\frac{\alpha_0^3}{6} \langle(U_1 \bar{U}_1)_0 T_0 (U_1 \bar{U}_1)_0\rangle \\
\nonumber
& +\frac{\alpha_0^3}{12} \langle(U_1 \bar{U}_1)_0 (U_1 \bar{U}_1)_0 \rangle T_0 -\frac{\alpha_0^3 N}{3} \langle\bar{T}_0 (U_1 \bar{U}_1)_0 (U_1 \bar{U}_1)_0\rangle \\
\nonumber
& -\frac{\alpha_0^3 N}{6} \langle(U_1 \bar{U}_1)_0 \bar{T}_0 (U_1 \bar{U}_1)_0\rangle +\frac{\alpha_0^3 N}{2} \langle\bar{T}_0\rangle \langle(U_1 \bar{U}_1)_0 (U_1 \bar{U}_1)_0\rangle \\
\nonumber
& = -\frac{\alpha_0^3 N^3}{6} \frac{-2E_6+3E_2 E_4-E_2^3}{1728} +\frac{\alpha_0^3}{12} \langle(\partial U_1 \bar{U}_1)_0 (U_1 \bar{U}_1)_0\rangle \\
& -\frac{\alpha_0^3}{12} \langle(U_1 \bar{U}_1)_0 (\partial U_1 \bar{U}_1)_0\rangle -\frac{\alpha_0^3 N}{2} \langle\bar{T}_0 (U_1 \bar{U}_1)_0 (U_1 \bar{U}_1)_0\rangle \\
\nonumber
& +\frac{\alpha_0^3 N}{2} \langle\bar{T}_0\rangle \langle(U_1 \bar{U}_1)_0 (U_1 \bar{U}_1)_0\rangle +\frac{\alpha_0^3 N}{6} \langle(U_1 \bar{U}_1)_0 (\partial U_1 \bar{U}_1)_0\rangle \\
\nonumber
& = -\frac{\alpha_0^3 N^3}{6} \frac{-2E_6+3E_2 E_4-E_2^3}{1728} \\
\nonumber
& -\alpha_0^3 N \sum_{m>0} \frac{m^2 q^m}{(1-q^m)^2} U_{1,-m} U_{1,m} -\frac{\alpha_0^3 N^2}{2} \sum_{m>0} \frac{m^3 q^m}{(1-q^m)^2} \\
\nonumber
& -\frac{\alpha_0^3 (N-1)}{6} \sum_{m>0} m^2 U_{1,-m} U_{1,m} +\frac{\alpha_0^3 N(N-1)}{6} \sum_{m>0} \frac{m^3 q^m}{1-q^m}
\end{align}
Adding everything, we finally find
\begin{align}
\nonumber
(\log \mathcal{H}_q)_3 & = \frac{\alpha_0}{3} \phi_{4,0} -\frac{\alpha_0^2}{2} \sum_{m>0} \frac{m(1+q^m)}{1-q^m} (U_{1,-m} T_m + T_{-m} U_{1,m}) \\
\nonumber
& +\frac{\alpha_0^3 (N+2)}{12} \sum_{m>0} m^2 U_{1,-m} U_{1,m} -\frac{\alpha_0^3 N}{4} \sum_{m>0} \frac{m^2 (1+q^m)^2}{(1-q^m)^2} U_{1,-m} U_{1,m} \\
& +\frac{\alpha_0 E_2}{12} T_0 -\alpha_0^2 N \frac{E_4-E_2^2}{288} U_{1,0} -\alpha_0^2 \sum_{m>0} \frac{m^2 q^m}{1-q^m} U_{1,0} \\
\nonumber
& +\alpha_0^3 N^3 \frac{2E_6-3E_2 E_4+E_2^3}{10368} +\frac{\alpha_0^3 N(N-1)}{6} \sum_{m>0} \frac{m^3 q^m}{1-q^m} \\
\nonumber
& -\frac{\alpha_0^3 N^2}{2} \sum_{m>0} \frac{m^3 q^m}{(1-q^m)^2} -\frac{N\alpha_0(1+\alpha_0^2+N\alpha_0^2) E_4}{1440} -\frac{N\alpha_0 E_2^2}{576}
\end{align}

\subsection{Fourth order ILW Hamiltonian}
\begin{align}
\langle \mathbf{r}^{(4)} \rangle & = \frac{\alpha_0}{4} \phi_{5,0} -\frac{\alpha_0 E_2}{8} \phi_{3,0} +\frac{\alpha_0 E_2^2}{192} U_{1,0} +\frac{\alpha_0(1+\alpha_0^2+N\alpha_0^2)E_4}{480} U_{1,0} \\
\frac{1}{2} \langle \mathbf{r}^{(1)} \mathbf{r}^{(3)} \rangle_{sc} & = \frac{1}{2} \left( \langle \mathbf{r}^{(1)} \mathbf{r}^{(3)} \rangle - \langle \mathbf{r}^{(1)} \rangle \langle \mathbf{r}^{(3)} \rangle + \langle \mathbf{r}^{(3)} \mathbf{r}^{(1)} \rangle - \langle \mathbf{r}^{(3)} \rangle \langle \mathbf{r}^{(1)} \rangle\right) \\
\nonumber
& = 2\alpha_0^2 N \langle \bar{T}_0^2 \rangle_c T_0 +\frac{\alpha_0^2}{2} \langle (U_1 \bar{J})_0 (\phi_3 \bar{J})_0 \rangle +\frac{\alpha_0^2}{2} \langle (\phi_3 \bar{J})_0 (U_1 \bar{J})_0 \rangle \\
\nonumber
& +\frac{\alpha_0^2}{6} \langle (U_1 \bar{J})_0 (U_1 (\bar{J}(\bar{J}\bar{J})))_0 \rangle +\frac{\alpha_0^2}{6} \langle (U_1 (\bar{J}(\bar{J}\bar{J})))_0 (U_1 \bar{J})_0 \rangle \\
\nonumber
& +\frac{\alpha_0^2 N^2}{12} \langle \bar{T}_0 (\bar{J}(\bar{J}(\bar{J}\bar{J})))_0 \rangle_c -\frac{\alpha_0^2(1+\alpha_0^2+N\alpha_0^2)N^2}{12} \langle \bar{T}_0 (\partial \bar{J} \partial \bar{J})_0 \rangle_c \\
\nonumber
& = \frac{\alpha_0^2}{2} \sum_{m>0} \frac{m(1+q^m)}{1-q^m} (U_{1,-m} \phi_{3,m} + \phi_{3,-m} U_{1,m}) \\
\nonumber
& -\frac{\alpha_0^2 E_2}{12} \sum_{m>0} \frac{m(1+q^m)}{1-q^m} U_{1,-m} U_{1,m} \\
\nonumber
& + \alpha_0^2 N \frac{E_4-E_2^2}{144} T_0 +2\alpha_0^2 \sum_{m>0} \frac{m^2 q^m}{1-q^m} T_0 \\
\nonumber
& +\frac{\alpha_0^2}{6} (1+\alpha_0^2+N\alpha_0^2) N \sum_{m>0} \frac{m^4 q^m}{1-q^m} -\frac{\alpha_0^2 N E_2}{12} \sum_{m>0} \frac{m^2 q^m}{1-q^m} \\
\nonumber
& +\frac{\alpha_0^2 N^2}{3456}(-E_2 E_4 + E_2^3) -\frac{\alpha_0^2(1+\alpha_0^2+N\alpha_0^2)N^2}{4320} (E_6-E_2 E_4) \\
\frac{1}{2} \langle \mathbf{r}^{(2)2} \rangle_{c} & = \frac{1}{2} \langle \mathbf{r}^{(2)2} \rangle - \frac{1}{2} \langle \mathbf{r}^{(2)} \rangle^2 \\
\nonumber
& = +\frac{\alpha_0^2}{2} \langle (T \bar{J})_0 (T \bar{J})_0 \rangle +\frac{\alpha_0^2}{2} \langle (U_1 \bar{T})_0 (U_1 \bar{T})_0 \rangle_c +\frac{\alpha_0^2 N^2}{72} \langle (\bar{J}(\bar{J}\bar{J}))_0 (\bar{J}(\bar{J}\bar{J}))_0 \rangle \\
\nonumber
& = +\frac{\alpha_0^2}{2} \sum_{m>0} \frac{m(1+q^m)}{1-q^m} T_{-m} T_m +\alpha_0^2 \sum_{m>0} \frac{m^2 q^m}{1-q^m} T_0 \\
\nonumber
& +\frac{\alpha_0^2}{2} \sum_{m>0} \frac{m(m^2-E_2)(1+q^m)}{12(1-q^m)} U_{1,-m} U_{1,m} \\
\nonumber
& +\frac{\alpha_0^2 N^2}{72} \frac{-2E_6-3E_2 E_4+5E_2^3}{720} +\frac{\alpha_0^2 N}{2} \sum_{m>0} \frac{m^2 q^m(m^2-E_2)}{12(1-q^m)} \\
\nonumber
& +\frac{\alpha_0^2 N(1+\alpha_0^2-\alpha_0^2 N^2)}{24} \sum_{m>0} \frac{m^4 q^m}{1-q^m} +\frac{\alpha_0^2}{2} \frac{E_4-E_2^2}{288} U_{1,0}^2 \\
\nonumber
\frac{1}{2} \langle \mathbf{r}^{(1)2} \mathbf{r}^{(2)} \rangle_{sc} & = \left( \frac{1}{6} \langle \mathbf{r}^{(1)2} \mathbf{r}^{(2)} \rangle - \frac{1}{4} \langle \mathbf{r}^{(1)2} \rangle \langle \mathbf{r}^{(2)} \rangle - \frac{1}{4} \langle \mathbf{r}^{(1)} \rangle \langle \mathbf{r}^{(1)} \mathbf{r}^{(2)} \rangle + \frac{1}{3} \langle \mathbf{r}^{(1)} \rangle^2\langle \mathbf{r}^{(2)} \rangle \right) \\
\nonumber
& + \left( \frac{1}{6} \langle \mathbf{r}^{(1)} \mathbf{r}^{(2)} \mathbf{r}^{(1)} \rangle - \frac{1}{4} \langle \mathbf{r}^{(1)} \mathbf{r}^{(2)} \rangle \langle \mathbf{r}^{(1)} \rangle - \frac{1}{4} \langle \mathbf{r}^{(1)} \rangle \langle \mathbf{r}^{(2)} \mathbf{r}^{(1)} \rangle + \frac{1}{3} \langle \mathbf{r}^{(1)} \rangle \langle \mathbf{r}^{(2)} \rangle \langle \mathbf{r}^{(1)} \rangle \right) \\
& + \left( \frac{1}{6} \langle \mathbf{r}^{(2)} \mathbf{r}^{(1)2} \rangle - \frac{1}{4} \langle \mathbf{r}^{(2)} \mathbf{r}^{(1)} \rangle \langle \mathbf{r}^{(1)} \rangle - \frac{1}{4} \langle \mathbf{r}^{(2)} \rangle \langle \mathbf{r}^{(1)2} \rangle + \frac{1}{3} \langle \mathbf{r}^{(2)} \rangle \langle \mathbf{r}^{(1)} \rangle^2 \right) \\
\nonumber
& = -\frac{\alpha_0^3}{12} \sum_{m>0} m^2 (T_{-m} U_{1,m}+U_{1,-m} T_m) +\frac{\alpha_0^3}{6} \sum_{m>0} \frac{m^3 q^m}{1-q^m} U_{1,0} \\
\nonumber
& -\frac{\alpha_0^3}{6} \sum_{m>0} m^2 (U_{1,-m} T_m + T_{-m} U_{1,m}) +\frac{\alpha_0^3}{3} \sum_{m>0} \frac{m^3 q^m}{1-q^m} U_{1,0} \\
\nonumber
& +\alpha_0^3 N \sum_{m>0} \frac{m^2 q^m}{(1-q^m)^2} (U_{1,-m} T_m + T_{-m} U_{1,m}) + \alpha_0^3 N \sum_{m>0} \frac{m^3 q^m}{(1-q^m)^2} U_{1,0} \\
\nonumber
& +\frac{\alpha_0^3 N}{6} \sum_{m>0} m^2 (U_{1,-m} T_m + T_{-m} U_{1,m}) -\frac{\alpha_0^3 N}{3} \sum_{m>0} \frac{m^3 q^m}{1-q^m} U_{1,0} \\
\nonumber
& +\frac{\alpha_0^3 N^2}{2} \frac{-2E_6+3E_2 E_4-E_2^3}{1728} U_{1,0} \\
\nonumber
& +\frac{\alpha_0^3}{12} \sum_{m,n>0} (U_{1,-m} U_{1,-n} U_{1,m+n} + U_{1,-m-n} U_{1,m} U_{1,n} ) \times \\
\nonumber
& \times \Big( \frac{mn(2+2q^{m+n}+q^m+q^n)}{(1-q^m)(1-q^n)} + \frac{m(m+n)(2q^m+1+2q^{m+n}+q^{2m+n})}{(1-q^m)(1-q^{m+n})} \\
\nonumber
& + \frac{n(m+n)(2q^n+1+2q^{m+n}+q^{m+2n})}{(1-q^n)(1-q^{m+n})} \Big) \\
\nonumber
& +\alpha_0^3 U_{1,0} \sum_{m>0} \frac{m^2 q^m}{(1-q^m)^2} U_{1,-m} U_{1,m} +\frac{\alpha_0^3 N}{2} U_{1,0} \sum_{m>0} \frac{m^3 q^m}{(1-q^m)^2} \\
\nonumber
& +\frac{\alpha_0^3}{6} U_{1,0} \sum_{m>0} m^2 U_{1,-m} U_{1,m} -\frac{\alpha_0^3 N}{6} U_{1,0} \sum_{m>0} \frac{m^3 q^m}{1-q^m} \\
\nonumber
\frac{1}{24} \langle \mathbf{r}^{(1)4} \rangle & = \Big( \frac{1}{24} \langle \mathbf{r}^{(1)4} \rangle - \frac{1}{12} \langle \mathbf{r}^{(1)3} \rangle \langle \mathbf{r}^{(1)} \rangle - \frac{1}{12} \langle \mathbf{r}^{(1)} \rangle \langle \mathbf{r}^{(1)3} \rangle - \frac{1}{8} \langle \mathbf{r}^{(1)2} \rangle^2 \\
& + \frac{1}{6} \langle \mathbf{r}^{(1)} \rangle^2 \langle \mathbf{r}^{(1)2} \rangle + \frac{1}{6} \langle \mathbf{r}^{(1)} \rangle \langle \mathbf{r}^{(1)2} \rangle \langle \mathbf{r}^{(1)} \rangle + \frac{1}{6} \langle \mathbf{r}^{(1)2} \rangle \langle \mathbf{r}^{(1)} \rangle^2 - \frac{1}{4} \langle \mathbf{r}^{(1)} \rangle^4 \Big) \\
\nonumber
& = \frac{\alpha_0^4}{24} \sum_{m>0} \frac{m^3(1+q^m)}{1-q^m} U_{1,-m} U_{1,m} -\frac{\alpha_0^4 N}{6} \sum_{m>0} \frac{m^3(1+q^m)}{1-q^m} U_{1,-m} U_{1,m} \\
\nonumber
& +\frac{\alpha_0^4 N^2}{2} \sum_{m>0} \frac{m^3 q^m(1+q^m)}{(1-q^m)^3} U_{1,-m} U_{1,m} +\frac{\alpha_0^4 N^2}{24} \sum_{m>0} \frac{m^3(1+q^m)}{1-q^m} U_{1,-m} U_{1,m} \\
\nonumber
& +\frac{\alpha_0^4 N}{24} \sum_{m>0} \frac{m^4 q^m}{1-q^m} -\frac{\alpha_0^4 N^2}{12} \sum_{m>0} \frac{m^4 q^m}{1-q^m} +\frac{\alpha_0^4 N^2}{6} \sum_{m>0} \frac{m^4 q^m}{(1-q^m)^2} \\
\nonumber
& +\frac{\alpha_0^4 N^2}{24} \sum_{m>0} \frac{m^4 q^m (1+2q^m)}{(1-q^m)^2} +\frac{\alpha_0^4 N^3}{24} \sum_{m>0} \frac{m^4 q^m}{1-q^m} -\frac{\alpha_0^4 N^3}{6} \sum_{m>0} \frac{m^4 q^m}{(1-q^m)^2} \\
\nonumber
& +\frac{\alpha_0^4 N^3}{4} \sum_{m>0} \frac{m^4 q^m(1+q^m)}{(1-q^m)^3} +\frac{\alpha_0^4 N^4}{24} \frac{3 E_4^2 - 8 E_2 E_6 + 6 E_2^2 E_4 - E_2^4}{6912}
\end{align}
Where we used the following commutation relations:
\begin{equation}
\left[ U_{1,m}, \phi_{3,-m} \right] = \frac{m^3}{6} N(1+\alpha_0^2+N\alpha_0^2) + 2m T_0
\end{equation}
and
\begin{equation}
\left[ \phi_{3,0}, U_{1,m} \right] = 2(\partial T)_m = -2m T_m.
\end{equation}
Adding everything, we find the fourth ILW Hamiltonian
\begin{align}
\nonumber
(\log \mathcal{H}_q)_4 & = +\frac{\alpha_0}{4} \phi_{5,0} +\frac{\alpha_0^2}{2} \sum_{m>0} \frac{m(1+q^m)}{1-q^m} (U_{1,-m} \phi_{3,m} + \phi_{3,-m} U_{1,m}) \\
\nonumber
& +\frac{\alpha_0^2}{2} \sum_{m>0} \frac{m(1+q^m)}{1-q^m} T_{-m} T_m +\frac{\alpha_0^3 (2N-3)}{12} \sum_{m>0} m^2 (U_{1,-m} T_m + T_{-m} U_{1,m}) \\
\nonumber
& +\alpha_0^3 N \sum_{m>0} \frac{m^2 q^m}{(1-q^m)^2} (U_{1,-m} T_m + T_{-m} U_{1,m}) \\
\nonumber
& +\frac{\alpha_0^3}{12} \sum_{m,n>0} (U_{1,-m} U_{1,-n} U_{1,m+n} + U_{1,-m-n} U_{1,m} U_{1,n} ) \times \\
\nonumber
& \times \Big( \frac{mn(2+2q^{m+n}+q^m+q^n)}{(1-q^m)(1-q^n)} + \frac{m(m+n)(2q^m+1+2q^{m+n}+q^{2m+n})}{(1-q^m)(1-q^{m+n})} \\
\nonumber
& + \frac{n(m+n)(2q^n+1+2q^{m+n}+q^{m+2n})}{(1-q^n)(1-q^{m+n})} \Big) \\
\nonumber
& +\alpha_0^3 \sum_{m>0} \frac{m^2 q^m}{(1-q^m)^2} U_{1,-m} U_{1,m} U_{1,0} +\frac{\alpha_0^3}{6} \sum_{m>0} m^2 U_{1,-m} U_{1,m} U_{1,0} \\
\nonumber
& +\frac{\alpha_0^4 N^2}{2} \sum_{m>0} \frac{m^3 q^m(1+q^m)}{(1-q^m)^3} U_{1,-m} U_{1,m} \\
\nonumber
& +\frac{\alpha_0^2(1+\alpha_0^2-4\alpha_0^2 N+\alpha_0^2 N^2)}{24} \sum_{m>0} \frac{m^3(1+q^m)}{1-q^m} U_{1,-m} U_{1,m} \\
& -\frac{\alpha_0 E_2}{8} \phi_{3,0} -\frac{\alpha_0^2 E_2}{8} \sum_{m>0} \frac{m(1+q^m)}{1-q^m} U_{1,-m} U_{1,m} \\
\nonumber
& +\alpha_0^2 N \frac{E_4-E_2^2}{144} T_0 +3\alpha_0^2 \sum_{m>0} \frac{m^2 q^m}{1-q^m} T_0 +\frac{\alpha_0^2}{2} \frac{E_4-E_2^2}{288} U_{1,0}^2 \\
\nonumber
& +\frac{3\alpha_0^3 N}{2} \sum_{m>0} \frac{m^3 q^m}{(1-q^m)^2} U_{1,0} -\frac{\alpha_0^3 (N-1)}{2} \sum_{m>0} \frac{m^3 q^m}{1-q^m} U_{1,0} \\
\nonumber
& +\frac{\alpha_0^3 N^2}{2} \frac{-2E_6+3E_2 E_4-E_2^3}{1728} U_{1,0} +\frac{\alpha_0 E_2^2}{192} U_{1,0} +\frac{\alpha_0(1+\alpha_0^2+N\alpha_0^2)E_4}{480} U_{1,0} \\
\nonumber
& +\frac{\alpha_0^4 N^3}{4} \sum_{m>0} \frac{m^4 q^m(1+q^m)}{(1-q^m)^3} +\frac{\alpha_0^4 N^2}{24} \sum_{m>0} \frac{m^4 q^m (1+2q^m)}{(1-q^m)^2} \\
\nonumber
& -\frac{\alpha_0^4 N^2(N-1)}{6} \sum_{m>0} \frac{m^4 q^m}{(1-q^m)^2} -\frac{\alpha_0^2 N E_2}{8} \sum_{m>0} \frac{m^2 q^m}{1-q^m} \\
\nonumber
& +\frac{\alpha_0^2 N(3+3\alpha_0^2+\alpha_0^2 N)}{12} \sum_{m>0} \frac{m^4 q^m}{1-q^m} \\
\nonumber
& +\frac{\alpha_0^2 N^2}{3456}(-E_2 E_4 + E_2^3) -\frac{\alpha_0^2(1+\alpha_0^2+N\alpha_0^2)N^2}{4320} (E_6-E_2 E_4) \\
\nonumber
& +\frac{\alpha_0^2 N^2}{72} \frac{-2E_6-3E_2 E_4+5E_2^3}{720} +\frac{\alpha_0^4 N^4}{24} \frac{3 E_4^2 - 8 E_2 E_6 + 6 E_2^2 E_4 - E_2^4}{6912}
\end{align}

\subsection{Lowest weight expectation value}
\label{secilwexplw}
Here we list the eigenvalues of lowest ILW Hamiltonians when acting on the lowest weight state. We have
\begin{align}
(\log \mathcal{H}_q)_{1,hw} & = -\alpha_0 \left(-u_2 + \frac{u_1^2}{2}-\frac{N}{24} \right) + \frac{\alpha_0 N E_2}{24} \\
\nonumber
(\log \mathcal{H}_q)_{2,hw} & = \frac{\alpha_0}{2} \left( u_3 - u_1 u_2 + \frac{1}{3} u_1^3 - \frac{u_1}{12} \right) - \frac{\alpha_0 E_2}{24} u_1 \\
& + \alpha_0^2 N^2 \frac{E_4 - E_2^2}{576} + \frac{\alpha_0^2 N}{2} \sum_{m>0} \frac{m^2 q^m}{1-q^m} \\
\nonumber
(\log \mathcal{H}_q)_{3,hw} & = \frac{\alpha_0}{3} \left( \frac{\alpha_0^2 N^2}{480}-\frac{\alpha_0^2 N}{160}-\frac{7 N}{960}+u_1^2 u_2-\frac{u_1^4}{4}+\frac{u_1^2}{8}-u_1 u_3-\frac{u_2^2}{2}-\frac{u_2}{4}+u_4 \right) \\
\nonumber
& +\frac{\alpha_0 E_2}{12} \left( -u_2 + \frac{u_1^2}{2} - \frac{N}{24} \right) \\
& -\alpha_0^2 N \frac{E_4-E_2^2}{288} u_1 -\alpha_0^2 \sum_{m>0} \frac{m^2 q^m}{1-q^m} u_1 \\
\nonumber
& +\alpha_0^3 N^3 \frac{2E_6-3E_2 E_4+E_2^3}{10368} +\frac{\alpha_0^3 N(N-1)}{6} \frac{E_4-1}{240} \\
\nonumber
& -\frac{\alpha_0^3 N^2}{2} \sum_{m>0} \frac{m^3 q^m}{(1-q^m)^2} -\frac{N\alpha_0(1+\alpha_0^2+N\alpha_0^2) E_4}{1440} -\frac{N\alpha_0 E_2^2}{576} \\
\nonumber
(\log \mathcal{H}_q)_{4,hw} & = +\frac{\alpha_0}{4} \Big( u_5 - u_1 u_4 - u_2 u_3  + u_1^2 u_3 + u_1 u_2^2 - u_1^3 u_2 + \frac{u_1^5}{5} \\
\nonumber
& - \frac{u_3}{2} + \frac{u_1 u_2}{2} - \frac{u_1^3}{6} - \frac{(2\alpha_0^2 N-6\alpha_0^2-7)u_1}{240} \Big) \\
\nonumber
& -\frac{\alpha_0 E_2}{8} \left( u_3 - u_1 u_2 + \frac{u_1^3}{3} - \frac{u_1}{12} \right) \\
\nonumber
& +\alpha_0^2 N \frac{E_4-E_2^2}{144} \left( -u_2 + \frac{u_1^2}{2} - \frac{N}{24} \right) \\
\nonumber
& +3\alpha_0^2 \sum_{m>0} \frac{m^2 q^m}{1-q^m} \left( -u_2 + \frac{u_1^2}{2} - \frac{N}{24} \right) \\
& +\frac{\alpha_0^2}{2} \frac{E_4-E_2^2}{288} u_1^2 \\
\nonumber
& +\frac{3\alpha_0^3 N}{2} \sum_{m>0} \frac{m^3 q^m}{(1-q^m)^2} u_1 -\frac{\alpha_0^3 (N-1)}{2} \frac{E_4-1}{240} u_1 \\
\nonumber
& +\frac{\alpha_0^3 N^2}{2} \frac{-2E_6+3E_2 E_4-E_2^3}{1728} u_1 +\frac{\alpha_0 E_2^2}{192} u_1 +\frac{\alpha_0(1+\alpha_0^2+N\alpha_0^2)E_4}{480} u_1 \\
\nonumber
& +\frac{\alpha_0^4 N^3}{4} \sum_{m>0} \frac{m^4 q^m(1+q^m)}{(1-q^m)^3} +\frac{\alpha_0^4 N^2}{24} \sum_{m>0} \frac{m^4 q^m (1+2q^m)}{(1-q^m)^2} \\
\nonumber
& -\frac{\alpha_0^4 N^2(N-1)}{6} \sum_{m>0} \frac{m^4 q^m}{(1-q^m)^2} -\frac{\alpha_0^2 N E_2}{8} \sum_{m>0} \frac{m^2 q^m}{1-q^m} \\
\nonumber
& +\frac{\alpha_0^2 N(3+3\alpha_0^2+\alpha_0^2 N)}{12} \sum_{m>0} \frac{m^4 q^m}{1-q^m} \\
\nonumber
& +\frac{\alpha_0^2 N^2}{3456}(-E_2 E_4 + E_2^3) -\frac{\alpha_0^2(1+\alpha_0^2+N\alpha_0^2)N^2}{4320} (E_6-E_2 E_4) \\
\nonumber
& +\frac{\alpha_0^2 N^2}{72} \frac{-2E_6-3E_2 E_4+5E_2^3}{720} +\frac{\alpha_0^4 N^4}{24} \frac{3 E_4^2 - 8 E_2 E_6 + 6 E_2^2 E_4 - E_2^4}{6912}
\end{align}
The expectation values of the power sums of Yangian Bethe roots $\mathcal{O}_k$ as defined in \label{yangrootpowersum} that we used in this calculation are given by
\begin{align}
\langle \mathcal{O}_0 \rangle_q & = \sum_{m>0} \mathfrak{d}_{-m} = \frac{1-E_2}{24} \\
\langle \mathcal{O}_1 \rangle_q & = \frac{\alpha_0}{2} \sum_{m>0} (1-\mathfrak{d}_m+\mathfrak{d}_{-m}) \mathfrak{d}_{-m} = \frac{\alpha_0(1-E_2)}{48} - \frac{\alpha_0}{2} \sum_{k>0} \frac{k^2 q^k}{1-q^k} \\
\langle \mathcal{O}_0^2 \rangle_{qc} & = \sum_{m>0} \mathfrak{d}_m \mathfrak{d}_{-m} = \frac{E_4-E_2^2}{288} \\
\nonumber
\langle \mathcal{O}_2 \rangle_q & = \sum_{m>0} \left[ \frac{7+4\alpha_0^2}{12} (\mathfrak{d}_{-m}^3 -2 \mathfrak{d}_{-m}^2 \mathfrak{d}_m + \mathfrak{d}_{-m} \mathfrak{d}_m^2) + \frac{\alpha_0^2}{2} \mathfrak{d}_{-m}^2 - \frac{1+\alpha_0^2}{2} \mathfrak{d}_{-m} \mathfrak{d}_m - \frac{1-2\alpha_0^2}{12} \mathfrak{d}_{-m} \right] \\
& = \frac{16\alpha_0^2-17}{2880}+\frac{(1-2\alpha_0^2) E_2}{288}+\frac{(1+2\alpha_0^2) E_4}{1440}+\frac{E_2^2}{576}-\frac{\alpha_0^2}{2} \sum_{k>0} \frac{k^2 q^k}{1-q^k} \\
\langle \mathcal{O}_0 \mathcal{O}_1 \rangle_{qc} & = \frac{\alpha_0}{4} \sum_{m>0} (\mathfrak{d}_{-m}^2 \mathfrak{d}_m - \mathfrak{d}_{-m} \mathfrak{d}_m^2 + \mathfrak{d}_{-m} \mathfrak{d}_m) = \frac{\alpha_0(E_4-E_2^2)}{576} -\frac{\alpha_0}{2} \sum_{k>0} \frac{k^3 q^k}{(1-q^k)^2} \\
\langle \mathcal{O}_0^3 \rangle_{qc} & = \sum_{m>0} (\mathfrak{d}_{-m}^2 \mathfrak{d}_m + \mathfrak{d}_{-m} \mathfrak{d}_m^2) = \frac{-2E_6+3E_2 E_4-E_2^3}{1728} \\
\nonumber
\langle \mathcal{O}_3 \rangle_q & = -\frac{\alpha_0(4\alpha_0^2+17)}{1920} + \frac{\alpha_0 E_2}{192} + \frac{\alpha_0(1+2\alpha_0^2)E_4}{960} + \frac{\alpha_0 E_2^2}{384} \\
& + \frac{\alpha_0(1-2\alpha_0^2+E_2)}{8} \sum_{k>0} \frac{k^2 q^k}{1-q^k} - \frac{\alpha_0(1+\alpha_0^2)}{4} \sum_{k>0} \frac{k^4 q^k}{1-q^k} \\
\nonumber
\langle \mathcal{O}_0 \mathcal{O}_2 \rangle_{qc} & = \sum_{m>0} \Big[ \frac{7+4\alpha_0^2}{12} (\mathfrak{d}_{-m}^3 \mathfrak{d}_m - 2 \mathfrak{d}_{-m}^2 \mathfrak{d}_m^2 + \mathfrak{d}_{-m} \mathfrak{d}_m^3) \\
& - \frac{1-\alpha_0^2}{2} \mathfrak{d}_{-m}^2 \mathfrak{d}_m - \frac{1+\alpha_0^2}{2} \mathfrak{d}_{-m} \mathfrak{d}_m^2 - \frac{1-2\alpha_0^2}{12} \mathfrak{d}_{-m} \mathfrak{d}_m \Big] \\
\nonumber
& = \frac{-4E_6-E_2 E_4+5E_2^3}{17280} - \frac{\alpha_0^2(E_6-E_2 E_4)}{2160} - \frac{(1-2\alpha_0^2)(E_4-E_2^2)}{3456} - \frac{\alpha_0^2}{2} \sum_{k>0} \frac{k^3 q^k}{(1-q^k)^2} \\
\nonumber
\langle \mathcal{O}_1^2 \rangle_{qc} & = \sum_{m>0} \Big[ \frac{4+3\alpha_0^2}{12} (\mathfrak{d}_{-m}^3 \mathfrak{d}_m - 2 \mathfrak{d}_{-m}^2 \mathfrak{d}_m^2 + \mathfrak{d}_{-m} \mathfrak{d}_m^3) \\
& - \frac{2-3\alpha_0^2}{2} \mathfrak{d}_{-m}^2 \mathfrak{d}_m - \frac{2+3\alpha_0^2}{2} \mathfrak{d}_{-m} \mathfrak{d}_m^2 + \frac{\alpha_0^2}{4} \mathfrak{d}_{-m} \mathfrak{d}_m \Big] \\
\nonumber
& = \frac{-2E_6-3E_2 E_4+5E_2^3}{25920} + \frac{\alpha_0^2(E_4-E_2^2)}{1152} + \frac{\alpha_0^2(-E_6+E_2 E_4)}{2880} - \frac{\alpha_0^2}{2} \sum_{k>0} \frac{k^3 q^k}{(1-q^k)^2} \\
\nonumber
\langle \mathcal{O}_0^2 \mathcal{O}_1 \rangle_{qc} & = \frac{\alpha_0}{2} \sum_{m>0} (\mathfrak{d}_{-m}^3 \mathfrak{d}_m - \mathfrak{d}_{-m} \mathfrak{d}_m^3 + \mathfrak{d}_{-m}^2 \mathfrak{d}_m + \mathfrak{d}_{-m} \mathfrak{d}_m^2) \\
& = -\frac{\alpha_0(2E_6-3E_2 E_4+E_2^3)}{3456} - \frac{\alpha_0}{2} \sum_{k>0} \frac{k^4 q^k(1+q^k)}{(1-q^k)^3} \\
\langle \mathcal{O}_0^4 \rangle_{qc} & = \sum_{m>0} (\mathfrak{d}_{-m}^3 \mathfrak{d}_m + 4 \mathfrak{d}_{-m}^2 \mathfrak{d}_m^2 + \mathfrak{d}_{-m} \mathfrak{d}_m^3) = \frac{3E_4^2-8E_2 E_6+6E_2^2 E_4-E_2^4}{6912}
\end{align}

\section{Level $1$ test of FJMM}

In this appendix we make another test of Feigin-Jimbo-Miwa-Mukhin formula. In the main text we were calculating the individual terms of the asymptotic expansions at $u \to \infty$ to all orders in $q$, i.e. our formulas were perturbative in $u$ but exact in $q$. On the other hand, the matrix elements of $R$ in the Fock representations can be evaluated as exact functions of $u$ (which are up to an overall normalization just rational functions).

The test that we want to do here will be to calculate the exact eigenvalue of $\mathcal{H}_q(u)$ on level $1$ excited state $\ket{\Box}$ in free boson Fock space (i.e. $N = 1$). The result is expressed as a sum over Young diagrams labeling states in the auxiliary space, but the summand will have slightly different structure than in FJMM formula. The strategy is the same one that we followed in section \ref{highestweightILWspectrum} when calculating the vacuum eigenvalue of $\mathcal{H}_q(u)$: we need to evaluate
\begin{equation}
\frac{1}{\sum_\lambda q^{|\lambda|}} \sum_\lambda q^{|\lambda|} \bra{\Box} \otimes \bra{\lambda} \mathbf{R}(u) \ket{\Box} \otimes \ket{\lambda},
\end{equation}
therefore it is enough to find the spectrum of the operator
\begin{equation}
\label{boxrbox}
\bra{\Box} \mathbf{R}(u) \ket{\Box}
\end{equation}
acting on the auxiliary Fock space.

\subsection{Yangian algebra from $\mathcal{R}$-matrix and useful identity}

Matrix elements of $\mathcal{R}$-matrix between simple Fock states such as \eqref{boxrbox} are exactly what leads to Yangian currents such as $\psi(u), e(u)$ and $f(u)$ \cite{Prochazka:2019dvu,Litvinov:2020zeq}. In this section we will review this construction mainly to fix the notation and also derive an identity \eqref{litidentity} following \cite{Litvinov:2020zeq}. This identity is exactly what allows us to understand the spectrum of \eqref{boxrbox}.

We start with $(N=1,\bar{N}=1)$ $\mathcal{R}$-matrix normalized such that it acts on the vacuum as identity and at level one we have
\begin{align}
\label{level1rmatrixelements}
\mathcal{R}(u) \ket{\Box} \otimes \ket{0} & = \frac{u}{u+\epsilon_3} \ket{\Box} \otimes \ket{0} + \frac{\epsilon_3}{u+\epsilon_3} \ket{0} \otimes \ket{\Box} \\
\mathcal{R}(u) \ket{0} \otimes \ket{\Box} & = \frac{\epsilon_3}{u+\epsilon_3} \ket{\Box} \otimes \ket{0} + \frac{u}{u+\epsilon_3} \ket{0} \otimes \ket{\Box} \\
\bra{\Box} \otimes \bra{0} \mathcal{R}(u) & = \frac{u}{u+\epsilon_3} \bra{\Box} \otimes \bra{0} + \frac{\epsilon_3}{u+\epsilon_3} \bra{0} \otimes \bra{\Box} \\
\bra{0} \otimes \bra{\Box} \mathcal{R}(u) & = \frac{\epsilon_3}{u+\epsilon_3} \bra{\Box} \otimes \bra{0} + \frac{u}{u+\epsilon_3} \bra{0} \otimes \bra{\Box}
\end{align}
As explained in \cite{Prochazka:2019dvu}, these matrix elements can be evaluated easily from the definition \eqref{instrdef}.

In the following, we will study the matrix elements of $\mathcal{R}$-matrix with the auxiliary space being one of a single free boson (Heisenberg algebra Fock space) while the quantum space will be a general one. We need to choose which of the two spaces we choose as the auxiliary space (the space where we take the matrix elements) and which one is the quantum space. We will follow the convention of the main text, taking the right (barred) space as the auxiliary space and the left space as the quantum space. This is the convention used in \cite{Litvinov:2020zeq} and the opposite of the one used in \cite{Prochazka:2019dvu}. For calculation of \eqref{boxrbox} it would actually be more convenient to use the opposite convention, but with the current choice we can more easily compare to the main text. Both conventions are exchanged by taking $u \to -u$ or $\mathcal{R}(u) \to \mathcal{R}(u)^{-1}$ so we can easily obtain results for the other choice.

We can now define (see sections 3.5 and 3.6 in \cite{Prochazka:2019dvu})
\begin{align}
\mathcal{H}(u) & = \bra{0}_A \mathcal{R}_{QA}(u) \ket{0}_A \\
\mathcal{E}(u) & = \bra{0}_A \mathcal{R}_{QA}(u) \ket{\Box}_A \\
\mathcal{F}(u) & = \bra{\Box}_A \mathcal{R}_{QA}(u) \ket{0}_A
\end{align}
and
\begin{align}
e(u) & = \mathcal{H}(u)^{-1} \mathcal{E}(u) \\
f(u) & = \mathcal{F}(u) \mathcal{H}(u)^{-1}
\end{align}
which differs from the conventions of \cite{Prochazka:2019dvu} only by an overall constant. Consider now three spaces, auxiliary spaces $H_A$ and $H_B$ and the quantum space $H_Q$. The Yang-Baxter equation for such three spaces reads
\begin{equation}
\mathcal{R}_{QA}(u) \mathcal{R}_{QB}(v) \mathcal{R}_{AB}(v-u) = \mathcal{R}_{AB}(v-u) \mathcal{R}_{QB}(v) \mathcal{R}_{QA}(u).
\end{equation}

\paragraph{Level $0 \to 0$ relations}
Taking the vacuum-to-vacuum matrix element in auxiliary Fock spaces we find immediately
\begin{multline}
\bra{0}_A \bra{0}_B \mathcal{R}_{QA}(u) \mathcal{R}_{QB}(v) \mathcal{R}_{AB}(v-u) \ket{0}_A \ket{0}_B = \\
= \bra{0}_A \bra{0}_B \mathcal{R}_{AB}(v-u) \mathcal{R}_{QB}(v) \mathcal{R}_{QA}(u) \ket{0}_A \ket{0}_B
\end{multline}
or in other words
\begin{equation}
\bra{0}_A \mathcal{R}_{QA}(u) \ket{0}_A \bra{0}_B \mathcal{R}_{QB}(v) \ket{0}_B = \bra{0}_B \mathcal{R}_{QB}(v) \ket{0}_B \bra{0}_A \mathcal{R}_{QA}(u) \ket{0}_A
\end{equation}
which is by definition
\begin{equation}
\mathcal{H}(u) \mathcal{H}(v) = \mathcal{H}(v) \mathcal{H}(u).
\end{equation}
We used the fact that the $\mathcal{R}$-matrix used in this section is normalized such that it preserves the vacuum state,
\begin{align}
\mathcal{R}_{AB}(u) \ket{0}_A \ket{0}_B & = \ket{0}_A \ket{0}_B \\
\bra{0}_A \bra{0}_B \mathcal{R}_{AB}(u) & = \bra{0}_A \bra{0}_B.
\end{align}
By taking the vacuum-to-vacuum matrix element in the auxiliary space we found a generating function of Yangian Hamiltonians, discovering one of the integrable structures of $\mathcal{W}_{1+\infty}$. Let us now turn to the ladder operators.

\paragraph{Level $0 \leftrightarrow 1$ relations}
Taking matrix elements between the vacuum and level $1$ state in the auxiliary space (4 matrix elements), we find $2$ equations for the creation operator
\begin{align}
\bra{\bullet,\bullet} \mathcal{R}_{QA}(u) \mathcal{R}_{QB}(v) \mathcal{R}_{AB}(v-u) \ket{\Box,\bullet} & = \bra{\bullet,\bullet} \mathcal{R}_{AB}(v-u) \mathcal{R}_{QB}(v) \mathcal{R}_{QA}(u) \ket{\Box,\bullet} \\
\bra{\bullet,\bullet} \mathcal{R}_{QA}(u) \mathcal{R}_{QB}(v) \mathcal{R}_{AB}(v-u) \ket{\bullet,\Box} & = \bra{\bullet,\bullet} \mathcal{R}_{AB}(v-u) \mathcal{R}_{QB}(v) \mathcal{R}_{QA}(u) \ket{\bullet,\Box}
\end{align}
or explicitly
\begin{align}
\label{ybaherel}
(u-v) \mathcal{E}(u) \mathcal{H}(v) - \epsilon_3 \mathcal{H}(u) \mathcal{E}(v) & = (u-v-\epsilon_3) \mathcal{H}(v) \mathcal{E}(u) \\
\label{ybelevel11eqn3}
-\epsilon_3 \mathcal{E}(u) \mathcal{H}(v) + (u-v) \mathcal{H}(u) \mathcal{E}(v) & = (u-v-\epsilon_3) \mathcal{E}(v) \mathcal{H}(u).
\end{align}
These two equations are not independent: taking for example the first equation as it is and the same equation with $u \leftrightarrow v$ exchanged, we can eliminate from these two equations $\mathcal{H}(v) \mathcal{E}(u)$ and the resulting quantities satisfy the second equation. In fact, we have four different ways of ordering $\mathcal{H}$ and $\mathcal{E}$ evaluated at $u$ and $v$ so there are four different equations with one of these quantities missing. Multiplying the first equation in \eqref{ybaherel} by $\mathcal{H}(u)^{-1}$ from the left and using the commutativity of $\mathcal{H}$ as well as the definition of $e(u)$ we can write equivalently
\begin{equation}
(u-v) e(u) \mathcal{H}(v) - \epsilon_3 \mathcal{E}(v) = (u-v-\epsilon_3) \mathcal{H}(v) e(u).
\end{equation}
Multiplying this from the left by $\mathcal{H}(v)^{-1}$, we find
\begin{equation}
(u-v-\epsilon_3) e(u) = (u-v) \mathcal{H}(v)^{-1} e(u) \mathcal{H}(v) - \epsilon_3 e(v).
\end{equation}
Sending $v \to u-\epsilon_3$,
\begin{equation}
e(u+\epsilon_3) \mathcal{H}(u) = \mathcal{H}(u) e(u) = \mathcal{E}(u).
\end{equation}
In particular, we find
\begin{equation}
e(u) = \mathcal{H}(u)^{-1} \mathcal{E}(u) = \mathcal{E}(u-\epsilon_3) \mathcal{H}(u-\epsilon_3)^{-1}
\end{equation}
For the annihilation operators, we have the following $2$ equations
\begin{align}
\bra{\Box,\bullet} \mathcal{R}_{QA}(u) \mathcal{R}_{QB}(v) \mathcal{R}_{AB}(v-u) \ket{\bullet,\bullet} & = \bra{\Box,\bullet} \mathcal{R}_{AB}(v-u) \mathcal{R}_{QB}(v) \mathcal{R}_{QA}(u) \ket{\bullet,\bullet} \\
\bra{\bullet,\Box} \mathcal{R}_{QA}(u) \mathcal{R}_{QB}(v) \mathcal{R}_{AB}(v-u) \ket{\bullet,\bullet} & = \bra{\bullet,\Box} \mathcal{R}_{AB}(v-u) \mathcal{R}_{QB}(v) \mathcal{R}_{QA}(u) \ket{\bullet,\bullet}
\end{align}
or explicitly
\begin{align}
\label{ybelevel11eqn2}
(u-v-\epsilon_3) \mathcal{F}(u) \mathcal{H}(v) & = (u-v) \mathcal{H}(v) \mathcal{F}(u) - \epsilon_3 \mathcal{F}(v) \mathcal{H}(u) \\
(u-v-\epsilon_3) \mathcal{H}(u) \mathcal{F}(v) & = -\epsilon_3 \mathcal{H}(v) \mathcal{F}(u) + (u-v) \mathcal{F}(v) \mathcal{H}(u)
\end{align}

\paragraph{Level $1 \to 1$ relations}
Taking the matrix elements between level $1$ states in the auxiliary space (4 matrix elements in total), we find equations
\begin{align}
\bra{\Box,\bullet} \mathcal{R}_{QA}(u) \mathcal{R}_{QB}(v) \mathcal{R}_{AB}(v-u) \ket{\Box,\bullet} & = \bra{\Box,\bullet} \mathcal{R}_{AB}(v-u) \mathcal{R}_{QB}(v) \mathcal{R}_{QA}(u) \ket{\Box,\bullet} \\
\bra{\bullet,\Box} \mathcal{R}_{QA}(u) \mathcal{R}_{QB}(v) \mathcal{R}_{AB}(v-u) \ket{\Box,\bullet} & = \bra{\bullet,\Box} \mathcal{R}_{AB}(v-u) \mathcal{R}_{QB}(v) \mathcal{R}_{QA}(u) \ket{\Box,\bullet} \\
\bra{\Box,\bullet} \mathcal{R}_{QA}(u) \mathcal{R}_{QB}(v) \mathcal{R}_{AB}(v-u) \ket{\bullet,\Box} & = \bra{\Box,\bullet} \mathcal{R}_{AB}(v-u) \mathcal{R}_{QB}(v) \mathcal{R}_{QA}(u) \ket{\bullet,\Box} \\
\bra{\bullet,\Box} \mathcal{R}_{QA}(u) \mathcal{R}_{QB}(v) \mathcal{R}_{AB}(v-u) \ket{\bullet,\Box} & = \bra{\bullet,\Box} \mathcal{R}_{AB}(v-u) \mathcal{R}_{QB}(v) \mathcal{R}_{QA}(u) \ket{\bullet,\Box}
\end{align}
Using now the explicit expressions for matrix elements of $\mathcal{R}_{AB}$ \eqref{level1rmatrixelements}, we simplify this to
\begin{align}
(u-v) \left[ \mathcal{H}(v), \mathcal{L}_{\Box,\Box}(u) \right] & = -\epsilon_3 \left( \mathcal{F}(u) \mathcal{E}(v) - \mathcal{F}(v) \mathcal{E}(u) \right) \\
(u-v) \left[ \mathcal{E}(u), \mathcal{F}(v) \right] & = -\epsilon_3 \left( \mathcal{H}(v) \mathcal{L}_{\Box,\Box}(u) - \mathcal{H}(u) \mathcal{L}_{\Box,\Box}(v) \right) \\
\label{ybelevel11eqn1}
(u-v) \left[ \mathcal{E}(v), \mathcal{F}(u) \right] & = -\epsilon_3 \left( \mathcal{L}_{\Box,\Box}(u) \mathcal{H}(v) - \mathcal{L}_{\Box,\Box}(v) \mathcal{H}(u) \right) \\
(u-v) \left[ \mathcal{H}(u), \mathcal{L}_{\Box,\Box}(v) \right] & = -\epsilon_3 \left( \mathcal{E}(v) \mathcal{F}(u) - \mathcal{E}(u) \mathcal{F}(v) \right)
\end{align}
where
\begin{equation}
\mathcal{L}_{\Box,\Box}(u) \equiv \bra{\Box}_A \mathcal{R}_{QA}(u) \ket{\Box}_A.
\end{equation}
Let us now construct the $\psi$ operator following \cite{Litvinov:2020zeq}. We start with the equation \eqref{ybelevel11eqn1} and write it as
\begin{equation}
(u-v) \mathcal{E}(v) \mathcal{H}^{-1}(v) \mathcal{H}(v) \mathcal{F}(u) - (u-v) \mathcal{F}(u) \mathcal{H}^{-1}(u) \mathcal{H}(u) \mathcal{E}(v) = -\epsilon_3 \left( \mathcal{L}_{\Box,\Box}(u) \mathcal{H}(v) - \mathcal{L}_{\Box,\Box}(v) \mathcal{H}(u) \right)
\end{equation}
We now use \eqref{ybelevel11eqn2} and \eqref{ybelevel11eqn3} and find
\begin{multline}
-\epsilon_3 \mathcal{L}_{\Box,\Box}(u) \mathcal{H}(v) + \epsilon_3 \mathcal{L}_{\Box,\Box}(v) \mathcal{H}(u) = \epsilon_3 \mathcal{E}(v) \mathcal{H}^{-1}(v) \mathcal{F}(v) \mathcal{H}(u) - \epsilon_3 \mathcal{F}(u) \mathcal{H}^{-1}(u) \mathcal{E}(u) \mathcal{H}(v) \\
+ (u-v-\epsilon_3) \mathcal{E}(v) \mathcal{H}^{-1}(v) \mathcal{F}(u) \mathcal{H}(v) - (u-v-\epsilon_3) \mathcal{F}(u) \mathcal{H}^{-1}(u) \mathcal{E}(v) \mathcal{H}(u)
\end{multline}
Multiplying by $\mathcal{H}^{-1}(u) \mathcal{H}^{-1}(v)$ from the right and using the definition of $e(u)$ and $f(u)$,
\begin{multline}
-\epsilon_3 \mathcal{L}_{\Box,\Box}(u) \mathcal{H}^{-1}(u) + \epsilon_3 \mathcal{L}_{\Box,\Box}(v) \mathcal{H}^{-1}(v) = \epsilon_3 \mathcal{E}(v) \mathcal{H}^{-1}(v) \mathcal{F}(v) \mathcal{H}^{-1}(v) - \epsilon_3 \mathcal{F}(u) \mathcal{H}^{-1}(u) \mathcal{E}(u) \mathcal{H}^{-1}(u) \\
+ (u-v-\epsilon_3) \mathcal{E}(v) \mathcal{H}^{-1}(v) \mathcal{F}(u) \mathcal{H}^{-1}(u) - (u-v-\epsilon_3) \mathcal{F}(u) \mathcal{H}^{-1}(u) \mathcal{E}(v) \mathcal{H}^{-1}(v)
\end{multline}
Expressing the terms proportional to $(u-v-\epsilon_3)$ in terms of $f(u)$ and $e(u+\epsilon_3)$ and rearranging the result, we see that
\begin{multline}
-\frac{\epsilon_3}{u-v-\epsilon_3} \Big( \mathcal{L}_{\Box,\Box}(u) \mathcal{H}^{-1}(u) - \mathcal{L}_{\Box,\Box}(v) \mathcal{H}^{-1}(v)  + \mathcal{E}(v) \mathcal{H}^{-1}(v) \mathcal{F}(v) \mathcal{H}^{-1}(v) \\
- \mathcal{F}(u) \mathcal{H}^{-1}(u) \mathcal{E}(u) \mathcal{H}^{-1}(u) \Big) = \left[ e(v+\epsilon_3), f(u) \right].
\end{multline}
Shifting $v \to v-\epsilon_3$ (and renaming $u$ and $v$) we finally find
\begin{equation}
(u-v) \left[ e(u), f(v) \right] = -\epsilon_3 \left( \psi(u) - \tilde{\psi}(v) \right)
\end{equation}
where
\begin{align}
\label{psileqn1}
\psi(u+\epsilon_3) \equiv \mathcal{L}_{\Box,\Box}(u) \mathcal{H}^{-1}(u) - \mathcal{E}(u) \mathcal{H}^{-1}(u) \mathcal{F}(u) \mathcal{H}^{-1}(u) \\
\label{psileqn2}
\tilde{\psi}(u) \equiv \mathcal{L}_{\Box,\Box}(u) \mathcal{H}^{-1}(u) - \mathcal{F}(u) \mathcal{H}^{-1}(u) \mathcal{E}(u) \mathcal{H}^{-1}(u)
\end{align}
Choosing $u = v$, the left-hand side vanishes while the form of the right-hand side implies a consistency relation
\begin{equation}
\psi(u) = \tilde{\psi}(u)
\end{equation}
which can equivalently be written as
\begin{equation}
\label{litidentity}
e(u) f(u-\epsilon_3) - f(u) e(u+\epsilon_3) = \mathcal{L}_{\Box,\Box}(u-\epsilon_3) \mathcal{H}^{-1}(u-\epsilon_3) -\mathcal{L}_{\Box,\Box}(u) \mathcal{H}^{-1}(u).
\end{equation}

\subsection{Level $1$ eigenvalue of $\mathcal{H}_q(u)$}
Let us now restrict to $N = 1 = \bar{N}$. The spectrum of $\mathcal{H}(u)$ and $\psi(u)$ are given by the usual Yangian expressions
\begin{equation}
\mathcal{H}(u) \leftrightarrow \prod_{\Box\in\lambda} \frac{u-\epsilon_{\Box}}{u-\epsilon_{\Box}+\epsilon_3}
\end{equation}
and
\begin{equation}
\psi(u) \leftrightarrow \psi_\lambda(u) \equiv \frac{u-\epsilon_3}{u} \times \prod_{\Box\in\lambda} \frac{(u-\epsilon_{\Box}+\epsilon_1)(u-\epsilon_{\Box}+\epsilon_2)(u-\epsilon_{\Box}+\epsilon_3)}{(u-\epsilon_{\Box}-\epsilon_1)(u-\epsilon_{\Box}-\epsilon_2)(u-\epsilon_{\Box}-\epsilon_3)}
\end{equation}
where $\lambda$ is a Young diagram labeling the common eigenstate of $\mathcal{H}(u)$ and $\psi(u)$ \cite{Prochazka:2019dvu}.

The relations that we just found allow us to calculate the matrix elements of $\mathcal{L}_{\Box,\Box}(u)$: \eqref{psileqn1}-\eqref{psileqn2} can be written as
\begin{align}
\mathcal{L}_{\Box,\Box}(u) & = \left[ \psi(u) + f(u) e(u+\epsilon_3) \right] \mathcal{H}(u) \\
& = \left[ \psi(u+\epsilon_3) + e(u+\epsilon_3) f(u) \right] \mathcal{H}(u)
\end{align}
Taking the diagonal matrix element of the first formula between the eigenstate $\ket{\lambda}$ of $\psi(u)$, we find
\begin{align}
\label{ldiagel}
\nonumber
\bra{\lambda} \mathcal{L}_{\Box,\Box}(u) \ket{\lambda} & = \left( \psi_\lambda(u) + \# \sum_{\Box\in\lambda^+} \frac{F(\lambda+\Box\to\lambda) E(\lambda\to\lambda+\Box)}{(u-\epsilon_\Box)(u-\epsilon_\Box+\epsilon_3)} \right) \times \prod_{\Box\in\lambda} \frac{u-\epsilon_{\Box}}{u-\epsilon_{\Box}+\epsilon_3} \\
& = \left( \psi_\lambda(u) - \epsilon_3 \sum_{\Box\in\lambda^+} \frac{\res_{u\to\epsilon_\Box} \psi_\lambda(u)}{(u-\epsilon_\Box)(u-\epsilon_\Box+\epsilon_3)} \right) \times \prod_{\Box\in\lambda} \frac{u-\epsilon_{\Box}}{u-\epsilon_{\Box}+\epsilon_3}
\end{align}
In the second line we used the fact that the product of creation and annihilation amplitudes can be expressed in terms of the residue of the generating function $\psi_\lambda(u)$ \cite{Prochazka:2015deb}. Analogously, the second formula implies
\begin{align}
\nonumber
\bra{\lambda} \mathcal{L}_{\Box,\Box}(u) \ket{\lambda} & = \left( \psi_\lambda(u+\epsilon_3) + \# \sum_{\Box\in\lambda^-} \frac{F(\lambda\to\lambda-\Box) E(\lambda-\Box\to\lambda)}{(u-\epsilon_\Box)(u-\epsilon_\Box+\epsilon_3)} \right) \times \prod_{\Box\in\lambda} \frac{u-\epsilon_{\Box}}{u-\epsilon_{\Box}+\epsilon_3} \\
& = \left( \psi_\lambda(u+\epsilon_3) + \epsilon_3 \sum_{\Box\in\lambda^-} \frac{\res_{u\to\epsilon_\Box} \psi_\lambda(u)}{(u-\epsilon_\Box)(u-\epsilon_\Box+\epsilon_3)} \right) \times \prod_{\Box\in\lambda} \frac{u-\epsilon_{\Box}}{u-\epsilon_{\Box}+\epsilon_3}.
\end{align}
We can now use the partial fraction expansion
\begin{equation}
\frac{1}{(u-\epsilon_\Box)(u-\epsilon_\Box+\epsilon_3)} = \frac{1}{\epsilon_3} \frac{1}{u-\epsilon_\Box} - \frac{1}{\epsilon_3} \frac{1}{u-\epsilon_\Box+\epsilon_3}
\end{equation}
as well as the fact that $\psi_\lambda(u)$ is determined once we know the positions of its simple poles and the corresponding residues,
\begin{equation}
\psi_\lambda(u) = 1 + \sum_{\Box\in\lambda^+} \frac{\res_{u\to\epsilon_\Box} \psi_\lambda(u)}{u-\epsilon_\Box} + \sum_{\Box\in\lambda^-} \frac{\res_{u\to\epsilon_\Box} \psi_\lambda(u)}{u-\epsilon_\Box}.
\end{equation}
Here we used the shell-formula \cite{Prochazka:2015deb}, the fact that $\psi_\lambda(u)$ has only simple poles and these poles are at positions of boxes which can be added or removed consistently with the Young diagram rules. We also used the fact that the value of $\psi_\lambda(u)$ at infinity is equal to $1$. Plugging this into \eqref{ldiagel}, we finally find
\begin{align}
\bra{\lambda} \mathcal{L}_{\Box,\Box}(u) \ket{\lambda} & = \left( 1 + \sum_{\Box\in\lambda^+} \frac{\res_{u\to\epsilon_\Box} \psi_\lambda(u)}{u+\epsilon_3-\epsilon_\Box} + \sum_{\Box\in\lambda^-} \frac{\res_{u\to\epsilon_\Box} \psi_\lambda(u)}{u-\epsilon_\Box} \right) \times \prod_{\Box\in\lambda} \frac{u-\epsilon_{\Box}}{u-\epsilon_{\Box}+\epsilon_3}.
\end{align}
Using the second formula for $\mathcal{L}_{\Box,\Box}(u)$ would lead to the same result. We see that the diagonal matrix element of $\mathcal{L}_{\Box,\Box}(u)$ between eigenstates of Yangian Hamiltonians is up to a contribution from $\mathcal{H}(u)$ eigenvalue given by decomposing $\psi_\lambda(u)$ into partial fractions and shifting the poles associated to addable boxes by $-\epsilon_3$ while leaving the poles associated to removable boxes intact. We can now write the formula for $\mathcal{H}_q(u)$ eigenvalue at level $1$ as
\begin{equation}
\mathcal{H}_q(u) \leftrightarrow \frac{1}{\sum_\lambda q^{|\lambda|}} \sum_\lambda q^{|\lambda|} \left( 1 + \sum_{\Box\in\lambda^+} \frac{\res_{u\to\epsilon_\Box} \psi_\lambda(u)}{u-\epsilon_\Box+\epsilon_3} + \sum_{\Box\in\lambda^-} \frac{\res_{u\to\epsilon_\Box} \psi_\lambda(u)}{u-\epsilon_\Box} \right) \times \prod_{\Box\in\lambda} \frac{u-\epsilon_{\Box}}{u-\epsilon_{\Box}+\epsilon_3}
\end{equation}
which is a closed form formula valid exactly to all orders in $u$.

Now we can compare this eigenvalue with the FJMM formula. More precisely, we calculate
\begin{equation}
X = \bra{\Box} \mathcal{H}_q(u) \ket{\Box} \times \frac{\sum_\lambda q^{|\lambda|} \times \prod_{\Box\in\lambda} \frac{u-\epsilon_{\Box}}{u-\epsilon_{\Box}+\epsilon_3}}{\bra{0} \mathcal{H}_q(u) \ket{0}}
\end{equation}
in two different ways (the second factor is introduced to fix the normalization). From the derivation in this appendix, we have
\begin{equation}
\label{boxtestx1}
X = \sum_\lambda q^{|\lambda|} \left( 1 + \sum_{\Box\in\lambda^+} \frac{\res_{u\to\epsilon_\Box} \psi_\lambda(u)}{u-\epsilon_\Box+\epsilon_3} + \sum_{\Box\in\lambda^-} \frac{\res_{u\to\epsilon_\Box} \psi_\lambda(u)}{u-\epsilon_\Box} \right) \times \prod_{\Box\in\lambda} \frac{u-\epsilon_{\Box}}{u-\epsilon_{\Box}+\epsilon_3}
\end{equation}
On the other hand, from BAE we see that there is only one Bethe root equal to
\begin{equation}
x = -\frac{\epsilon_3 q}{1-q}
\end{equation}
and the FJMM formula takes the form
\begin{align}
\label{boxtestx2}
X & = \frac{u-x}{u-x+\epsilon_3} \sum_\lambda q^{|\lambda|} \prod_{\Box\in\lambda} \frac{u-\epsilon_\Box}{u-\epsilon_\Box+\epsilon_3} \varphi(u-x-\epsilon_\Box+\epsilon_3) \\
& = \sum_\lambda q^{|\lambda|} \psi_\lambda\left(u+\frac{\epsilon_3}{1-q}\right) \prod_{\Box\in\lambda} \frac{u-\epsilon_\Box}{u-\epsilon_\Box+\epsilon_3}
\end{align}
Both formulas are exact to all orders in $u$ and can be compared perturbatively in small $q$ expansion. We checked that they agree to order $\mathcal{O}(q^{27})$. It is interesting that the effect of splitting the poles corresponding to addable and removable boxes and shifting the addable boxes by $-\epsilon_3$ in \eqref{boxtestx1} is equivalent to translating the argument of $\psi_\lambda(u)$ by $q$-dependent Bethe root as in \eqref{boxtestx2}.

\section{Expansions of ILW Hamiltonians in the local limit}

\subsection{ILW Hamiltonians with rational matrix elements}
Here we summarize the ILW conserved quantities obtained from $(\log\mathcal{H})_j$ by change of normalization and more importantly subtraction of the transcendental and central terms:
\begin{align}
\mathcal{A}_1 & = T_0 \\
\mathcal{A}_2 & = -\frac{1}{2} \phi_{3,0} - \frac{\alpha_0}{2} \sum_{m>0} m \frac{1+q^m}{1-q^m} U_{1,-m} U_{1,m} \\
\nonumber
\mathcal{A}_3 & = -\frac{1}{3} \phi_{4,0} +\frac{\alpha_0}{2} \sum_{m>0} \frac{m(1+q^m)}{1-q^m} (U_{1,-m} T_m + T_{-m} U_{1,m}) \\
& -\frac{\alpha_0^2 (N+2)}{12} \sum_{m>0} m^2 U_{1,-m} U_{1,m} +\frac{\alpha_0^2 N}{4} \sum_{m>0} \frac{m^2 (1+q^m)^2}{(1-q^m)^2} U_{1,-m} U_{1,m} \\
\nonumber
\mathcal{A}_4 & = -\frac{1}{4} \phi_{5,0} -\frac{\alpha_0}{2} \sum_{m>0} \frac{m(1+q^m)}{1-q^m} (U_{1,-m} \phi_{3,m} + \phi_{3,-m} U_{1,m}) \\
\nonumber
& -\frac{\alpha_0}{2} \sum_{m>0} \frac{m(1+q^m)}{1-q^m} T_{-m} T_m -\frac{\alpha_0^2 (2N-3)}{12} \sum_{m>0} m^2 (U_{1,-m} T_m + T_{-m} U_{1,m}) \\
\nonumber
& -\alpha_0^2 N \sum_{m>0} \frac{m^2 q^m}{(1-q^m)^2} (U_{1,-m} T_m + T_{-m} U_{1,m}) \\
\nonumber
& -\frac{\alpha_0^2}{12} \sum_{m_1,m_2>0} (U_{1,-m_1} U_{1,-m_2} U_{1,m_1+m_2} + U_{1,-m_1-m_2} U_{1,m_1} U_{1,m_2} ) \times \\
& \times \Big( 2\mathfrak{d}_{m_1} \mathfrak{d}_{m_2} + \mathfrak{d}_{-m_1}\mathfrak{d}_{m_2} + \mathfrak{d}_{m_1}\mathfrak{d}_{-m_2} + 2\mathfrak{d}_{-m_1}\mathfrak{d}_{-m_2} \\
\nonumber
& + \mathfrak{d}_{m_1} \mathfrak{d}_{m_1+m_2} + 2\mathfrak{d}_{-m_1} \mathfrak{d}_{m_1+m_2} + 2\mathfrak{d}_{m_1} \mathfrak{d}_{-m_1-m_2} + \mathfrak{d}_{-m_1} \mathfrak{d}_{-m_1-m_2} \\
\nonumber
& + \mathfrak{d}_{m_2} \mathfrak{d}_{m_1+m_2} + 2\mathfrak{d}_{-m_2} \mathfrak{d}_{m_1+m_2} + 2 \mathfrak{d}_{m_2} \mathfrak{d}_{-m_1-m_2} + \mathfrak{d}_{-m_2} \mathfrak{d}_{-m_1-m_2} \Big) \\
\nonumber
& -\alpha_0^2 \sum_{m>0} \frac{m^2 q^m}{(1-q^m)^2} U_{1,-m} U_{1,m} U_{1,0} -\frac{\alpha_0^2}{6} \sum_{m>0} m^2 U_{1,-m} U_{1,m} U_{1,0} \\
\nonumber
& -\frac{\alpha_0^3 N^2}{2} \sum_{m>0} \frac{m^3 q^m(1+q^m)}{(1-q^m)^3} U_{1,-m} U_{1,m} \\
\nonumber
& -\frac{\alpha_0(1+\alpha_0^2-4\alpha_0^2 N+\alpha_0^2 N^2)}{24} \sum_{m>0} \frac{m^3(1+q^m)}{1-q^m} U_{1,-m} U_{1,m} \\
\nonumber
\mathcal{A}_5 & = -\frac{1}{5} \phi_{6,0} -\frac{\alpha_0}{2} \sum_{m>0} \frac{m (1+q^m)}{1-q^m} (U_{1,-m} \tilde{\phi}_{4,m} + \tilde{\phi}_{4,-m} U_{1,m}) \\
\nonumber
& +\frac{\alpha_0}{2} \sum_{m>0} \frac{m(1+q^m)}{1-q^m} (T_{-m} \phi_{3,m} + \phi_{3,-m} T_m) \\
\nonumber
& +\frac{\alpha_0^2 (N-2)}{6} \sum_{m>0} m^2 (U_{1,-m} \phi_{3,m} + \phi_{3,-m} U_{1,m}) \\
\nonumber
& +\frac{\alpha_0^2 (N-2)}{6} \sum_{m>0} m^2 T_{-m} T_m +\alpha_0^2 N \sum_{m>0} \frac{m^2 q^m}{(1-q^m)^2} T_{-m} T_m \\
\nonumber
& +\frac{\alpha_0}{12} \sum_{m>0} \frac{m^3 (1+q^m)}{1-q^m} (U_{1,-m} T_m + T_{-m} U_{1,m}) \\
\nonumber
& +\frac{(N-1)(N-2) \alpha_0^3}{24} \sum_{m>0} \frac{m^3(1+q^m)}{1-q^m} (U_{1,-m} T_m + T_{-m} U_{1,m}) \\
\nonumber
& +\frac{\alpha_0^3 N^2}{2} \sum_{m>0} \frac{m^3 q^m(1+q^m)}{(1-q^m)^3} (U_{1,-m} T_m + T_{-m} U_{1,m}) \\
& +\alpha_0^2 N \sum_{m>0} \frac{m^2 q^m}{(1-q^m)^2} (U_{1,-m} \phi_{3,m} + \phi_{3,-m} U_{1,m}) \\
\nonumber
& +\frac{\alpha_0^4 N^3}{6} \sum_{m>0} \frac{m^4 q^m(1+4q^m+q^{2m})}{(1-q^m)^4} U_{1,-m} U_{1,m} \\
\nonumber
& +\frac{\alpha_0^2 N(1+\alpha_0^2+N\alpha_0^2)}{36} \sum_{m>0} \frac{m^4(1+4q^m+q^{2m})}{(1-q^m)^2} U_{1,-m} U_{1,m} \\
\nonumber
& +\frac{\alpha_0^2 N}{4} \sum_{m>0} \frac{m^4 q^m}{(1-q^m)^2} U_{1,-m} U_{1,m} \\
\nonumber
& +\frac{\alpha_0^4 N(N^2-4N+1)}{12} \sum_{m>0} \frac{m^4 q^m}{(1-q^m)^2} U_{1,-m} U_{1,m} \\
\nonumber
& +\frac{\alpha_0^2}{72} (3N+\alpha_0^2+2N\alpha_0^2+N^2\alpha_0^2) \sum_{m>0} m^4 U_{1,-m} U_{1,m} \\
\nonumber
& +\frac{\alpha_0^4 (N-1)^3}{120} \sum_{m>0} m^4 U_{1,-m} U_{1,m} \\
\nonumber
& +\frac{\alpha_0^3 N}{6} \mathcal{I}_{4,3} +\frac{\alpha_0^4 (N-1)}{240} \mathcal{I}_{5,1} +\frac{\alpha_0^3}{24} \mathcal{I}_{5,2} +\frac{\alpha_0^3 N}{144} \mathcal{I}_{5,3} \\
\nonumber
& -\frac{\alpha_0^3 (N-1)}{24} \mathcal{I}_{5,5} +\frac{\alpha_0^2}{3} \mathcal{I}_{5,4} +\frac{\alpha_0^2}{6} \mathcal{I}_{5,6}
\end{align}
where
\begin{align}
\nonumber
\mathcal{I}_{4,3} & = \frac{3}{2} \sum_{m_1,m_2>0} \left(U_{1,-m_1-m_2} U_{1,m_1} U_{1,m_2} + U_{1,-m_1} U_{1,-m_2} U_{1,m_1+m_2} \right) \times \\
\nonumber
& \times \Big( \mathfrak{d}_{m_1} \mathfrak{d}_{-m_1} \mathfrak{d}_{m_2} + \mathfrak{d}_{m_1} \mathfrak{d}_{-m_1} \mathfrak{d}_{-m_2} + \mathfrak{d}_{m_1} \mathfrak{d}_{m_2} \mathfrak{d}_{-m_2} + \mathfrak{d}_{-m_1} \mathfrak{d}_{m_2} \mathfrak{d}_{-m_2} + \mathfrak{d}_{m_1} \mathfrak{d}_{-m_1} \mathfrak{d}_{m_1+m_2} \\
\nonumber
& + \mathfrak{d}_{m_1} \mathfrak{d}_{-m_1} \mathfrak{d}_{-m_1-m_2} + \mathfrak{d}_{m_2} \mathfrak{d}_{-m_2} \mathfrak{d}_{m_1+m_2} + \mathfrak{d}_{m_2} \mathfrak{d}_{-m_2} \mathfrak{d}_{-m_1-m_2} + \mathfrak{d}_{m_1} \mathfrak{d}_{m_1+m_2} \mathfrak{d}_{-m_1-m_2} \\
& + \mathfrak{d}_{-m_1} \mathfrak{d}_{m_1+m_2} \mathfrak{d}_{-m_1-m_2} + \mathfrak{d}_{m_2} \mathfrak{d}_{m_1+m_2} \mathfrak{d}_{-m_1-m_2} + \mathfrak{d}_{-m_2} \mathfrak{d}_{m_1+m_2} \mathfrak{d}_{-m_1-m_2} \Big) \\
\nonumber
& + 6 U_{1,0} \sum_{m>0} (\mathfrak{d}_m^2 \mathfrak{d}_{-m} + \mathfrak{d}_m \mathfrak{d}_{-m}^2) U_{1,-m} U_{1,m} \\
\mathcal{I}_{5,1} & = -8N \sum_{m>0} \frac{m^4(2+q^m+2q^{2m})}{(1-q^m)^2} U_{1,-m} U_{1,m} \\
\nonumber
\mathcal{I}_{5,2} & = -\sum_{m_1,m_2>0} (m_1^2+m_1 m_2+m_2^2) (\mathfrak{d}_{m_1+m_2} + \mathfrak{d}_{-m_1-m_2} + \mathfrak{d}_{m_1} + \mathfrak{d}_{-m_1} + \mathfrak{d}_{m_2} + \mathfrak{d}_{-m_2}) \times \\
& \times (U_{1,-m_1} U_{1,-m_2} U_{1,m_1+m_2} + U_{1,-m_1-m_2} U_{1,m_1} U_{1,m_2})
 \\
\nonumber
& - 2N \sum_{m>0} \frac{m^3(1+q^m)}{1-q^m} (T_{-m} U_{1,m} + U_{1,-m} T_m) - 4 U_{1,0} \sum_{m>0} \frac{m^3(1+q^m)}{1-q^m} U_{1,-m} U_{1,m} \\
\nonumber
\mathcal{I}_{5,3} & = 6\sum_{m_1,m_2>0} \Big( \mathfrak{d}_{m_1} \mathfrak{d}_{m_2} \mathfrak{d}_{m_1+m_2} + \mathfrak{d}_{-m_1} \mathfrak{d}_{m_2} \mathfrak{d}_{m_1+m_2} + \mathfrak{d}_{m_1} \mathfrak{d}_{-m_2} \mathfrak{d}_{m_1+m_2} + \mathfrak{d}_{-m_1} \mathfrak{d}_{m_2} \mathfrak{d}_{-m_1-m_2} \\
\nonumber
& + \mathfrak{d}_{m_1} \mathfrak{d}_{-m_2} \mathfrak{d}_{-m_1-m_2} + \mathfrak{d}_{-m_1} \mathfrak{d}_{-m_2} \mathfrak{d}_{-m_1-m_2} + 3 \mathfrak{d}_{-m_1} \mathfrak{d}_{-m_2} \mathfrak{d}_{m_1+m_2} + 3 \mathfrak{d}_{m_1} \mathfrak{d}_{m_2} \mathfrak{d}_{-m_1-m_2} \Big) \times \\
& \times \left( U_{1,-m_1} U_{1,-m_2} U_{1,m_1+m_2} + U_{1,-m_1-m_2} U_{1,m_1} U_{1,m_2} \right) \\
\nonumber
\mathcal{I}_{5,4} & = \sum_{m_1,m_2>0} \Big( \mathfrak{d}_{m_1+m_2} \mathfrak{d}_{m_2} + 2\mathfrak{d}_{m_1+m_2} \mathfrak{d}_{-m_2} + 2\mathfrak{d}_{-m_1-m_2} \mathfrak{d}_{m_2} + \mathfrak{d}_{-m_1-m_2} \mathfrak{d}_{-m_2} \Big) \times \\
\nonumber
& \times \Big(U_{1,-m_2} T_{-m_1} U_{1,m_1+m_2} + U_{1,-m_1-m_2} T_{m_1} U_{1,m_2} \Big) \\
& + \sum_{m_1,m_2>0} \Big( \mathfrak{d}_{m_1} \mathfrak{d}_{m_2} + \frac{1}{2} \mathfrak{d}_{m_1} \mathfrak{d}_{-m_2} + \frac{1}{2} \mathfrak{d}_{-m_1} \mathfrak{d}_{m_2} + \mathfrak{d}_{-m_1} \mathfrak{d}_{-m_2} \Big) \times \\
\nonumber
& \times \Big( T_{-m_1-m_2} U_{1,m_1} U_{1,m_2} + U_{1,-m_1} U_{1,-m_2} T_{m_1+m_2} \Big) \\
\nonumber
& + \sum_{m>0} \left( \frac{6m^2 q^m}{(1-q^m)^2} + m^2 \right) U_{1,-m} T_0 U_{1,m} \\
\nonumber
\mathcal{I}_{5,5} & = \frac{1}{2} \sum_{m_1,m_2>0} (U_{1,-m_1-m_2} U_{1,m_1} U_{1,m_2} + U_{1,-m_1} U_{1,-m_2} U_{1,m_1+m_2}) \times \\
\nonumber
& \times \Big[ -2m_1^2 \mathfrak{d}_{-m_1} - 2m_2^2 \mathfrak{d}_{-m_2} -8m_1 m_2 \mathfrak{d}_{m_1} - 8m_1 m_2 \mathfrak{d}_{m_2} - 8m_2^2 \mathfrak{d}_{m_1} - 8m_1^2 \mathfrak{d}_{m_2} \\
& + 7m_1^2 m_2 + 7m_1 m_2^2 - 2(m_1-m_2)^2 \mathfrak{d}_{m_1+m_2} \Big] \\
\nonumber
& - 2 U_{1,0} \sum_{m>0} \frac{m^3(1+q^m)}{1-q^m} U_{1,-m} U_{1,m} \\
\nonumber
\mathcal{I}_{5,6} & = \sum_{m_1,m_2>0} \Bigg[ \frac{m_2 (m_1+m_2) (1+2q^{m_2}+2q^{m_1+m_2}+q^{m_1+2m_2})}{(1-q^{m_2})(1-q^{m_1+m_2})} \times \\
\nonumber
& \times \Big( T_{-m_1-m_2} U_{1,m_1} U_{1,m_2} + U_{1,-m_1} U_{1,-m_2} T_{m_1+m_2} \\
\nonumber
& + U_{1,-m_1-m_2} U_{1,m_1} T_{m_2} + T_{-m_2} U_{1,-m_1} U_{1,m_1+m_2} \Big) \\
\nonumber
& + \frac{m_1 m_2 (2+q^{m_1}+q^{m_2}+2q^{m_1+m_2})}{(1-q^{m_1})(1-q^{m_2})} (U_{1,-m_1-m_2} U_{1,m_1} T_{m_2} + T_{-m_2} U_{1,-m_1} U_{1,m_1+m_2}) \\
\nonumber
& - 3 \frac{m_1 m_2 q^{m_1} (1+q^{m_2})}{(1-q^{m_1})(1-q^{m_1+m_2})} \Big(U_{1,-m_1-m_2} \left[ U_{1,m_1}, T_{m_2} \right] + \left[ T_{-m_2}, U_{1,-m_1} \right] U_{1,m_1+m_2}\Big) \Bigg] \\
& - \frac{1}{6} \sum_{m>0} m^4 U_{1,-m} U_{1,m} + \frac{1}{60} \sum_{m>0} U_{1,-m} U_{1,m} \\
\nonumber
& + 6 U_{1,0} \sum_{m>0} \frac{m^2 q^m}{(1-q^m)^2} (U_{1,-m} T_m + T_{-m} U_{1,m}) + U_{1,0} \sum_{m>0} m^2 (U_{1,-m} T_m + T_{-m} U_{1,m})
\end{align}
These quantities have matrix elements which are manifestly rational functions of $q$ with poles only at roots of unity, i.e. there are no transcendental terms such as the Eisenstein series. Their spectra in terms of solutions of Bethe ansatz equations are given by:
\begin{align}
\mathcal{A}_1 & \leadsto p_0 + hw. \\
\mathcal{A}_2 & \leadsto p_1 - \frac{\alpha_0}{2} p_0 + hw. \\
\mathcal{A}_3 & \leadsto p_2 - \alpha_0 p_1 + \frac{1+4\alpha_0^2}{12} p_0 + hw. \\
\mathcal{A}_4 & \leadsto p_3 - \frac{3\alpha_0}{2} p_2 + \frac{1+4\alpha_0^2}{4} p_1 - \frac{\alpha_0(1+2\alpha_0^2)}{8} p_0 + hw. \\
\mathcal{A}_5 & \leadsto p_4 - 2\alpha_0 p_3 + \frac{1+4\alpha_0^2}{2} p_2 - \frac{\alpha_0 (1+2\alpha_0^2)}{2} p_1 + \frac{9+114\alpha_0^2-4\alpha_0^2 N+144\alpha_0^4}{720} p_0 + hw.
\end{align}

\subsection{Local expansion of conserved quantities}
\label{applocal}
Here we list the expansion of $\mathcal{A}_j$ as $q \to 1$:
\begin{align}
\nonumber
\mathcal{A}_2 & \sim -\frac{\alpha_0}{1-q} \sum_{m>0} U_{1,-m} U_{1,m} + \Big[ -\frac{1}{2} \phi_{3,0} + \frac{\alpha_0}{2} \sum_{m>0} U_{1,-m} U_{1,m} \Big] \\
& + (1-q) \frac{\alpha_0}{12} \sum_{m>0} (1-m^2) U_{1,-m} U_{1,m} + O((1-q)^2) \\
\nonumber
\mathcal{A}_3 & \sim \frac{N \alpha_0^2}{(1-q)^2} \sum_{m>0} U_{1,-m} U_{1,m} + \frac{1}{1-q} \Big[ \alpha_0 \sum_{m>0} (U_{1,-m} T_m + T_{-m} U_{1,m}) - N \alpha_0^2 \sum_{m>0} U_{1,-m} U_{1,m} \Big] \\
\nonumber
& + \Big[ -\frac{1}{3} \phi_{4,0} - \frac{\alpha_0}{2} \sum_{m>0} (U_{1,-m} T_m + T_{-m} U_{1,m}) - \frac{\alpha_0^2(N+2)}{12} \sum_{m>0} m^2 U_{1,-m} U_{1,m} \\
& + \frac{N \alpha_0^2}{12} \sum_{m>0} (1+2m^2) U_{1,-m} U_{1,m} \Big] + O((1-q)^1) \\
\nonumber
\mathcal{A}_4 & \sim -\frac{N^2 \alpha_0^3}{(1-q)^3} \sum_{m>0} U_{1,-m} U_{1,m} \\
\nonumber
& + \frac{1}{(1-q)^2} \Big[ -N \alpha_0^2 \sum_{m>0} (U_{1,-m} T_m + T_{-m} U_{1,m}) - \alpha_0^2 U_{1,0} \sum_{m>0} U_{1,-m} U_{1,m} \\
\nonumber
& - \frac{3\alpha_0^2}{2} \sum_{m_1,m_2>0} (U_{1,-m_1} U_{1,-m_2} U_{1,m_1+m_2} + U_{1,-m_1-m_2} U_{1,m_1} U_{1,m_2}) + \frac{3N^2 \alpha_0^3}{2} \sum_{m>0} U_{1,-m} U_{1,m} \Big] \\
\nonumber
& + \frac{1}{1-q} \Big[ -\alpha_0 \sum_{m>0} (U_{1,-m} \phi_{3,m} + \phi_{3,-m} U_{1,m}) - \alpha_0 \sum_{m>0} T_{-m} T_m \\
\nonumber
& + N \alpha_0^2 \sum_{m>0} (U_{1,-m} T_m + T_{-m} U_{1,m}) - \frac{\alpha_0(1+\alpha_0^2-4\alpha_0^2 N+\alpha_0^2 N^2)}{12} \sum_{m>0} m^2 U_{1,-m} U_{1,m} \\
\nonumber
& + \frac{3\alpha_0^2}{2} \sum_{m_1,m_2>0} (U_{1,-m_1} U_{1,-m_2} U_{1,m_1+m_2} + U_{1,-m_1-m_2} U_{1,m_1} U_{1,m_2}) \\
& + \alpha_0^2 U_{1,0} \sum_{m>0} U_{1,-m} U_{1,m} - \frac{N^2 \alpha_0^3}{2} \sum_{m>0} U_{1,-m} U_{1,m} \Big] + \mathcal{O}((1-q)^0) \\
\nonumber
\mathcal{A}_5 & \sim +\frac{\alpha_0^4 N^3}{(1-q)^4} \sum_{m>0} U_{1,-m} U_{1,m} \\
\nonumber
& +\frac{1}{(1-q)^3} \Big[ \alpha_0^3 N^2 \sum_{m>0} (T_{-m} U_{1,m} + U_{1,-m} T_m) +2\alpha_0^3 N U_{1,0} \sum_{m>0} U_{1,-m} U_{1,m} \\
\nonumber
& +\frac{7\alpha_0^3 N}{2} \sum_{m_1,m_2>0} (U_{1,-m_1} U_{1,-m_2} U_{1,m_1+m_2} + U_{1,-m_1-m_2} U_{1,m_1} U_{1,m_2}) \\
\nonumber
& -2\alpha_0^4 N^3 \sum_{m>0} U_{1,-m} U_{1,m} \Big] \\
\nonumber
& \frac{1}{(1-q)^2} \Big[ +2\alpha_0^2 \sum_{m_1,m_2>0} (T_{-m_1-m_2} U_{1,m_1} U_{1,m_2} + U_{1,-m_1} U_{1,-m_2} T_{m_1+m_2}) \\
\nonumber
& +2\alpha_0^2 \sum_{m_1,m_2>0} (U_{1,-m_2} T_{-m_1} U_{1,m_1+m_2} + U_{1,-m_1-m_2} T_{m_1} U_{1,m_2}) \\
& +2\alpha_0^2 \sum_{m_1,m_2>0} (T_{-m_2} U_{1,-m_1} U_{1,m_1+m_2} + U_{1,-m_1-m_2} U_{1,m_1} T_{m_2}) \\
\nonumber
& -\frac{21\alpha_0^3 N}{4} \sum_{m_1,m_2>0} (U_{1,-m_1} U_{1,-m_2} U_{1,m_1+m_2} + U_{1,-m_1-m_2} U_{1,m_1} U_{1,m_2}) \\
\nonumber
& +2\alpha_0^2 \sum_{m>0} U_{1,-m} T_0 U_{1,m} +\alpha_0^2 N \sum_{m>0} (U_{1,-m} \phi_{3,m} + \phi_{3,-m} U_{1,m}) \\
\nonumber
& +\alpha_0^2 N \sum_{m>0} T_{-m} T_m -3\alpha_0^3 N U_{1,0} \sum_{m>0} U_{1,-m} U_{1,m} -\frac{\alpha_0^2}{3} \sum_{m>0} m^2 U_{1,-m} U_{1,m} \\
\nonumber
& +\alpha_0^2 U_{1,0} \sum_{m>0} (U_{1,-m} T_m + T_{-m} U_{1,m}) -\frac{3\alpha_0^3 N^2}{2} \sum_{m>0} (U_{1,-m} T_m + T_{-m} U_{1,m}) \\
\nonumber
& +\frac{5\alpha_0^2 N}{12} \sum_{m>0} m^2 U_{1,-m} U_{1,m} +\frac{5\alpha_0^4 N}{12} \sum_{m>0} m^2 U_{1,-m} U_{1,m} \\
\nonumber
& -\frac{\alpha_0^4 N^2}{3} \sum_{m>0} m^2 U_{1,-m} U_{1,m} +\frac{\alpha_0^4 N^3}{12} \sum_{m>0} m^2 U_{1,-m} U_{1,m} \\
\nonumber
& +\frac{\alpha_0^2}{3} \sum_{m>0} U_{1,-m} U_{1,m} +\frac{7\alpha_0^4 N^3}{6} \sum_{m>0} U_{1,-m} U_{1,m} \Big] + \mathcal{O}((1-q)^{-1})
\end{align}
The linear combinations of $\mathcal{A}_j$ that are finite in $q \to 1$ limit
\begin{align}
\mathcal{B}_1 & \equiv \mathcal{A}_1 \\
\mathcal{B}_2 & \equiv \lim_{q \to 1} \left[ (1-q) \mathcal{A}_2 \right] \\
\mathcal{B}_3 & \equiv \lim_{q \to 1} \left[ (1-q) \mathcal{A}_3 + \alpha_0 N \mathcal{A}_2 \right] \\
\mathcal{B}_4 & \equiv \lim_{q \to 1} \left[ (1-q)^2 \mathcal{A}_4 + (1-q) \alpha_0 N \mathcal{A}_3 \right] \\
\mathcal{B}_5 & \equiv \lim_{q \to 1} \left[ (1-q)^2 \mathcal{A}_5 + (1-q) \frac{7\alpha_0 N}{3} \mathcal{A}_4 + \frac{4\alpha_0^2 N^2}{3} \mathcal{A}_3 + \frac{\alpha_0^3 N^3}{6} \mathcal{A}_2 - \frac{\alpha_0^2 N u_1}{3} \mathcal{A}_2 \right].
\end{align}
In terms of the mode operators, these are given by
\begin{align}
\mathcal{B}_1 & = T_0 + hw. \\
\mathcal{B}_2 & = -\frac{\alpha_0}{2} (U_1 U_1)_0 + hw. \\
\mathcal{B}_3 & = -\frac{\alpha_0 N}{2} W_{3,0} - \alpha_0 U_{1,0} T_0 + \frac{\alpha_0}{3N} (U_1 (U_1 U_1))_0 - \frac{\alpha_0^2 N}{4} (U_1 U_1)_0 + hw. \\
\mathcal{B}_4 & = -\frac{\alpha_0^2}{2} (U_1 (U_1 U_1))_0 + \alpha_0^2 U_{1,0} (U_1 U_1)_0 + \frac{\alpha_0^3 N^2}{4} (U_1 U_1)_0 + hw. \\
\nonumber
\mathcal{B}_5 & = -\frac{4\alpha_0^2 N^2}{9} \left( W_{4,0} - \frac{3(N-3)(\alpha_0^2 N^2+3\alpha_0^2 N-9)}{2N(5\alpha_0^2 N^3-5\alpha_0^2 N-5N-17)} (T_{\infty} T_{\infty})_0 \right) \\
\nonumber
& + \frac{\alpha_0^2}{2N} \left( (U_1(U_1(U_1 U_1)))_0 - N (U_1^\prime U_1^\prime)_0 \right) \\
\nonumber
& + \frac{2\alpha_0^2 N}{3} T_0^2 - \alpha_0^2 T_0 (U_1 U_1)_0 \\
& + \frac{3\alpha_0^2 N}{2} U_{1,0} W_{3,0} - \frac{\alpha_0^3 N^3}{12} W_{3,0} \\
\nonumber
& - \frac{\alpha_0^2}{N} U_{1,0} (U_1 (U_1 U_1))_0 - \frac{19\alpha_0^3 N}{36} (U_1 (U_1 U_1))_0 \\
\nonumber
& + 2\alpha_0^2 U_{1,0}^2 T_0 - \frac{\alpha_0^3 N^2}{6} U_{1,0} T_0 - \frac{\alpha_0^2 N}{4} T_0 \\
\nonumber
& + \frac{4\alpha_0^3 N}{3} U_{1,0} (U_1 U_1)_0 + \frac{\alpha_0^2 (36+7\alpha_0^2 N^3)}{72} (U_1 U_1)_0 + hw.
\end{align}
Their spectra in terms of Bethe roots are given by
\begin{align}
\mathcal{B}_1 & \leadsto p_{0,0} + hw. \\
\mathcal{B}_2 & \leadsto p_{1,-1} + hw. \\
\mathcal{B}_3 & \leadsto p_{2,-1} + \alpha_0 N p_{1,0} - \alpha_0 p_{1,-1} - \frac{\alpha_0^2 N}{2} p_{0,0} + hw. \\
\mathcal{B}_4 & \leadsto p_{3,-2} + \alpha_0 N p_{2,-1} + \frac{\alpha_0^2 N}{2} p_{1,-1} + hw. \\
\nonumber
\mathcal{B}_5 & \leadsto p_{4,-2} + \frac{7\alpha_0 N}{3} p_{3,-1} + \frac{4\alpha_0^2 N^2}{3} p_{2,0} - 2\alpha_0 p_{3,-2} - \frac{7\alpha_0^2 N}{2} p_{2,-1} \\
\nonumber
& + \frac{\alpha_0^3 N^2(N-8)}{6} p_{1,0} - \frac{\alpha_0^2 N}{3} u_1 p_{1,0} + \frac{\alpha_0 N(1+4\alpha_0^2)}{12} p_{1,-1} \\
& + \frac{\alpha_0^2 N^2(4+16\alpha_0^2-3\alpha_0^2 N)}{36} p_{0,0} + \frac{\alpha_0^3 N}{6} u_1 p_{0,0}
\end{align}

\section{Solutions of Bethe ansatz equations}

\subsection{Lee-Yang, $\Delta=0$, level $3$}

\subsubsection{Resultant}
\label{appleeyangreslvl3}
The factor of resultant which gives all $15$ physical roots corresponding to $5$ states at level $3$ in $\Delta=0$ representation of Lee-Yang is of the form
\begin{equation}
\sum_{j=0}^{15} \beta_j x^j
\end{equation}
with
\begin{align}
\nonumber
\beta_{15} & = (q-1)^{10} (q+1)^2 (q^2+q+1)^3 \\
\nonumber
\beta_{14} & = -3 (q-1)^9 (q+1) (q^2+q+1)^2 (33 q^4+43 q^3+28 q^2+13 q+3) \\
\nonumber
\beta_{13} & = +2 (q-1)^8 (q^2+q+1) (2233 q^8+5861 q^7+7678 q^6+6703 q^5+4118 q^4 \\
\nonumber
& +1915 q^3+622 q^2+65 q-35) \\
\nonumber
\beta_{12} & = -6 (q-1)^7 (20307 q^{11}+60224 q^{10}+99023 q^9+108981 q^8+87774 q^7+55299 q^6 \\
\nonumber
& +26595 q^5+8604 q^4+951 q^3-687 q^2-434 q-77) \\
\nonumber
\beta_{11} & = +3 (q-1)^6 (749783 q^{12}+1744223 q^{11}+2365345 q^{10}+2212756 q^9 +1574487 q^8 \\
\nonumber
& +880113 q^7+336042 q^6+35985 q^5-44217 q^4-37388 q^3-16847 q^2-3361 q+839) \\
\nonumber
\beta_{10} & = -9 (q-1)^5 (3308679 q^{13}+5646690 q^{12}+6135516 q^{11}+4858939 q^{10} \\
\nonumber
& +2992059 q^9 +1255144 q^8+131899 q^7-313931 q^6-287540 q^5-159815 q^4 \\
& -52583 q^3-6224 q^2+5010 q+781) \\
\nonumber
\beta_9 & = +2 (q-1)^4 (146080348 q^{14}+161581637 q^{13}+135923944 q^{12}+92019809 q^{11} \\
\nonumber
& +42057082 q^{10}+1378169 q^9-15764754 q^8-17393766 q^7-11748774 q^6-4558375 q^5 \\
\nonumber
& -907046 q^4+94313 q^3+226048 q^2+113861 q-23456) \\
\nonumber
\beta_8 & = -12 (q-1)^3 (180333417 q^{15}+94388201 q^{14}+61100585 q^{13}+37723050 q^{12} \\
\nonumber
& -47534 q^{11}-19473224 q^{10}-24531861 q^9-17036862 q^8-9470139 q^7-1708392 q^6 \\
\nonumber
& +40666 q^5+598774 q^4+313866 q^3+155033 q^2-55000 q+300) \\
\nonumber
\beta_7 & = +8 (q-1)^2 (1525470224 q^{16}-67057602 q^{15}+58007498 q^{14}+64468027 q^{13} \\
\nonumber
& -127940362 q^{12} -159580082 q^{11}-206469540 q^{10}-63924311 q^9-23482788 q^8 \\
\nonumber
& +977641 q^7+9511632 q^6 +7954486 q^5+2154422 q^4+568675 q^3-139810 q^2 \\
\nonumber
& -191490 q+40100) \\
\nonumber
\beta_6 & = -48 (q-1) (1091911035 q^{17}-655148860 q^{16}-7896247 q^{15}+25788450 q^{14} \\
\nonumber
& -29750149 q^{13}-132857401 q^{12}-117530061 q^{11}+20865566 q^{10}+14616293 q^9 \\
\nonumber
& -2642898 q^8+15160759 q^7+2887639 q^6+1689999 q^5-275210 q^4+195565 q^3 \\
\nonumber
& -378795 q^2+125350 q-12875) \\
\nonumber
\beta_5 & = +288 (590434004 q^{18}-679495775 q^{17}+105467922 q^{16}+57781429 q^{15} \\
\nonumber
& +59562920 q^{14} -175016546 q^{13}+30903476 q^{12}+1357527 q^{11}+33182072 q^{10} \\
\nonumber
& -9897385 q^9+7510916 q^8 +292398 q^7+1384190 q^6-780377 q^5+92840 q^4 \\
\nonumber
& -172070 q^3 +82650 q^2-16625 q+1250) \\
\nonumber
\beta_4 & = -1728 q^3 (236868899 q^{15}-167106540 q^{14}-27475861 q^{13}+30085783 q^{12} \\
\nonumber
& +71194720 q^{11} -93267345 q^{10}+26586077 q^9-34251559 q^8+17589658 q^7 \\
\nonumber
& -3235715 q^6-1579579 q^5 +967928 q^4+321965 q^3-16556 q^2+5065 q+100) \\
\nonumber
\beta_3 & = +10368 q^6 (68241566 q^{12}-18895570 q^{11}-20786934 q^{10}+7199369 q^9 \\
\nonumber
& +33561740 q^8 -22673859 q^7+12236878 q^6-12682141 q^5+3702336 q^4 \\
\nonumber
& -429295 q^3-709332 q^2+75072 q-25990) \\
\nonumber
\beta_2 & = -62208 q^9 (13326500 q^9+1552000 q^8-5598873 q^7-169064 q^6+7716992 q^5 \\
\nonumber
& -2044760 q^4+2513825 q^3 -1187532 q^2+408704 q+15928) \\
\nonumber
\beta_1 & = +8211456 q^{12} (71676 q^6+32140 q^5-37217 q^4-23402 q^3+31535 q^2-268 q+4268) \\
\nonumber
\beta_0 & = -8671297536 q^{15} (q-1) (22 q^2+37 q+22)
\end{align}
This expression is very long, but we should remember that it captures all the higher spin charges of all five physical states at level $3$ for any value of the twist parameter $q$.

\subsubsection{Yangian limit}
\label{appleeyangyanglvl3}
The $q$-expansion of the physical solutions around $q=0$ is as follows:
\begin{align}
& \begin{cases}
-10 - \frac{24}{13}q + \mathcal{O}(q^2) \\
-5 - \frac{264}{65}q^2 + \mathcal{O}(q^3) \\
-\frac{528}{25}q^3 + \frac{2376}{1625}q^4 + \mathcal{O}(q^5)
\end{cases} \\
& \begin{cases}
-5 - \frac{22}{7}q - + \mathcal{O}(q^2) \\
\frac{132}{5}q^3 + \frac{6138}{25}q^4 + \mathcal{O}(q^5) \\
2 - \frac{90}{7}q + \mathcal{O}(q^2)
\end{cases} \\
& \begin{cases}
-5 - \frac{39}{14}q + \mathcal{O}(q^2) \\
\frac{66}{5}q^3 - \frac{136176}{2275} q^4 + \mathcal{O}(q^5) \\
3 + \frac{165}{26}q + \mathcal{O}(q^2)
\end{cases} \\
& \begin{cases}
-12 q^3 - \frac{1410}{7}q^4 + \mathcal{O}(q^5) \\
2 + \frac{30}{7}q + \mathcal{O}(q^2) \\
3 - 30q + \mathcal{O}(q^2)
\end{cases} \\
& \begin{cases}
-6q^3 + 45q^4 + \mathcal{O}(q^5) \\
2 - 20q^2 + \mathcal{O}(q^3) \\
4 + 10q + \mathcal{O}(q^2)
\end{cases}
\end{align}

\subsubsection{Local limit $q \to 1$}
\label{appleeyangloclvl3}
On the other hand, the expansion around the local point is more interesting. The five solutions have expansion around $q = 1$ as follows:
\begin{align}
& \begin{cases}
0 - \frac{88}{81}(1-q) + \frac{1496}{6561}(1-q)^2 + \mathcal{O}((1-q)^3) \\
3 - \frac{148}{81}(1-q) - \frac{74}{81}(1-q)^2 + \mathcal{O}((1-q)^3) \\
6 - \frac{88}{81}(1-q) - \frac{8624}{6561}(1-q)^2 + \mathcal{O}((1-q)^3)
\end{cases} \\
& \begin{cases}
-\frac{6}{1-q} + 6 - \frac{10}{3}(1-q) - \frac{5}{3}(1-q)^2 + \mathcal{O}((1-q)^3) \\
\frac{6+\sqrt{14}}{2} - \frac{11}{12}(1-q) - \frac{11(168+5\sqrt{14})}{4032}(1-q)^2 + \mathcal{O}((1-q)^3) \\
\frac{6-\sqrt{14}}{2} - \frac{11}{12}(1-q) - \frac{11(168-5\sqrt{14})}{4032}(1-q)^2 + \mathcal{O}((1-q)^3)
\end{cases} \\
& \begin{cases}
-\frac{6}{1-q} -2^{\frac{2}{3}}5^{\frac{1}{3}} (1-q)^{-\frac{1}{3}} + 6 + \mathcal{O}((1-q)^{\frac{1}{3}}) \\
-\frac{6}{1-q} -2^{\frac{2}{3}}5^{\frac{1}{3}} e^{-\frac{2\pi i}{3}} (1-q)^{-\frac{1}{3}} + 6 + \mathcal{O}((1-q)^{\frac{1}{3}}) \\
-\frac{6}{1-q} -2^{\frac{2}{3}}5^{\frac{1}{3}} e^{\frac{2\pi i}{3}} (1-q)^{-\frac{1}{3}} + 6 + \mathcal{O}((1-q)^{\frac{1}{3}})
\end{cases} \\
& \begin{cases}
-\frac{6}{1-q} +3\times 5^{\frac{1}{4}} (1-q)^{-\frac{1}{2}} - \frac{19\sqrt{5}}{18} + \mathcal{O}((1-q)^{\frac{1}{2}}) \\
-\frac{6}{1-q} -3\times 5^{\frac{1}{4}} (1-q)^{-\frac{1}{2}} - \frac{19\sqrt{5}}{18} + \mathcal{O}((1-q)^{\frac{1}{2}}) \\
-\frac{6}{1-q} -\frac{2(4\sqrt{5}-27)}{9} + \mathcal{O}((1-q))
\end{cases} \\
& \begin{cases}
-\frac{6}{1-q} +3i\times 5^{\frac{1}{4}} (1-q)^{-\frac{1}{2}} + \frac{19\sqrt{5}}{18} + \mathcal{O}((1-q)^{\frac{1}{2}}) \\
-\frac{6}{1-q} -3i\times 5^{\frac{1}{4}} (1-q)^{-\frac{1}{2}} + \frac{19\sqrt{5}}{18} + \mathcal{O}((1-q)^{\frac{1}{2}}) \\
-\frac{6}{1-q} +\frac{2(4\sqrt{5}+27)}{9} + \mathcal{O}((1-q))
\end{cases}
\end{align}

\bibliographystyle{JHEP}
\bibliography{winffjmm}

\end{document}